\documentclass[onecolumn, 12 pt, doublespace, fullpage, letterpaper]{report}

\usepackage{amsmath,amsfonts,bm}

\def\eqref#1{equation~\ref{#1}}

\def\1{\bm{1}}

\def\vb{{\bm{b}}}

\def\vx{{\bm{x}}}

\DeclareMathAlphabet{\mathsfit}{\encodingdefault}{\sfdefault}{m}{sl}
\SetMathAlphabet{\mathsfit}{bold}{\encodingdefault}{\sfdefault}{bx}{n}

\def\gA{{\mathcal{A}}}
\def\gB{{\mathcal{B}}}

\def\gD{{\mathcal{D}}}

\def\gL{{\mathcal{L}}}

\def\gO{{\mathcal{O}}}

\def\gR{{\mathcal{R}}}

\def\gT{{\mathcal{T}}}

\def\gX{{\mathcal{X}}}
\def\gY{{\mathcal{Y}}}

\newcommand{\R}{\mathbb{R}}

\usepackage{amssymb}
\usepackage{textcomp}
\usepackage{graphicx}
\usepackage{graphics}
\usepackage{epsfig}
\usepackage{epstopdf}
\usepackage{float}
\usepackage{color}
\usepackage{mathtools}
\usepackage{latexsym,amsfonts}
\usepackage{amsthm}
\usepackage{url}
\usepackage{longtable}
\usepackage[figuresright]{rotating}
\usepackage{listings}
\usepackage{etoolbox}
\usepackage{sectsty}
\usepackage{datetime}
\usepackage{multirow}
\usepackage{booktabs}
\usepackage{hyperref}
\usepackage{enumitem}
\usepackage[vlined,ruled]{algorithm2e}
\usepackage{xcolor}
\usepackage{Definitions}
\usepackage{wrapfig}
\usepackage{multicol}
\usepackage{lipsum}
\usepackage{sidecap}
\usepackage{cleveref}
\usepackage{setspace}
\usepackage{subfigure}
\usepackage{tabularx}

\definecolor{deepmagenta}{rgb}{0.8, 0.0, 0.8}
\usepackage{letltxmacro}
\renewcommand{\eqref}{Eq.~\ref}

\definecolor{mydarkblue}{rgb}{0,0.08,0.45}
\hypersetup{
 colorlinks=true,
 citecolor=mydarkblue,
 linkcolor=black}

\newcommand{\angstrom}{\mbox{\normalfont\AA}}
\newcommand{\irow}[1]{
  \begin{smallmatrix}(#1)\end{smallmatrix}%
}

\newdateformat{monthyeardate}{%
  \monthname[\THEMONTH], \THEYEAR}

\chapterfont{\fontsize{14}{15}\selectfont}
\sectionfont{\fontsize{14}{15}\selectfont}
\subsectionfont{\fontsize{14}{15}\selectfont}
\subsubsectionfont{\fontsize{14}{15}\selectfont}

\usepackage{fancyhdr}       
\fancyheadoffset[L]{0.25in}

\makeatletter
\patchcmd{\@makechapterhead}{50\p@}{20pt}{}{}
\patchcmd{\@makeschapterhead}{50\p@}{20pt}{}{}
\makeatother

\fancypagestyle{plain}{
\fancyhf{} 
\fancyhead[C]{\thepage} 

}

   \usepackage{ifthen,xkeyval,xfor,amsgen}
   \usepackage[acronym,toc, nogroupskip]{glossaries}
   \newglossary[slg]{symbols}{syi}{sbl}{List of Symbols}
 
   \makeglossaries

\newglossaryentry{symb:Pi}{
name=$\pi$, type=symbols,
description=A mathematical constant whose value is the ratio of any circle's circumference to its diameter,
sort=symbolpi
}

\newglossaryentry{symb:Phi}{
name=$\varphi$, type=symbols,
description=An angle,
sort=symbolphi
}

\newglossaryentry{symb:Lambda}{
name=$\lambda$, type=symbols,
description=Lambda indicates usually an eigenvalue in linear algebra,
sort=symbollambda
}

\newacronym{ngs}{NGS}{Next-generation sequencing}
\newacronym{ont}{ONT}{Oxford Nanopore Technology}
\newacronym{bilstm}{Bi-LSTM}{Bi-directional Long Short-Term Memory}
\newacronym{dctw}{DCTW}{Deep Canonical Time Warping}
\newacronym{bdctw}{BDCTW}{BiLSTM-extended Deep Canonical Time Warping}
\newacronym{dnns}{DNNs}{Deep Neural Networks}
\newacronym{cca}{CCA}{Canonical Correlation Analysis}
\newacronym{dtw}{DTW}{Dynamic Time Warping}
\newacronym{svd}{SVD}{singular value decomposition}

\usepackage{hyperref}

\usepackage[lmargin=1.5in, rmargin=1in, vmargin=1in, headsep=0.083334in]{geometry}
\newcommand{\mathsym}[1]{{}}
\newcommand{\unicode}[1]{{}}
\renewcommand{\thechapter}{\arabic{chapter}}
\renewcommand\bibname{\centering BIBLIOGRAPHY}

\newenvironment{benumerate}[1]{
    \let\oldItem\item
    \def\item{\addtocounter{enumi}{-2}\oldItem}
    \begin{enumerate}
    \setcounter{enumi}{#1}
    \addtocounter{enumi}{1}
}{
    \end{enumerate}
}

\begin{document}


\vspace{2pt}
\thispagestyle{empty}
\addvspace{10mm}

\begin{center}

{\textbf{{\large Towards Structured Prediction in Bioinformatics with Deep Learning}}}\vfill 
{Ph.D. Dissertation by}\\
{Yu Li}\vfill

{ In Partial Fulfillment of the Requirements}\\[12pt]
{ For the Degree of}\\[12pt]
{Doctor of Philosophy} \vfill
{King Abdullah University of Science and Technology }\\
{Thuwal, Kingdom of Saudi Arabia}
\vfill
{\monthyeardate\today}

\end{center}

\newpage

\begin{center}

\end{center}

\begin{center}

{ \textbf{{\large EXAMINATION COMMITTEE PAGE}}}\\\vspace{1cm}

\end{center}
\noindent{The dissertation of Yu Li is approved by the examination committee}
\addcontentsline{toc}{chapter}{Examination Committee Page}

\vspace{4\baselineskip}

\begin{onehalfspacing}
\noindent{Committee Chairperson: Xin Gao}\\
Committee Members: Robert Hoehndorf, Stefan Arold, \textcolor{black}{Jian Ma}\vfill

\end{onehalfspacing}


\newpage
\addcontentsline{toc}{chapter}{Copyright}
\vspace*{\fill}
\begin{center}
{ \copyright \monthyeardate\today}\\
{Yu Li}\\
{All Rights Reserved}
\end{center}

\begin{center}

\end{center}

\begin{center}
{{\bf\fontsize{14pt}{14.5pt}\selectfont \uppercase{ABSTRACT}}}
\end{center}

\doublespacing
\addcontentsline{toc}{chapter}{Abstract}

\begin{center}
{{\fontsize{14pt}{14.5pt}\selectfont {Towards Structured Prediction in Bioinformatics with Deep Learning\\
Yu Li}}}
\end{center}

Using machine learning, especially deep learning, to facilitate biological research is a fascinating research direction. However, in addition to the standard classification or regression problems, whose outputs are simple vectors or scalars, in bioinformatics, we often need to predict more complex structured targets, such as 2D images and 3D molecular structures. The above complex prediction tasks are referred to as structured prediction. Structured prediction is more complicated than the traditional classification but has much broader applications, especially in bioinformatics, considering the fact that most of the original bioinformatics problems have complex output objects.

Due to the properties of those structured prediction problems, such as having problem-specific constraints and dependency within the labeling space, the straightforward application of existing deep learning models on the problems can lead to unsatisfactory results. In this dissertation, we argue that the following two ideas can help resolve a wide range of structured prediction problems in bioinformatics. Firstly, we can combine deep learning with other classic algorithms, such as probabilistic graphical models, which model the problem structure explicitly. Secondly, we can design and train problem-specific deep learning architectures or methods by considering the structured labeling space and problem constraints, either explicitly or implicitly. We demonstrate our ideas with six projects from four bioinformatics subfields, including sequencing analysis, structure prediction, function annotation, and network analysis. The structured outputs cover 1D electrical signals, 2D images, 3D structures, hierarchical labeling, and heterogeneous networks. With the help of the above ideas, all of our methods can achieve state-of-the-art performance on the corresponding problems. 

The success of these projects motivates us to extend our work towards other more challenging but important problems, such as health-care problems, which can directly benefit people's health and wellness. We thus conclude this thesis by discussing such future works, and the potential challenges and opportunities.






\begin{center}
•
\end{center}
\begin{center}

{\bf\fontsize{14pt}{14.5pt}\selectfont \uppercase{Acknowledgements}}\\\vspace{1cm}
\end{center}

\addcontentsline{toc}{chapter}{Acknowledgements}

\vspace{-1cm}

{\setstretch{1.25}
I want to thank all the people who made this dissertation possible. First and foremost, I am tremendously grateful for my adviser, Professor Xin Gao, for his continuous support and guidance throughout my Ph.D. journey. It was him who led me to the computational world. He gave me the freedom to work on a variety of problems and the opportunities to collaborate with a number of top-tier researchers all over the world. Without him, my academic career would be much less successful.

Special thanks to my committee members, Professor Robert Hoehndorf, Professor Stefan Arold, and Professor \textcolor{black}{Jian Ma}, for taking their precious time to review this thesis and provide valuable comments.

I want to thank all the SFB members, especially, Sheng Wang, Ramzan Umarov, Lizhong Ding, Renmin Han, Zhenzhen Zou, Zhongxiao Li, Siyuan Chen, Wenkai Han, Hiroyuki Kuwahara, Trisevgeni Rapakoulia, Haoyang Li, Zhihao Xia, Vasiliki Kordopati, Adil Salhi, Christophe Van Neste, and Xuefeng Cui. I enjoyed working and discussing with you. Beyond the work included in this thesis, I had the pleasure of working with many other students and professors in KAUST, including Professor Shuyu Sun, Tao Zhang, Yiteng Li, Professor Mo Li, Chongwei Bi, Professor Xiangliang Zhang, Professor Vladimir Bajic, Rabab Al-omairy, Hatem Ltaief, and Professor David E. Keyes.

My Ph.D. study would be much less fruitful without working with the top-tier researchers in other universities and institutes. I am grateful to Professor Xuhui Huang, Jordy Homing Lam, Lizhe Zhu, and Jinping Lei, Professor Wei Chen, Professor Le Song, Xinshi Chen, Hanjun Dai, Professor Andrey Rzhetsky, Gengjie Jia, Professor Russ Altman, Professor Maojun Yang, Sensen Zhang, Maofei Chen, Professor Wenning Wang, Dongdong Wang, Professor Zhi-Hua Zhou, Fan Xu, and Huiluo Cao. It is my great pleasure to work with you on those wonderful projects.

Much thanks to my friends, Dapeng Liu, Jin Ren, Haneen Mohammed, Yang Feng, Zihao Wang, Piao Yu, and Hassan Irshad Bhatti, met in KAUST. It is you who made my life here in Saudi Arabia more enjoyable. 

Great thanks to King Abdullah, KAUST, and Saudi Arabia for providing the scholarship and a unique environment to support my research. I enjoyed the fantastic Ph.D. journey here greatly. 

Finally, I would like to dedicate this thesis to my parents, grandmother, and the entire family for their generous and unconditional support all over the years. 

}



\begin{onehalfspacing}

\renewcommand{\contentsname}{\centerline{\textbf{{\large TABLE OF CONTENTS}}}}
\tableofcontents
\cleardoublepage



\cleardoublepage
\addcontentsline{toc}{chapter}{\listfigurename} 
\renewcommand*\listfigurename{\centerline{LIST OF FIGURES}} 
\listoffigures

\cleardoublepage
\addcontentsline{toc}{chapter}{\listtablename}
\renewcommand*\listtablename{\centerline{LIST OF TABLES}} 
\listoftables

\end{onehalfspacing}

\chapter{Introduction}
\label{chapter_background}


\section{Motivation}
Biological and biomedical research is fascinating and critical for directly improving the wellness of all human beings. Machine learning, especially deep learning, can potentially benefit biological research greatly, considering that it has achieved great successes in other fields \cite{li2019deep}, such as computer vision and natural language processing. Usually, when we develop deep learning methods for solving the standard computational problems, the output of the deep learning model is a vector for classification problems or a scalar for regression problems. However, sometimes, especially when handling computational problems in bioinformatics, we need to predict much more complex targets, such as time-course electrical signals, 2D images, 3D molecular structures, and interaction graphs, whose output space contains structures. In other words, there are multiple variables in the output space, and these variables may be dependent, instead of being independent of each other in the standard classification or regression problems. The above complex prediction task is referred as \textbf{structured prediction} \cite{bakir2007predicting}. Structured prediction is much more general and difficult than simple classification. It has much wider application scenarios, especially in bioinformatics, considering that most of the original bioinformatics problems are coupled with the complicated real-life biological problems, with complex output objects. In this dissertation, we focus on tackling structured prediction in bioinformatics with deep learning.

In this chapter, from the next section, we will introduce the background of deep learning (Section \ref{chpt1_sec:dl}) and structured prediction (Section \ref{chpt1_sec:sp}), surveying the existing computational methods (Section \ref{chapter2_sec:bio_methods}) and pointing out challenges for solving the structured prediction problems in bioinformatics with deep learning (Section \ref{chapter2_sec:challenges}). Then, after discussing the limitations of the existing methods (Section \ref{chapter2_sec:limitations}), we present our ideas for resolving those challenges (Section \ref{chapter2_sec:ideas}), improving deep learning methods' performance on the problems. Finally, we give a detailed overview of the rest of this thesis (Section \ref{chpt1_sec:overview}).

\section{Deep Learning}
\label{chpt1_sec:dl}
Since AlexNet \cite{RN69}, deep learning methods have achieved great successes across different fields \cite{RN4}, including bioinformatics \cite{li2019deep}. Two key factors contribute to the success of deep leaning. Firstly, the model architectures are highly flexible, including both feature extractors and classifiers. When we train the models in an end-to-end fashion, such models allow the data to determine which information in the original input is important to the final prediction. Using features determined by the data, instead of the predefined hand-crafted ones, we are more likely to achieve impressive prediction performance. Secondly, the availability of a huge amount of scientific and industrial data has made it possible to train the complex models, without getting stuck in over-fitting. Regarding specific deep learning models, there are several different types of them, which are suitable for different kinds of data and computational problems. For example, convolutional neural networks (CNNs) \cite{RN69} are suitable for image processing while recurrent neural networks (RNNs) \cite{chang2018antisymmetricrnn} or attention networks \cite{vaswani2017attention} are suitable for natural language processing. In addition to supervised learning, researchers have designed deep generative models to conduct unsupervised learning, such as generative adversarial networks (GANs) \cite{RN1150} and variational autoencoders (VAEs) \cite{li2019deep}. In terms of the successful applications of deep learning in bioinformatics, researchers have used it to perform sequence analysis \cite{xia2018deerect,umarov2019promoter}, structure prediction and reconstruction \cite{chen2020rna,li2018dlbi}, biomolecular property and function prediction \cite{RN140,lam2019deep}, \textit{et al.}.
Because of the great expression capability of deep learning models, deep learning methods usually can achieve a better performance than the shallow methods on standard classification problems, as long as we can handle the overfitting issue properly. However, to take advantage of its power, we need a large amount of training data, which may not be available in the biological field. Furthermore, how to incorporate prior knowledge and constraints of biological problems into the deep learning methods remains to be an open research topic. Failing to consider the prior knowledge or constraints can lead to invalid outputs and inferior performance. In Section \ref{chapter2_sec:limitations}, we will further discuss the limitations of directly applying deep learning models to resolve the computational problems in bioinformatics, especially the complicated structured prediction problems.

\section{Structured Prediction}
\label{chpt1_sec:sp}
In this section, we give a short introduction to structured prediction \cite{bakir2007predicting}. We first give a relatively formal definition of structured prediction. Then, we distinguish the term ``structured prediction'' from ``structure prediction'' in bioinformatics. Finally, we introduce the commonly used methods in the machine learning field for tackling structured prediction problems.

\subsection{Definition}
\label{sub:sp_def}
Usually, the output of a standard supervised machine learning model is a vector or a scalar. The vector can represent the predicted probability of the input object belonging to different classes. And the scalar can be the predicted value of a regression problem. However, in real-world applications, we often need to tackle problems that are much more complicated. For example, in bioinformatics, we need to handle at least the following tasks:

\begin{itemize}
    \item The outputs can be sequences, such as DNA sequences or 1D electrical signals. Notice that the values at different locations on the sequence may be dependent. 
    \item The problem targets are 2D images. In bioinformatics, we sometimes need to perform image denoising or super-resolution tasks.
    \item The outputs of the model can be molecular secondary or 3D structures. Predicting or determining molecular structures is one of the most important tasks in bioinformatics.
    \item The labeling space has a hierarchical structure, instead of the plain, unstructured labeling space in most classification problems. The two famous hierarchical labeling systems in bioinformatics are Gene Ontology (GO) system \cite{RN27} and Enzyme Commission number (EC number) system \cite{RN140}.
    \item Within the labeling space, an object belongs to more than one class, which is usually referred to as ``multi-label classification'' \cite{zou2019mldeepre}. A typical example from biology is that an enzyme can be a multi-functional enzyme, being able to catalyze more than one reaction in our body.
    \item We want to predict a graph, which represents the interaction between different objects in a bio-system.
\end{itemize}

Notice that in the above problems, not only is the output of the model much more complicated than a vector or a scale, but different parts of the solution are interdependent, which makes the problem even harder.

Although there is no formal definition of structured prediction, we use the following statement \cite{deshwal2019learning} in this dissertation:

\begin{definition}
For a structured input space $\gX$, a structured output space $\gY$, and a loss function $L: \gX \times \gY \times \gY \mapsto \gR^{+}$, in structured prediction problems, we have a $L(x, y^\prime, y^*)$ as the loss associated with the input $x$, the predicted $y^\prime$, and the true $y^*$. Each structured output $y \in \gY$ can be decomposed into $d$ discrete/continuous variables $v_1, v_2, ..., v_d$ and each decomposed variable $v_i$ can take the value for a set $C(v_i)$. 
\end{definition}

\begin{figure}[!hbpt]
\centering
\includegraphics[width=0.7\textwidth]{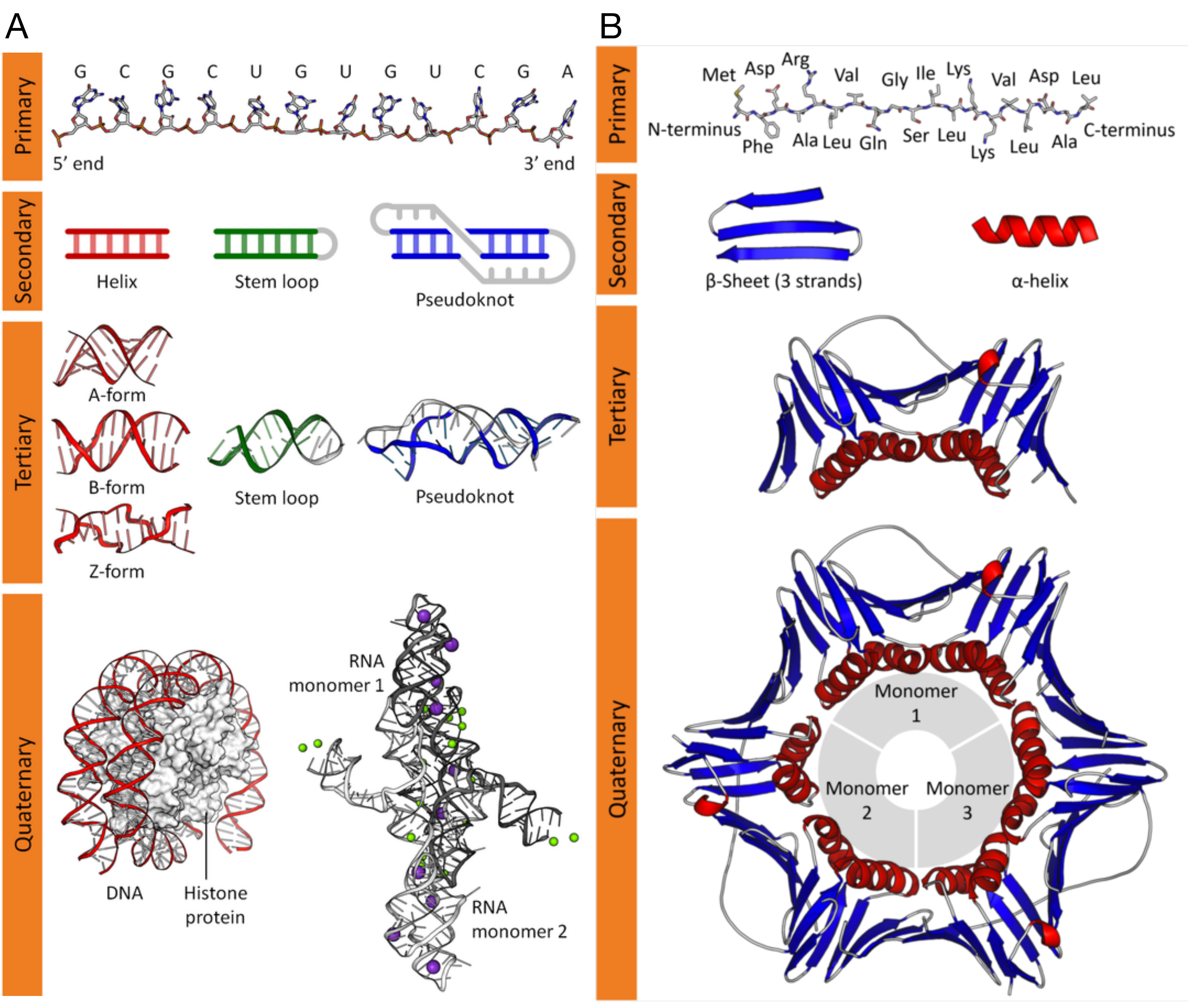}
\caption{(A) Nucleic acid structure. (B) Protein structure.}
\label{fig:structure_cha2}
\end{figure}

\subsection{Structured Prediction VS Structure Prediction}
Although these two terms are very similar, and both of them define a set of problems, they are from different fields. ``Structured prediction'' is from the machine learning field, which has been discussed in detail in Section \ref{sub:sp_def}. ``Structure prediction'' is from the bioinformatics field, which refers to the problem of predicting macromolecular higher-order structures given primary structures (the sequences). The images from Wikipedia (Figure~\ref{fig:structure_cha2}) show the different levels of structures in protein and nucleic acid. Researchers in bioinformatics are also interested in structures of other bio-entities, including chromosome. As illustrated in Figure~\ref{fig:structure_cha2}, the targets of structure prediction are usually complex objects, which means that most of the structure prediction problems belong to structured prediction.

\subsection{Related Works on Structured Prediction from Machine Learning Field}
From the machine learning aspect, the previous methods for tackling structured prediction can be classified into the following categories. Firstly, the earliest attempt to address structured output is to use probabilistic graphical models (PGMs) \cite{lafferty2001conditional}. Although PGMs have strong modeling power, in practice, due to the limit of computational power, researchers tended to use weak graphical models with pairwise or small clique potentials on the output \cite{gygli2017deep} in the first decade of the 21st century. Such a strategy can work for relatively simple problems; however, it encounters bottlenecks when dealing with complex problems because it fails to learn the complicated relationship between different random variables.
Secondly, researchers have also tried to involve features from the output space into the margin-based methods, \textit{i.e.,} support vector machines (SVMs) \cite{tsochantaridis2004support,bakir2007predicting}. Such methods have indeed achieved successes in sequence labeling \cite{tsochantaridis2004support}. However, they have not been applied to more complex problems, such as image processing. Thirdly, people also tried to use energy models \cite{bakir2007predicting,belanger2016structured} to resolve structured prediction. Such methods score joint configurations of the input and different structured outputs. The output with the lowest joint energy score is the final prediction. Despite the success of such methods, how to perform inference efficiently remains to be a difficult problem. Recently, people have tried to use gradient descent \cite{belanger2016structured,belanger2017end}, adversarial networks \cite{tu2018learning}, and reinforcement learning (search) \cite{gygli2017deep} to do inference. Finally, there is also a trend of learning neural networks with differential algorithms \cite{chen2020rna}. If there are non-differentiable operators within the algorithm, such as the max operator, people will turn the non-differentiable operators into differentiable ones, with relaxations and regularizers \cite{cuturi2017soft}. Then, the downstream algorithm can be trained together with the upstream deep learning model. Differentiable dynamic programming \cite{mensch2018differentiable} is a typical example.

\section{Existing Prediction Methods in Bioinformatics}
\label{chapter2_sec:bio_methods}
The computational methods in bioinformatics to perform prediction can be roughly divided into the following categories. 

The most traditional and popular kind is based on similarity-search, either on the raw sequence level or on the hand-crafted feature level. If it is on the raw sequence level, then, it is based on sequence alignment \cite{RN95}. If it is on the feature level, the algorithm is very similar to k-nearest neighbors (KNNs). For example, if we want to predict the function of a newly discovered enzyme, following this idea, we can use the enzyme sequence to search against an enzyme database, finding out the annotated enzyme with the highest sequence similarity against the new enzyme and transferring the old annotation to it \cite{RN140}. On the other hand, because the methods are based on similarity, they are unable to handle new queries without homologs. Furthermore, for the very complicated targets, such as 2D images or graphs, it is challenging to build such databases and define the similarity. Consequently, the similarity-based methods are usually not suitable for handling structured prediction. 

Secondly, researchers have tried to use shallow learning with hand-crafted features to tackle the prediction problems in bioinformatics. Taking the enzyme function prediction as an example again, we can extract some features from the raw sequence, such as the frequency of each residual type, and use standard shallow methods, such as SVMs, to perform the prediction. Such a strategy is usually adopted before the surging of deep learning. Similar to the similarity-based approaches, it is not suitable for structured prediction neither. Firstly, the hand-crafted features are usually sub-optimal for representing the input. Secondly, the standard shallow learning methods cannot deal with complicated objects with structures in the output space. Although in the machine learning field, people have tried to incorporate information from labeling space, such as structured SVM \cite{tsochantaridis2004support}, researchers have seldom used those methods in bioinformatics due to the complexity of problems in this field.

People have indeed tried to handle the structured prediction problems in bioinformatics. However, the solutions are usually based on classic algorithms in computer science, such as dynamic programming and Bayesian inference, instead of cutting-edge learning algorithms. Using RNA secondary structure prediction as an example \cite{chen2020rna}, we want to predict the pairing information between different bases within the RNA sequence in this task. The traditional methods would define a specific energy value for each pair and then enumerate all the possible RNA secondary structure patterns, identifying the particular pattern with the lowest summarized energy of all the pairs with dynamic programming. Despite being reasonable solutions for handling the structured prediction problems, such methods rely heavily on the predefined energy value and optimization. Involving optimization algorithms with high time complexity in the inference step, such as dynamic programming and expectation-maximization (EM), can make the methods very slow. The limitations of such classic algorithms would be further discussed in Sections \ref{chapter2_sec:limitations}.

Regarding deep learning methods, people have applied deep learning models to solve computational problems in biology. However, the covered scenarios largely overlap with shallow learning ones. People usually formulate the computational problem into a supervised learning problem and then utilize the most suitable deep learning model to solve it \cite{li2019deep}. Under most circumstances, the outputs of such methods are vectors or scalars, limiting the power of deep learning to solve real-life problems. 
In fact, seldom did people manage to handle the complicated structured prediction problems with deep learning in bioinformatics. Firstly, the problems are very challenging. Moreover, the existing deep learning models have limitations for handling the structured prediction in bioinformatics, which makes the direct application unsuitable. 
We will discuss the challenges of using deep learning to resolve the structured prediction problems in detail in Section \ref{chapter2_sec:challenges}. The limitations of the previous deep learning methods for solving those problems will be discussed further in Section \ref{chapter2_sec:limitations}.

\section{Challenges of Structure Prediction Problems in Bioinformatics}
\label{chapter2_sec:challenges}
\subsection{Data Problems}
The biggest challenge of tackling the structured prediction in bioinformatics is the data. First of all, the training data are almost always insufficient for such problems in this field. The commonly used training dataset in the computer vision field, ImageNet, contains more than 10 million images. In contrast, in bioinformatics, for example, we only have around 20K sequences for the enzyme function prediction task. Regarding the RNA secondary structure prediction problem, we only have about 30K training RNAs. For the protein-RNA interaction project, we only have roughly 500 interaction complexes.
Furthermore, the data can be biased and imbalanced. In the disease gene prioritization project, where we want to predict a heterogeneous network, the negatives samples are much more than the positive samples. More specifically, because such a network is usually sparse, the number of non-edges is much larger than that of edges.
Even worse, sometimes, we do not have the training data. For instance, we do not have the ground-truth super-resolution structure images for the structure super-resolution project, in which we want to surpass the limitation of optical microscopy, because there is no trivial experimental way to obtain them. Without such images, we cannot train a deep learning model as those images are the training targets for the model.

\subsection{Tremendous Search Space and Output Dimension}
\label{chapter2_sub:dimension}
As we have discussed, for the structured prediction problems, the targets are complex objects. Such complicated objects can have an enormous dimension and tremendous search space, which makes the problem computational prohibitive. Taking the image recognition problem as a baseline, we know that currently, the most intricate image recognition problem from ImageNet has 1000 classes, which means the output of the solution deep learning model is a 1000D vector. On the contrary, for the large-field structure super-resolution task, we may need to reconstruct an image whose dimension is 2K by 3.2K. For the RNA secondary structure prediction task, we should investigate the pairwise potential between each pair within the sequence, which means the output dimension can be $L$ by $L$, where $L$ is the sequence length and can be as large as 1800. Regarding the interaction between protein and RNA, since we are modeling two 3D objects at the same time, if we want to find the optimal configuration with the lowest binding energy, the search space is virtually infinite.

\subsection{Problem Structure and Prior Knowledge}
\label{chapter2_sub:structure}
The essence of structured perdition is to deal with the problem structure and problem-specific constraints. Sometimes, those constraints are explicit. For example, in the Nanopore modeling project, we know a scale mismatch (8-10 times) between the raw input sequences and outputted electrical signals. The model should perform internal warping to handle the mismatch; otherwise, the outputted signals would be invalid. In the enzyme function annotation project, we know that the labeling space has a hierarchical structure. Failing to consider that can lead to a wrong prediction, which is not self-consistent.
Regarding the RNA secondary structure prediction, it is known that only specific pairs are allowed, while the other pairs are not allowed as they are not physically stable. In terms of the structure super-resolution project, we know that the physical process behind this problem can be described with a known mathematical model. The prediction from our method should be consistent with that mathematical model. Sometimes, the constraints can be implicit. For example, in the protein-RNA interaction project, the model should be able to learn and identify the specific local physiochemical environment and spatial conformation, which is favorable to a particular RNA chemical group. Regarding the disease gene prioritization project, although we do not define hard constraints for such a structured prediction problem, the model should take both the topological information in the network and the side information of each node into consideration. 
In fact, the problem structure is not just an obstacle. From the other perspective, such constraints are the prior knowledge we know about the problem. If we can incorporate such knowledge into the algorithm design, we can potentially reduce the data size requirement for training the deep learning model.

\subsection{Interpretability} 
\label{chapter2_sub:interpretability}
As we know, deep learning models are often criticized for acting like a black-box. Sometimes, if we only care about the prediction performance, we can tolerate such a black-box model. However, under some circumstances in bioinformatics, when we care about how we obtained the results, we cannot allow a black-box model even if it is fast and accurate. Taking the structure super-resolution project as an example, as we discussed in Section \ref{chapter2_sub:structure}, we know the physical process and the mathematical modeling behind the problem clearly (more discussion in Chapter \ref{chapter_dlbi}). If we did not consider the physical process explicitly when designing the solution, the biologists would not believe in it and use it, even if the proposed approach may be consistent with the modeling implicitly.

\section{Limitations of the Existing methods}
\label{chapter2_sec:limitations}
Because of the properties and challenges of structured prediction problems in bioinformatics, as we have discussed above, the previously existing methods may have the following limitations when we use them to tackle the problems.

\subsection{Existing Deep Learning Models}
As we have discussed above, the biggest challenge for resolving structured prediction problems in bioinformatics is the lack of data. At the same time, the existing deep learning models are exactly very data-hungry. To train them, we need to prepare a large amount of annotated data, which are usually unavailable in the bioinformatics field. This data requirement of deep learning models limits their application severely, especially for handling the structured prediction problems. Secondly, the existing deep learning models, including CNNs, RNNs, and attention networks, are not explicitly designed to consider the problem-specific constraints. People usually train those models with a tremendous amount of data, hoping they can learn the constraints and distributions implicitly from the data themselves. Such a strategy would not work for the problems in bioinformatics. When we are designing the models, failing to consider the problem-specific constraints can lead to invalid outputs, which are inconsistent with the biological and physical principles. Furthermore, for some structured problems, in which we care about how we obtained the results, it is not suitable to apply the existing black-box deep learning models onto those problems directly, as we discussed in Section \ref{chapter2_sub:interpretability}.


\subsection{Existing Structured Prediction Methods}
Regarding the existing structured prediction methods in the machine learning field, most of them are not scalable to handle the bioinformatics problems. As we have discussed in Section \ref{chapter2_sub:dimension}, the outputs of structured prediction problems in bioinformatics can have a dimension of 1000 by 1000. The traditional PGMs, structured SVM and energy models would encounter severe time and space complexity issues when handling such high dimension outputs. That is why they have seldom been applied to solve complex real-life problems after they were proposed a long time ago, although they can provide rigorous mathematical modeling and decent theoretical guarantee. Furthermore, even if we can overcome the scalability issue, it is not trivial to use them to deal with biological problems. Those problems have their unique properties and problem-specific constraints. How to translate the constraints into mathematical formulations and integrate such formulations into the existing machine learning frameworks remains a problem.
Utilizing those methods from the machine learning field to resolve the problems in bioinformatics, we need to have deep understandings of both the biological problems and the algorithms, making proper customization and optimization to fit the tasks.

\subsection{Existing Prediction Methods in Bioinformatics}
As we have discussed in Section \ref{chapter2_sec:bio_methods}, most of the existing computational methods in bioinformatics, such as the similarity-based methods, shallow learning methods, and directly applied deep learning methods, are not suitable for structured prediction problems. Regarding those classic algorithms, such as dynamic programming and Bayesian inference, which can handle structured prediction, they are usually very slow because a high time-complexity optimization algorithm is involved in the inference step within such methods. Furthermore, often, there is no learning process in the approaches. Consequently, they do not utilize the information from the annotated data properly and take advantage of representation learning. Those algorithms are at the opposite extreme against the deep learning models, only considering the problem-specific constraints and prior knowledge. Given the success of the learning algorithms, it is undesirable to continue omitting the information from the annotated data. However, people in this field usually regard those classic algorithms as the standard algorithms for solving structured prediction problems. They would follow the idea of those algorithms and only make marginal refinement for the methods, such as refining the potential function. 
Consequently, the newly proposed methods can only achieve marginal improvement regarding performance. Seldom did people consider how to reformulate the problems and redesign the algorithms, especially from the learning aspect.

\section{Our Ideas}
\label{chapter2_sec:ideas}
To resolve the above challenges and obstacles, we used the following ideas when designing our methods for solving the structured prediction problems in bioinformatics. 

Firstly, we used various techniques to deal with the data problem. To handle the data insufficiency problem, for example, in the structure super-resolution project (Chapter \ref{chapter_dlbi}), we utilized simulated data for training the deep learning model. In the RNA-protein interaction project (Chapter \ref{chapter_nucleicnet}), although we only had around 500 complexes, we zoomed in the granularity. By training the model with the grid point data, instead of the entire complex, we could significantly boost the training data size and force the model to focus on the local physiochemical information. The other techniques for handling the data problem, such as transfer learning and negative sampling, will be discussed in the following chapters in detail. 

Secondly, when designing the deep learning methods, we considered the problem structure and problem-specific constraints, either by developing new deep learning architectures or incorporating the labeling structure in the training process. For example, in the Nanopore sequencing modeling project (Chapter \ref{chapter_ds}), we proposed a new deep learning architecture, which incorporates a canonical time warping module and can handle the scale difference problem automatically. In the RNA secondary structure prediction project (Chapter \ref{chapter_e2efold}), we proposed an integrated deep learning model with the constraints embedded in the architecture. The output from such a model can satisfy the constraints of the problem directly. In the disease gene prioritization project (Chapter \ref{chapter_pgcn}), the proposed method, based on graph convolutional neural networks, can consider the topological structure of the heterogeneous graph effectively.
Regarding the training process, for example, in the enzyme function prediction project (Chapter \ref{chapter_deepre}), we hierarchically trained multiple deep learning models, following the hierarchical architecture of the labeling space. We also used hierarchical transfer learning to simulate the information flow in the labeling space. In the structure super-resolution project (Chapter \ref{chapter_dlbi}), we incorporated the perceptual loss, which measures the high-level structure and texture difference, into the loss function. In fact, considering the problem-specific constraints when designing methods can help us alleviate the data deficiency issue implicitly since it can reduce the search space of the original problem. In the reduced search space, it is likely to train a biased model towards the desired distribution with less training data. 

Finally, we tried to combine deep learning with classic algorithms, such as canonical time warping, constrained optimization, and PGMs. Those classic algorithms can provide relatively rigorous mathematical modeling and incorporate the constraints into the deep learning models. As we have discussed above, deep learning models and classic algorithms are in two extremes. Deep learning relies heavily on the data without considering the problem structure explicitly. The classic algorithms are mainly based on our prior knowledge about the problem structure while they do not fully utilize the annotated data. The proper integration of such two kinds of methods can leverage the power of both the data and the prior knowledge. We will demonstrate this idea in the Nanopore sequencing modeling project (Chapter \ref{chapter_ds}), the structure super-resolution project (Chapter \ref{chapter_dlbi}), and the RNA secondary structure prediction project (Chapter \ref{chapter_e2efold}). In the protein-RNA interaction project (Chapter \ref{chapter_nucleicnet}), we also tried to integrate deep learning with PGMs. However, compared to the other three projects, the integration in this project is relatively loose.


We will give six concrete examples of how to resolve the structured prediction problems in bioinformatics with the above ideas from the next chapter.

\section{Thesis Overview}
\label{chpt1_sec:overview}

\subsection{Relationship between the Involved Projects}

\begin{figure}[!t]
\centerline{\includegraphics[width=0.65\textwidth]{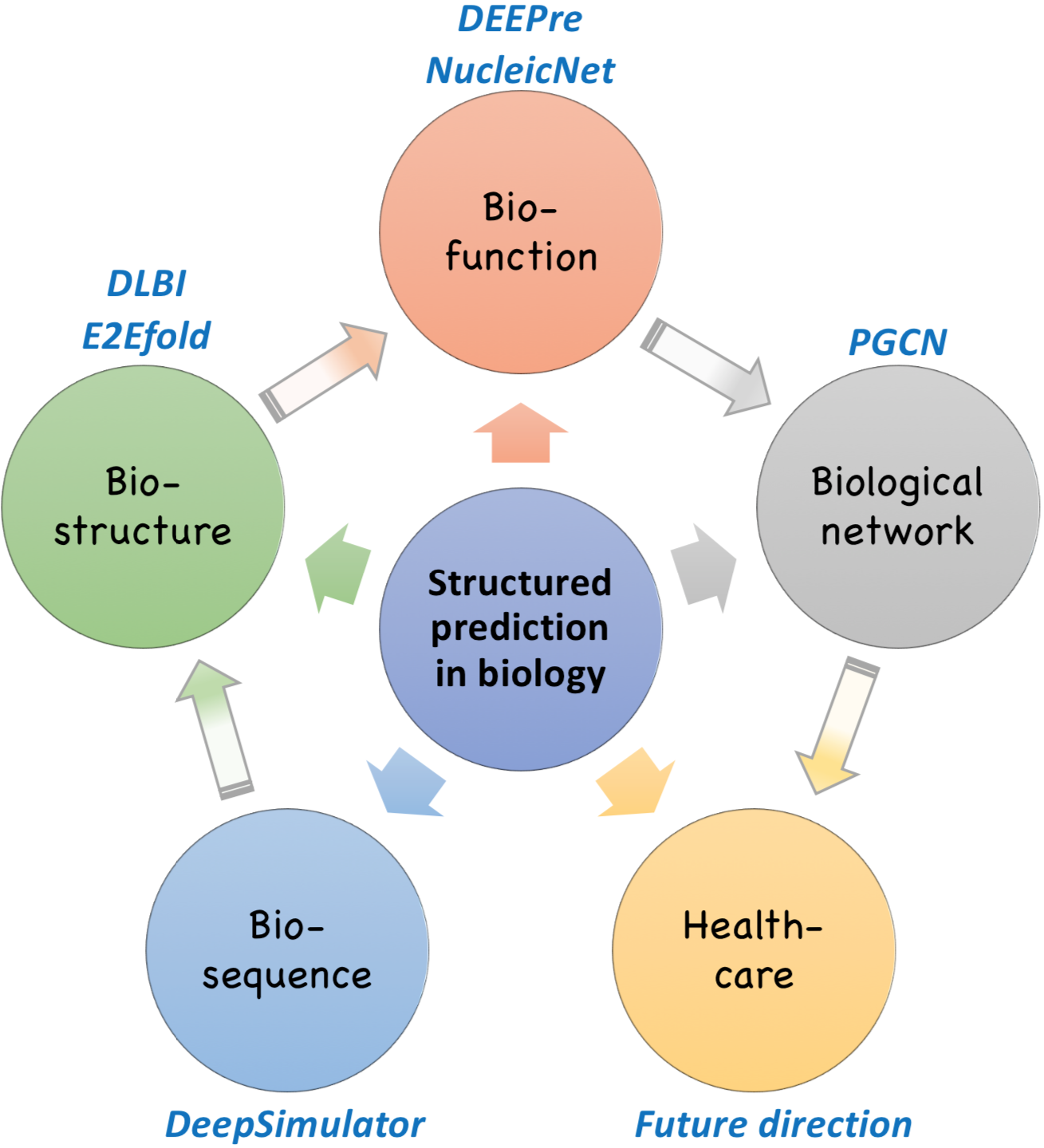}}
\caption{Overview of the projects involved in this thesis from the bioinformatics aspect.} 
\label{fig:thesis_overview}
\end{figure}

Before discussing each project in detail, we want to show the relationship between the six projects involved in this thesis from the bioinformatics aspect. In bioinformatics, there is a well-known paradigm. Molecular sequences, which are usually represented by the combinations of different alphabets, can partially determine their 3D structures. After folding into 3D structures, molecules can interact with other biomolecules to perform their functions. Multiple functional biomolecules form biological pathways or bio-systems, which ensure our body to operate correctly. The deficiency of a critical functional molecule or part of the bio-system can lead to diseases. This paradigm suggests that computational biological research can be divided into five different scales: sequence, structure, function, system, and diseases, as shown in Figure~\ref{fig:thesis_overview}. The six projects involved in this thesis are related to the structured prediction problems from the first four scales. We will discuss the problems from the health-care scale in the concluding chapter.

More specifically, on the sequence level, we discuss the DeepSimulator project \cite{li2018deepsimulator,li2020deepsimulator1}. In this project, we proposed the first deep learning-based simulator, modeling and mimicking the entire pipeline of Nanopore sequencing. On the structure level, we discuss two projects, DLBI \cite{li2018dlbi} and E2Efold \cite{chen2020rna}. As for DLBI, we developed a deep learning guided Bayesian inference framework for reconstructing super-resolved structures from super-resolution fluorescence microscopy data. Regarding E2Efold, we designed a new deep learning architecture, which has the unrolled algorithm embedded in the network, for predicting the RNA secondary structure. On the function level, we discuss two projects, NucleicNet \cite{lam2019deep} and DEEPre \cite{RN140,zou2019mldeepre}. As for NucleicNet, we proposed a deep learning framework for predicting the RNA binding preference landscape on the RNA-binding protein surface. Regrading DEEPre, we built a new tool for annotating the detailed enzyme function hierarchically. On the system level, we present PGCN \cite{li2019pgcn}, which can predict and prioritize the disease genes. 
The examples cover almost all the possible structured outputs, including 1D electrical signals, 2D images, 3D structures, hierarchical labeling, and heterogeneous networks.
This dissertation is related to eight papers, including seven published ones and one preprint. Appendix \ref{appendix_publications} presents the complete list of my publications during the Ph.D. study, including 30 publications and five preprints or papers under review.

\subsection{Summary}
To sum up, the contributions of this thesis are as follows:
\begin{itemize}
    \item In this chapter, Chapter \ref{chapter_background}, we discussed the background of deep learning and structured prediction, pinpointing the challenges for solving the structured prediction problems in bioinformatics with deep learning. Then, we explained our instructive ideas for handling the problems and challenges.
    \item From Chapter \ref{chapter_ds} to Chapter \ref{chapter_pgcn}, we use six examples to substantiate our ideas for solving the structured prediction problems in bioinformatics. In Chapter \ref{chapter_ds}, we focus on the sequence scale in Figure~\ref{fig:thesis_overview}, discussing the DeepSimulator project, in which we utilize deep learning to model the 1D electrical signals in Nanopore sequencing.
    \item In Chapter \ref{chapter_dlbi} and \ref{chapter_e2efold}, we pay attention to the problems in the structure scale, discussing DLBI and E2Efold, respectively, showing how to use deep learning to reconstruct and predict biological structures. 
    \item In Chapter \ref{chapter_nucleicnet} and \ref{chapter_deepre}, we discuss the projects in the function scale, presenting NucleicNet and DEEPre. In the former project, we illustrate how to use deep learning to predict the interaction detail between two 3D biomolecules. In the latter one, we build a tool based on deep learning to annotate the detailed function of enzymes in a hierarchical and multi-labeling way.
    \item In Chapter \ref{chapter_pgcn}, we go to the system scale, presenting the PGCN project, and showing how to use deep learning to aggregate the topological information from biological networks and then predict disease genes.
    \item After demonstrating the effectiveness of our idea, we want to extend our work towards more challenging but important problems, such as the ones in health-care, which can directly benefit people's health and wellness. In Chapter \ref{chapter_conclusion}, we conclude this thesis by discussing such future works and the potential challenges as well as opportunities.
\end{itemize}

\chapter{DeepSimulator: A Deep Simulator for Nanopore Sequencing}
\label{chapter_ds}

\section{Chapter Introduction}
Next-generation sequencing (NGS) technologies allow researchers to sequence DNA and RNA in a high-throughput manner, which have facilitated numerous breakthroughs in genomics, transcriptomics, and epigenomics \cite{metzker2010sequencing,RN135,RN249,shi2016long}. The most popular NGS technologies on the market include Illumina, PacBio and Nanopore. Unlike the other sequencing technologies, Nanopore, whose core component is the pore chemistry that contains a voltage-biased membrane embedded with nanopores, would detect the electrical current signal changes when DNA or RNA molecules are forced to pass through the pore by voltage. Inputting the detected signals to a basecaller specifically designed for Nanopore, one can obtain the nucleotide sequence reads. Benefited from the underlying design, Nanopore sequencing owns the advantages of long-reads \cite{byrne2017nanopore}, point-of-care \cite{lu2016oxford}, and PCR-free \cite{simpson2017detecting}, which enable \textit{de novo} genome or transcriptome \textcolor{black}{assembly} with repetitive regions, field real-time analysis, and direct epigenetic detection, respectively.

Along with the rapid development in Nanopore sequencing, the downstream data analytical methods and tools have also been rapidly emerging. For example, Graphmap \cite{RN136}, Minimap2 \cite{RN137} and MashMap2 \cite{RN138} were designed to map the Nanopore data to the genome. Canu \cite{RN139} and Racon \cite{racon} were created to assemble long and noisy reads produced by Nanopore. It is foreseeable that an even larger number of methods and tools would be developed in the near future. Therefore, it is quite important to benchmark those new methods using either empirical data (i.e., experimentally obtained) or simulated data \cite{RN129}. Although it is essential that one should finally run the method on the empirical data, the empirical data are sometimes difficult and expensive to obtain, with unknown ground truth. On the contrary, the simulated data can be easily obtained at a low cost, and its ground truth can be under full control. These features allow the simulated data to serve as the cornerstone to benchmark new methods.

\begin{figure}[!hbpt]
\centerline{\includegraphics[width=0.9\textwidth]{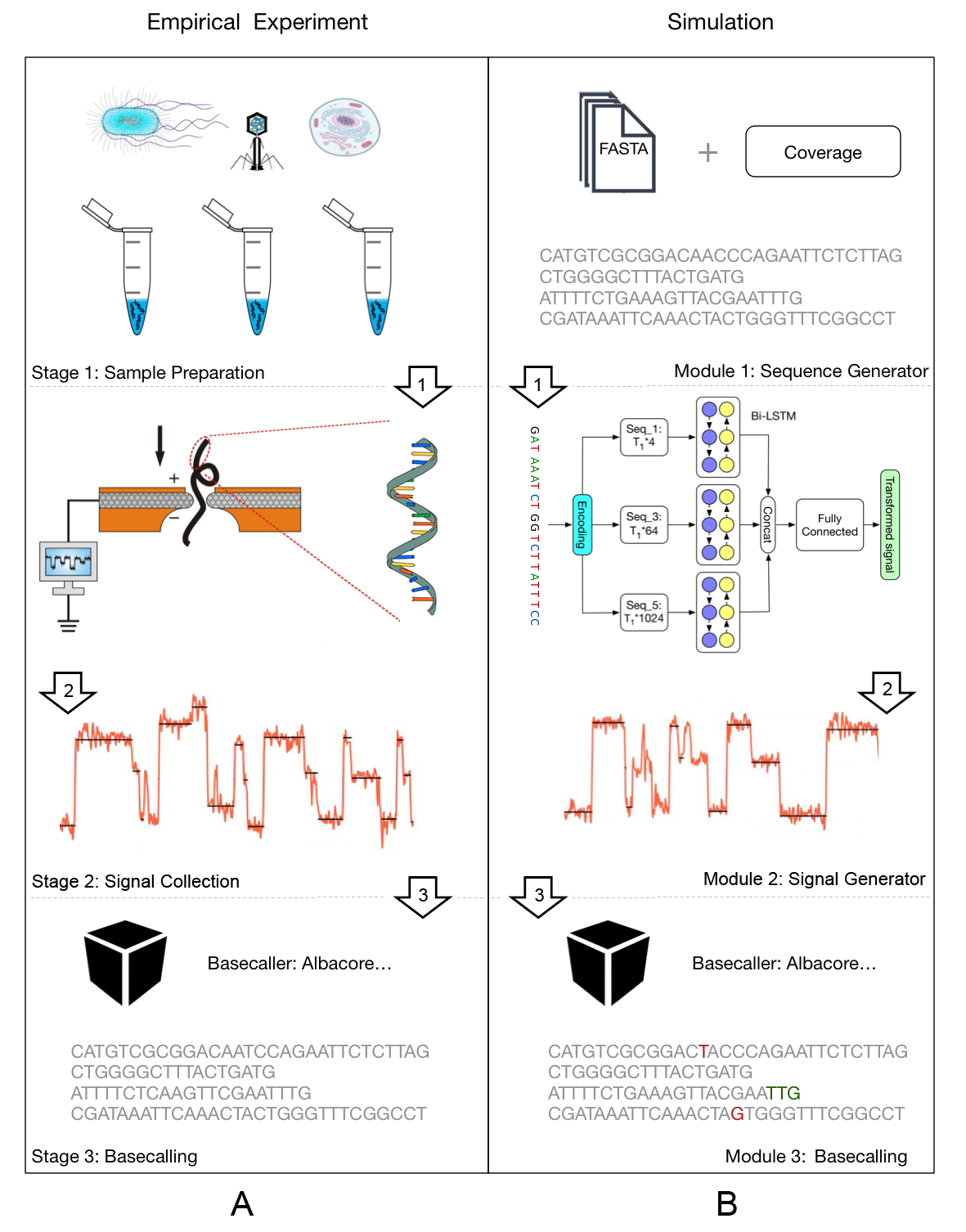}}
\caption{ (A) The Nanopore sequencing procedure. (B) The main workflow of DeepSimulator. 
} \label{fig:overview_ds}
\end{figure}

Despite the existence of more than twenty simulators for NGS technologies \cite{RN129}, there were only three simulators created for the Nanopore sequencing before our method, namely ReadSim \cite{RN132}, SiLiCO \cite{RN128}, and NanoSim \cite{RN127}. Although there are some differences between the three simulators, they share the same property of generating simulated data utilizing the input nucleotide sequence and the explicit \textit{profiles}\footnote{Here the profiles refer to a set of parameters, such as insertion and deletion rates, substitution rates, read lengths, error rates and quality scores. For instance, ReadSim uses the fixed profile; SiLiCO uses the user provided profile; and NanoSim uses the user provided empirical data to learn the profile which would be used in the simulation stage.} with a statistical model. However, those simulators do not truly capture the complex nature of the Nanopore sequencing procedure, which contains multiple stages including sample preparation, current signal collection, and basecalling (Figure~\ref{fig:overview_ds}(A)). More importantly, the current signal is the essence of Nanopore sequencing, yet there was no such simulator that attempted to mimic the signal generation step before our tool was developed.




Instead of following the commonly adapted scenario of designing a simulator from the statistical aspect, we tackle the problem from a different angle, proposing a novel simulator, DeepSimulator, that is designed more naturally for Nanopore sequencing. To run the simulator, the user just need to input a reference genome or assembled contigs, specifying the coverage or the number of reads. The sequence would first go through a preprocessing stage, which produces several shorter sequences, satisfying the input coverage requirement and the read length distribution of real Nanopore reads. Then, those sequences would pass through the signal generation module, which contains the pore model component and the signal repeating component. The pore model component is used to model the expected current signal of a given $k$-mer ($k$ usually equals to 5 or 6 and here we use 5-mer without loss of generality), which is followed by the signal repeating component to produce the simulated current signals. These simulated signals are similar to the real signals in both strength and scale. Finally, the simulated signal would go through Albacore or Guppy, the Oxford Nanopore Technology (ONT) official basecaller, to produce the final simulated reads.



Obviously, the core component of DeepSimulator is the pore model in the
signal generation module. All the official pore models\footnote{https://github.com/nanoporetech/kmer\_models} before ours were context-independent, which assigned each 5-mer a fixed value for the expected current signal regardless of its location on the nucleotide sequence. In order to further polish our simulator, we propose a novel context-dependent pore model, taking advantage of deep learning techniques, which have shown great potential in bioinformatics \cite{RN140,RN141}. Nonetheless, it is not straightforward to train the deep learning model because of the fact that the current signal is usually 8-10 times longer than the nucleotide sequence. To conquer this difficulty, we propose a novel deep learning strategy, BiLSTM-extended Deep Canonical Time Warping (BDCTW), which combines bi-directional long short-term memory (Bi-LSTM) \cite{graves2005framewise} with deep canonical time warping (DCTW) \cite{trigeorgis2016deep} to solve the scale difference issue.

From the structured prediction point of view, in this project, we model the expected electrical signals as well as the base-called reads in Nanopore experiments. The outputs are 1D signals and sequences, which can be context-dependent. We use Bi-LSTM to model such dependency and problem structure. Moreover, as discussed above, the inputs and outputs of the model can have a scale difference (8-10 times). To train such a model, we need to warp the inputs and outputs when calculating the loss function, which can be time-consuming. So we use a novel deep learning architecture with the canonical time warping algorithm embedded in the model. The entire network can be trained jointly and efficiently, as shown in Figure~\ref{fig:dctw}. Next, we explain the technical details and the performance of our method in detail.

\begin{figure*}[!hbpt]
\centering
\includegraphics[width=0.9\textwidth]{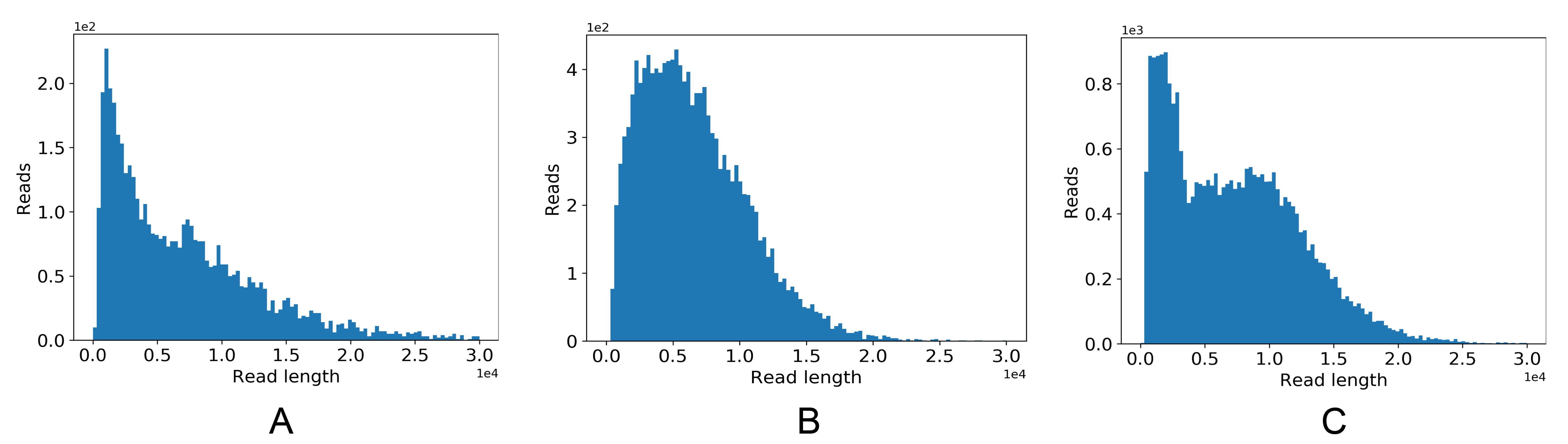}
\caption{The three common read length distribution patterns in Nanopore sequencing. 
} \label{fig_dis1}
\end{figure*}

\section{Methods}

\subsection{Main Workflow}
The main workflow of our DeepSimulator is shown in Figure~\ref{fig:overview_ds}. Unlike the previous simulators \cite{RN127,RN128} that only simulate the final reads from statistical models, our simulator attempts to mimic the entire pipeline of Nanopore sequencing. There are three main stages in Nanopore sequencing. The first stage is sample preparation which results in the nucleotide specimen used in the experiment. After obtaining the specimen, the next stage is to measure the electrical current signals of the nucleotide sequences using a Nanopore sequencing device, such as the MinION. These collected signals are usually stored in a FAST5 file. Finally, we obtain the reads by applying a basecaller to the current signals. Correspondingly, DeepSimulator has three modules. The first module is the sequence generator. Providing the whole genome or the assembled contigs, as well as the desired coverage requirement, DeepSimulator generates relatively short sequences, which satisfy the coverage requirement and the length distribution of Nanopore reads. The read length distribution is described in Section \ref{read_length}. Then, those generated sequences are fed into the second module, namely the signal generation module. As the core module of DeepSimulator, it is used to generate the simulated current signals which aim to approximate the current signals produced by the MinION. There are two components within this module: the pore model component and the signal simulation component. The pore model component takes as input a nucleotide sequence and outputs the context-dependent expected current signal for each 5-mer in the sequence, which is discussed in detail in Section \ref{pore_model}. The signal simulation component repeats an expected signal several times at each position based on the signal repeat time distribution and then adds random noise to produce the simulated current signals. This component is discussed in Section \ref{rep_time_section}. The last module of DeepSimulator is the commonly used basecallers.

Notice that during the entire simulating process, we do not explicitly introduce mismatches and indels (insertions and deletions), which is usually performed in the statistical simulators \cite{RN127,RN128} directly at the read-level. Instead, we try to mimic the current signal produced by Nanopore sequencing as similar as possible, making the basecaller introduce mismatches and indels by itself. Thus, the mismatches and indels in our method are implicitly introduced at the signal-level, which is more reasonable and closer to the real-world situation.



\subsection{Sequence Generation}\label{read_length}


The first module of our simulator is the sequence generator. Given the user-specified reference genome or assembled contigs, as well as the desired coverage or the number of reads, the sequence generation module randomly chooses a starting position on the genome or contigs to produce the relatively short sequences, which satisfy the coverage requirement and the length distribution of the experimental Nanopore reads.

As discussed in the previous papers \cite{RN127,RN128}, the read length of Nanopore sequencing is not very straightforward to model. Many factors, such as the experimental purpose and the experimenter's experience, would influence the read length distribution greatly. By investigating the dataset published by Nanoporetech and datasets provided by our collaborators (in Section \ref{datasets}), we found that the distribution of the read length could be categorized into three patterns by using DBSCAN \cite{DBSCAN} as the clustering method and histogram intersection \cite{Intersect} as the distance metric (Figure~\ref{fig_dis1}). For the first pattern shown in Figure~\ref{fig_dis1}(A), we used an exponential distribution to fit it (e.g., reads from the human genome). For the second pattern shown in the Figure~\ref{fig_dis1}(B), we used a beta distribution to fit it (e.g., reads from the \emph{E. coli} genome). For the last pattern shown in Figure~\ref{fig_dis1}(C), it was not easy to fit it using a single distribution (e.g., reads from the lambda phage genome). To deal with this pattern, we used a mixture distribution with two gamma distributions to fit it. When using the simulator, the users can choose either of the three patterns. Alternatively, the user can also specify the other distribution patterns for the read length.


\subsection{Context-dependent Pore Model}\label{pore_model}

Given a nucleotide sequence, the first step to simulate its corresponding electrical current signals (i.e., raw signal) is the transformation to its expected current signals via the pore model. In this subsection, we first formulate the problem of building the pore model, followed by the proposed solution, BiLSTM-extended Deep Canonical Time Warping (BDCTW). We divide BDCTW into three parts: general framework of deep canonical time warping, feature representation, and neural network architecture. Finally, we introduce our context-dependent pore model.

\vspace{\baselineskip}
\noindent {\bf Problem formulation}

A pore model is defined as the correspondence between the expected current signal and the 5-mer nucleotide sequence that is in the pore at the same time \cite{deamer2016three}. The pore model prediction problem is formulated as follows: given an input nucleotide sequence $X = x_{1}, x_{2}, \ldots, x_{T_{1}}$ with $T_{1}$ nucleotides where $x_{i}$ is a 4-state nucleotide base that can take one of the four values from $\{\text{A,T,C,G}\}$ for DNA or $\{\text{A,U,C,G}\}$ for RNA, we need to predict the corresponding expected electrical current signals $Y = y_{1}, y_{2}, \ldots, y_{T_{1}-4}$, where $y_{i}$ is the predicted expected current signal of a 5-mer starting from position $i$ in $X$ (e.g., ``ACGTT").

Here, we propose a novel method for building the pore model in consideration of the contextual information. Specifically, our method learns the context-dependent (or position-specific) pore model $Y^{dep}$ with length $T_1-4$ for the nucleotide sequence $X$ with length $T_1$ from the raw signals (i.e., the observed electrical current signals from a Nanopore sequencing device) $\hat{Y}$ with length $T_2$.

There are three challenges for learning the context-dependent pore model.
\begin{itemize}
\item \textbf{Scale difference}. Since the frequency of the electrical current measurements (taken at 4000 Hz) is about 8-10 times faster than the speed at which the single-strand nucleotide sequence passes through the pore (the translocation speed is around 450 bases per second for Rapid Kit, for example) \cite{stoiber2017basecrawller}, the temporal scale difference between the raw signals $\hat{Y}$ and the nucleotide sequence $X$ is large.
\item \textbf{Dimensionality difference}. The feature space dimensionality is different between $X$ and $\hat{Y}$, due to the fact that $\hat{Y}$ is a one-dimensional electrical current signal sequence whereas $X$ is a nucleotide sequence with the feature dimension being at least four. Usually, in order to preserve the original sequence information, one-hot encoding is commonly used \cite{graves2013generating} and thus four-dimension is needed to encode the four nucleotide bases.
\item \textbf{Complex non-linear correlation}. The measurement of the raw signals $\hat{Y}$ is under a noisy sequencing environment because of voltage changes, noise and interactions between nanopore channels, etc \cite{david2016nanocall}. Thus, the relationship between $X$ and $\hat{Y}$ is very complex, having high-order or non-linear correlation.
\end{itemize}
\begin{figure}[!hbpt]
\centering
\includegraphics[width=0.9\textwidth]{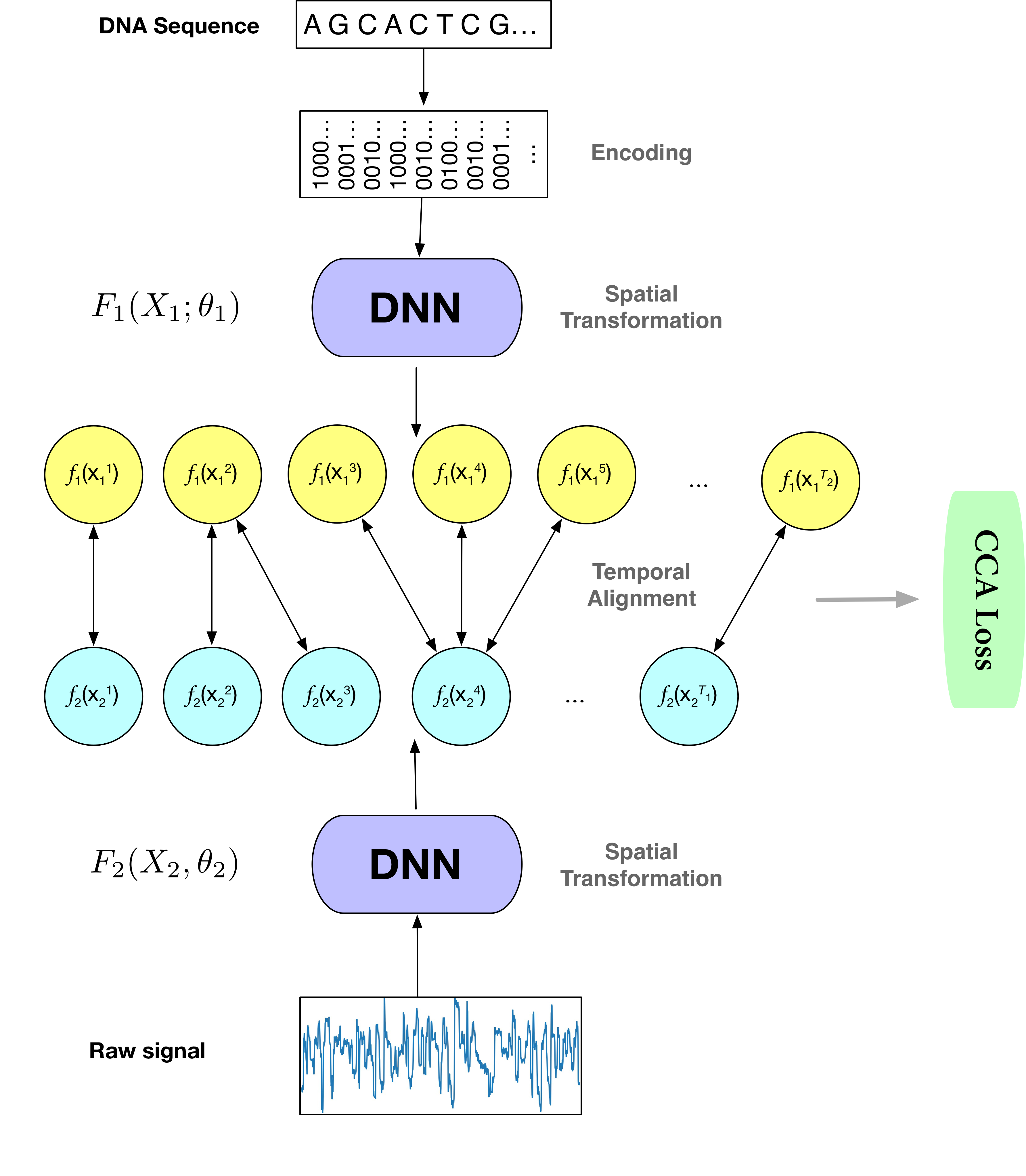}
\caption{Illustration of the deep canonical time warping architecture with two deep neural networks.
}
\label{fig:dctw}
\end{figure}

\vspace{\baselineskip}
\noindent {\bf General framework of deep canonical time warping}

The goal of deep canonical time warping (DCTW) is to discover a hierarchical or recurrent non-linear relationship between two input linearly structured data sets $X_{1}$ and $X_2$ with different lengths $T_{1},T_{2}$ and feature dimensionality $d_{1},d_{2}$ (i.e., $X_{i} \in \mathbb{R}^{d_{i} \times T_{i}}$) \cite{trigeorgis2016deep}. That is, DCTW simultaneously performs spatial transformation and temporal alignment between the two input data sequences. In our case, the two inputs are the nucleotide sequence $X$ and the observed electrical current signal sequence $\hat{Y}$. As shown in Figure~\ref{fig:dctw}, after DCTW, the transformed features from $X$ and $\hat{Y}$ are not only temporally aligned with each other, but also maximally correlated. To this end, let us consider that $Y_{i}=F_{i}(X_{i}; \theta_{i})$ representing the activation function of the final layer of the corresponding deep neural network (DNN) for $X_{i}$, which has $d$ maximally correlated units where $d \leqslant \text{min}(d_{1}, d_{2})$. Such an operation reduces the input data samples to the same feature dimension and then performs a maximal correlation analysis, which essentially resembles the classical canonical correlation analysis (CCA) \cite{akaike1976canonical}. Consequently, we try to optimize the following objective function,
\begin{gather}
\text{argmin}_{\substack{\theta_{1},\theta_{2},\Delta_{1},\Delta_{2}}} || F_{1}(X_{1}; \theta_{1}) \Delta_{1} - F_{2}(X_{2}; \theta_{2}) \Delta_{2} ||_{F}^{2} \nonumber \\
\text{subject to: }
F_{i}(X_{i}; \theta_{i}) \Delta_{i}\textbf{1}_{T}=\textbf{0}_{d}, \nonumber \\
F_{i}(X_{i}; \theta_{i}) \Delta_{i} \Delta_{i}^\top F_{i}(X_{i}; \theta_{i})^\top = \textbf{I}_{d}, \nonumber \\
F_{1}(X_{1}; \theta_{1}) \Delta_{1} \Delta_{2}^\top F_{2}(X_{2}; \theta_{2})^\top = \textbf{D}_{d}, \nonumber \\
\Delta_{i} \in \{0, 1\}^{T_{i} \times T}, i = \{1, 2\}
\label{eq:objective}
\end{gather}
\noindent where $X_1=X$ and $X_2=\hat{Y}$. $T_1$, $T_2$ and $T$ are the length of $X$, $\hat{Y}$, and the final alignment, respectively. $\Delta_{i}$ are the binary selection matrices that encode the alignment paths for $X_i$. That is, $\Delta_1$ and $\Delta_2$ remap the nucleotide sequence $X$ with length $T_1$ and raw signals $\hat{Y}$ with length $T_2$ to a common temporal scale $T$. \textbf{D} is a diagonal matrix. \textbf{I} is the identity matrix. And \textbf{1} (\textbf{0}) is an appropriate dimensionality vector of all 1's (0's).

Such an objective function can be solved via alternating optimization \cite{trigeorgis2016deep}. Specifically, given the final layer output $F_{i}(X_{i}; \theta_{i})$, we employ dynamic time warping (DTW) \cite{fastdtw} to obtain the optimal warping matrices $\Delta_{i}$ which temporally align the input sequence $X_i$ and the final alignment. After obtaining the warping matrices $\Delta_{i}$ via DTW, we infer the maximally correlated nonlinear transformation on the temporally aligned input features $F_{i}(X_{i}; \theta_{i})$ by maximizing the following function,
\begin{gather}
\text{corr} (F_{1}(X_{1}; \theta_{1}) \Delta_{1},F_{2}(X_{2}; \theta_{2}) \Delta_{2})=||\textbf{K}_{DCTW}||_*,
\label{eq:correlation}
\end{gather}
\noindent where $||.||_*$ is the nuclear norm, $\textbf{K}_{DCTW} = \hat{\Sigma}_{11}^{-1/2} \hat{\Sigma}_{12} \hat{\Sigma}_{22}^{-1/2} $ is the kernel matrix of DCTW, $\hat{\Sigma}_{ij}=\frac{1}{T-1}F_{i}(X_{i}; \theta_{i}) \Delta_{i} \textbf{C}_{T} \Delta_{j}^\top F_{j}(X_{j}; \theta_{j})^\top  $ denotes the empirical covariance between the transformed data sets, where $\textbf{C}_T$ is the centering matrix, $\textbf{C}_{T}=\textbf{I}-\frac{1}{T}\textbf{11}^\top$.

The gradient of the objective function $||\textbf{K}_{DCTW}||_*$ with respect to the activation layer of one neural network, such as $Y_{1} = F_{1}(X_{1}; \theta_{1}) $, can be calculated as
\begin{gather}
\frac{\partial ||\textbf{K}_{DCTW}||_*}{\partial Y_{1}} = \frac{1}{T-1} (\textbf{F}^{(pos)} - \textbf{F}^{(neg)}), \nonumber \\
\textbf{F}^{(pos)}= \hat{\Sigma}_{11}^{-1/2}\textbf{UV}^\top\hat{\Sigma}_{22}^{-1/2}Y_{2}\Delta_{2}\textbf{C}_{T}, \nonumber \\
\textbf{F}^{(neg)}= \hat{\Sigma}_{11}^{-1/2}\textbf{USU}^\top\hat{\Sigma}_{11}^{-1/2}Y_{1}\Delta_{1}\textbf{C}_{T},
\label{eq:gradient}
\end{gather}
\noindent where $\textbf{USV}^\top = \textbf{K}_{DCTW}$ is the singular value decomposition (SVD) of the kernel matrix $\textbf{K}_{DCTW}$. By employing this equation as the subgradient, we can optimize the parameters $\theta_{i}$ in each neural network via back-propagation.

\begin{figure}[!hbpt]
\centering
\includegraphics[width=0.9\textwidth]{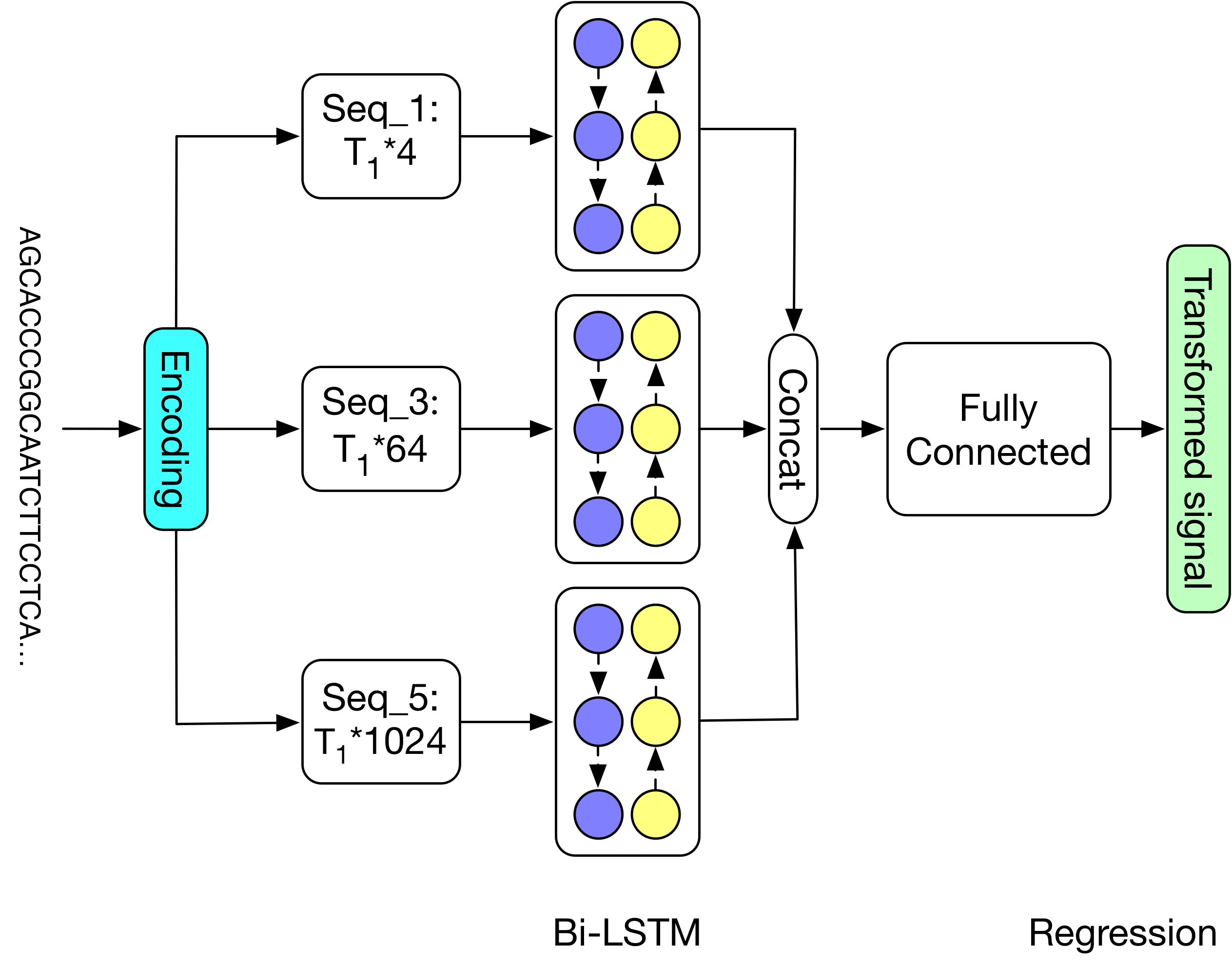}
\caption{Detailed architecture of the deep neural network in deep canonical time warping for feature mapping of the input nucleotide sequence. 
}
\label{fig:bilstm}
\end{figure}

Since the electrical current signal of a 5-mer could be influenced by the surrounding sequences, we extend the feature function $F_{1}(X_{1}; \theta_{1})$ in the original DCTW with bi-directional long short-term memory (Bi-LSTM) \cite{bovza2017deepnano} to incorporate the contextual information. The DNN architecture in Figure~\ref{fig:dctw} is further elucidated in Figure~\ref{fig:bilstm}, which is introduced in detail in the following paragraphs.

\vspace{\baselineskip}
\noindent {\bf Feature representation}
\label{feature_representation}

To preserve the original sequence information, we use one-hot encoding as the representation of the nucleotide sequence $X$. When a nucleotide sequence passes through the nanopore, each 5-mer inside the pore will cause a change in the magnitude of the electrical current. Thus, instead of just considering one nucleotide ($4^1=4$ combinations) at position $t$, we encode the 3-mer ($4^3=64$ combinations) and the 5-mer ($4^5=1024$ combinations) centered at $t$ as well. Specifically, we use one 1 and ($4^\text{k}-1$) 0's to represent each $k$-mer ($ k \in \{1,3,5\}$). Then, for each nucleotide sequence $X$ with length $T_1$, the one-hot encoding would produce three feature matrices with dimensions $T_1 \times 4 $, $T_1 \times 64 $, and $T_1 \times 1024 $, respectively. Each row in the feature matrix represents a specific position and each column represents the appearance of a certain $k$-mer.

\vspace{\baselineskip}
\noindent {\bf Neural network architecture}
\label{nn_archi}

To simplify our model architecture, we use an identical transformation as the feature mapping to deal with the raw signal data. That is, we set $F_{2}(X_{2}; \theta_{2})=\hat{Y}$. For the other feature mapping function $F_{1}(X_{1}; \theta_{1})$ for the nucleotide sequence, we use the Bi-LSTM architecture. Specifically, as shown in Figure~\ref{fig:bilstm}, for each feature matrix, we use a Bi-LSTM block to obtain the hidden representation, with $50$ forward LSTM cells and $50$ backward LSTM cells. After concatenating the obtained hidden representations of different feature matrices, we feed it into a fully-connected layer with $200$ nodes, which is followed by a regression layer. All the weights are initialized using the Xavier method. To avoid overfitting, we utilize weight decay with the coefficient as $1\mathrm{e}^{-4}$. We choose Adam \cite{RN102} as the optimizer with the learning rate $1\mathrm{e}^{-4}$. Deploying batch normalization \cite{RN72} to accelerate training, we set the batch size as 64 during training. The deep neural network model is implemented using Tensorflow \cite{RN98} and can converge within 6 hours with the help of two Pascal Titan X cards.

\vspace{\baselineskip}
\noindent {\bf Context-dependent pore model}

The deep neural network in deep canonical time warping for feature mapping of the input nucleotide sequence (Figure~\ref{fig:bilstm}) becomes the context-dependent pore model after training. To use it, the pore model first uses one-hot vector encoding of \textit{k}-mers, where \textit{k}={1, 3, 5}, to encode the input sequence. The encodings then go through BiLSTM layers, fully-connected layers as well as the final regression layer to generate the expected electrical signals. 


\subsection{Signal Simulation}\label{rep_time_section}


After obtaining the expected current signals of a given nucleotide sequence, the second step of simulating its corresponding electrical current signals is to repeat the signal at each position and add random noise. It is well-known that during sequencing, the raw signal acquisition speed is much faster than the DNA or RNA moving speed, causing a certain 5-mer being measured multiple times. Thus, to convert the expected signals produced by the pore model to the electrical current signals which can be put into a basecaller, we need to repeat a certain position on the expected signal several times. Similar to the read length, we manage to model the repeat time using a mixture alpha distribution. When running the simulator, the repeat time would be drawn from the distribution for each position on the expected signal, generating the simulated current signal by repeating that position for a certain number of times. It should also be noted that the raw signals are extremely noisy due to the complicated sequencing environment \cite{david2016nanocall}. Therefore, we add Gaussian noise with the user-defined variance parameter to each position of the simulated signals.

\begin{figure}[!hbpt]
\centerline{\includegraphics[width=0.8\textwidth]{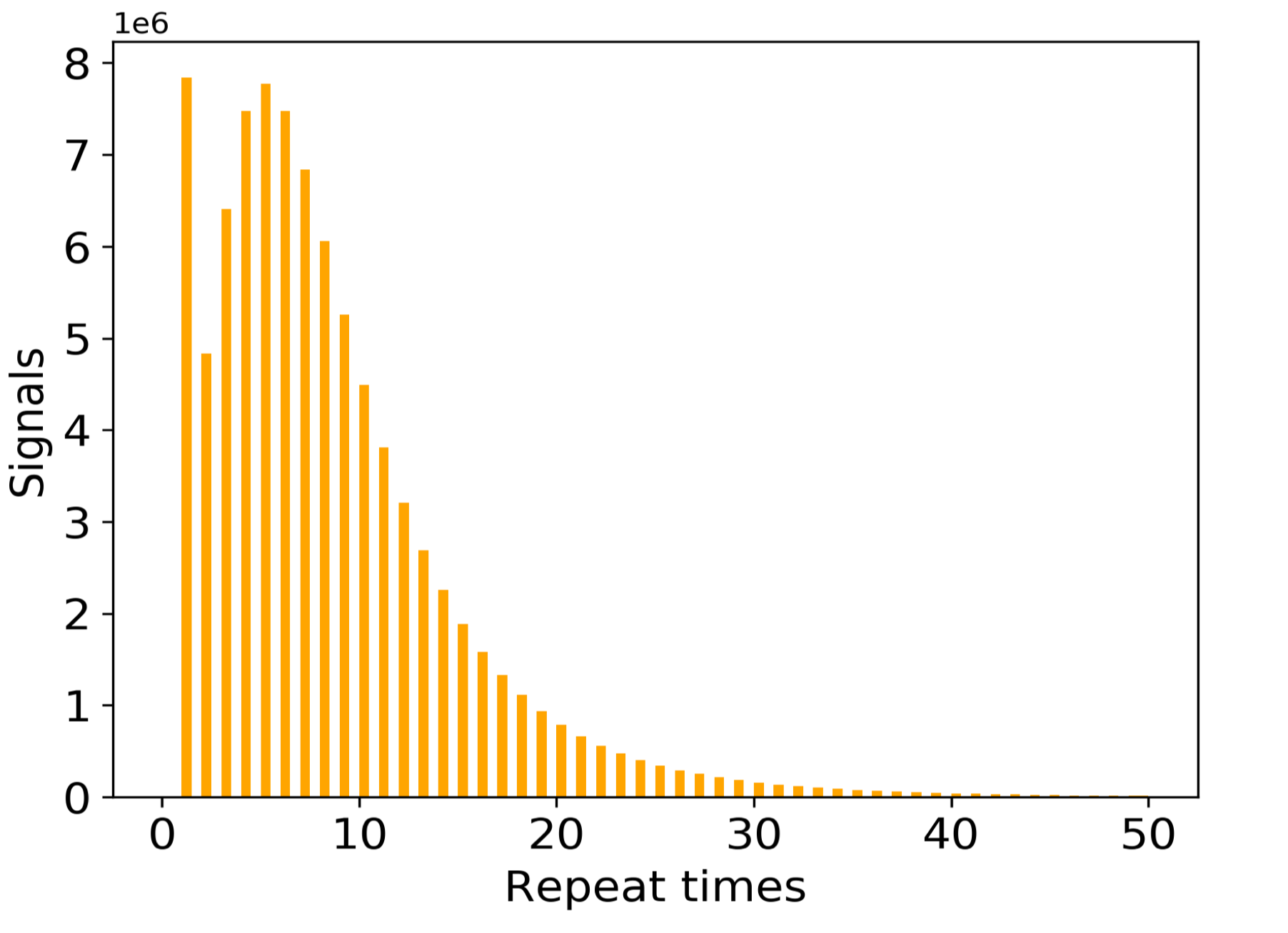}}
\caption{The distribution of the signal repeat times of 5-mer nucleotides.} \label{fig_dis}
\end{figure}


The main difficulty of this step is to get the statistics of the repeat time, as shown in Figure~\ref{fig_dis}. Currently, it is almost impossible to get the precise repeat time of a certain 5-mer, but it is possible to obtain the approximate repeat time statistics. Here we show the four basic steps for obtaining the statistics. (i) Taking as input the reference genome, raw signals produced by the MinION, and the basecalled reads from Albacore, we first map the reads on to the reference genome by Minimap \cite{minimap_miniasm}, which would mark out the ground truth (at least approximate) sequence that corresponds to the raw signal. (ii) With the ground truth sequence, we can get the expected signal of each 5-mer in the sequence using the context-independent pore model. (iii) We then apply dynamic time warping (DTW) \cite{fastdtw} to map the raw signal and the expected signal, which is based on the fact that those two signals should have similar shapes. (iv) Based on the mapping, we can find out the repeat time from the raw signal positions that correspond to each expected signal position. Performing the above procedure on a large dataset, we can get a stable statistic of the repeat time. We then fit the distribution as a mixture model.

\section{Results}

We comprehensively evaluated each of the three modules in DeepSimulator. In summary, the results in this section show that (i) the length distribution of the simulated reads satisfies the empirical read length distribution; (ii) the signals generated by our context-dependent pore model are more similar to the experimental signals than the signals generated by the official context-independent pore model; and (iii) the final reads generated by DeepSimulator with the default parameter have almost the same profile as the experimental data. We finally show that DeepSimulator can benefit the development of tools or methods in \textit{de novo} assembly and low coverage SNP detection.

\subsection{Experimental Setting}

\noindent {\bf Datasets}\label{datasets}

Four Nanopore sequencing datasets from different species were used in this paper: ranging from the in-house datasets lambda phage, \textit{E.coli} K-12 sub-strain MG1655, \textit{Pandoraea pnomenusa} strain 6399, to the public available human data. The three in-house datasets were prepared and sequenced by Prof. Lachlan Coin's lab at University of Queensland. In particular, all the samples were sequenced on the MinION device with 1D ligation kits on R9.4 flow cells (SQK-LSK108 protocol). The publicly available human dataset is the human chromosome 21 from the Nanopore WGS Consortium \cite{Jain128835}. The samples in this dataset were sequenced from the NA12878 human genome reference on the Oxford Nanopore MinION using 1D ligation kits (450 bp/s) with R9.4 flow cells. The Nanopore raw signal datasets in the FAST5 format were downloaded from nanopore-wgs-consortium\footnote{http://s3.amazonaws.com/nanopore-human-wgs/rel3-fast5-chr21.part03.tar}. The reference genomes of the four datasets were downloaded from NCBI\footnote{https://www.ncbi.nlm.nih.gov/nuccore/J02459, https://www.ncbi.nlm.nih.gov/nuccore/U00096, https://www.ncbi.nlm.nih.gov/nuccore/JTCR01000000, https://www.ncbi.nlm.nih.gov/nuccore/NC\_000021}.

The context-dependent pore model of the second module in DeepSimulator was trained on the \textit{Pandoraea pnomenusa} dataset. To construct the dataset used in Section \ref{signal_setion}, which is used to check the performance of the pore models, we randomly sampled 700 reads from each of remaining three species to form a dataset containing 2100 reads.


In addition to the four species for which we have both the reference genome and the empirical experimental data, we also included another extremely small genome, mitochondria, for which we only have the reference genome\footnote{https://www.ncbi.nlm.nih.gov/nuccore/AY172335}. We used the \textit{E.coli} K-12 genome, the lambda phage genome, and the mitochondrial genome to perform the assembly experiments in Section \ref{ass_section}. Finally, the mitochondrial genome and lambda phage genome were used for the single nucleotide polymorphisms (SNP) calling experiments in Section \ref{snp_section}.

\subsection{Read Length Distribution}
As mentioned in Section \ref{read_length}, for an input genome sequence, DeepSimulator generates reads whose length distribution satisfies the empirical length distribution.
\textcolor{black}{In order to find the distributions of the Nanopore sequencing reads, we applied the DBSCAN clustering algorithm with histogram intersection as the distance metric to the datasets, which found three distinguished patterns from the data. We used three distributions, beta distribution, exponential distribution and the mixed gamma distribution to fit the three patterns. The three distributions are thus provided as options in DeepSimulator.}
In general, the mixed gamma distribution is often the most suitable length distribution. As a result, we set it as the default length distribution pattern. In addition to that, considering the property of different sequencing tasks, some biological experiments may be designed on purpose so that the read length distribution would satisfy a predefined distribution. In order to simulate this case, we also provide the interface for the user-defined read length distributions. The distributions of the length of the simulated reads by DeepSimulator on human, \textit{E.coli} K-12 sub-strain MG1655, and lambda phage are very similar to that of the experimental reads. \textcolor{black}{SiLiCO and Nanosim also investigated the read length distribution fitting problem. More detailed discussion of their methods could be found in \cite{RN127,RN128}.}

\subsection{Simulated Signals} \label{signal_setion}
\begin{figure}[!hbpt]
\centerline{\includegraphics[width=0.8\textwidth]{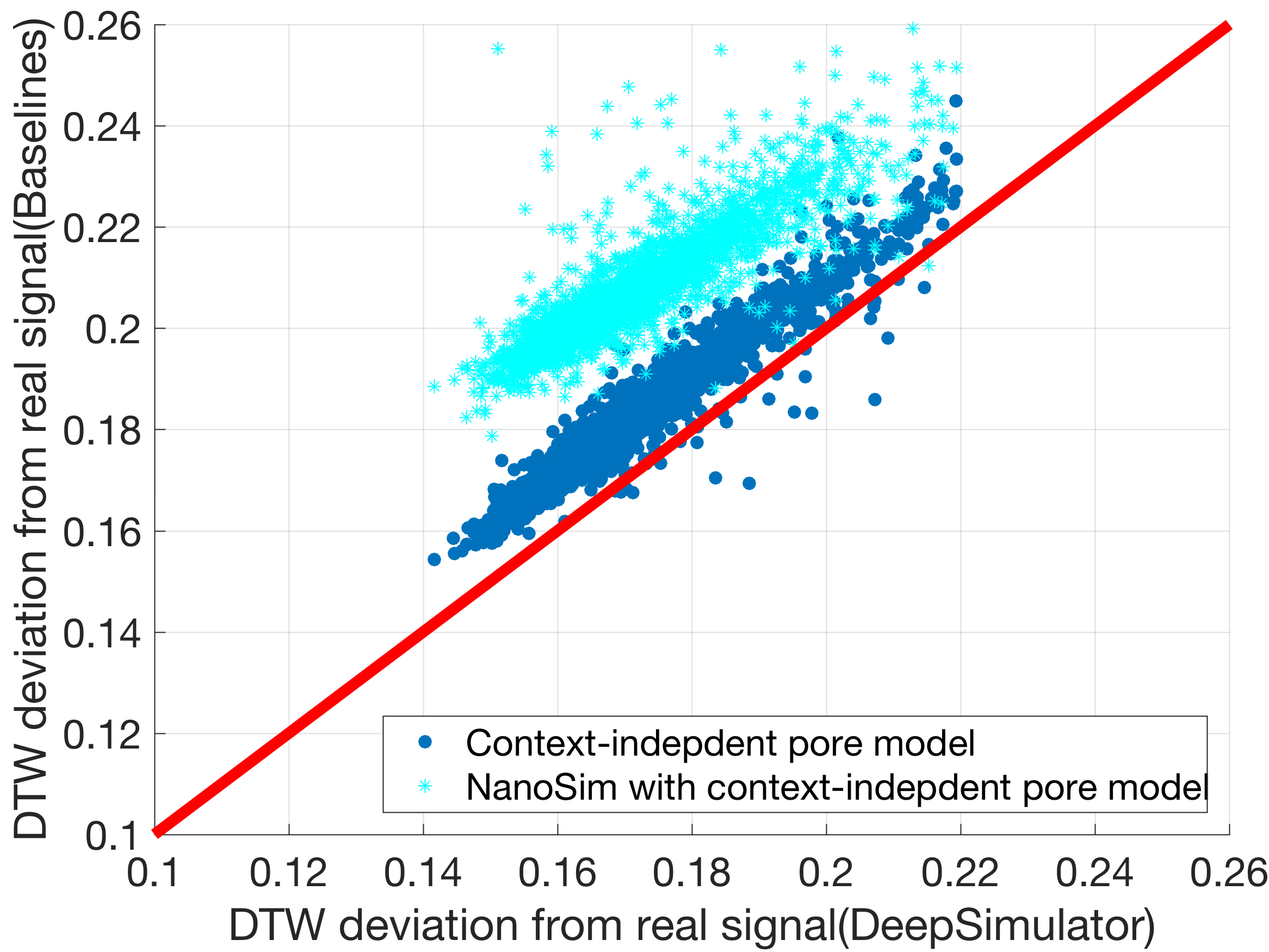}}
\caption{Comparison of the context-dependent pore model component of DeepSimulator with the context-independent pore model on the signal-level. 
} \label{fig_sig}
\end{figure}

To check the signal-level similarity between the simulated signals generated by DeepSimulator and the experimental ones produced by the MinION (i.e., the raw signals), we employed dynamic time warping (DTW) \cite{fastdtw} which is the standard way of checking the difference between two signals. We tested the performance on the randomly selected 2100 reads from lambda phage, \textit{E.coli} K-12 sub-strain MG1655, and human (as described in Section \ref{datasets}). The average deviation between the simulated signals and the raw signals is 0.175. We also performed the same analysis using the official content-independent pore model followed by the same signal repeat component used in DeepSimulator to obtain the context-independent simulated signals. Using the same set of reads, the average deviation of the context-independent signals to the raw ones is 0.185, which is about 5.7\% higher than that of DeepSimulator. Furthermore, we performed another experiment on the reads generated by NanoSim \cite{RN127} to derive the simulated signals by the context-independent pore model. The average deviation of the NanoSim signals to the raw ones is 0.210, which is 20\% higher than that of DeepSimulator. Figure~\ref{fig_sig} shows the comparison of the deviation scores of the DeepSimulator signals and that of the context independent signals as well as that of the NanoSim signals for the 2100 reads. Notice that DeepSimulator was trained solely on \textit{Pandoraea pnomenusa} and tested on the three other species, which demonstrates the generality of our model.




\subsection{Simulated Reads}
\textcolor{black}{The read-level outputs are also of significant importance for sequence level analysis. This section further investigates whether DeepSimulator can simulate reads with the same profile as the real reads from the Nanopore sequencing.} For the read-level outputs, we provided a parameter interface in DeepSimulator, which can be adjusted continuously so that the user could control the final read basecalling accuracy as well as the indel ratio. Internally, the parameters change the noise and the signal repeat time distribution, which are the two factors that affect the read profile greatly. To check the read profile of the simulated reads, for a given input ground truth sequence, we ran DeepSimulator to obtain the simulated read. Performing BLAST \cite{RN95} between the simulated read and the ground truth read, we can calculate the profiles such as the accuracy, mismatch number, and gap numbers. According to our experiment, the output reads of DeepSimulator can have a basecalling accuracy ranging from 83\% to 97\%. Table \ref{read_cmp} shows the profile of the real reads and the profiles of DeepSimulator reads using four typical parameter settings. In addition, we also checked the profile of the reads generated from the official context-independent pore model, whose output is extended using the noise-free repeat time distribution and further basecalled using Albacore, which is shown in the third column of Table \ref{read_cmp}. \textcolor{black}{Due to the modularization of DeepSimulator, we know the ground truth of each read from the Sequence Generator module. As a result, we can run BLAST and obtain the exact profile. As for the reads from other baseline methods, of which it is difficult to determine the ground truth, we performed a global mapping of the reads to first find the regions of the reference genome that are the most similar to the reads, followed by a BLAST analysis to approximate the true profile.}

\begin{table*}
\caption{The profiles of different types of simulated Nanopore reads.
}
\label{read_cmp}
\begin{tabular}{ |p{2cm}|p{1.5cm}|p{1.5cm}|p{1.5cm}|p{1.5cm}|p{1.5cm}|p{1.5cm}|p{1.5cm}|}
 \hline
 Criteria&Real data& OPM & DS (noise free) & DS (high acc) & DS (med acc) & DS (low acc) & NanoSim \\
 \hline
 Accuracy  & 88.49\textcolor{black}{\%} & 95.99\textcolor{black}{\%} & 97.01\textcolor{black}{\%} & 92.96\textcolor{black}{\%} & 88.78\textcolor{black}{\%} & 83.45\textcolor{black}{\%} & 83.80\textcolor{black}{\%} \\
 \hline
 Mismatch & 2.88\textcolor{black}{\%} & 1.24\textcolor{black}{\%}  & 0.94\textcolor{black}{\%} & 1.87\textcolor{black}{\%} & 2.74\textcolor{black}{\%} & 4.36\textcolor{black}{\%} & 4.51\textcolor{black}{\%} \\
 \hline
 Gap open & 5.38\textcolor{black}{\%} & 2.21\textcolor{black}{\%}  & 1.69\textcolor{black}{\%} & 3.63\textcolor{black}{\%} & 5.28\textcolor{black}{\%} & 7.08\textcolor{black}{\%} & 7.31\textcolor{black}{\%} \\
 \hline
 Gap total & 8.62\textcolor{black}{\%} & 2.77\textcolor{black}{\%} & 2.04\textcolor{black}{\%} & 5.17\textcolor{black}{\%} & 8.48\textcolor{black}{\%} & 12.19\textcolor{black}{\%} & 11.69\textcolor{black}{\%} \\
 \hline
\end{tabular}
\end{table*}

\begin{figure*}[!t]
  \centering
  \subfigure[\textit{E.coli} (simulated)]{
    \includegraphics[width=0.35\textwidth]{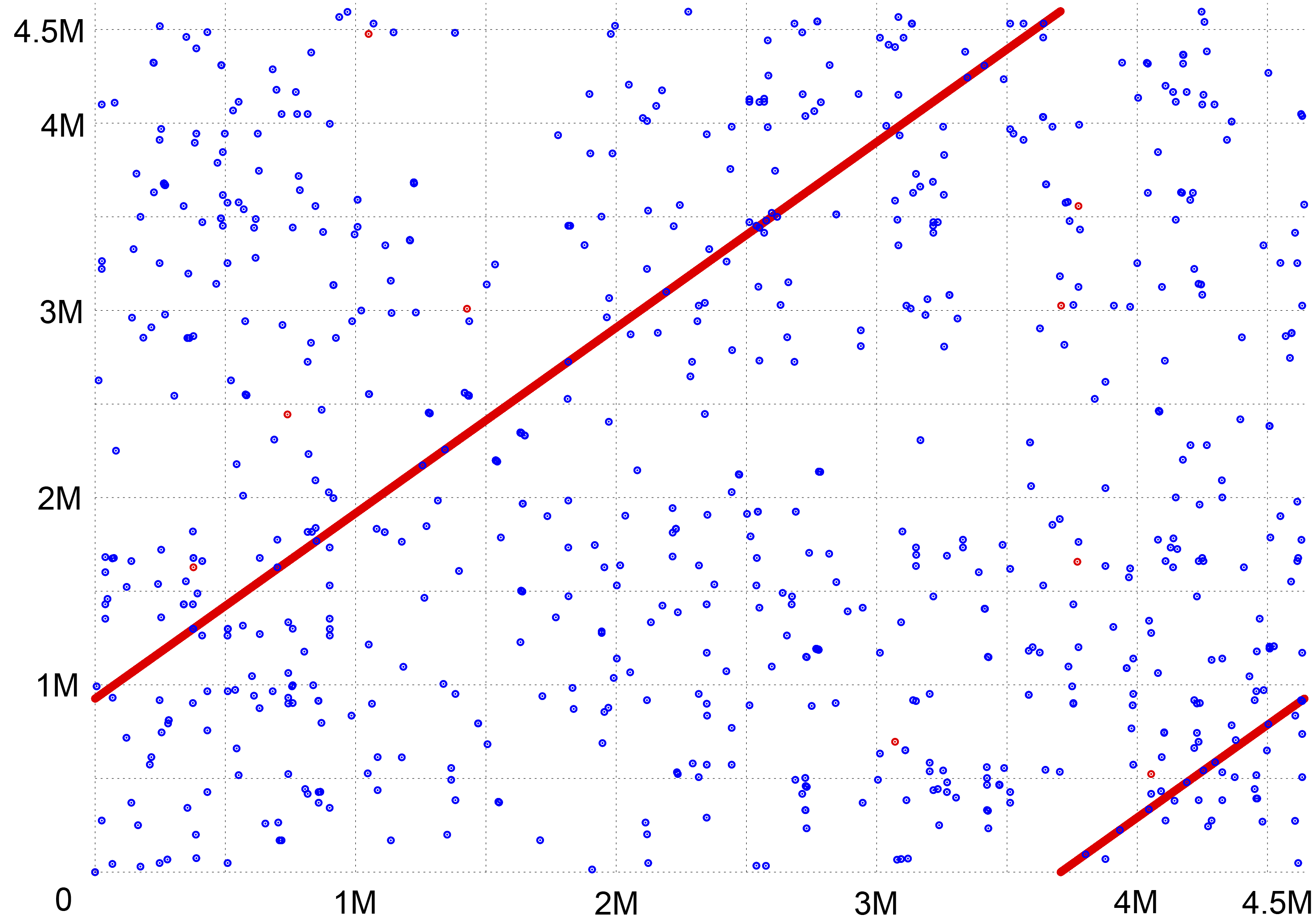}
  }
  \subfigure[\textit{E.coli} (empirical)]{
    \includegraphics[width=0.35\textwidth]{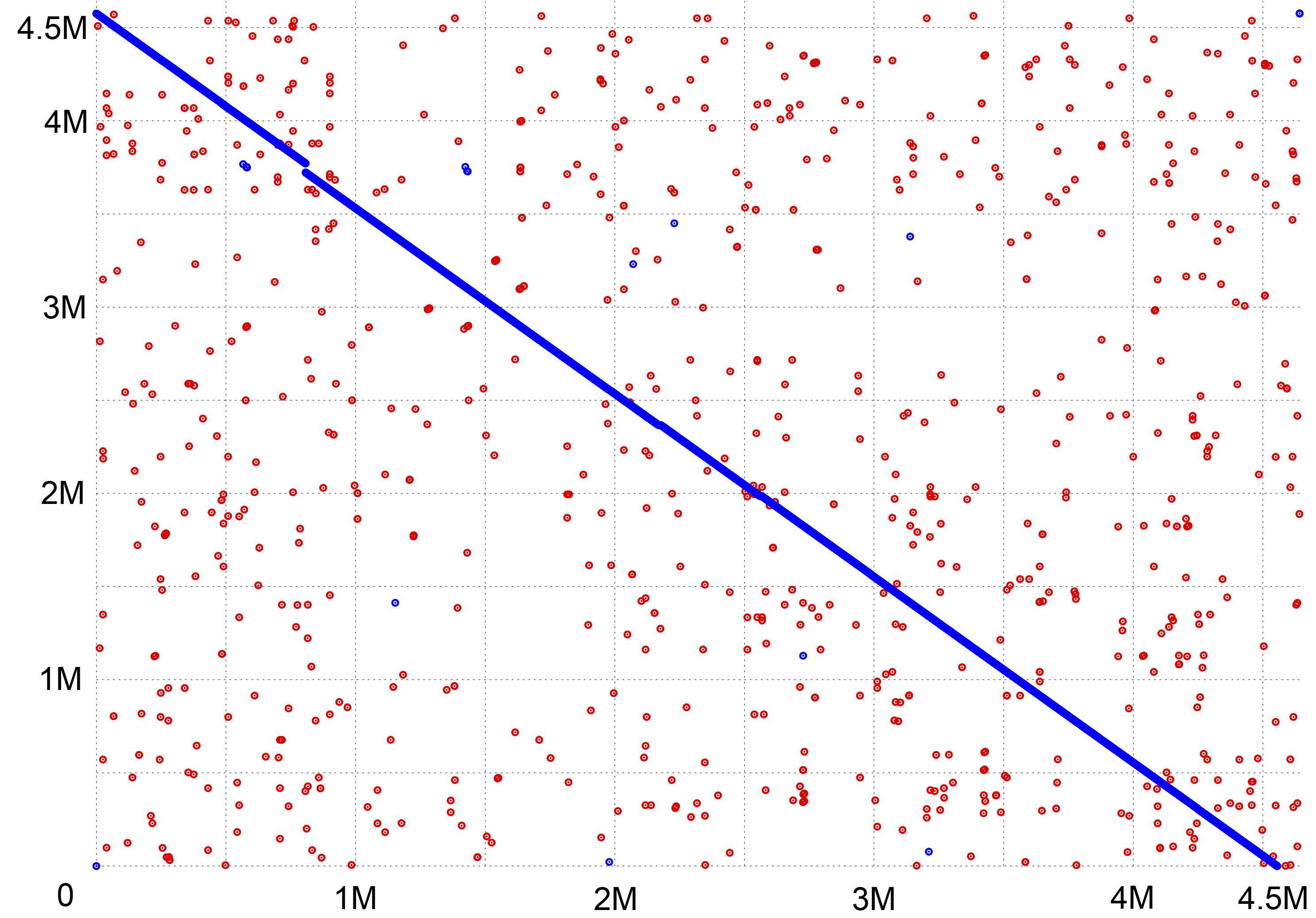}
  }
    \subfigure[Lambda phage (simulated)]{
    \includegraphics[width=0.35\textwidth]{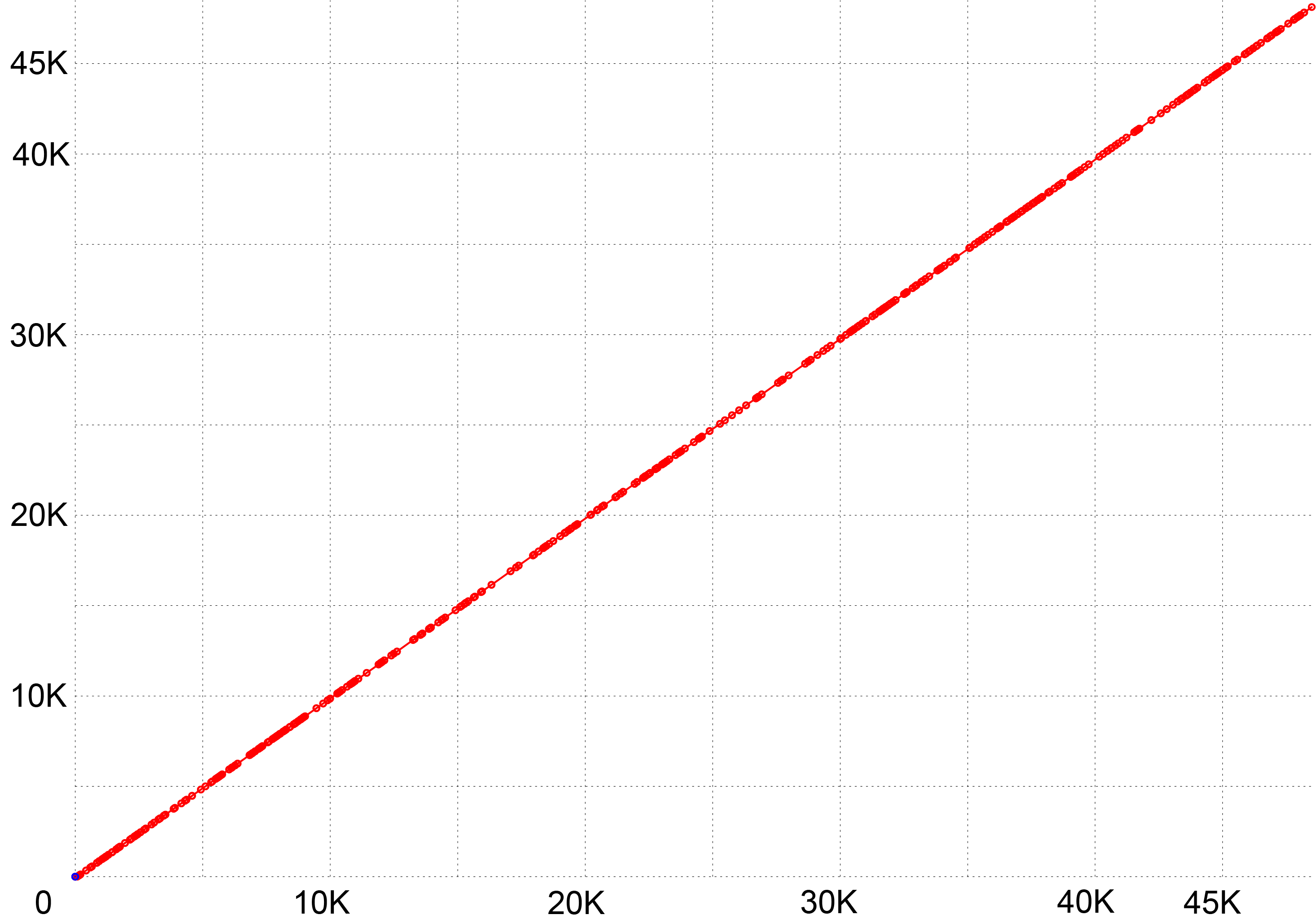}
  }
    \subfigure[Lambda phage (empirical)]{
    \includegraphics[width=0.35\textwidth]{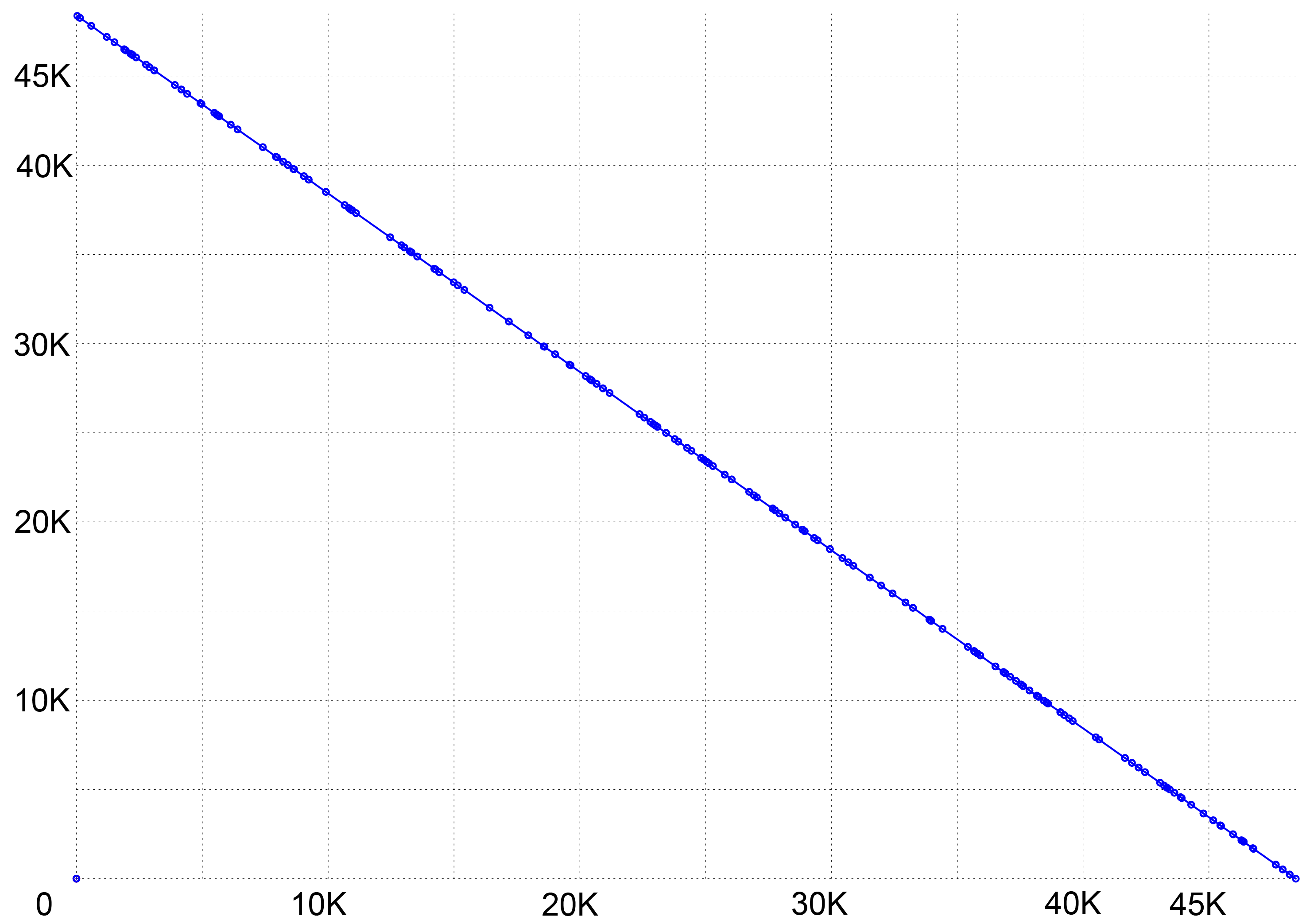}
  }
  \caption{Mummer plots comparing the reference genome on the x-axis with the assembled genome on the y-axis. 
  }
  \label{fig_ass}
\end{figure*}

\subsection{Applications of DeepSimulator}

\noindent {\bf \textit{De novo} Assembly}\label{ass_section}

Because of long reads, Nanopore sequencing has higher potential in genome assembly than the other short-reads sequencing technologies \cite{cao2017scaffolding}. Thus, one of the main applications for Nanopore sequencing is \textit{de novo} assembly. We used two widely recognized \textit{de novo} assembly pipelines, Canu \cite{RN139} and Miniasm \cite{minimap_miniasm} with Racon \cite{racon}, to perform such a task on two different sets of simulated reads generated by DeepSimulator from the \textit{E.coli} K-12 genome and the lambda phage genome, respectively. Both experiments succeeded in assembling the simulated reads into one contig. The comparison between the assemblies and the reference genome was plotted using MUMmer \cite{RN238}, as shown in Figure~\ref{fig_ass}(A, C). As a comparison, we also show the assembly results of \textit{E.coli} K-12 and lambda phage using the empirical data (Figure~\ref{fig_ass}(B, D)). It is clear that the results of the empirical data show similar patterns as the results of the simulated data. In addition to the relatively large genome, \textit{E.coli} K-12, which is 4.6 Mbp, and a small genome, lambda phage, which is 48 Kbp, we also performed another experiment on an extremely small genome, the mitochondrial genome (16 Kbp). Miniasm with Racon also succeeded in assembling the simulated reads into one contig.
%

\vspace{\baselineskip}
\noindent {\bf Low Coverage SNP Detection}\label{snp_section}

Single nucleotide polymorphisms (SNPs) are found to be involved in the etiology of many human diseases. For example, hundreds of SNPs in the mitochondrial DNA (mtDNA) have been linked to aging-related diseases \cite{stewart2015dynamics,RN278}. Despite the importance of the complete haplotyping of the mitochondrial genome, the current methods, which are designed for detecting mitochondrial mutations from a population of cells, would perform massively parallel sequencing of short DNA fragments, having difficulty in performing the complete haplotyping. On the other hand, the Nanopore sequencing, which has the potential of performing the long-read single-molecular sequencing of mtDNA, may overcome the hurdle. Under this circumstance, mimicking the ideal single molecular Nanopore sequencing scenarios, we conducted experiments on the success rate of SNPs detection with respect to sequencing coverage, using the simulated reads from DeepSimulator.

Considering the basecalling accuracy of the Nanopore sequencing, although the current basecalling accuracy is not high enough (around 86\% to 88\%), theoretically, we can consider those errors as random errors instead of systematic errors, and the consensus analysis could help us get rid of such random noise and detect the systematic variants which are caused by SNPs.

\begin{figure}[!hbpt]
\centerline{\includegraphics[width=0.9\textwidth]{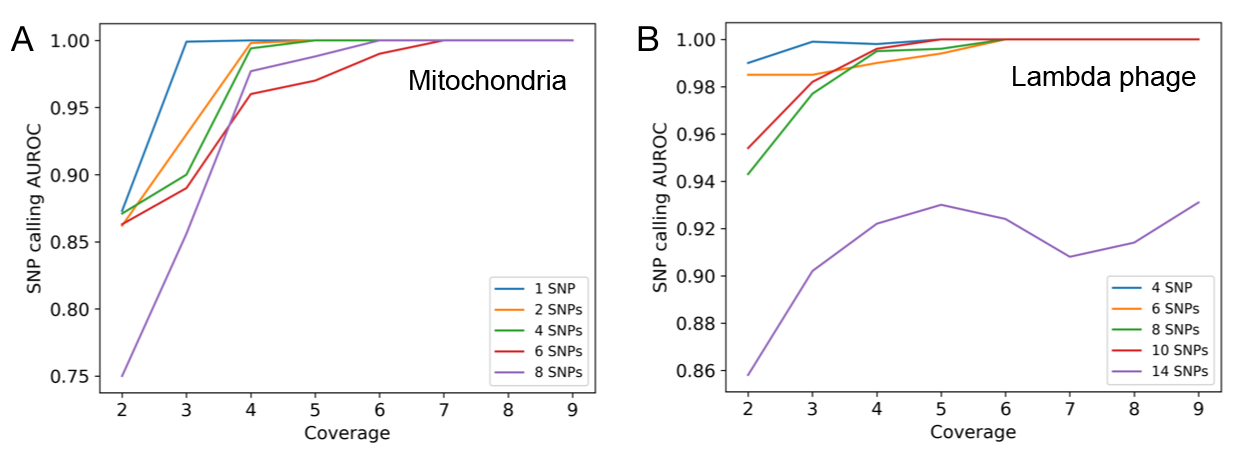}}
\caption{The relationship between the SNP detection performance and the coverage as well as the number of introduced SNPs on the simulated reads.
} \label{fig_snp}
\end{figure}

The results are shown in Figure~\ref{fig_snp}. On the simulated data of mitochondrial genome, we could detect SNPs when the coverage is above 6$\times$ using the standard pipeline of samtools \cite{samtools} and bcftools \cite{bcftools} (Figure~\ref{fig_snp}(A)), which is consistent with the conclusion in \cite{btt512}. As the number of the implanted SNPs increases, the coverage should increase to ensure all the SNPs to be successfully called. Figure~\ref{fig_snp}(B) shows the same analysis on the lambda phage genome, which shares the similar pattern as the mitochondrial experiment. In summary, the detection of the SNPs would become more difficult as the number of SNPs increases. Our experiments demonstrate that in general, 6$\times$ coverage would be enough to detect a small number of SNPs.

\section{Discussion}

In this chapter, we proposed DeepSimulator, the first Nanopore simulator that aims at mimicking the entire procedure of Nanopore sequencing. Unlike the previous simulators which only simulate the reads from the statistical patterns of the real data, DeepSimulator simulates both the raw electrical current signals and nucleotide reads.



There are three advantages of DeepSimulator. First of all, our pipeline is highly modularized, which is easier to be customized by users. For example, the users can use another basecaller, to replace Albacore, to obtain the reads with the profile of that basecaller. Secondly, because of the modularization, compared with other simulators, it is more likely for our simulator to keep up with the rapid development of the Nanopore sequencing technology. If one step of the Nanopore sequencing pipeline is updated, we can also update the corresponding module without changing the entire pipeline completely. 
Thirdly, in addition to the final simulated reads, we are also able to obtain the simulated electrical current signals, which are very useful for the development of basecallers and for the benchmarking of signal-level read mappers.



There are two potential applications of DeepSimulator. On one hand, DeepSimulator can generate benchmark datasets to evaluate the newly developed methods for Nanopore sequencing data analysis. Unlike the empirical datasets whose ground truth is difficult to obtain, DeepSimulator can be fully controlled, which makes it a practical complement to the empirical data. On the other hand, as shown in the SNP detection experiments, it can act as a guidance to the empirical experiment by simulating the ideal situation.


As for this project, we show an example of using deep learning to tackle structured prediction problem in sequence analysis. We proposed a new deep learning architecture, BDCTW, which combines deep learning with the CTW algorithm, to model the dependency in the 1D electrical signals and sequences as well as the scale difference between the inputs and the outputs. The CTW algorithm was embedded in the deep learning model. As a result, the entire model can be trained in an end-to-end fashion, which is more likely to approximate the actual distribution of the data. 
In the next chapter, we will show an example of using deep learning to determine the super-resolved bio-entity structures, which are represented by 2D images.









\chapter{DLBI: Deep Learning Guided Bayesian Inference for Structure Reconstruction of Super-resolution Fluorescence Microscopy}
\label{chapter_dlbi}

\section{Chapter Introduction}
Fluorescence microscopy with a resolution beyond the diffraction limit of light (i.e., super-resolution) has played an important role in biological sciences. The application of super-resolution fluorescence microscope techniques to living-cell imaging promises dynamic information on complex biological structures with nanometer-scale resolution.

Recent development of fluorescence microscopy takes advantages of both the development of optical theories and computational methods. Living cell stimulated emission depletion (STED) \cite{hein2008stimu}, reversible saturable optical linear fluorescence transitions (RESOLFT) \cite{a2007wide}, and structured illumination microscopy (SIM) \cite{gustafsson2005nonlinear} mainly focus on the innovation of instruments, which requires sophisticated, expensive optical setups and specialized expertise for accurate optical alignment. The time-series analysis based on localization microscopy techniques, such as photoactivatable localization microscopy (PALM) \cite{hess2006ultra} and stochastic optical reconstruction microscopy (STORM) \cite{rust2006sub}, is mainly based on the computational methods, which build a super-resolution image from the localized positions of single molecules in a large number of images. Though compared with STED, RESOLFT and SIM, PALM and STORM do not need specialized microscopes, the localization techniques of PALM and STORM require the fluorescence emission from individual fluorophores to not overlap with each other, leading to long imaging time and increased damage to live samples \cite{lippincott2009putting}. More recent methods \cite{holden2011daostorm,huang2011simultaneous,quan2011high,zhu2012faster} alleviate the long exposure problem by developing multiple-fluorophore fitting techniques to allow relatively dense fluorescent data, but still do not solve the problem completely.

Bayesian-based time-series analysis of high-density fluorescent images \cite{cox2012bayesian,xu2015bayesian,xu2016live} further pushes the limit. By using data from overlapping fluorophores as well as information from blinking and bleaching events, it extends the super-resolution imaging to the large-field imaging of living cells. Despite its potential to resolve ultrastructures and fast cellular dynamics in living cells, several bottlenecks still remain. The state-of-the-art methods, such as Bayesian analysis of the blinking and bleaching (i.e., the 3B analysis) \cite{cox2012bayesian}, are computationally expensive, and may cause artificial thinning and thickening of structures due to local sampling. Significant improvements on runtime and accuracy have been achieved by single molecule-guided Bayesian localization microscopy (SIMBA) \cite{xu2016live} with the introduction of dual-channel fluorescent imaging and single molecule-guided Bayesian inference. However, the enhanced process is severely limited by the specialized class of proteins.

Deep learning has accomplished great success in super-resolution imaging \cite{ledig2016photo,kim2016accurate,lim2017enhanced}. Among different deep learning architectures, the generative adversarial network (GAN) \cite{RN1150} achieved the state-of-the-art performance on single image super-resolution (SISR) \cite{ledig2016photo}. However, there are two fundamental differences between the SISR and super-resolution fluorescence microscopy. First, the input of SISR is a downsampled (i.e., low-resolution) image of a static high-resolution image and the expected output is the original image, whereas the input of super-resolution fluorescence microscopy is a time-series of low-resolution fluorescent images and the output is the high-resolution image containing estimated locations of the fluorophores (i.e., the reconstructed structure). Second, the nature of SISR ensures that there are readily a huge amount of existing data to train deep learning models, whereas for fluorescence microscopy, there are only limited time-series datasets. Furthermore, most of these datasets do not have the ground-truth high-resolution images, which makes supervised deep learning infeasible.

In this chapter, we discuss a novel deep learning guided Bayesian inference framework, DLBI, for structure reconstruction of high-resolution fluorescent microscopy. Our framework combines the strength of stochastic simulation, deep learning and statistical inference. 
In particular, the stochastic simulation module simulates time-series low-resolution images from high-resolution images based on experimentally calibrated parameters of fluorophores and stochastic modeling, which provides supervised training data for deep learning models. The deep learning module takes the simulated time-series low-resolution images as inputs, captures the underlying distribution that generates the ground-truth super-resolutions images by exploring local features and correlation along time-axis of the low-resolution images, and outputs a predicted high-resolution image. To achieve this goal, we develop a generative adversarial network (GAN) in which a generator network and a discriminator network contest with each other. The generator network tries to learn the distribution of the high-resolution images in a multi-scale manner, whereas the discriminator network tries to discriminate the ground-truth images and the images produced by the generator network. In order to capture the deep features in the images, we further ease the degradation issue by integrating residual networks \cite{RN7} into our GAN model, where degradation means that stacking more network layers does not lead to better accuracy. The high-resolution image produced by the deep learning module is often very close to the ground-truth image. However, it can still contain some artifacts, and more importantly, lacks the physical meaning. Thus, we develop the Bayesian inference module to take the predicted high-resolution image from deep learning, run Bayesian inference from the initial locations of fluorophores in the predicted image, and predict a more accurate high-resolution image.


\section{Methods}
\begin{figure}[!hpbt]
\centering
\includegraphics[width=1.0\textwidth]{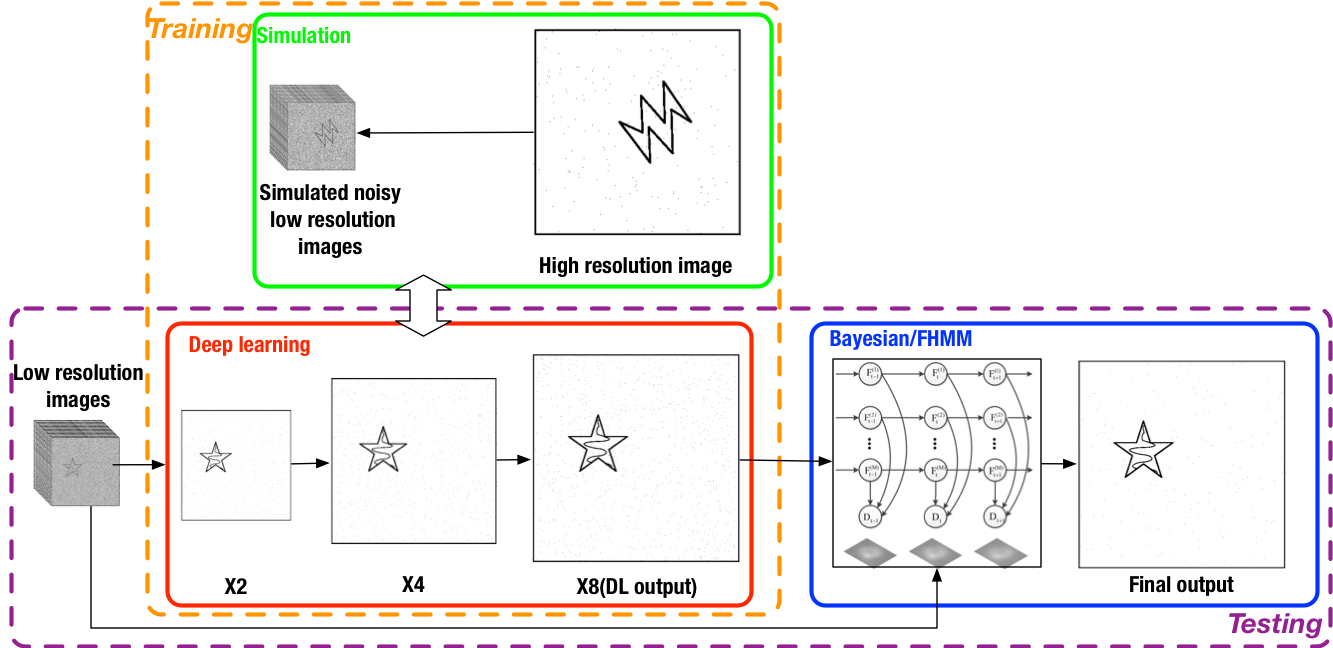}
\caption{The overall workflow of DLBI. 
}
\label{fig:workflow}
\end{figure}

As shown in Figure~\ref{fig:workflow}, DLBI contains three modules: (i) stochastic simulation (Section \ref{sec:2.1}), (ii) deep neural networks (Section \ref{sec:2.2}), and (iii) Bayesian inference (Section \ref{sec:2.3}).

Although deep learning has proved its great superiority in various fields, it has not been used for fluorescent microscopy image analysis. One of the possible reasons is the lack of supervised training data, which means the number of time-series low-resolution image datasets is limited and even for the existing datasets, the ground-truth high-resolution images are often unknown. Here, a stochastic simulation based on the experimentally calibrated parameters is designed to solve this issue, without the need of collecting a massive amount of real fluorescent images. This empowers our deep neural networks to effectively learn the latent structures under the low-resolution, high-noise and stochastic fluorescing conditions. The primitive super-resolution images produced by deep neural networks still contain artifacts and lack physical meaning, we finally develop a Bayesian inference module based on the mechanism of fluorophore switching to produce high-confident images.

Our method combines the strength of deep learning and statistical inference, where deep learning captures the underlying distribution that generates the training super-resolution images by exploring local features and correlation along time-axis, and statistical inference removes artifacts and refines the ultrastructure extracted by deep learning, and further endues physical meaning to the final image.

\subsection{The Stochastic Simulation Module}
\label{sec:2.1}
The input of our simulation module is a high-resolution image that depicts the distribution of the fluorophores and the output is a time-series of low-resolution fluorescent images with different fluorescing states. 

\begin{figure}[!b]
\centering
\includegraphics[width=0.9\textwidth]{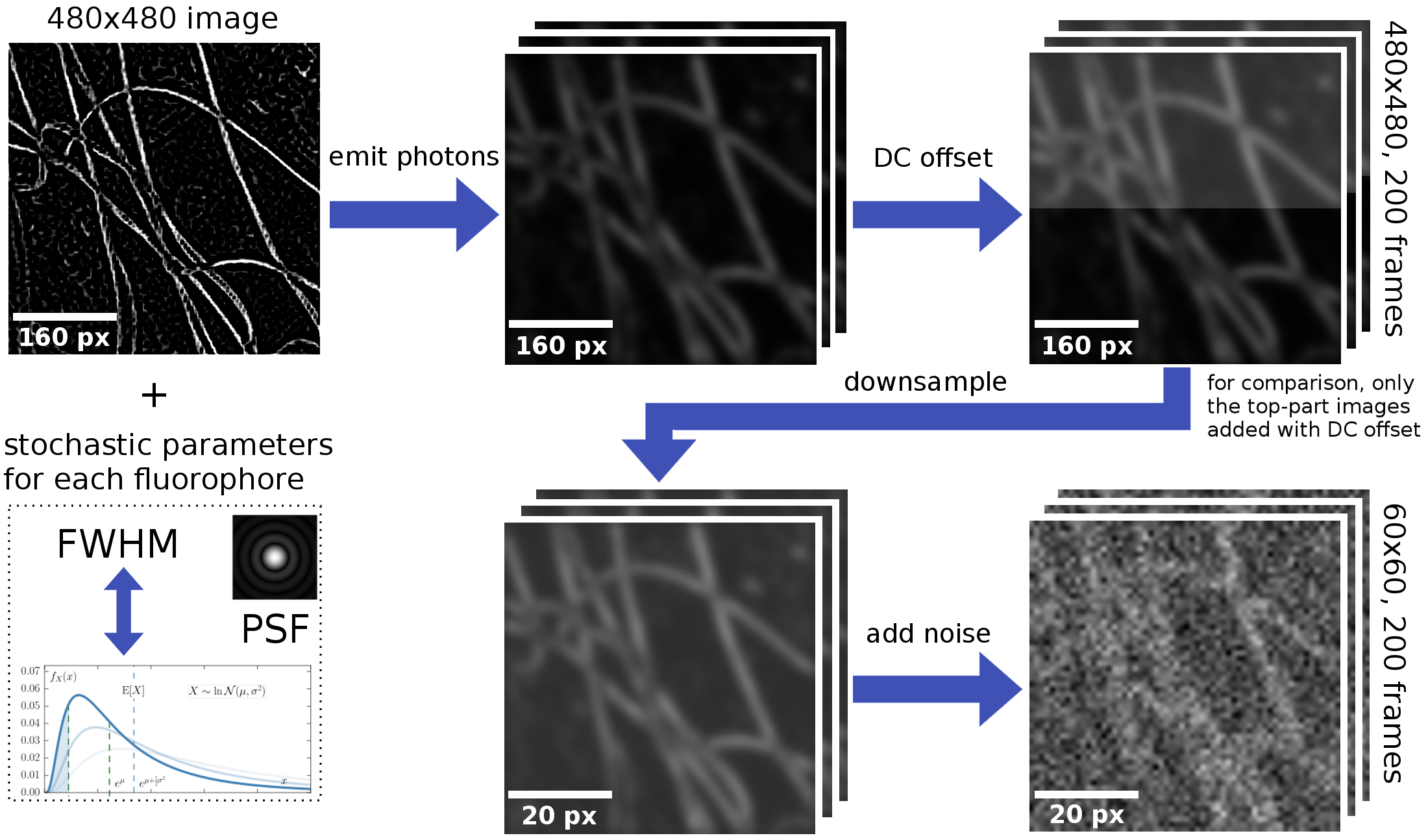}
\caption{The workflow of stochastic simulation. 
}
\label{fig:simu}
\end{figure}

In our simulation, Laplace-filtered natural images and sketches are used as the ground-truth high-resolution images that contain the fluorophore distribution. If a gray-scale image is given, the depicted shapes are considered as the distribution of fluorophores and each pixel value on the image is considered as the density of fluorophores at the location. We then create a number of simulated fluorophores that are distributed according to the distribution and the densities. For each fluorophore, it switches according to a Markov model, i.e., among states of emitting (activated), not emitting (inactivated), and bleached. The emitting state means that the fluorophore emits photons and a spot according to the point spread function (PSF) is depicted on the canvas. All the spots of the emitting fluorophores thus result in a high-resolution fluorescent image. Applying the Markov model on the initial high-resolution image generates a time-series of high-resolution images. After adding the background to the high-resolution images, they are downsampled to low-resolution images and noise is finally added. Figure~\ref{fig:simu} summarizes the stochastic simulation procedure.

Here, the success of simulation relies on three factors: (i) the principal of the linear optical system, (ii) experimentally calibrated parameters of fluorophores, and (iii) stochastic modeling.

\vspace{\baselineskip}
\noindent {\bf Linear Optics}

A fluorescence microscope is considered as a linear optical system, in which the superposition principle is valid, i.e., Image(Obj1 + Obj2) = Image(Obj1) + Image(Obj2). The behavior of fluorophores is considered invariant to mutual interaction. Therefore, for high-density fluorescent images, the pixel density can be directly calculated from the light emitted from its surrounding fluorophores. 

When a fluorophore is activated, an observable spot can be recorded by the sensor, the shape of which is called the point spread function (PSF). Considering the limitation of sensor capability, the PSF of an isotropic point source is often approximated as a Gaussian function:
\begin{footnotesize}
\begin{eqnarray}
   I(x, y) = I_0\exp(-\frac{1}{2\sigma^2}((x-x_0)^2 + (y-y_0)^2)),\label{eq:psf}
\end{eqnarray}
\end{footnotesize}
where $\sigma$ is calculated from the fluorophore in the specimen that specifies the width of the PSF, $I_0$ is the peak intensity and is proportional to the photon emission rate and the single-frame acquisition time, $(x_0, y_0)$ is the location of the fluorophore.

While PSF describes the shape, the full width at half maximum (FWHM) describes the distinguishability. It is defined to be the half width of the maximum amplitude of PSF. If PSF is modeled as a Gaussian function, the relationship between FWHM and $\sigma$ is given by
\begin{footnotesize}
\begin{eqnarray}
   \mathrm{FWHM} = 2\sqrt{2\ln2}\;\sigma \approx 2.355\;\sigma. \label{eq:fwhm}
\end{eqnarray}
\end{footnotesize}

Considering the probability of linear optics, a high-density fluorescent image is composed by PSFs of the fluorophores.

\vspace{\baselineskip}
\noindent {\bf Calibrated Parameters of Fluorophores}

In most imaging systems, the characteristics of a fluorescent protein can be calibrated by experimental techniques. With all the calibrated parameters, it is not difficult to describe and simulate the fluorescent switching of a specialized protein.

\begin{figure}[!hpbt]
\centering
\includegraphics[width=0.8\textwidth]{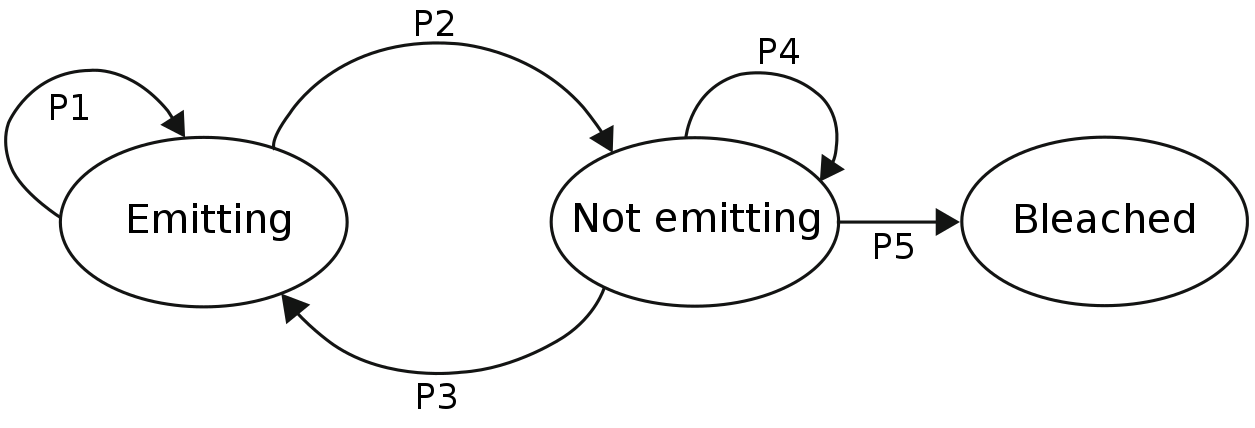}
\caption{The Markov model describing state transition of a fluorophore.}
\label{fig:trans}
\end{figure}

The first characteristic of a fluorophore is its switching probability. A fluorophore always transfers among three states, emitting, not emitting and bleached, which can be specified by a Markov model (Figure~\ref{fig:trans}). If the fluorophore transfers from not emitting to bleached, it will not emit any photon anymore. As linear optics, each fluorophore's transitions are assumed to be independent.

The second characteristic of a fluorophore is its PSF. When a real-world fluorophore is activated, the emitted photons and its corresponding PSF will not stay unchanged over time. The stochasticity of the PSF and photon strength describes the characteristics of a fluorescent protein. To simulate the fluorescence, we should not ignore these properties. Fortunately, the related parameters can be well-calibrated. The PSF and FWHM of a fluorescent protein can be measured in low molecule density. In an instrument for PALM or STORM, the PSF of the microscope can be measured by acquiring image frames, fitting the fluorescent spots parameter, normalizing and then averaging the aligned single-molecule images. The distribution of FWHM can be obtained from statistical analysis. The principle of linear optics ensures that the parameters measured in single-molecule conditions is also applicable to high-density conditions.

In our simulation, a log-normal distribution \cite{cox2012bayesian,zhu2012faster} is used to approximate the experimentally measured single fluorophore photon number distribution. Firstly, a table of fluorophore's experimentally calibrated FWHM parameters is used to initialize the PSF table in our simulation, according to Eq.\ref{eq:psf} and Eq.\ref{eq:fwhm}. Then for each fluorophore recorded in the high-resolution image, the state of the current image frame is calculated according to the transfer table [P1, P2, P3, P4, P5] (Figure~\ref{fig:trans}) and a random PSF shape is produced if the corresponding fluorophore is at the ``emitting'' state. This procedure is repeated for each fluorophore, which results in the final fluorescent image.

\vspace{\baselineskip}
\noindent {\bf Stochastic Modeling}

The illumination of real-world objects is different at different time. In general, the illumination change of real-world objects can be suppressed by high-pass filtering with a large Gaussian kernel. However, this operation will sharpen the random noise and cannot remove the background (or DC offset\footnote{DC offset, DC bias or DC component denotes the mean value of a signal. If the mean amplitude is zero, there is no DC offset. For most microscopy, the DC offset can be calibrated but cannot be completely removed.}). To make our simulation more realistic, several stochastic factors are introduced. First, for a series of simulated fluorescent images, a background value calculated from the multiplication between a random strength factor and the average image intensity is added to the fluorescent images to simulate the DC offset. For the same time-series, the strength factor remains unchanged but the background strength changes with the image intensity. Second, the high-resolution fluorescent image is downsampled and random Gaussian noise is added to the low-resolution image. Here, the noise is also stochastic for different time-series and close to the noise strength that is measured from the real-world microscopy.

The default setting of our simulation takes a $480\times480$ pixel high-resolution image as the input and simulates 200 frames of $60\times60$ pixel (i.e., $8\times$ binned) low-resolution images.

\subsection{The Deep Learning Module}
\label{sec:2.2}
\begin{figure}[!hpbt]
\centering
\includegraphics[width=0.9\textwidth]{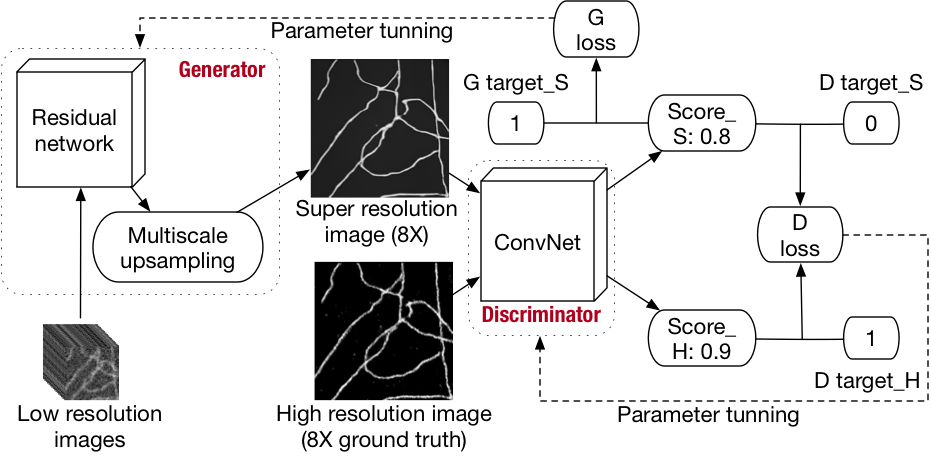}
\caption{Overview of our deep learning framework. 
} 
\label{fig:gan}
\end{figure}

We build a deep residual network under the generative adversarial network (GAN) framework \cite{RN1150,ledig2016photo} to estimate the primitive super-resolution image $I^{SR}$ (the latent structure features) from time-series of low-resolution fluorescent images $ \mathcal{T}=\{I^{FL}_{k}\}_{k=1,...,K}$. Instead of building just one generative model, our approach builds a pair of models, a generator model, $\mathbf{G}$, which produces the estimation of the underling structure of the training images, and a discriminator model, $\mathbf{D}$, which is trained to distinguish the reconstructed super-resolution image from the ground-truth one. Figure~\ref{fig:gan} demonstrates the overview of our deep learning framework.

\vspace{\baselineskip}
\noindent {\bf Basic Concepts}

The goal of training a generator neural network is to obtain the optimized parameters, $\theta_G$, for the generating function, $G$, with the minimum difference between the output super-resolution image, $I^{SR}$, and ground-truth, $I^{HR}$:
\begin{footnotesize}
\begin{eqnarray}
  \hat{\theta}_G=\mathop{\arg\min}\limits_{\theta_G}\frac{1}{N}\sum_{n=1}^{N}{l^{SR}(G(\mathcal{T}_n, \theta_G), I^{HR}_n)},
\end{eqnarray}
\end{footnotesize}
where $G(\mathcal{T}_n, \theta_G)$ is the generated super-resolution image by $\mathbf{G}$ for the $n$th training sample, $N$ is the number of training images, and $l^{SR}$ is a loss function that will be specified later.

For the discriminator network $\mathbf{D}$, $D(x)$ represents the probability of the data being the real high-resolution image rather than from $\mathbf{G}$. When training $\mathbf{D}$, we try to maximize its ability to differentiate ground-truth from the generated image, to force $\mathbf{G}$ to learn better details. When training $G$, we try to minimize $\log(1-D(G(\mathcal{T}_n, \theta_G), \theta_D))$, which is the log likelihood of $\mathbf{D}$ being able to tell that the image generated by $\mathbf{G}$ is not ground-truth. That is, we minimax the following function:
\begin{footnotesize}
\begin{eqnarray}
\mathop{\min}\limits_{\theta_G}\mathop{\max}\limits_{\theta_D}\mathbb{E}_{I^{HR}\sim p_{train}(I^{HR})}[\log(D(I^{HR}, \theta_D))]\\ \notag
\vspace{5em}+\mathbb{E}_{I^{HR}\sim p_{G}(\mathcal{T})}[\log(1-D(G(\mathcal{T}, \theta_G), \theta_D))].
\end{eqnarray}
\end{footnotesize}
In this way, we force the generator to optimize the generative loss, which is composed of perceptual loss, content loss and adversarial loss (more details of the loss function will be introduced later).

\vspace{\baselineskip}
\noindent {\bf Model Architecture}

Our network is specialized for the analysis of time-series images through: (1) 3D filters in the neural network that take all the image frames into consideration, which extracts the time dependent information naturally, (2) two specifically designed modules in the generator residual network, i.e., Monte Carlo dropout \cite{mcdropout} and denoise shortcut, to cope with the stochastic switching of fluorophores and random noise, and (3) a novel incremental multi-scale architecture and parameter tuning scheme, which is designed to suppress the error accumulation in large upscaling factor neural networks.

\begin{figure}[!hpbt]
\centering
\includegraphics[width=0.8\textwidth]{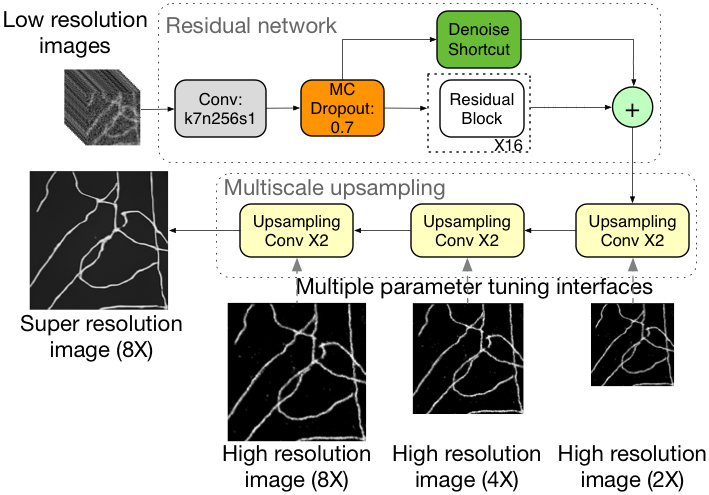}
\caption{Architecture of the generator network, which is composed of a residual network component and a multi-scale upsampling component. 
}
\label{fig:genor}
\end{figure}

Figure~\ref{fig:genor} illustrates the entire architecture of the generator model. The input is time-series low-resolution images. We first use a convolutional layer with the filter size as 7 by 7\footnote{Here, the filter size depends on the FWHM of a PSF. Generally, a fluorescence microscope produces low-resolution images with PSF spanning $3\sim7$ pixels. The specially designed filter size can balance between the computational time and physical meaning.}, which is larger than the commonly used filter, to capture meaningful features of the input fluorescence microscope images. The Monte Carlo dropout layer, which dropouts some pixels from the input feature maps during both training and testing, is applied to the output of the first layer to suppress noise. To further alleviate the noise issue, we use another technique, the denoise shortcut. It is similar to the identical shortcut in the residual network block. However, instead of being exactly the same as the input, we set each channel of the input feature map as the average of all the channels. The denoise shortcut is added to the output of the convolutional layer, which is after 16 residual blocks, element-wisely. After this feature map extraction process, we use a pixel shuffle layer combined with the convolutional layer to increase the dimensionality of the image gradually (upsampling Conv X2 in Figure~\ref{fig:genor}).

Here we adopt a novel multi-scale tuning procedure to stabilize the $8\times$ images. As shown in Figure~\ref{fig:genor}, our generator can output and thus calculate the training error of multi-scale super-resolution images, ranging from $2\times$ to $8\times$, which means that our model has multiple training interfaces for backpropagation. Thus during training, we use the $2\times$, $4\times$, $8\times$ high-resolution ground-truth images to tune the model simultaneously to ensure that the dimensionality of the images increases smoothly and gradually without introducing too much fake detail.

For the discriminator network, we adopt the traditional convolutional neural network, which contains eight convolutional layers, one residual block and one sigmoid layer. The convolutional layers increase the number of channels gradually to 2048 and then decrease it using 1 by 1 filters. Those convolutional layers are followed by a residual block, which further increases the model ability of extracting features.

\label{sec:training}
\vspace{\baselineskip}
\noindent {\bf Model Training and Testing}

GAN is known to be difficult to train \cite{salimans2016improved}. We use the following techniques to obtain stable models. For the generator model, we do not train GAN immediately after initialization. Instead, we pretrain the model. During the pretrain process, we minimize the mean squared error between the super-resolution image and the ground-truth, i.e., with the pixel-wise MSE loss as
\begin{footnotesize}
\begin{eqnarray}
  l^{SR}_{MSE_\mu} = \frac{1}{\mu^2WH}\sum_{x}^{\mu W}\sum_{y}^{\mu H}{(G(\mathcal{T}, \theta_{G_\mu})-I^{HR}_{x,y})^2},
\end{eqnarray}
\end{footnotesize}
\noindent where $W$ is the width of the low-resolution image, $H$ is the height of the low-resolution image, and $\mu=2,4,8$ is the upscaling factor. During pretraining, we optimize $l^{SR}_{MSE_8}$, $l^{SR}_{MSE_4}$, $l^{SR}_{MSE_2}$ simultaneously, instead of optimizing the sum of them.

Only after the model has been well-pretrained do we start training the GAN. During that process, we also use VGG19 \cite{simonyan2014very} to calculate the perceptual loss \cite{johnson2016perceptual} and use Adam optimizer \cite{RN102} with learning rate decay as the optimizer. When feeding an image to the VGG model, we resize the image to fulfill the dimensionality requirement:
\begin{footnotesize}
\begin{eqnarray}
  l^{SR}_{VGG_\mu} = \sum_{i=1}^{V}{(VGG(G(\mathcal{T}, \theta_{G_\mu}))_i-VGG(I^{HR})_i)^2},
\end{eqnarray}
\end{footnotesize}
where $V$ is the dimensionality of the VGG embedding output.

During final tuning, we simultaneously optimize the $2\times$, $4\times$, and $8\times$ upscaling by the generative loss:
\begin{footnotesize}
\begin{eqnarray}
  l^{SR}_{GAN_\mu} = 0.4*l^{SR}_{MSE_\mu} + 10^{-6}*l^{SR}_{VGG_\mu},
\end{eqnarray}
\end{footnotesize}
and
\begin{footnotesize}
\begin{eqnarray}
  l^{SR}_{GAN_8} = 0.5*l^{SR}_{MSE_8} + 10^{-3}*l^{SR}_{ADV_8} + 10^{-6}*l^{SR}_{VGG_8},
\end{eqnarray}
\end{footnotesize}
where $\mu=2,4$ and the $8\times$ upscaling has an additional term, the adversarial loss $l^{SR}_{ADV_8} = \sum_{n=1}^{N}{\log(1-D(G(\mathcal{T}_n, \theta_G), \theta_D))}$. For the discriminator network, we use the following loss function:
\begin{footnotesize}
\begin{eqnarray}
  l^{SR}_{DIS} = \sum_{n=1}^{N}{\log(D(G(\mathcal{T}_n, \theta_G), \theta_D))} + \sum_{n=1}^{N}{\log(1-D(I^{HR}_n, \theta_D))}.
\end{eqnarray}
\end{footnotesize}

During testing, for the same input time-series images, we run the model multiple times to get a series of super-resolution images. Because of the Monte Carlo dropout layer in the generator model, all of the super-resolution images are not identical. We then compute the average of these images as the final prediction, with another map showing the p-value of each pixel. We use Tensorflow combined with TensorLayer \cite{dong2017tensorlayer} to implement the deep learning module. Trained on a workstation with one Pascal Titan X, the model gets converged in around 8 hours.

\subsection{The Bayesian Inference Module}
\label{sec:2.3}
Our Bayesian inference module takes both the time-series low-resolution images and the primitive super-resolution image produced by the deep learning module as inputs, and generates a set of optimized fluorophore locations, which are further interpreted as a high-confident super-resolution image. Since the deep learning module has already depicted the ultrastructures in the image, we use these structures as the initialization of the fluorophore locations, re-sampling with a random punishment against artifacts. For each pixel, we re-sample the fluorophore intensity by $\sqrt{I_{x,y}}$ and the location by $(x,y)\pm rand(x,y)$, where $I_{x,y}$ is the pixel value in the image produced by deep learning, $rand(x,y)$ is limited in $\pm8$. In this way, the extremely high illumination can be suppressed and fake structures will be re-estimated.

\vspace{\baselineskip}
\noindent {\bf Basic Concepts}

As shown in Figure~\ref{fig:trans}, a fluorophore has three states: emitting (light), not emitting and bleached. In classic Bayesian-based time-series analysis, the switching procedure of fluorophores is modeled by Bayesian inference, i.e., given an observed region $R$, deciding whether there is a fluorophore ($F$) or not ($N$) by
\begin{footnotesize}
\begin{eqnarray}
  \frac{P(F|R)}{P(N|R)} = \frac{P(R|F)P(F)}{P(R|N)P(N)},	\label{eq:fn}
\end{eqnarray}
\end{footnotesize}
\noindent where $P(F)$ and $P(N)$ are constants which are based on experimental prior, $P(R|F)$ is the probability of the observed data region $R$ given the location of the fluorophore, $P(R|N)$ is the probability of the observed data region $R$ if there is no fluorophore, which can be calculated by integrating all the probability of observing pixels given the noise model.

\begin{figure}[!t]
\centering
\includegraphics[width=0.8\textwidth]{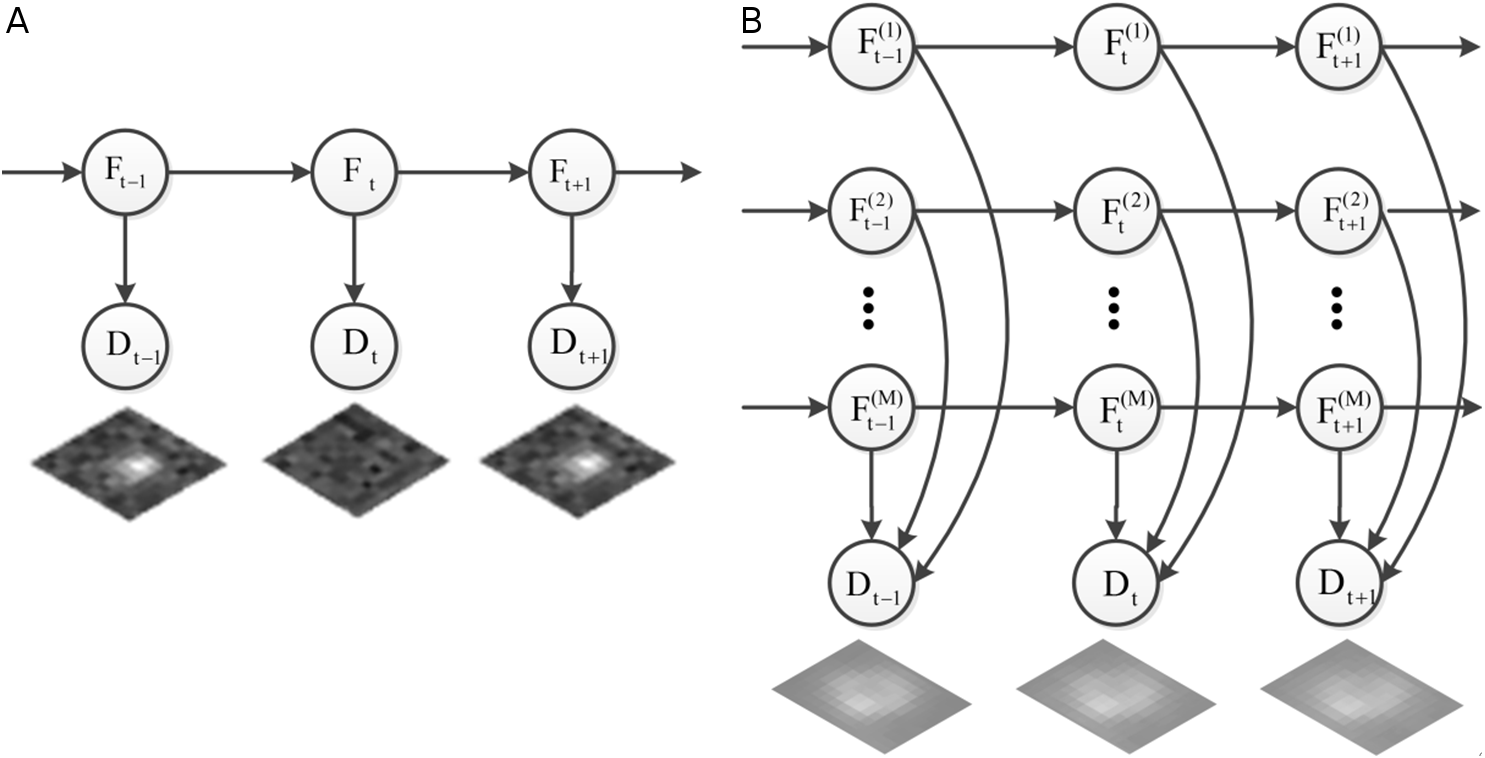}
\caption{The Bayesian inference model used for fluorophore switching. 
}
\label{fig:fhmm}
\end{figure}

For a single fluorophore, the switching procedure can be modeled by a hidden Markov model (HMM) \cite{rabiner1989tutorial}, as shown in Figure~\ref{fig:fhmm}(A). However, for high-density fluorophores, each fluorophore transfers the state independently with a stable probability \cite{cox2012bayesian} and all the fluorophores together can be modeled by a factorial hidden Markov model (FHMM) \cite{ghahramani1996factorial}, as shown in Figure~\ref{fig:fhmm}(B), which has been used and proved in \cite{xu2016live}.

\vspace{\baselineskip}
\noindent {\bf Refining Results with Physical Meaning}

Although HMM and FHMM are capable of modeling the fluorophore switching process, they are localization-guided, which often ignore the global information, and are computationally expensive to learn. Thus, we initialize the fluorophores' locations by using the image generated by deep learning and use Bayesian inference to further refine the results.

We apply the FHMM model to deal with high-density fluorescent microscopy. The parameters of FHMM are estimated by the expectation-maximization (EM) algorithm: 
\begin{footnotesize}
\begin{eqnarray}
  Q\left( {{\phi^{new}}|\phi} \right) = E\left\{ {\log P\left( {\left\{ {{F_t},{D_t}} \right\}|{\phi^{new}}} \right)|\phi,\left\{ {{D_t}} \right\}} \right\},	 \label{eq:fhmm-em}
\end{eqnarray}
\end{footnotesize}
\noindent where the observation sequence has $T$ frames, $\{D_t\}$, $t = 1,...,T$. The hidden states are $\{F_t\}$, where each fluorophore has three possible states in the model. $Q$ is a function of the fluorophore parameters $\phi^{new}$ given the current parameter estimation and the observation sequence $\{D_t\}$. The procedure iterates between a step that fixes the current parameters and computes posterior probabilities over the hidden states (the E-step), and a step that uses these probabilities to maximize the expected log likelihood of the observations as a function of the parameters (the M-step).

In the E-step, we fix the fluorophore parameters in the model and utilize the hybrid of Markov chain Monte Carlo and forward algorithm to sample the initial model. When a new fluorophore is determined, we take samples of this fluorophore using the forward filtering backward sampling algorithm \cite{godsill2004monte}. Thus, the sampled image sequence contains this fluorophore. In the M-step, we optimize the fluorophore parameters and find the maximum a posteriori (MAP) fluorophore positions using the conjugate gradient. Then, based on already known positions of fluorophores, the surrounding fluorophores with high probability are expanded. The final super-resolution image is obtained by iterating these two steps until convergence.

\section{Results}
\subsection{Experimental Setting}

\noindent {\bf Training Deep Learning}

To train our deep learning module, the stochastic simulation module was used to simulate time-series low-resolution images from 12000 gray-scale high-resolution images. These images were downloaded from two databases: (i) 4000 natural images were downloaded from ILSVRC \cite{ILSVRC15} and Laplace filtered, and (ii) 8000 sketches were downloaded from the Sketchy Database \cite{sangkloy2016sketchy}. Note that our simulation is a generic method, which does not depend on the type of the input images. Thus any gray-scale image can be interpreted as the fluorophore distribution and used to generate the corresponding time-series low-resolution images.

To initialize all the weights of the deep learning models, we used the random normal initializer with the mean as $0$ and standard deviation as $0.02$. As for the Monte Carlo dropout layer, we set the keep ratio as $0.8$. In terms of the Adam optimizer \cite{RN102}, we set the learning rate as $1*10^{-4}$ and the $beta\_1$, which is the exponential decay rate for the first moment estimates, as $0.9$. During training, we set the batch size as $8$, the initialization training epoch as $2$ and the GAN training epoch as $40$. When performing the real GAN training, we utilized the learning rate decay technique, reducing the learning rate by half every $10$ epochs.

\begin{figure*}[!hbpt]
\centering
\includegraphics[width=0.81\textwidth]{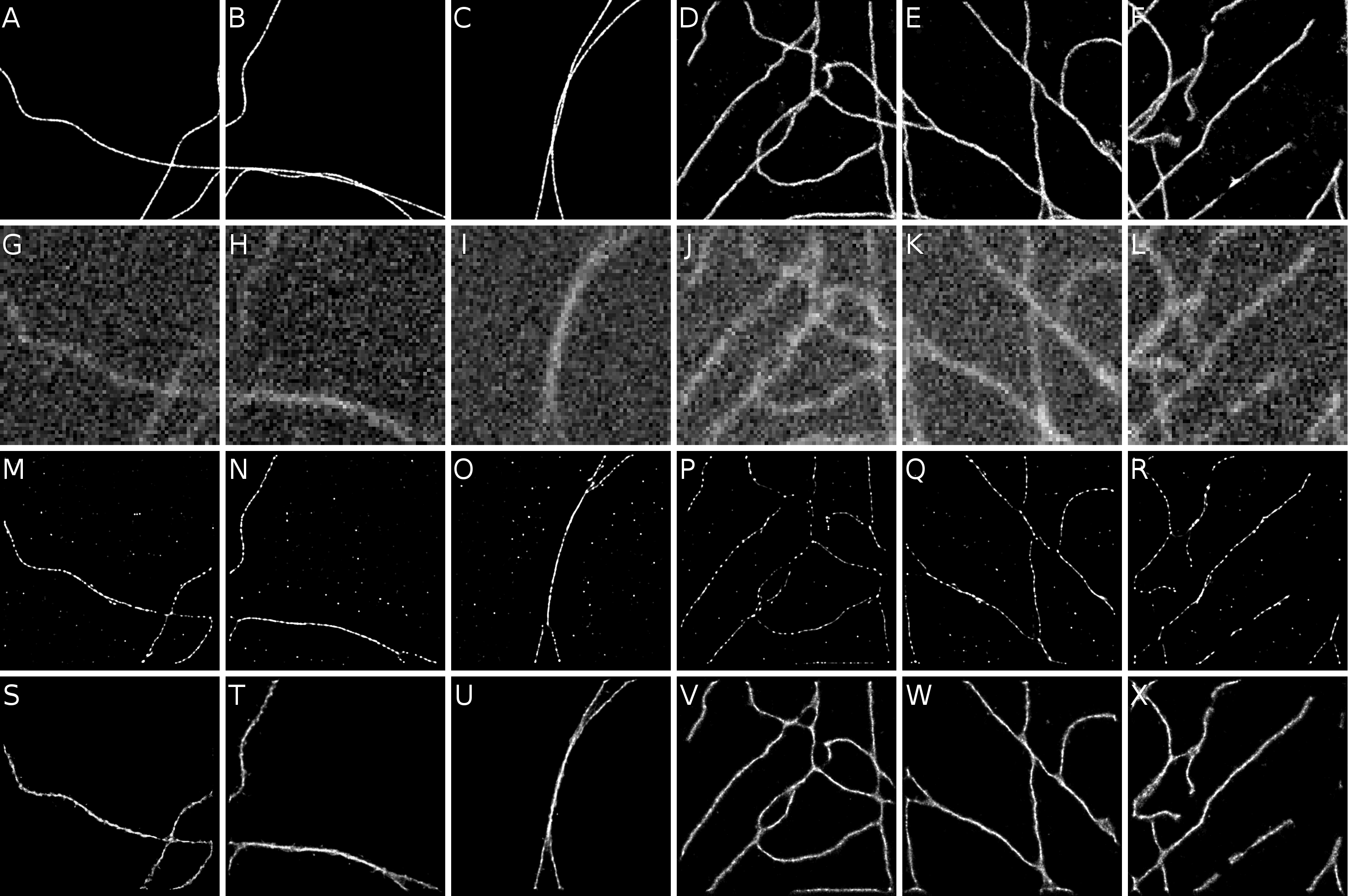}
\caption{Visualization of the ground-truth high-resolution images, representative low-resolution input images, the reconstruction results of the 3B analysis, and the results of our method on three representative areas of each simulated dataset.
}
\label{fig:simu_result}
\end{figure*}

\vspace{\baselineskip}
\noindent {\bf Evaluation Datasets}

Two simulated datasets and three real-world datasets were used to evaluate the performance of the proposed method. Simulated datasets were used due to the availability of ground-truth.

The first two datasets are simulated datasets, for which the ground-truth (i.e., high-resolution images) was downloaded from the Single-Molecule Localization Microscopy (SMLM) challenge\footnote{http://bigwww.epfl.ch/smlm/datasets/} \cite{sage2015quantitative}. The two datasets correspond to two structures: MT0.N1.HD (abbr. MT) and Tubulin ConjAL647 (abbr. Tub). For each structure, single molecule positions were downloaded and then transformed to fluorophore densities according to Section \ref{sec:2.1}. For simulation, the  photo-convertible fluorescent protein (PCFP) mEos3.2 \cite{zhang2012rational} and its associated PSF, FWHM and state transfer table were used. For the convenience of calculation, we cropped the large-field structure into four separate areas, each with $480\times480$ pixels ($1 px = 20 nm$). For each high-resolution image, 200 frames of low-resolution fluorescent images were generated, each with $60\times60$ pixels.

The third dataset is a real-world dataset, which was used in recent work \cite{xu2016live}. The actin was labeled with mEos3.2\footnote{For the convenience of cellular labeling and instrument setup, here all the experiments were carried out by mEos3.2.} in U2OS cells (abbr. Actin1) and taken with an exposure time of 50 ms per image frame. The actin network is highly dynamic and exhibits different subtype structures criss-crossing at various distances and angles, including stress fibers and bundles with different sizes and diameters. The dataset has 200 frames of high-density fluorescent images, each with $249\times395$ pixels ($1 px = 160 nm$) in the green channel. This is a good benchmark set that has been well tested which can compare our method with SIMBA \cite{xu2016live}, a recent Bayesian approach based on dual-channel imaging and photo-convertible fluorescent proteins.

Two other real-world datasets labeled with mEos3.2 were also used. One is an actin cytoskeleton network (abbr. Actin2), which was labeled and taken under a similar exposure condition with Actin1, but was completely new and had not been used before this work. The other one is an Endoplasmic reticulum structure (abbr. ER), which has a more complex structure. It is a type of organelle that forms an interconnected network of flattened, membrane-enclosed sacs or tubes known as cisternae, which exhibits different circular-structures and connections at different scales. For the ER dataset, the exposure time is 6.7 ms per frame. The resolution of each image in Actin2 is $263\times337$ pixels ($1 px = 160 nm$) and that in ER is $256\times170$ pixels ($1 px = 100 nm$). Both datasets have 200 frames of high-density fluorescent images and the same photographing parameters as Actin1. These datasets were used to demonstrate the power of our method in diverse ultrastructures.

Since the 3B analysis \cite{cox2012bayesian} is one of the most widely used high-density fluorescent super-resolution techniques, which can deal with high temporal and spatial resolutions \cite{lidke2012super,cox2012bayesian}, it was chosen to compare with our method.

\subsection{Performance on Simulated Datasets}

\noindent {\bf Visual Performance}

Figure~\ref{fig:simu_result} shows the visualization of the ground-truth high-resolution images, representative low-resolution input images, the reconstruction results of the 3B analysis, and the results of our method on the simulated datasets.

As shown in Figure~\ref{fig:simu_result}, the ground-truth images have very clear structures while the low-resolution image frames are very blurry and noisy ($8\times$ downsampled). To reconstruct the ultrastructures, we ran the 3B analysis with 240 iterations and ran our Bayesian inference module after the deep learning module with 60 iterations. In each iteration, the Bayesian inference module of our method searches four neighbor points for each fluorophore, whereas the 3B analysis takes isolated estimation strategy. Thus the difference in iteration numbers is comparable. Due to the high computational expense of the 3B analysis, each $60\times60$ image was subdivided into nine overlapped subareas for multi-core process, whereas for our method, the entire image was processed by a single CPU core.

Clearly, the reconstructions of our method are very similar to the ground-truth in terms of smoothness, continuity, and thickness. On the other hand, the reconstructions of the 3B analysis consist of a number of interrupted short lines and points with thin structures. In general, two conclusions can be drawn from the visual inspection.

First, DLBI discovered much more natural structures than the 3B analysis. For example, in the bottom part of Figure~\ref{fig:simu_result}(B), there are two lines overlapping with each other and a bifurcation at the tail. Due to the very low resolution in the input time-series images (e.g., Figure~\ref{fig:simu_result}(H)), neither DLBI nor the 3B analysis was able to recover the overlapping structure. However, DLBI reconstructed the proper thickness of that structure (Figure~\ref{fig:simu_result}(T)), whereas the 3B analysis only recovered a very thin line structure (Figure~\ref{fig:simu_result}(N)). Moreover, the bifurcation structure was reconstructed naturally by DLBI. Similar conclusions can be drawn on the more complex structures in the Tub dataset (columns 4-6 in Figure~\ref{fig:simu_result}).

Second, DLBI discovered much more latent structures than the 3B analysis. The Tub dataset consists of a lot of lines (tubulins) with diverse curvature degrees (Figure~\ref{fig:simu_result}(D),(E),(F)). The reconstructions of the 3B analysis successfully revealed most of the tubulin structures but left the crossing parts interrupted (Figure~\ref{fig:simu_result}(P),(Q),(R)). As a comparison, the reconstruction results of DLBI recovered both the line-like tubulin structures and most of the crossing parts accurately (Figure~\ref{fig:simu_result}(V),(W),(X)).

\vspace{\baselineskip}
\noindent {\bf Quantitative Performance}

For single-molecule super-resolution fluorescence microscopy, the quantitative performance has been measured by assessing the localization accuracy of single-emitters in each frame \cite{ram2006beyond,small2009theoretical,huang2011simultaneous}. For high-density super-resolution fluorescence microscopy, the entire time-series are analyzed and the production is the probability map of the locations of fluorophores.

\begin{table}[hbpt]
\centering
\caption{Performance comparison between the 3B analysis and DLBI on the simulated datasets.
}
\label{table:psnr}
\begin{tabular}{c|c|c c c c |c c c c }
\hline
\multicolumn{2}{c|}{\multirow{2}{*}{Datasets}} &\multicolumn{4}{c|}{MT0.N1.HD}&\multicolumn{4}{c}{Tubulin ConjAL647}       \\ \cline{3-10}  %
\multicolumn{2}{c|}{\multirow{2}{*}{}}  &   01   &   02   &   03   & \multicolumn{1}{c|}{04}  &   01   &   02   &   03   &   04    \\ \hline
PSNR&\small 3B & 17.99 & 17.62 & 17.84 & 17.89 & 13.42 & 15.49 & 15.00 & 13.21   \\
(dB)&\small DLBI & \textbf{18.59} & \textbf{19.16} & \textbf{18.51} & \textbf{20.42} & \textbf{18.72} & \textbf{19.17} & \textbf{18.72} & \textbf{16.63}  \\ \hline

SSIM&\small 3B &  0.89 &  0.89 &  0.90 & 0.90 &  0.74 &  0.81 &  0.75  & 0.69 \\
&\small DLBI &  \textbf{0.92} &  \textbf{0.92} &  \textbf{0.93} &   \textbf{0.94} &  \textbf{0.82} &  \textbf{0.85} &  \textbf{0.80} &  \textbf{0.76}    \\ \hline
\end{tabular}
\end{table}

Since the ground-truth is known for the simulated datasets, here we use peak signal-to-noise ratio (PSNR) and structural similarity (SSIM) to measure the reconstruction performance, both of which are widely-used criteria for image reconstruction in computer vision. The performance of the 3B analysis and DLBI on the two simulated datasets are given in Table \ref{table:psnr}. Here, we denote the four areas of each dataset as ``01'', ``02'', ``03'' and ``04'', respectively. It can be seen that DLBI clearly outperforms the 3B analysis in terms of both PSNR and SSIM on all the areas of the two datasets.

\subsection{Performance on Real Datasets}
\begin{figure}[!hbpt]
\centering
\includegraphics[width=0.8\textwidth]{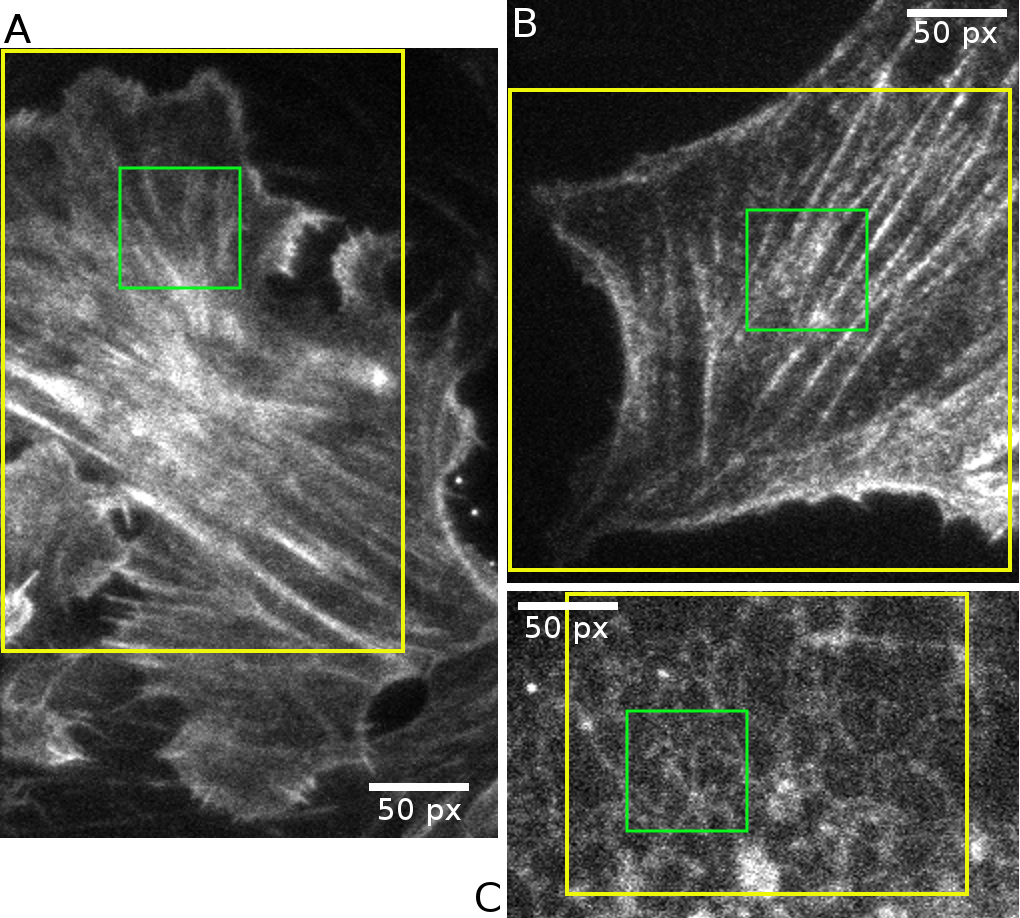}
\caption{The first frame of the time-series fluorescent images for each of the three real-world datasets: (A) Actin1, (B) Actin2, and (C) ER. 
}
\label{fig:realdata}
\end{figure}

Figure~\ref{fig:realdata} shows the first frame of the time-series fluorescent images for each of the three real-world datasets. Here we evaluate the performance of our method for both local-patch reconstruction (areas selected by green rectangles) and large-field reconstruction (areas selected by yellow rectangles).

\vspace{\baselineskip}
\noindent {\bf Local-patch Reconstruction}

\begin{figure}[!t]
\centering
\includegraphics[width=0.7\textwidth]{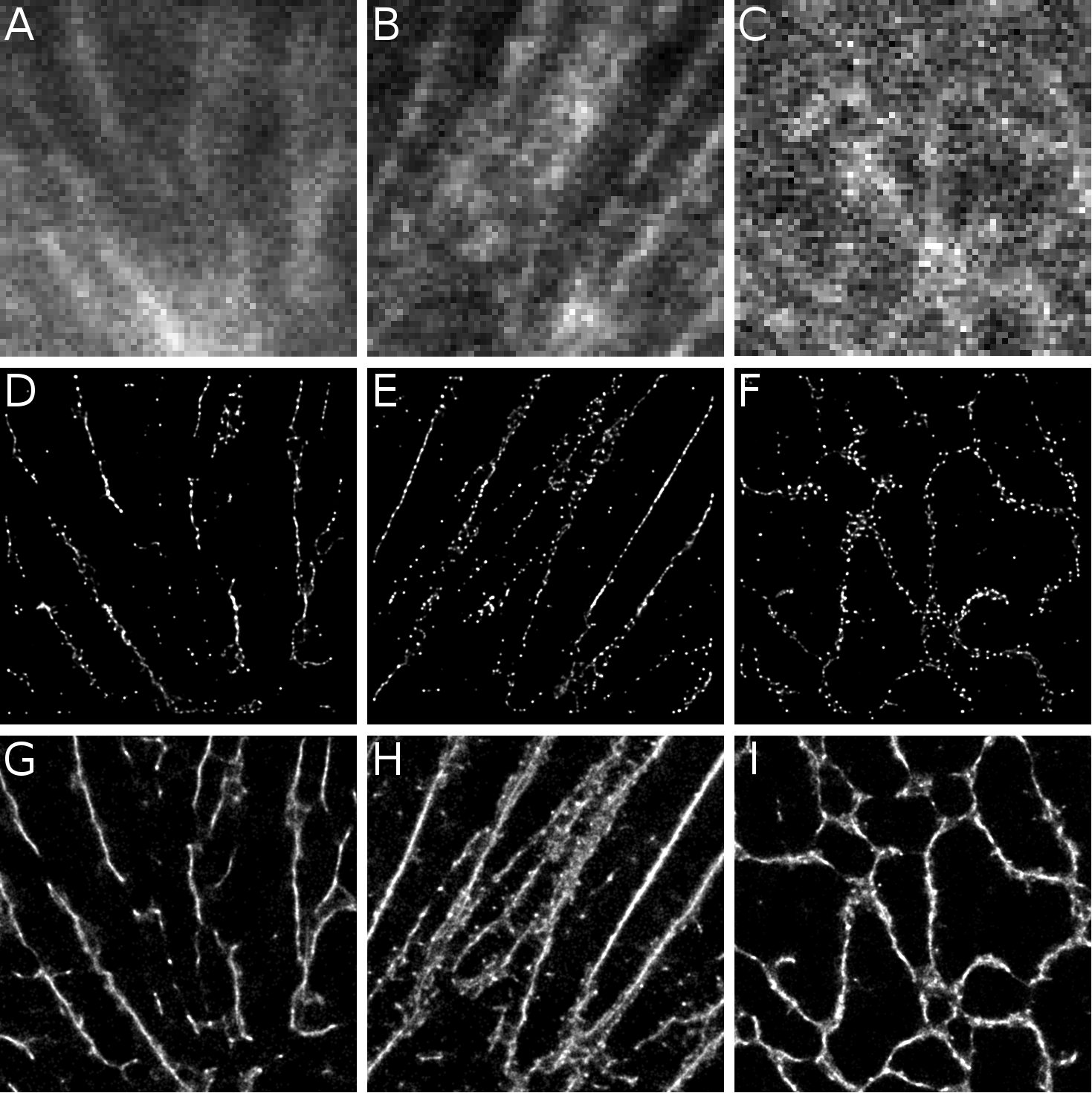}
\caption{Reconstructions of the local patches of the three real datasets. 
}
\label{fig:real_patch}
\end{figure}

Figure~\ref{fig:real_patch} shows the first frames of the low-resolution images of the three local-patches, and the reconstruction results of the 3B analysis and DLBI. The regions of interests of the selected patches are $(60,60,60,60)$, $(120,120,60,60)$ and $(60,60,60,60)$ for the three datasets, respectively\footnote{Region of interest is usually denoted as $(X,Y,W,H)$, where $(X,Y)$ are the coordinates of the top left point of the rectangle, $W$ is the width of the rectangle, and $H$ is the height of the rectangle.}. The temporal resolutions for the two actin datasets and the ER dataset were 10s and 1.34s respectively, according to the exposure time of the image frames.

It can be seen that the reconstruction results of the 3B analysis capture the main structures in the fluorescent images, but mainly consist of isolated high-illuminating spots, with details being interrupted (Figure~\ref{fig:real_patch}(D),(E),(F)). In contrast, the results of DLBI recover most of the latent ultrastructures, and the reconstructed structures have well-estimated fluorophore distribution and continuous depiction (Figure~\ref{fig:real_patch}(G),(H),(I)).

\begin{table}[hbpt]
\centering
\caption{Performance comparison between the 3B analysis and DLBI on the real datasets in terms of RSP and RSE with SQUIRREL. 
}
\label{table:squirrel}
\begin{tabular}{c|c c |c c |c c }
\hline
Dataset &\multicolumn{2}{c|}{Actin1-patch}&\multicolumn{2}{c|}{Actin2-patch}&\multicolumn{2}{c}{ER-patch}  \\ \hline
Criteria &   RSP   &   RSE &   RSP   &  RSE &   RSP   &   RSE        \\ \hline
 3B & 0.583 & 3915.096 & 0.770  & 2196.068 & 0.827 & 4077.037    \\
 DLBI & \textbf{0.721} & \textbf{3326.007} & \textbf{0.878} & \textbf{1648.919} & \textbf{0.916} & \textbf{2904.707}   \\ \hline
\end{tabular}
\end{table}

We further assessed the reconstruction quality of the 3B analysis and DLBI by SQUIRREL (super-resolution quantitative image rating and reporting of error locations) \cite{culley2018quantitative}. SQUIRREL compares the diffraction-limited image (the reference image) and the reconstructed equivalents to generate a quantitative map, in which two scores are calculated: the resolution-scaled Pearson coefficient (RSP) and the resolution-scaled error (RSE). The higher RSP and lower RSE values, the higher the image quality is. Table \ref{table:squirrel} shows the RSP and RSE scores for the 3B analysis and DLBI. It is clear that DLBI significantly outperforms the 3B analysis.

\vspace{\baselineskip}
\noindent {\bf From Deep Learning to Bayesian Inference}

Our method combines the strength of deep learning and statistical inference, where deep learning captures both local features in the images and the time-course correlation, and statistical inference removes artifacts from deep learning and enhances physical meaning to the final results. Conceptually, this is equivalent to using the power of deep learning to automatically and systematically explore and extract spatial and temporal features, and taking advantages of the explicit and rigorous mathematical foundation of probabilistic graphical models. Here we investigate the effectiveness of this combination.

\begin{figure}[!hbpt]
\centering
\includegraphics[width=0.8\textwidth]{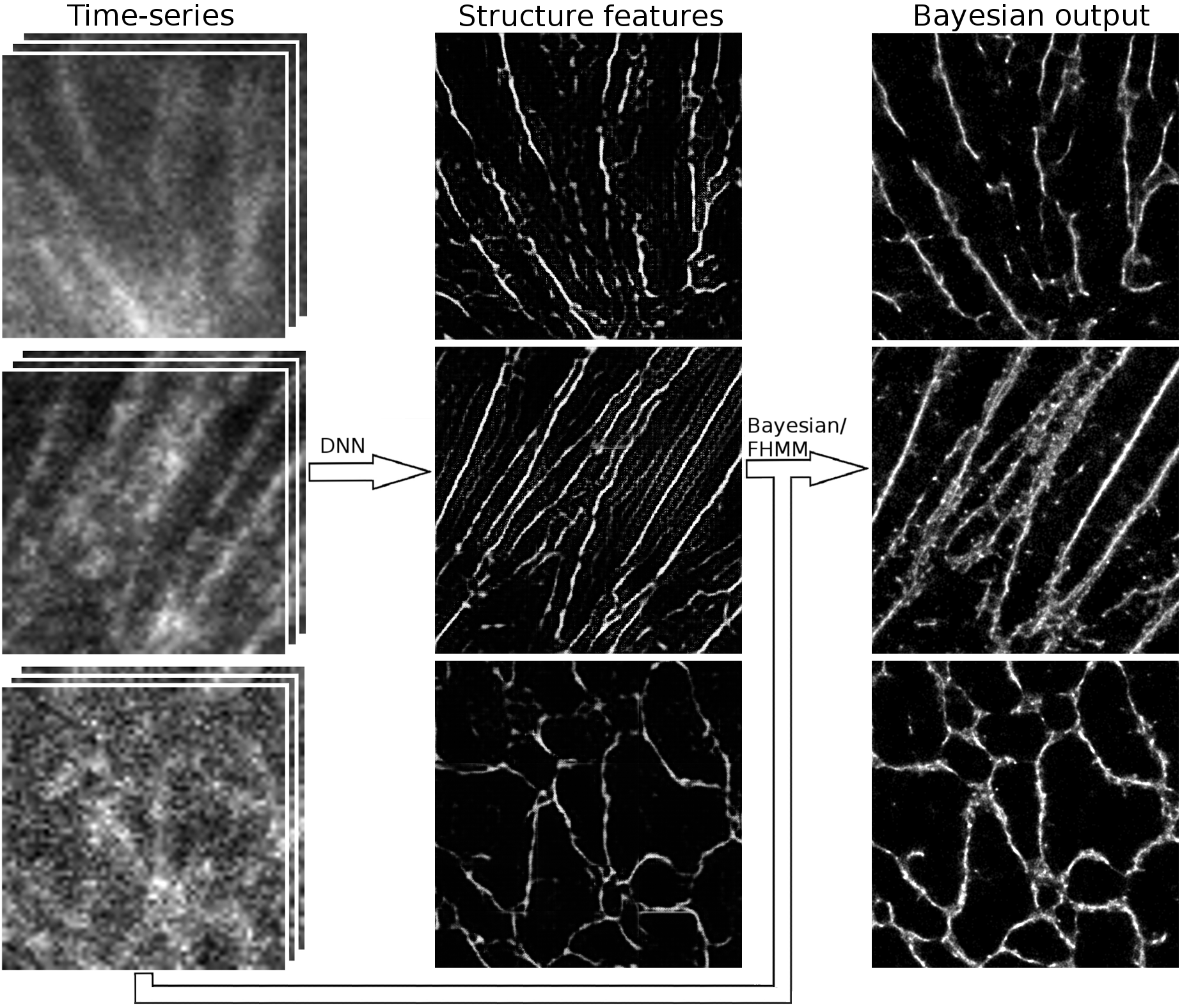}
\caption{Reconstructions of the three local patches of the three real datasets (the first column) by the deep learning module (the second column) and by deep learning guided Bayesian inference (the third column). }
\label{fig:dlbi}
\end{figure}

Figure~\ref{fig:dlbi} demonstrates the outputs of the deep learning module and the Bayesian inference module. It can be seen that the super-resolution images outputted from the deep learning module are very close to the final images from the Bayesian inference module, except for some artifacts and false structures. This is due to two reasons: (i) the abundance of training data provided by our simulation module, which are simulated under the real experimentally-calibrated parameters, enable deep learning to effectively learn spatial and temporal features; and (ii) the high diversity of biological structures is still a challenge, which causes the artifacts and false structures to be learned by deep learning.

After the deep learning module generates the super-resolution image, the Bayesian inference module uses both the original time-series low-resolution images and the deep learning image to statistically infer a ``false/true'' determination on each fluorophore location and produce the final image. In particular, the false structures are not directly rejected but used as seeds to search for true structures. Therefore, as shown in
Figure~\ref{fig:dlbi}, although the deep learning module outputted some unnatural structures for the Actin2 and ER datasets, these structures were further corrected by the Bayesian inference module.

\vspace{\baselineskip}
\noindent {\bf Runtime Analysis}

\begin{figure}[!hbpt]
\centering
\includegraphics[width=0.68\textwidth, height=0.44\textwidth]{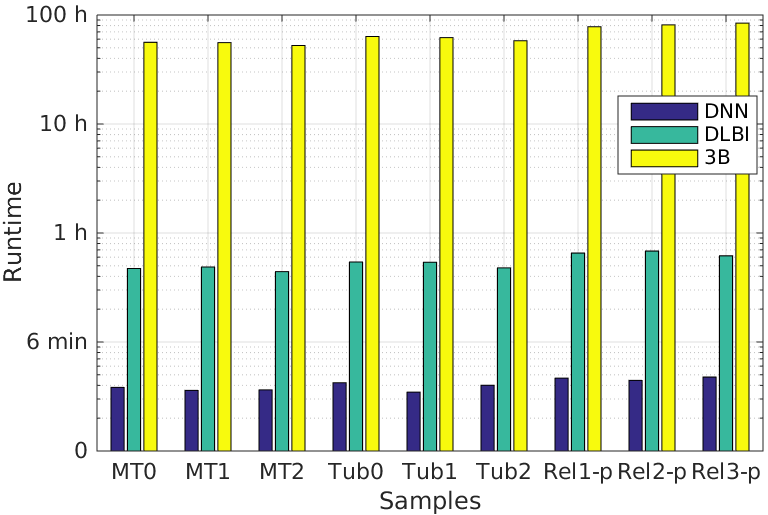}
\caption{Runtime comparison of the deep learning module (DNN), the entire DLBI pipeline (DLBI), and the 3B analysis (3B) 
}
\label{fig:runtime}
\end{figure}

After being trained, running the deep learning model is very computationally inexpensive. Furthermore, the results of deep learning provide a close-to-optimal initialization for Bayesian inference, which also significantly reduces trial-and-error and leads to faster convergence. Figure~\ref{fig:runtime} shows the runtime comparison of the deep learning module, the entire DLBI pipeline, and the 3B analysis on the nine reconstruction tasks (i.e., the six areas of the simulated datasets shown in Figure~\ref{fig:simu_result} and the three local patches of the real datasets shown in Figure~\ref{fig:real_patch}). It can be seen that the runtime for the deep learning module ranges between 1 to 3 minutes and that of DLBI ranges between 30 to 40 minutes. In contrast, the runtime for the 3B analysis is around 75 hours, which is more than 110 times higher than that for DLBI. Our results have demonstrated that the super-resolution images from the deep learning module alone is a good estimation to the ground-truth. Therefore, for users who value time and can compromise accuracy, the results from the deep learning module provide a good tradeoff, and thus a good estimation of the ground-truth.

\vspace{\baselineskip}
\noindent {\bf Large-field Reconstruction}

\begin{figure*}[!hbpt]
\centering
\includegraphics[width=0.75\textwidth]{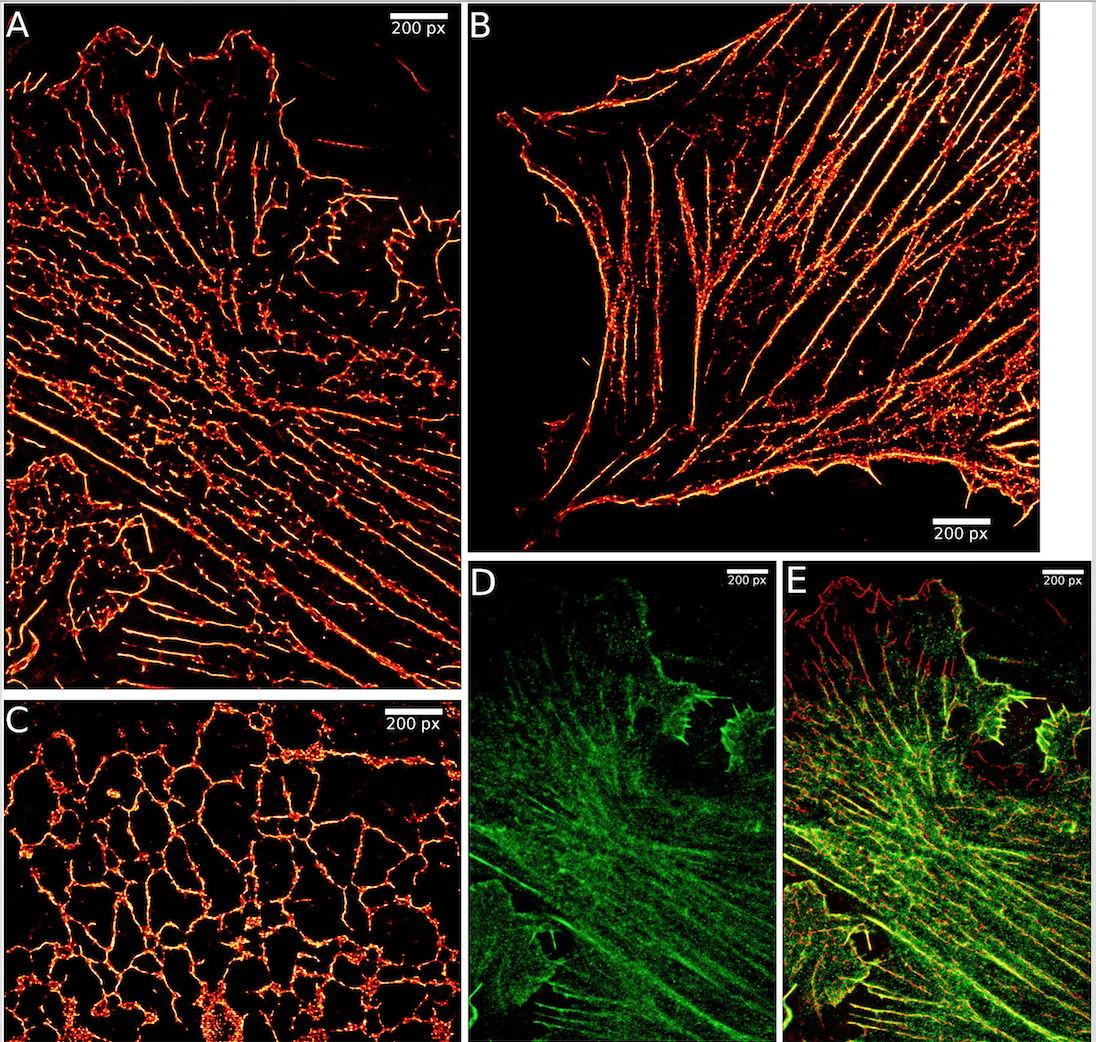}
\caption{Performance on large-field reconstructions of real datasets.
}
\label{fig:wholescale}
\end{figure*}

To analyze a dataset with 200 frames, each with about $200\times300$ pixels, it takes our method about $7\sim10$ hours on a single CPU core. Therefore, our method is able to achieve large-field reconstruction on the yellow areas shown in Figure~\ref{fig:realdata}. Figure~\ref{fig:wholescale} shows the large-field reconstruction images of the three real datasets. For the Actin1 dataset, the selected area is $200\times300$ pixels and the reconstructed super-resolution image is $1600\times2400$ pixels. For the Actin2 dataset, the selected area is $250\times240$ pixels and the reconstructed image is $2000\times1920$ pixels. And for the ER dataset, the selected area is $200\times150$ pixels and the reconstructed image is $1600\times1200$ pixels.

As shown in Figure~\ref{fig:wholescale}(A) and (B), the actin networks in the two datasets have been successfully recovered by DLBI. The thinning and thickening trends of the cytoskeleton have been clearly depicted, as well as the small latent structures, including actin filaments, actin bundles and ruffles. For the endoplasmic reticulum structure (Figure~\ref{fig:wholescale}(C)), the circular-structures and connections of the cytoskeleton have also been accurately reconstructed.

For the Actin1 dataset, the single-molecule reconstruction of the red channel is available (Figure~\ref{fig:wholescale}(D)). This reconstruction was produced by PALM \cite{hess2006ultra} using 20,000 frames, whereas the reconstruction image of DLBI (Figure~\ref{fig:wholescale}(A)) used only 200 frames. We further overlayed the image produced by DLBI with that of PALM to check how well they overlap (Figure~\ref{fig:wholescale}(E)). It is clear that the main structures of the two images almost perfectly agree with each other. In addition, our method was able to recover the latent structure on the top-left part which was not photographed by PALM due to out of range of views in dual-channel photographing. If we carefully check the original low-resolution fluorescent images, we could find that this predicted structure indeed exists, which is consistent with our reconstruction.

\section{Discussion}
In this chapter, we discussed a deep learning guided Bayesian inference method for the structure reconstruction of super-resolution fluorescence microscopy. 
When resolving this standard structured prediction problem, whose targets are 2D images, we encountered the following challenges. Firstly, the traditional Bayesian algorithm for solving this problem is very slow. If we want to reconstruct a mid-size patch with the 3B method, whose dimension is 480 by 480, it can take us 75 hours, which means this method is not practical for real-life usage. Secondly, the Bayesian methods can get stuck in local minimum easily as they do not consider the global structure information. Thirdly, it is not suitable to apply the existing deep learning models to this problem directly as there are no sufficient training data. Even worse, we do not have the super-resolved structure images, which are the training targets for the deep learning model.
Furthermore, deep learning models are usually criticized for being black-box models, while for this problem, we know clearly about the physical process behind it. If we built a black-box method, the biologists would not believe in it and thus would not use it. Finally, the inputted data are highly noisy, which are different from the data in the computer vision field. 
To handle this challenging structured prediction problem, we developed a framework, which combines the strength of stochastic simulation, deep learning, and Bayesian inference. The stochastic simulation module, which was built upon the physical process of this problem, can generate a large amount of training data for the deep learning module. The trained deep learning model can extract global structure features from the input low-resolution images, and its outputs can serve as the prior for the downstream Bayesian inference module. With the deep learning output as a good prior, the Bayesian inference module can converge much faster than the simple Bayesian methods. 
As we have discussed in Chapter \ref{chapter2_sec:ideas}, we used stochastic simulation based on the physical process behind this task to handle the data issue in this bioinformatics problem. We developed a new deep learning architecture to tackle the noise issue. We further combined deep learning with the traditional Bayesian algorithm to endue physical meaning to the deep learning outputs and refine the results. With those ideas, we can resolve this problem efficiently and accurately.



Essential, here, we show a typical example of using deep learning to tackle structured prediction in the structure determination field.
In the next chapter, we will show another example of predicting RNA secondary structures, whose targets are 2D contact maps.


\chapter{E2Efold: RNA Secondary Structure Prediction by Learning Unrolled Algorithms}
\label{chapter_e2efold}

\section{Chapter Introduction}



As hub molecules in the central dogma, RNAs play numerous roles in our bodies. Their functions include transportation, information-carrier, catalyzing, the scaffolding of other molecules. RNA structures largely determine RNA functions. For example, the unique structure of transportation RNAs (tRNAs) enables them to transport amino acids and elongate peptides during protein synthesis. To understand the RNA function, we should first investigate the RNA structure. As we know, understanding RNA structure is a very complicated topic. RNA can first form into a linear chain of the building block elements. The linear chain can fold into an RNA secondary structure, which can further fold into a 3D structure to perform functionality. In this chapter, we will focus on investigating the RNA secondary structure based on the RNA preliminary sequence information.

\begin{figure}[!t]
    \centering
    \includegraphics[width=1.0\textwidth]{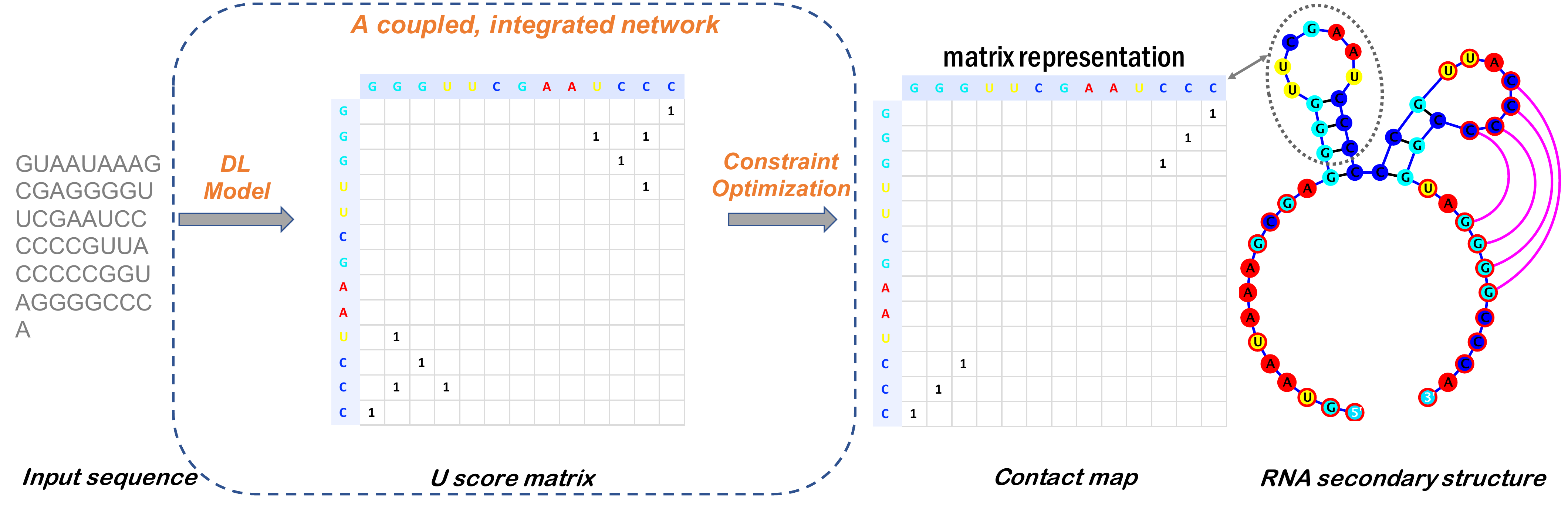}
    \caption{The overall idea of E2Efold. We solve the RNA secondary structure prediction problem as a translation with constraints problem.}
    \label{fig:e2efold_overall}
\end{figure}

The RNA secondary structure refers to the pairing information for each base within an RNA sequence, as illustrated at the most right of Figure~\ref{fig:e2efold_overall}. That is, for each base, we want to know whether it can bond with another base. And if it can bond, with which base it can bond. Because determining RNA secondary structure by experimental assays is laborious and expensive, computational approaches have emerged to complement the experimental results. Actually, Predicting the RNA secondary structure from the RNA sequence is one of the oldest problems in bioinformatics, which has been studied for more than 40 years. Almost all the previous methods, including those widely used methods, such as RNAstructure \cite{bellaousov2013rnastructure}, Vienna RNAfold \cite{lorenz2011viennarna} and UNAFold~\cite{markham2008unafold}, are based on dynamic programming (DP). Essentially, people assign a particular energy value to each pair and then use DP to identify the specific secondary structure pattern with the lowest summarized energy of all the pairs. However, the DP-based methods may have the following limitations. Firstly, DP can be slow, as its time complexity is $\gO(L^3)$. Secondly, the DP-based methods have reached the prediction performance bottleneck, with the F1 score being around 0.6. Furthermore, the canonical DP-based methods are unable to handle pseudoknot prediction, where there is a jump in the paring, as shown in Figure~\ref{fig:e2efold_overall}. In fact, pseudoknots make up roughly 1.4\% of base-pairs~\cite{mathews2006prediction}, and are overrepresented in functionally important regions~\cite{hajdin2013accurate,staple2005pseudoknots}. Furthermore, pseudoknots are present in around  40\% of the RNAs. They also assist folding into 3D structures~\cite{fechter2001novel} and thus should not be ignored.
To handle pseudoknot, some heuristic algorithms, such as HotKnots~\cite{andronescu2010improved} and Probknots~\cite{bellaousov2010probknot}, have been proposed, but their predictive accuracy and efficiency still need to be improved. Some researchers also tried to corporate DP with a learning algorithm. For example, ContraFold \cite{do2006contrafold} and ContextFold \cite{zakov2011rich} have been proposed for energy parameter estimation due to the increasing availability of known RNA structures, resulting in higher prediction accuracies. However, these methods still rely on the above DP-based algorithms for energy minimization. A recent deep learning method, CDPfold~\cite{zhang2019new}, applies convolutional neural networks to predict base-pairings, but it adopts the dot-bracket representation for RNA secondary structure, which can not represent pseudoknotted structures. Moreover, it requires a DP-based post-processing step whose computational complexity is prohibitive for sequences longer than a few hundreds.

Because the legendary idea for solving the problem has so many limitations, we solve the task from an entirely new angle, starting from problem formulation. Instead of considering it as an optimization problem, we creatively formulate the task as a translation with constraints problem. Given the RNA sequence, we translate it into a binary contact map matrix, which has a one-to-one mapping against a specific secondary structure (Figure~\ref{fig:e2efold_overall}). When resolving the translation problem, we also need to consider the constraints for the original task: 1) only specific pairs are allowed; 2) no sharp loops are allowed; 3) it is a matching problem. We should enforce those constraints into the outputted matrix to ensure the corresponding secondary structure satisfying the physical properties of the problem. Notice that such a formulation can cover the pseudoknot prediction naturally.

This classic structured prediction problem can have the following challenges. First of all, the available annotated data for this problem are very limited. The most commonly used benchmark dataset, ArchiveII \cite{sloma2016exact}, only contains around 4K data points. Even for the most comprehensive dataset, RNAStralign \cite{tan2017turbofold}, after we removing redundancy, there are just around 30K data points. Even worse, the search space and output dimension of this problem are very daunting. As shown in Figure~\ref{fig:e2efold_overall}, we need to investigate the pairwise potential between all the elements within the sequence, which means the output dimension of this problem can be $\gO(L^2)$, where $L$ is the sequence length. Regarding the RNA sequences included in this project, the longest one can go up to 1851 bases. Finally, we should incorporate the problem-specific constraints into the solution. Without considering those constraints and the problem structure, the outputs from the proposed method may not be physically possible and valid. 

\begin{figure}[!t]
    \centering
    \includegraphics[width=0.5\textwidth]{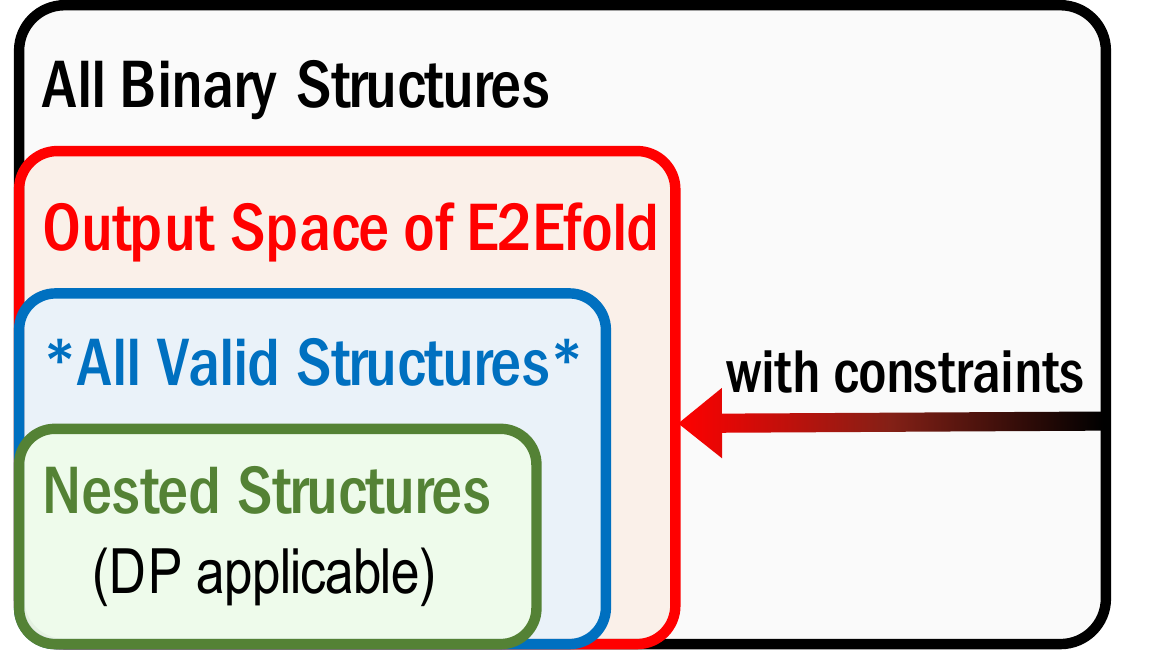}
      \caption{Output space of E2Efold.}
      \label{fig:space}
\end{figure}

To resolve such a challenging task of predicting RNA secondary structures from RNA sequences, in this chapter, we propose a new deep learning method, E2Efold. We present the overall idea of E2Efold in Figure~\ref{fig:e2efold_overall}. Firstly, we use a deep learning model to resolve a translation without constraints problem, whose outputs may not satisfy all the constraints. Then, we utilize constrained optimization to enforce the constraints and adjust the result from the first step, obtaining the final solution. However, the above two-step solution is not our ultimate solution. Eventually, we combine those two steps into one integrated deep learning model, proposing a new deep learning architecture, which has the constraints embedded in the network. Essentially, we unroll the gradient descent algorithm, which is the solution for the constrained optimization, to a specific time step, and then incorporate the unrolled algorithm into the integrated network, as shown in Figure~\ref{fig:e2efold_archi}. Given the RNA sequence, the new model can output a contact map in one step, which satisfies the constraints directly, without intermediate output. 
As we discussed extensively in this thesis, the above ideas can handle the challenges of this structured problem efficiently. 
By combining deep learning with constrained optimization, we can enforce the problem-specific constraints into the final results. Secondly, with the novel deep learning architecture, which couples the translation subnetwork with the unrolled gradient descent algorithm, we can achieve deep integration between the deep learning model and the classic algorithm. Such integration can embed the constraints into the deep learning architecture naturally.
As we incorporate the constraints into the deep learning model, the output space of the model can be reduced significantly, as shown in Figure~\ref{fig:space}. Moreover, with the embedded constraints, the deep learning model would bias towards the desired distribution that we want it to learn. Such introduced bias can reduce the requirement of the training data size. Below, we present the idea and performance of E2Efold in detail.

\section{Method}
\label{chapter5_sec:method}
In this section, we present the proposed method, E2Efold. We first formulate the problem in Section~\ref{sec:problem}. Then, we introduce the novel deep learning architecture in Section~\ref{chapter5_sec:e2efold}. Finally, we show how to train the model in an end-to-end fashion in Section~\ref{sec:training_e2efold}.

\subsection{RNA Secondary Structure Prediction Problem}
\label{sec:problem}
Regarding the translation with constraints problem formulation for solving the RNA secondary structure prediction task, the input is the ordered sequence of bases $\vx=(x_1,\ldots,x_L)$ and the output is the RNA secondary structure represented by a matrix $A^*\in \{0,1\}^{L\times L}$.
Hard constraints on the forming of an RNA secondary structure dictate that certain kinds of pairings cannot occur at all~\cite{steeg1993neural}. Formally, these constraints are:
\begin{center}
{\renewcommand{\arraystretch}{1.4}
\begin{tabular}{p{8pt}p{258pt}|p{90pt}}
    \hline
    (i) & Only three types of nucleotides combinations, $\gB:=\{AU,UA\}\cup\{GC,CG\}\cup\{GU,UG\}$, can form base-pairs.
    & $\forall i,j$, if $x_ix_j\notin \gB$, then $A_{ij}=0$. \\
    \hline
    (ii) &No sharp loops are allowed. & 
    $\forall|i-j|<4$, $A_{ij}=0$.
    \\
    \hline
    (iii)&  There is no overlap of pairs, i.e., it is a matching. & $\forall i,\sum_{j=1}^L A_{ij}\leq 1$.\\
    \hline
\end{tabular}}
\end{center}

(i) and (ii) prevent pairing of certain base-pairs based on their types and relative locations.
Incorporating these two constraints can help the model exclude lots of illegal pairs. 
(iii) 
is a global constraint among the entries of $A^*$. 

The space of all valid secondary structures contains all {\it symmetric} matrices $A\in\{0,1\}^{L\times L}$ that satisfy the above three constraints.
This space is much smaller than the space of all binary matrices $\{0,1\}^{L\times L}$. Therefore, if we could incorporate these constraints in our deep model, the reduced output space could help us train a better predictive model with less training data. We do this by using an unrolled algorithm as the inductive bias to design deep architecture.


\begin{figure}[!t]
    \centering
    \includegraphics[width=0.9\textwidth]{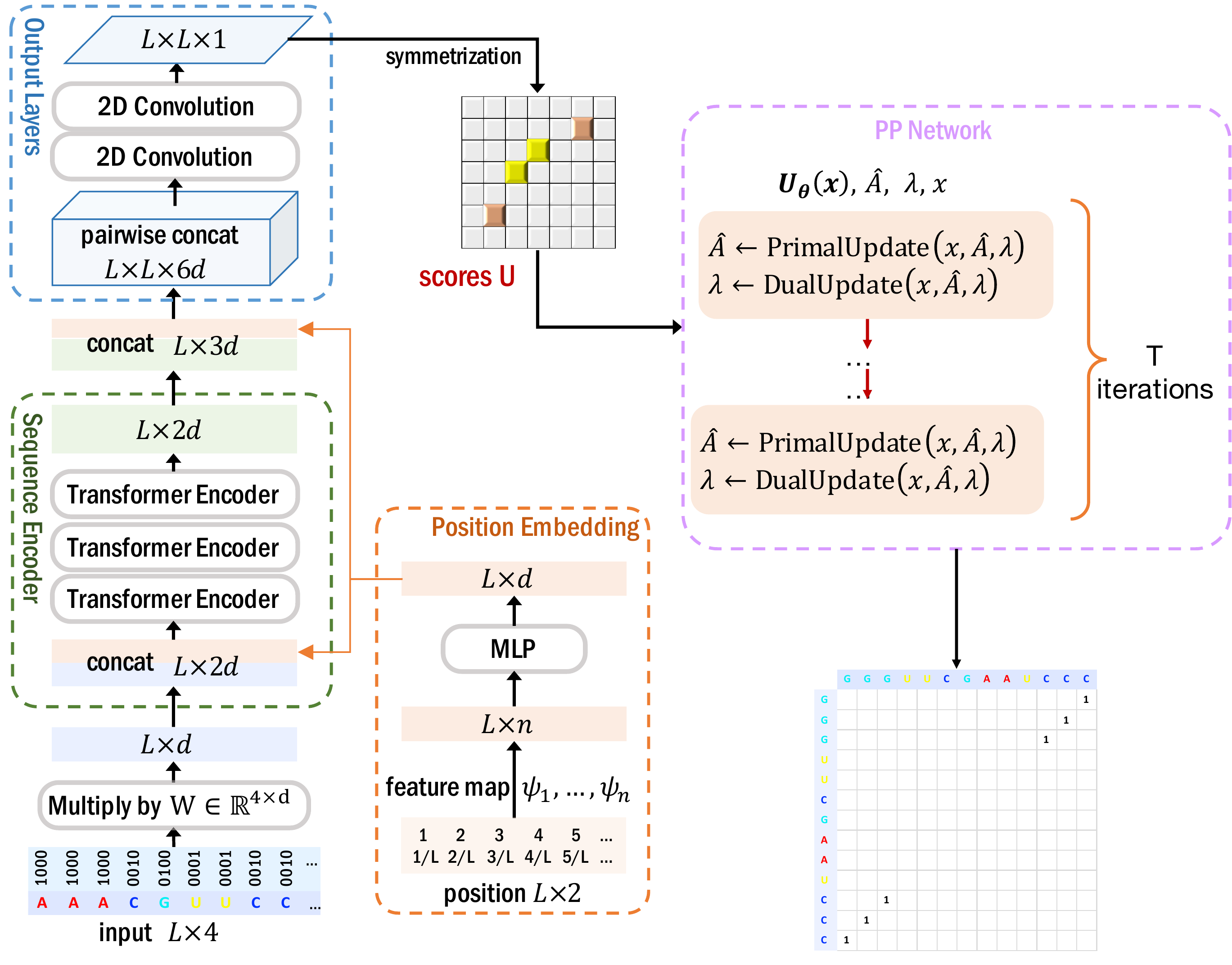}
    \caption{The E2Efold architecture. It contains two parts, the translation Deep Score Network and the Post-Processing Network. The latter subnetwork is based on the unrolled gradient descent algorithm, which can enforce the constraints.}
    \label{fig:e2efold_archi}
\end{figure}

\subsection{Deep Learning Model Based on Unrolled Algorithm}
\label{chapter5_sec:e2efold}

In the literature on feed-forward networks for structured prediction, most models are designed using traditional deep learning architectures. However, for RNA secondary structure prediction, directly using these architectures does not work well due to the limited amount of RNA data points and the hard constraints on forming an RNA secondary structure. These challenges motivate the design of our E2Efold deep model, which combines a {\it Deep Score Network} with a {\it Post-Processing Network} based on the unrolled gradient descent algorithm for solving a constrained optimization problem. We present the overview of the novel integrated network in Figure~\ref{fig:e2efold_archi}.

\subsubsection{Deep Score Network}

The first part of E2Efold is  a {\it Deep Score Network} $U_{\theta}(\vx)$ whose output is an $L\times L$ symmetric matrix. 
Each entry of this matrix, i.e., $U_{\theta}(\vx)_{ij}$, indicates the score of nucleotides $x_i$ and $x_j$ being paired. The $\vx$ input to the network here is the $L\times 4$ dimensional one-hot embedding.
The specific architecture of $U_{\theta}$ is shown in
Figure~\ref{fig:e2efold_archi} left part, before PP Network. It mainly consists of 

\begin{itemize}[leftmargin=*,nolistsep,nosep]
    \item a position embedding matrix $\bm{P}$ which distinguishes $\{x_i\}_{i=1}^L$ by their exact and relative positions:
    \begin{align}
        \bm{P}_i  =\textsc{MLP}\big(\psi_1(i), \ldots,\psi_{\ell}(i),\psi_{\ell+1}(i/L),\ldots,\psi_{n}(i/L)\big),
    \end{align}
    where $\{\psi_j\}$ is a set of $n$ feature maps such as $\sin(\cdot), \text{poly}(\cdot), \text{sigmoid}(\cdot)$, etc., and $\textsc{MLP}(\cdot)$ denotes multi-layer perceptions. Such position embedding idea has been used in natural language modeling such as BERT~\cite{devlin2018bert}, but we adapted for RNA sequence representation; 
    \item a stack of Transformer Encoders~\cite{vaswani2017attention} which encode the sequence information and the global dependency between nucleotides; 
    \item 2D Convolution layers~\cite{wang2017accurate} for outputting the pairwise scores.
\end{itemize}


With the representation power of neural networks, the hope is that we can learn an informative $U_{\theta}$ such that higher scoring entries in $U_\theta(\vx)$ correspond well to actual paired bases in RNA structure. 
Once the score matrix $U_\theta(\vx)$ is computed, a naive approach to use it is to choose an offset term $s\in \R$ (e.g., $s=0$) and let $A_{ij}=1$ if $U_{\theta}(\vx)_{ij}>s$.
However, such entry-wise independent predictions of $A_{ij}$ may result in a matrix $A$ that violates the constraints for a valid RNA secondary structure. Therefore, a post-processing step is needed to make sure the predicted $A$ is valid. This step could be carried out separately after $U_{\theta}$ is learned. But such decoupling of base-pair scoring and post-processing for constraints may lead to sub-optimal results, where the errors in these two stages can not be considered together and tuned together.
Instead, we will introduce a Post-Processing Network which can be trained end-to-end together with $U_{\theta}$ to enforce the constraints. 



\subsubsection{Post-Processing Network} 
The second part of E2Efold is a {\it Post-Processing Network} $\text{PP}_{\phi}$ which is an unrolled and parameterized algorithm for solving a constrained optimization problem. We first present how we formulate the post-processing step as a constrained optimization problem and the algorithm for solving it.  
After that, we show how we use the algorithm as a template to design deep architecture $\text{PP}_\phi$.

\label{sec:pp-copt}

\vspace{\baselineskip}
\noindent {\bf Post-Processing with Constrained Optimization}




\paragraph{Formulation of constrained optimization.}
Given the scores predicted by $U_{\theta}(\vx)$, we define the total score $\frac{1}{2}\sum_{i,j}(U_{\theta}(\vx)_{ij}-s)A_{ij}$ as the objective to maximize, where $s$ is an offset term. 
Clearly, without structure constraints, the optimal solution is to take $A_{ij}=1$ when $U_{\theta}(\vx)_{ij}>s$. Intuitively, the objective measures the covariation between the entries in the scoring matrix and the $A$ matrix. With constraints, the exact maximization becomes intractable. To make it tractable, we consider a convex relaxation of this discrete optimization to a continuous one by allowing $A_{ij}\in[0,1]$. Consequently, the solution space that we consider to optimize over is
$	\gA(\vx):=\cbr{A\in[0,1]^{L\times L} \mid A~\text{is symmetric and satisfies constraints (i)-(iii) in Section~\ref{sec:problem}}}.
$

To further simplify the search space, we define a nonlinear transformation $\gT$ on $\R^{L\times L}$ as
$	\gT(\hat{A}):= \textstyle{ \frac{1}{2}}\big(\hat{A}\circ \hat{A} + (\hat{A}\circ \hat{A})^\top \big)\circ M(\vx)$, where $\circ$ denotes element-wise multiplication. Matrix $M$ is defined as $M(\vx)_{ij}:=1$ if $x_ix_j\in\gB$ and also $|i-j|\geq4$, and $M(\vx)_{ij}:=0$ otherwise. From this definition, we can see that $M(\vx)$ encodes both constraint (i) and (ii). With transformation $\gT$, the resulting matrix is {non-negative}, {symmetric}, and {satisfies constraint }(i) and (ii). Hence, by defining $A:=\gT(\hat{A})$,
the solution space is simplified as
$    \gA(\vx)=\{ A = \gT(\hat{A})\mid \hat{A}\in\R^{L\times L}, A\mathbf{1}\leq \mathbf{1}\}.
$§

Finally, we introduce a $\ell_1$ penalty term $\|\hat{A}\|_1:=\sum_{i,j}|\hat{A}_{ij}|$ to make $A$ sparse and formulate the post-processing step as: ($\langle \cdot,\cdot\rangle$ denotes matrix inner product, i.e., sum of entry-wise multiplication)
				\begin{align*}
					\textstyle{\max_{\hat{A}\in \R^{L\times L}} \frac{1}{2}}\left\langle U_{\theta}(\vx)-s, A:=\gT(\hat{A})\right\rangle - \rho\|\hat{A}\|_1 \quad
					\text{s.t.}~~A\mathbf{1}\leq \mathbf{1}.
					\label{eq:pp-nonlinear}
				\end{align*}
The advantages of this formulation are that the variables $\hat{A}_{ij}$ are free variables in $\R$ and there are only $L$ inequality constraints $A\mathbf{1}\leq \mathbf{1}$. This system of linear inequalities can be replaced by a set of nonlinear equalities $\text{relu}(A\mathbf{1-\mathbf{1}})=\mathbf{0}$  so that the constrained problem can be easily transformed into an unconstrained problem by introducing 
a Lagrange multiplier $\boldsymbol{\lambda} \in \R_+^{L}$: 
\begin{align}
    \min_{\boldsymbol{\lambda} \geq \mathbf{0}} \max_{\hat{A}\in\R^{L\times L}}~ \underbrace{{\textstyle\frac{1}{2}}\langle U_{\theta}(\vx)-s,A \rangle - \langle \boldsymbol{\lambda}, \text{relu}(A\mathbf{1}-\mathbf{1})\rangle}_{f} - \rho\|\hat{A}\|_1.
\end{align}
\textbf{Algorithm for solving it.} We use proximal gradient for maximization and gradient descent for minimization. In each iteration, $\hat{A}$ and $\boldsymbol{\lambda}$ are updated alternatively by:
\begin{align}
    \text{gradient step:}\quad &\dot{A}_{t+1}\gets \hat{A}_t + \alpha\cdot \gamma_{\alpha}^t \cdot\hat{A}_t\circ M(\vx)\circ\Big(\partial f/\partial A_t+(\partial f/\partial A_t)^\top\Big),
    \label{eq:algo-start}
    \\
    & \text{where}~
    \begin{cases}
    {\partial f}/ {\partial  A_t}={\textstyle \frac{1}{2}} (U_{\theta}(\vx)-s)-\rbr{\boldsymbol{\lambda} \circ \text{sign}(A_t\mathbf{1}-\mathbf{1})}\mathbf{1}^\top, \\
    \text{sign}(c):=1~\text{when}~c>0~\text{and}~0~\text{otherwise},
    \end{cases}\label{eq:gradient_e2efold}
    \\
    \text{soft threshold:}\quad &\hat{A}_{t+1}\gets\text{relu}(|\dot{A}_{t+1}|-\rho\cdot \alpha\cdot \gamma_{\alpha}^t),
    \quad A_{t+1} \gets \gT(\hat{A}_{t+1}), \label{eq:algo-soft}\\
    \text{gradient step:}\quad & \boldsymbol{\lambda}_{t+1} \gets \boldsymbol{\lambda}_{t+1} + \beta\cdot \gamma_{\beta}^t\cdot \text{relu}(A_{t+1
    }\mathbf{1} - \mathbf{1}),
    \label{eq:algo-end}
\end{align}
where $\alpha$, $\beta$ are step sizes and $\gamma_{\alpha},\gamma_{\beta}$ are decaying coefficients.
When it converges at $T$, an approximate solution $Round\big(A_T = \gT(\hat{A}_T)\big)$ is obtained. 
With this algorithm operated on the learned $U_{\theta}(\vx)$, even if this step is disconnected to the training phase of $U_{\theta}(\vx)$, the final prediction works much better than
many other existing methods (as reported in Section \ref{sec:experiments}). 
Next, we introduce how to couple this post-processing step with the training of $U_{\theta}(\vx)$ to further improve the performance.


\vspace{\baselineskip}
\noindent {\bf Post-Processing Network via an Unrolled Algorithm}



We design a {\it Post-Processing Network}, denoted by $\text{PP}_{\phi}$, based on the above algorithm.
After it is defined, we can connect it with the deep score network $U_{\theta}$, as shown in Figure~\ref{fig:e2efold_archi}, and train them jointly in an end-to-end fashion, so that the training phase of $U_{\theta}(\vx)$ is aware of the post-processing step.

\begin{algorithm}[H]
		\DontPrintSemicolon
		\SetKwFunction{Grad}{Grad}
		\SetKwProg{Fn}{Function}{:}{end}
		\SetKwFor{uFor}{For}{do}{}
		\SetKwFor{ForPar}{For all}{do in parallel}{}
		\SetKwFunction{PP}{$\text{PP}_{\phi}$}
		\SetKwFunction{PPcell}{\rm$\text{PPcell}_{\phi}$}
		\parbox{\linewidth}{
		    {\it Parameters}  $\phi:= \{ 
		    {\color{blue}w}, {\color{blue}s}, {\color{blue}\alpha},{\color{blue}\beta},{\color{blue}\gamma_{\alpha}},{\color{blue}\gamma_\beta},{\color{blue}\rho}\}$\;
		    
            $U\gets \text{softsign}(U-{\color{blue}s})\circ U$\;
            $\hat{A}_0\gets \text{softsign}(U-{\color{blue}s})\circ \text{sigmoid}(U)$\;
            $A_0\gets \gT(\hat{A}_0)$;$\quad \boldsymbol{\lambda}_0\gets {\color{blue}w}\cdot\text{relu}(A_0\mathbf{1}-\mathbf{1})$\;
            
            \uFor{$t=0,\ldots,T-1$}{
                $\boldsymbol{\lambda}_{t+1},A_{t+1},\hat{A}_{t+1}$ = \PPcell{$U,M,\boldsymbol{\lambda}_t,A_t,\hat{A}_t,t$}\;
            }
            \KwRet $\{A_t\}_{t=1}^T$\;
		}
		\caption{Post-Processing Network $\text{PP}_{\phi}(U,M)$}
		\label{algo:pp}
\end{algorithm}

\begin{algorithm}[H]
		\DontPrintSemicolon
		\SetKwFunction{Grad}{Grad}
		\SetKwProg{Fn}{Function}{:}{end}
		\SetKwFor{uFor}{For}{do}{}
		\SetKwFor{ForPar}{For all}{do in parallel}{}
		\SetKwFunction{PP}{$\text{PP}_{\phi}$}
		\SetKwFunction{PPcell}{$\text{PPcell}_{\phi}$}
		\parbox{\linewidth}{
		\Fn{\PPcell{$U,M,\boldsymbol{\lambda},A,\hat{A},t$}}{
			$G \gets {\textstyle \frac{1}{2}} U-\rbr{\boldsymbol{\lambda} \circ \text{softsign}(A\mathbf{1}-\mathbf{1})}\mathbf{1}^\top$\;
			$\dot{A} \gets \hat{A}+ {\color{blue}\alpha} \cdot {\color{blue}\gamma_{\alpha}}^t \cdot \hat{A} \circ M\circ (G+G^\top)$\;
			$\hat{A} \gets \text{relu}(|\dot{A}| - {\color{blue}\rho}\cdot{\color{blue}\alpha}\cdot{\color{blue}\gamma_{\alpha}}^t)$\;			$\hat{A} \gets 1-\text{relu}(1-\hat{A})$ ~ [i.e.,$\min(\hat{A},1)$]\;
			$A\gets \gT(\hat{A})$; $\boldsymbol{\lambda} \gets \boldsymbol{\lambda} + {\color{blue}\beta} \cdot {\color{blue}\gamma_{\beta}}^t \cdot \text{relu}(A\mathbf{1} - \mathbf{1})$\;
			\KwRet $\boldsymbol{\lambda}, A,\hat{A}$
		}
        }
		\caption{Neural Cell $\text{PPcell}_\phi$}
		\label{algo:ppcell}
\end{algorithm}

The specific computation graph of $\text{PP}_{\phi}$ is given in Algorithm~\ref{algo:pp}, whose main component is a recurrent cell, namely, $\text{PPcell}_{\phi}$. The computation graph 
is almost the same as the iterative update from \eqref{eq:algo-start} to \eqref{eq:algo-end}, except for several modifications:
\begin{itemize}[leftmargin=*,nolistsep]
    \item {\it(learnable hyperparameters)} The hyperparameters including step sizes $\alpha,\beta$, decaying rate $\gamma_\alpha,\gamma_\beta$, sparsity coefficient $\rho$ and the offset term $s$ are treated as learnable parameters in ${\phi}$, so that there is no need to tune the hyperparameters by hand. They can be learned from data automatically.
    \item {\it(fixed \# iterations)} Instead of running the iterative updates until convergence, $\text{PPcell}_\phi$ is applied recursively for $T$ iterations where $T$ is a manually fixed number. This is why in Figure~\ref{fig:space} the output space of E2Efold is slightly larger than the true solution space.
    \item {\it(smoothed sign function)} Resulted from the gradient of $\text{relu}(\cdot)$, the update step in \eqref{eq:gradient_e2efold} contains a sign$(\cdot)$ function. However, to push gradient through $\text{PP}_{\phi}$, we require a differentiable update step. Therefore, we use a smoothed sign function defined as $\text{softsign}(c):=1/(1+\exp(-kc))$, where $k$ is a temperature. 
    \item {\it(clip $\hat{A}$)} An additional step, $\hat{A}\gets\min(\hat{A},1)$, is included to make the output $A_{t}$ at each iteration stay in the range $[0,1]^{L\times L}$. This is useful for computing the loss over intermediate results $\{A_{t}\}_{t=1}^T$, for which we will explain more in Section \ref{sec:training_e2efold}.
\end{itemize}
With these modifications, the Post-Processing Network $\text{PP}_{\phi}$ is a {tuning-free} and {differentiable} unrolled algorithm with {meaningful intermediate outputs}. Combining it with the deep score network, the final deep model is
\begin{align}
      \text{\bf E2Efold}:   \quad  \{A_t\}_{t=1}^T= \overbrace{\text{PP}_{\phi}(\underbrace{U_{\theta}(\vx)}_\text{\it Deep Score Network}, M(\vx))}^\text{\it Post-Process~Network}.
\end{align}

\subsection{End-to-End Training Algorithm}
\label{sec:training_e2efold}

Given a dataset $\gD$ containing examples of input-output pairs $(\vx,A^*)$, the training procedure of E2Efold is similar to standard gradient-based supervised learning. 
However,
for RNA secondary structure prediction problems, commonly used metrics for evaluating predictive performances are F1 score, precision and recall, which are non-differentiable.

\paragraph{Differentiable F1 loss.} To directly optimize these metrics, we mimic true positive (TP), false positive (FP), true negative (TN) and false negative (FN) by defining continuous functions on $[0,1]^{L\times L}$:
\begin{align*}
    \text{TP}=\langle A,A^* \rangle,\  \text{FP}= \langle A, 1-A^*\rangle,\
     \text{FN}=\langle 1-A,A^*\rangle,\ \text{TN}= \langle 1-A,1-A^*\rangle.
\end{align*}
Since $\text{F1} = 2\text{TP}/(2\text{TP} + \text{FP} + \text{FN})$, we define a loss function to mimic the negative of F1 score as:
\begin{align}
    \gL_{-\text{F1}}(A,A^*):={-2\langle A,A^*\rangle}/\rbr{2\langle A,A^*\rangle+\langle A,1-A^*\rangle+\langle 1-A,A^*\rangle}.
\end{align}
Assuming that $\sum_{ij}A^*_{ij}\neq 0$, this loss is well-defined and differentiable on $[0,1]^{L\times L}$. Precision and recall losses can be defined in a similar way, but we optimize F1 score in this work.

It is notable that this F1 loss takes advantages over other differentiable losses including $\ell_2$ and cross-entropy losses, because there are much more negative samples (i.e. $A_{ij}=0$) than positive samples (i.e. $A_{ij}=1$). A hand-tuned weight is needed to balance them while using $\ell_2$ or cross-entropy losses, but F1 loss handles this issue automatically, which can be useful for a number of other problems~\cite{RN1152,RN140}.

\paragraph{Overall loss function.} As noted earlier, E2Efold outputs a matrix $A_t\in[0,1]^{L\times L}$ in each iteration. This allows us to add auxiliary losses to regularize the intermediate results, guiding it to learn parameters which can generate a smooth solution trajectory. More specifically, we use an objective that depends on the entire trajectory of optimization:
\begin{align}
    \min_{\theta,\phi}\frac{1}{|\gD|}\sum_{(x,A^*)\in\gD} \frac{1}{T}\sum_{t=1}^{T} \gamma^{T-t}\gL_{-\text{F1}}(A_t, A^*),\label{eq:loss}
\end{align}
where $\{A_t\}_{t=1}^T= \text{PP}_{\phi}(U_{\theta}(\vx),M(\vx))$ and $\gamma\leq1$ is a discounting factor. Empirically, we find it very useful to pre-train $U_{\theta}$ using logistic regression loss. Also, it is helpful to add this additional loss to \eqref{eq:loss} as a regularization.

\section{Results}
\label{sec:experiments}

We compare E2Efold with the state-of-the-art methods in this RNA secondary structure prediction field on two benchmark datasets (Section~\ref{cha5_exp:dataset}). The experiments suggest that E2Efold can outperform all the previous methods significantly across different experimental settings, including cross-fold validation (Section~\ref{cha5_exp:rnastralign}), cross-dataset validation (Section~\ref{cha5_exp:archieveii}), and pseudoknot prediction (Section~\ref{cha5_exp:pseudoknot}). Among them, for the pseudoknot prediction, we can reach 29.7\% improvement over the previous methods regarding the F1 score on the RNAstralign dataset. Furthermore, our method is the fastest method for inferring the RNA secondary structure (Section~\ref{cha5_exp:time}). An ablation study was also conducted to show the necessity of pushing gradient through the post-processing step (Section~\ref{cha5_exp:rnastralign}). We further visualize the prediction in Section~\ref{cha5_exp:vis}. 


\begin{table}[!t]
\centering
\caption{Dataset Statistics}
    \resizebox{0.55\textwidth}{!}{
    \label{Tab:rnastralign_data}
    \begin{tabular}{@{}l@{}c@{}c@{\hspace{1mm}}|@{\hspace{1mm}}c@{}c@{}}
    \toprule
    \multirow{2}{*}{Type}& \multicolumn{2}{c}{ArchiveII} & \multicolumn{2}{c}{RNAStralign}\\
    \cmidrule(l){2-5} 
     &  length & \#samples & length & \#samples \\ 
     \hline
        All & 28$\sim$2968 & 3975 & 30$\sim$1851 & 30451\\ 
        \hline
        16SrRNA & 73$\sim$1995 & 110 & 54$\sim$1851 & 11620 \\ 
        5SrRNA & 102$\sim$135 & 1283 & 104$\sim$132 & 9385\\ 
        tRNA & 54$\sim$93 & 557 & 59$\sim$95 &  6443\\ 
        grp1 & 210$\sim$736 & 98 &163$\sim$615 & 1502 \\ 
        SRP & 28$\sim$533 & 928 & 30$\sim$553 &  468\\ 
      tmRNA & 102$\sim$437 & 462 & 102$\sim$437 & 572 \\ 
        RNaseP & 120$\sim$486 & 454 & 189$\sim$486 & 434\\ 
      telomerase & 382$\sim$559 & 37 & 382$\sim$559 & 37 \\ 
      23SrRNA & 242$\sim$2968 & 35 & - & -\\ 
      grp2 & 619$\sim$780 &  11 & - & - \\ 
      \bottomrule
     \end{tabular}}
\end{table}

\subsection{Experimental Setting}

\noindent {\bf Dataset}
\label{cha5_exp:dataset}

We used two benchmark datasets to evaluate the proposed method. The first dataset, ArchiveII \cite{sloma2016exact}, which contains 3975 RNA structures from 10 RNA types, is a widely used benchmark dataset for classical RNA folding methods. The second dataset, RNAStralign \cite{tan2017turbofold}, which is composed of 37149 structures from 8 RNA types, is one of the most comprehensive collections of RNA structures in the market. After removing redundant sequences and structures in the RNAStralign dataset, 30451 structures remain. We summarized the statistics of these two datasets in Table \ref{Tab:rnastralign_data}. In our experiments, we used the larger dataset, the RNAStralign dataset, for the cross-fold validation, while the smaller dataset, the ArchiveII dataset, for the cross-dataset validation. Regarding the cross-dataset validation, we trained the model on the RNAStralign and then applied the trained model on the ArchiveII directly without re-training. Such a validation simulates the real-life usage, which can test the generalization performance of the proposed method further.

\subsection{Experiments on RNAStralign}
\label{cha5_exp:rnastralign}

\noindent {\bf Overall Performance}

We divided the RNAStralign dataset into training, testing, and validation sets by stratified sampling so that each set contains all RNA types. We compared the performance of E2Efold against six existing methods, including { CDPfold}, { LinearFold}, { Mfold}, RNAstructure (ProbKnot), { RNAfold} and { CONTRAfold}.
Both E2Efold and CDPfold were learned from the same training/validation sets. For other methods, we directly used the provided packages or web-servers to predict the structures. 
We evaluated the F1 score, Precision and Recall for each sequence in the test set and reported the average values in Table \ref{Tab:overall_perf_rnastralign}. As suggested by \cite{mathews2019benchmark}, for a base pair ($i,j$), the following predictions were also considered as correct: ($i+1,j$), ($i-1,j$), ($i,j+1$), ($i,j-1$), so we also reported the metrics when one-position shift was allowed.

\begin{table}[!t]
\centering
\caption{Results on RNAStralign test set.  ``(S)'' indicates the results when one-position shift is allowed.}
\resizebox{0.7\textwidth}{!}{
 \begin{tabular}{@{}c c c cc@{\hspace{2pt}}c@{\hspace{3pt}}c@{}} 
 \toprule
 Method & Prec & Rec & F1  & Prec(S) & Rec(S) & F1(S) \\ 
 \midrule
  {\bf E2Efold} & {\bf 0.866} &	{\bf 0.788} & {\bf 0.821} & 	{\bf 0.880} &{\bf 	0.798} &	{\bf 0.833}   \\
  $U_{\theta}+\text{PP}$ & 0.755 & 0.712 & 0.721 & 0.782 &  0.737 &  0.752  \\
  CDPfold & 0.633  & 0.597  & 0.614 & 0.720  & 0.677 & 0.697      \\
  LinearFold & 0.620&0.606&0.609&0.635&0.622&0.624     \\
  Mfold & 0.450&0.398&0.420&0.463&0.409&0.433   \\
 RNAstructure & 0.537&0.568&0.550&0.559&0.592&0.573    \\
 RNAfold & 0.516&0.568&0.540&0.533&0.587&0.558   \\
 CONTRAfold & 0.608&0.663&0.633&0.624&0.681&0.650 \\
 \bottomrule
 \end{tabular}
 \label{Tab:overall_perf_rnastralign}
 }
\end{table}

\begin{figure}[h!]
    \centering
    \includegraphics[width=0.6\textwidth]{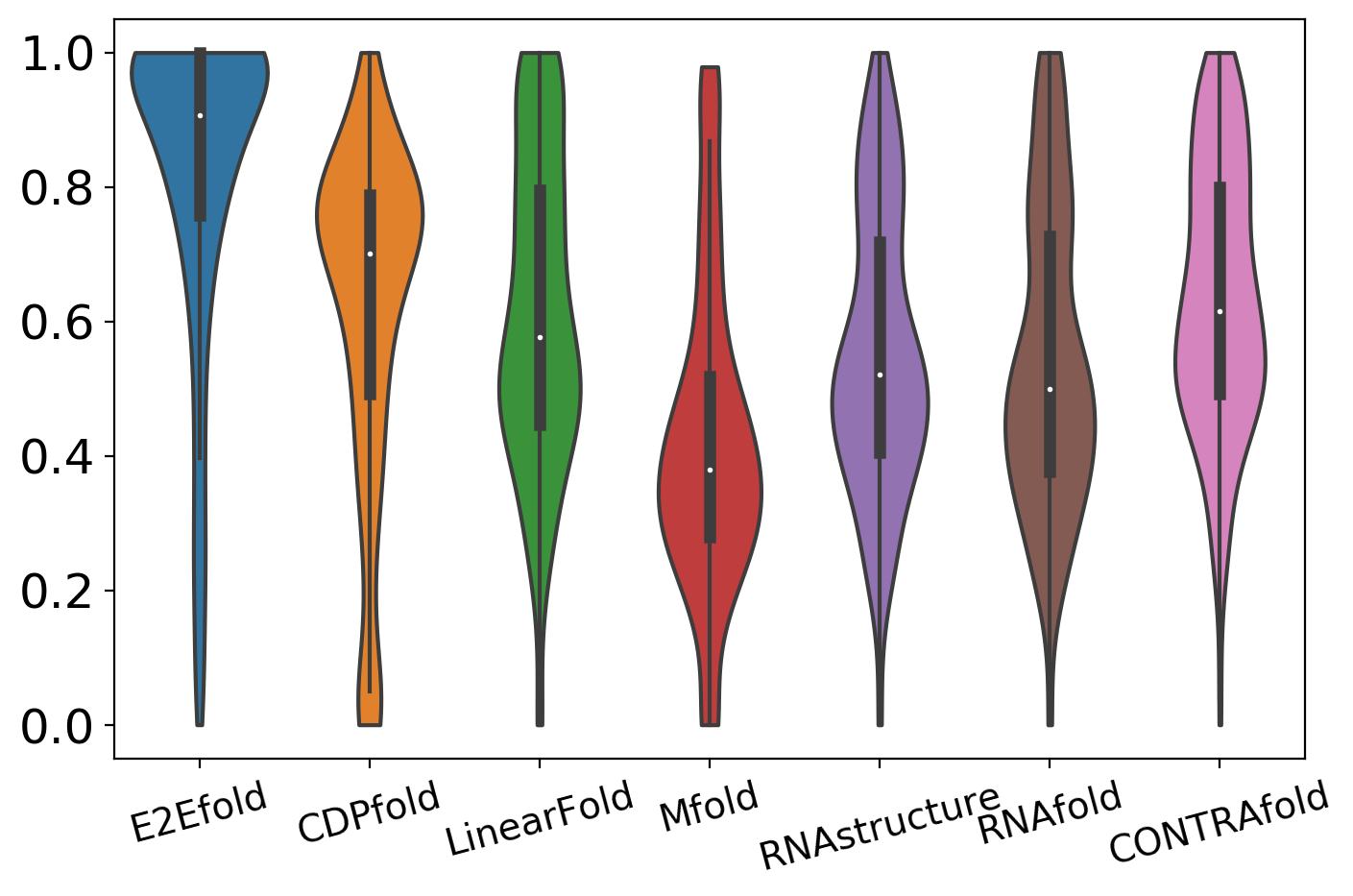}
    \caption{Distribution of F1 score.}
    \label{fig:overall_perf_rna_stralign}
\end{figure}



As shown in Table \ref{Tab:overall_perf_rnastralign}, traditional methods can achieve an averaged F1 score ranging from 0.433 to 0.624, which is consistent with the performance reported in their original papers. 
The two learning-based methods, CONTRAfold and CDPfold, can outperform classical methods with reasonable margin on some criteria. E2Efold, on the other hand, significantly outperforms all the previous methods across all criteria, with at least 20\% improvement. Notice that, for almost all the other methods, the recall is usually higher than precision, while for E2Efold, the precision is higher than recall. That can be the result of incorporating constraints during neural network training. 
We also show the distributions of F1 scores for each method in Figure~\ref{fig:overall_perf_rna_stralign}. As illustrated in the figure, E2Efold has consistently good performance in this experiment.

\vspace{\baselineskip}
\noindent {\bf Per-family Performance}
\label{cha5_exp:family}

To evaluate the detailed performance of E2Efold on different RNA types, we included the per-family F1 scores in Table \ref{tab:per-family}. Although the results are from just one single deep learning model, E2Efold can still outperform the other methods significantly in 16S rRNA, tRNA, 5S RNA, tmRNA, and telomerase. In the future, we can view this problem as a multi-task learning problem and further improve the performance by learning multiple models for different RNA families and an additional meta-classifier to predict which model to use for the input sequence. Furthermore, we can incorporate the energy function designed by the previous researchers, which is another source of prior knowledge, into the deep learning model to improve E2Efold's performance on SRP, Group I intron, and RNaseP further.

\begin{table}[!t]
\centering
\caption{RNAStralign: per-family performances}\label{tab:per-family}
 \begin{tabular}{c |c c|c c|c c|cc} 
  \toprule
  & \multicolumn{2}{c|}{16S rRNA} & \multicolumn{2}{c|}{tRNA} & \multicolumn{2}{c|}{5S RNA} & \multicolumn{2}{c}{SRP}  \\ 
 & F1 & F1(S) & F1 & F1(S) & F1 & F1(S) &F1 & F1(S) 
 \\
 \midrule
E2Efold & 0.783 & 0.795 & 0.917 & 0.939 & 0.906 & 0.936 & 0.550 & 0.614\\
LinearFold &0.493&0.504 & 0.734 & 0.739 & 0.713 & 0.738 & 0.618 & 0.648\\
Mfold & 0.362&0.373 & 0.662 & 0.675 & 0.356 & 0.367 & 0.350 & 0.378\\
RNAstructure &0.464&0.485 & 0.709 & 0.736 & 0.578 & 0.597 & 0.579 & 0.617\\
RNAfold & 0.430&0.449 & 0.695 & 0.706 & 0.592 & 0.612 & 0.617 & 0.651\\
CONTRAfold &0.529&0.546 & 0.758 & 0.765 & 0.717 & 0.740 & 0.563 & 0.596\\
 \toprule
 & \multicolumn{2}{c|}{tmRNA} & \multicolumn{2}{c|}{Group I intron} &\multicolumn{2}{c|}{RNaseP} & \multicolumn{2}{c|}{telomerase}\\
 & F1 & F1(S) & F1 & F1(S) & F1 & F1(S) &F1 & F1(S) \\
 \midrule
E2Efold & 0.588 & 0.653 & 0.387 & 0.428 & 0.565 & 0.604 & 0.954 & 0.961\\
LinearFold & 0.393 & 0.412 & 0.565 & 0.579 & 0.567 & 0.578 & 0.515 & 0.531\\
Mfold & 0.290 & 0.308 & 0.483 & 0.498 & 0.562 & 0.579 & 0.403 & 0.531\\
RNAstructure & 0.400 & 0.423 & 0.566 & 0.599 & 0.589 & 0.616 & 0.512 & 0.545\\
RNAfold & 0.411 & 0.430 & 0.589 & 0.599 & 0.544 & 0.563 & 0.471 & 0.496\\
CONTRAfold & 0.463 & 0.482 & 0.603 & 0.620 & 0.645 & 0.662 & 0.529 & 0.548\\
 \bottomrule
 \end{tabular}
\end{table}

\vspace{\baselineskip}
\noindent {\bf Ablation Study}

As we have discussed in Section~\ref{chapter5_sec:method}, we incorporate the deep learning model with the gradient descent algorithm deeply by designing a new deep learning architecture, which has the unrolled gradient descent algorithm embedded in the network architecture, to tackle this structured prediction problem. During training, we also differentiate the loss function through the unrolled algorithm to truly integrate the two steps into one model with such end-to-end training. To exam whether the deep integration is necessary for the performance of E2Efold, we further conducted an ablation study (Table \ref{Tab:overall_perf_rnastralign}). We compared the performance of E2Efold against the two-step solution, ``$U_{\theta}+\text{PP}$", where the post-processing step was disconnected from the training of the Deep Score Network ($U_{\theta}$). We applied the post-processing step (i.e., for solving augmented Lagrangian) after the Deep Score Network was learned (thus the notation ``$U_{\theta}+\text{PP}$" in Table~\ref{Tab:overall_perf_rnastralign}). Although ``$U_{\theta}+\text{PP}$" performs decently well, compared to the previous method, with the deep integration of deep learning method and the traditional algorithm, E2Efold can outperform the two-step solution significantly with 10\% performance improvement regarding the F1 score. This experiment demonstrates the effectiveness of our ideas for solving this structured prediction problem further.

\begin{table}[!t]
    \centering
    \caption{Performance comparison on ArchiveII}
    \resizebox{0.7\textwidth}{!}{
     \begin{tabular}{@{}c c c c c@{\hspace{2pt}}c c} 
     \toprule
     Method & Prec & Rec & F1 & Prec(S) & Rec(S) & F1(S)\\ 
     \midrule
      {\bf E2Efold} &{\bf 0.734}&{\bf0.66}&{\bf0.686}&{\bf0.758}&{\bf0.676}&{\bf0.704}\\
      CDPfold & 0.557 & 0.535 & 0.545 & 0.612 & 0.585 & 0.597   \\
      LinearFold & 0.641 & 0.617 & 0.621 & 0.668 & 0.644 & 0.647     \\
      Mfold & 0.428 & 0.383 & 0.401 & 0.450 & 0.403 & 0.421    \\
     RNAstructure & 0.563 & 0.615 & 0.585 & 0.590 & 0.645 & 0.613    \\
     RNAfold & 0.565 & 0.627 & 0.592 & 0.586  & 0.652 & 0.615    \\
     CONTRAfold & 0.607 & 0.679 & 0.638 & 0.629 & 0.705 & 0.662 \\
     \bottomrule
     \end{tabular}}
     \label{Tab:archiveii_perf}
\end{table}

\subsection{Test on ArchiveII without Re-training}
\label{cha5_exp:archieveii}
To mimic the real-world scenario where users want to predict the newly discovered RNAs' structures that may have a distribution different from the training dataset, 
we directly tested the model learned from the RNAStralign training set on the ArchiveII dataset, without re-training the model. To make the comparison fair, we excluded sequences that overlapped the RNAStralign dataset.
We then tested the model on sequences in ArchiveII that had overlapping RNA types (5SrRNA, 16SrRNA, etc.) with the RNAStralign dataset. Results are shown in Table \ref{Tab:archiveii_perf}. Understandably, the performance of classical methods which are not learning-based is consistent with that on RNAStralign. 
Both CDPfold and E2Efold have performance degradation on this cross-dataset validation because of the distribution shifting between training and testing sets. However, 
although the performance of E2Efold is not as good as that on RNAStralign, it is still better than all the other methods across different evaluation criteria. 
This experiment shows the real-world usefulness of E2Efold.

\begin{table}[!t]
\centering
\caption{Evaluation of pseudoknot prediction}
 \resizebox{0.5\textwidth}{!}{
 \begin{tabular}{@{}c c c c c c} 
 \toprule
 Method & Set F1 & TP & FP & TN & FN \\ 
 \midrule
  E2Efold &  0.710 & 1312 & 242 & 1271 & 0    \\
 {\small RNAstructure} & 0.472 & 1248 & 307 & 983 & 286  \\
 \bottomrule
 \end{tabular}}
 \label{Tab:pseudoknot_perf}
\end{table}

\subsection{Pseudoknot Prediction}
\label{cha5_exp:pseudoknot}
As we discussed in Section~\ref{chapter5_sec:method}, E2Efold can cover pseudoknot prediction. So, we evaluated the performance of E2Efold on such a prediction. As most of the previous methods are unable to predict pseudoknots while RNAstructure is the most famous one that can predict pseudoknots, we performed a head-to-head comparison between RNAstructure and E2Efold. As adopted in the literature \cite{bellaousov2013rnastructure}, we picked all the sequences containing pseudoknots and computed the averaged F1 score only on this set. Besides, we counted the number of pseudoknotted sequences that were predicted as pseudoknotted and reported this count as true positive (TP). Similarly, we reported TN, FP, and FN in Table~\ref{Tab:pseudoknot_perf} along with the F1 score. As shown in Table~\ref{Tab:pseudoknot_perf}, E2Efold can outperform RNAstructure significantly on the pseudoknot prediction, bringing around 25\% performance improvement regarding the F1 score. 


\begin{table}[!t]
    \centering
    \caption{Inference time on RNAStralign}
    \resizebox{0.6\textwidth}{!}{
    \begin{tabular}{l@{\hspace{-7mm}}rc@{}c@{}} 
     \toprule
     \multicolumn{2}{c}{Method} & Total run time & Time per seq\\ 
     \midrule
      \bf{E2Efold} & {\bf (Pytorch)} & {\bf 19m (GPU)} & {\bf 0.40s}      \\
      CDPfold&  (Pytorch) & 440m*32 threads & 300.107s   \\
      LinearFold&  (C) & 20m & 0.43s    \\
      Mfold&  (C) &  360m & 7.65s   \\
     RNAstructure&  (C) & 3 days & 142.02s   \\
     RNAfold&  (C) & 26m & 0.55s     \\
     CONTRAfold & (C) & 1 day & 30.58s     \\
     \bottomrule
     \end{tabular}}
     \label{Tab:inference_time}
\end{table}

\subsection{Inference Time Comparison}
\label{cha5_exp:time}
We recorded the running time of all algorithms for predicting RNA secondary structures on the RNAStralign test set and compared their performance regarding the inference time in Table \ref{Tab:inference_time}. 
As people realized that DP could be slow, all the previous methods, except for CDPfold, were implemented using C to reduce the running time.
LinearFold is the most efficient one among baselines because it uses beam pruning heuristic to accelerate DP. CDPfold, which achieves a higher F1 score than other baselines, however, is extremely slow due to its DP post-processing step.
Since we use a gradient-based algorithm, which is simple to design the Post-Processing Network, E2Efold is fast. On GPU, E2Efold has similar inference time as LinearFold, outputting one RNA secondary structure in 0.4 seconds.

\subsection{Visualization}
\label{cha5_exp:vis}
\begin{figure}[!t]
    \centering
    \includegraphics[width=1.0\textwidth]{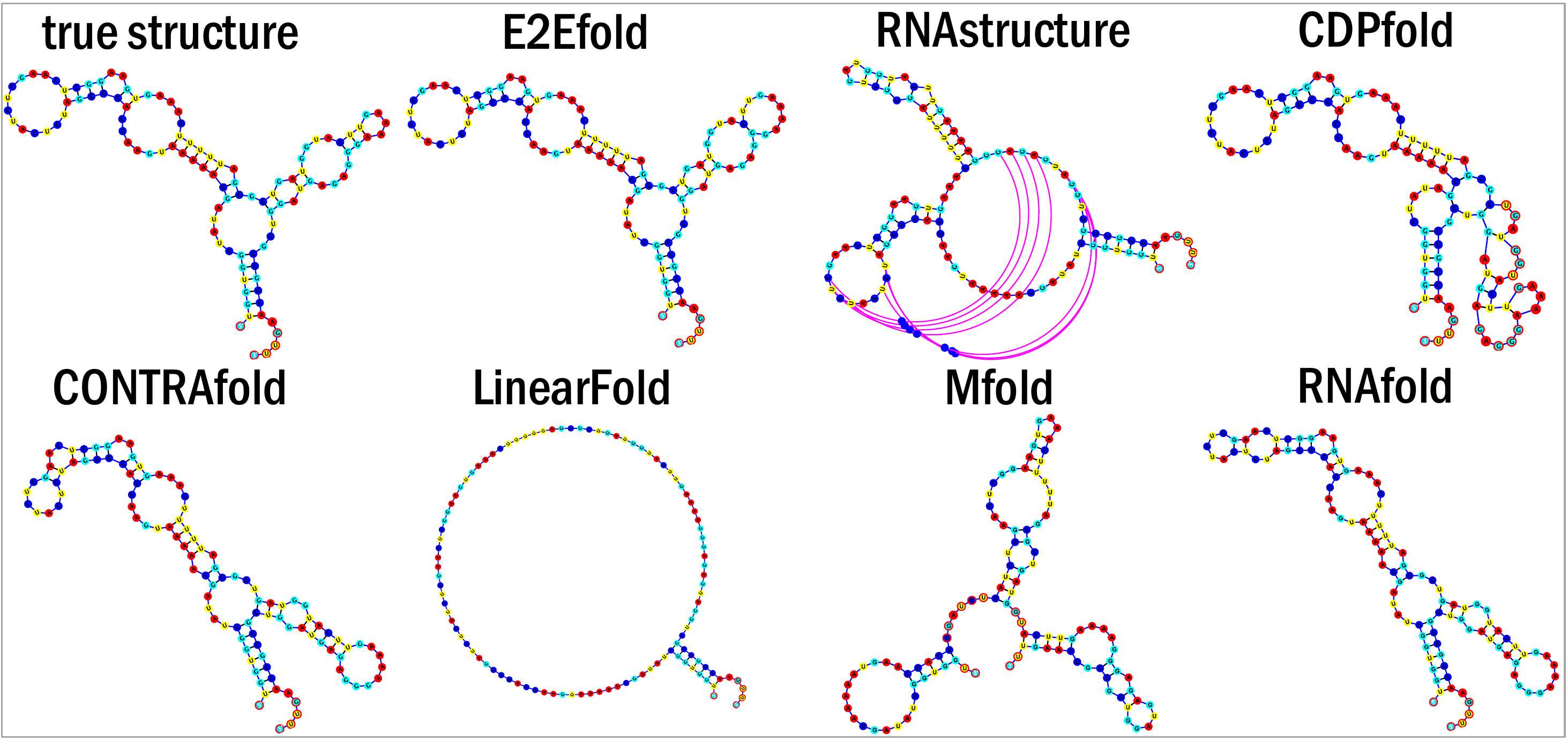}
    \caption{Visualization of a 5S rRNA, B01865.}
    \label{fig:visual_5s}
\end{figure}

\begin{figure}[!t]
    \centering
    \includegraphics[width=1.0\textwidth]{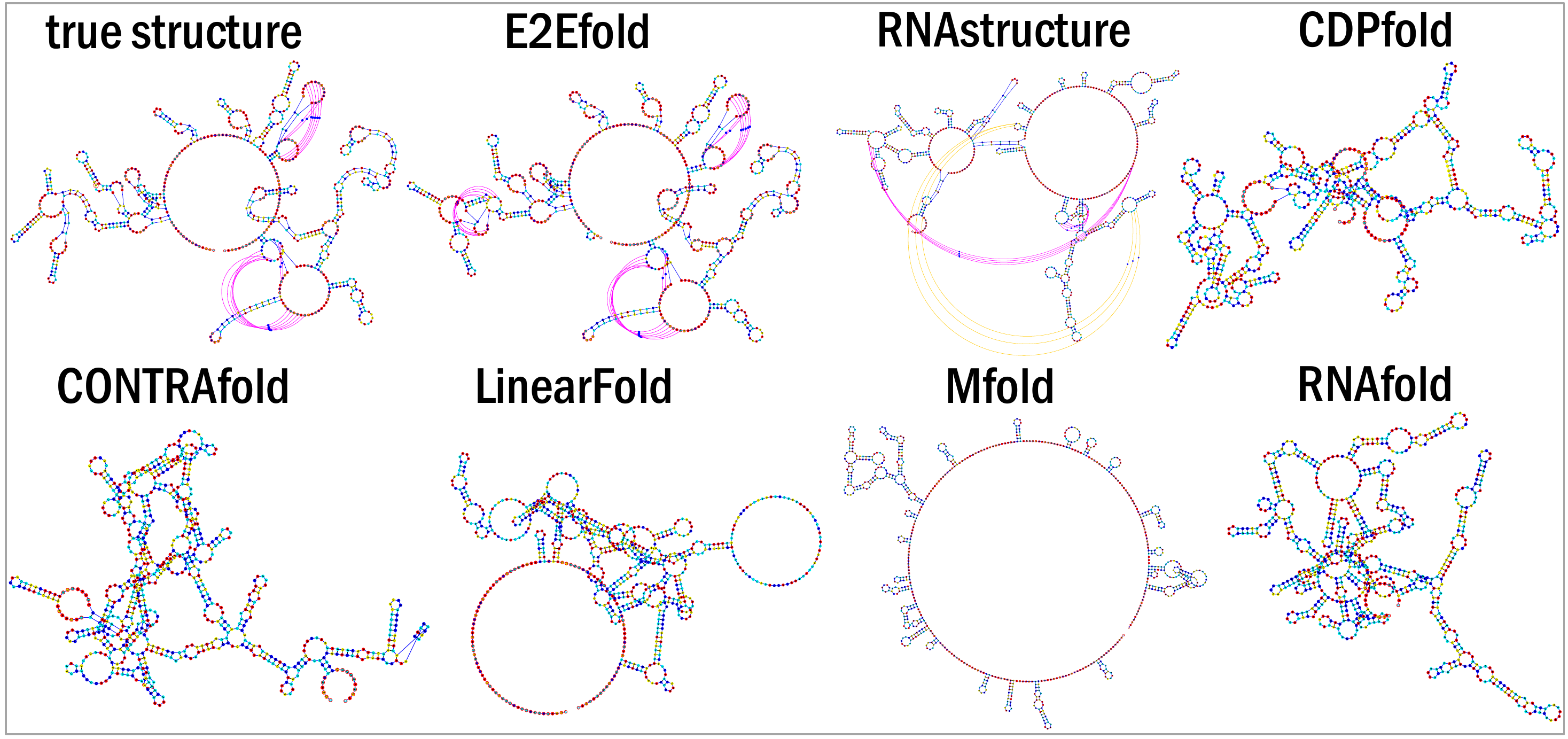}
    \caption{Visualization of a 16S rRNA, DQ170870.}
    \label{fig:visual}
\end{figure}

To further check whether E2Efold can make valid predictions, we randomly selected two RNA secondary structures, one simple and one hard, and visualized the predictions from different methods in Figure~\ref{fig:visual_5s} and ~\ref{fig:visual}, respectively. 
In these figures, {\color{deepmagenta}purple lines} indicate edges of pseudoknotted elements.
As shown in Figure~\ref{fig:visual_5s}, for this relatively simple structure, most of the methods can have relatively reasonable performance. Regarding E2Efold, it can predict a structure, which is identical to the ground truth structure. However, regarding the hard structure, as shown in Figure~\ref{fig:visual}, except for E2Efold, all the other methods even cannot predict the outline of the ground truth structure. As for E2Efold, although there are some minor errors in its prediction, the predicted structure is almost identical to the ground truth structure. 

\section{Discussion}
In this chapter, we discussed a new deep learning method, E2Efold, for predicting the RNA secondary structure. Instead of solving the task as an optimization problem, which is the legendary idea for resolving it, we creatively formulated it as a translation with constraints problem. Given the RNA sequence, we translate it into a binary contact map matrix, which has a one-to-one mapping against a specific RNA secondary structure, with the binary matrix satisfying some constraints. Those constraints are the prior knowledge we knew in advance about the task. This structured prediction problem can have the following challenges: 1) the lack of data; 2) tremendous output dimension and search space; 3) how to incorporate the problem-specific constraints into the method. To handle the above challenges, we used the following ideas. Firstly, we combined deep learning with constrained optimization. We used deep learning to model the complex mapping between the input sequences and the secondary structures. Then, we used constrained optimization to enforce the problem-specific constraints into the final results. As we have discussed in Section~\ref{chapter2_sec:ideas} and demonstrated here, such a combination can indeed incorporate the problem structures into the solution. Secondly, we designed a new deep learning architecture by coupling the Deep Score Network with the unrolled gradient descent algorithm, which achieved deep integration between the deep learning model and the classic algorithm. Such deep integration can resolve the above three challenges efficiently. First of all, the integration can embed the constraints into the deep learning architecture naturally.
Meanwhile, because we enforced the constraints into the deep learning model, the output space of the model can be reduced significantly. Furthermore, with the embedded constraints, the deep learning model would bias towards the desired distribution that we want it to learn. Such introduced bias can reduce the requirement of the training data size.  

With E2Efold, we illustrated a representative work of combining differentiable algorithm with deep learning to solve a classic 2D structured prediction problem in bioinformatics. In the next chapter, we will go from 2D to 3D, showcasing a framework for investigating the interactions between RNAs and RNA-binding proteins (RBPs).





\chapter{NucleicNet: A Deep Learning Framework to Predict Binding Preference of RNA Constituents on Protein Surface}
\label{chapter_nucleicnet}

\section{Chapter Introduction}
After transcription, mRNAs undergo a series of intertwining processes before being finally translated into functional proteins. These post-transcriptional regulations, which provide cells an extended option to fine-tune their proteomes, are in general mediated through interactions between RNAs and RNA-binding proteins (RBPs). In cells, RNAs are largely regulated by two modes of specific interactions -- either by direct recognition of RNA motifs on the RBP surface or by an indirect RNA-guided manner. In the former case, the RBP makes direct contact with the bases of RNA. For instance, the Pumilio/FBF (PUF) family can control translations via direct base-protein contact, e.g., with UGUR motifs on RNA transcripts \cite{RN1101}. In the latter case, the RBP interacts with backbone or non-Watson-Crick (WC) edges of the bases leaving WC-edges for target recognition. For example, in core enzymes of RNA interference (RNAi, e.g., Argonautes) and gene-editing complexes (e.g., CRISPR-Cas), selective loading of a guide-RNA (gRNA) into the RBP is a prerequisite to activate the enzyme; target D/RNA recognition is then mediated through the WC edges of gRNA while other parts of the gRNA remain in contact with the RBP. Therefore, deciphering the specificity and mechanisms in RNA-protein interactions is of fundamental importance to understanding the functions of RBPs, identifying RBPs, and designing RNAs for RBP recognition and regulation.

To approach systematic mapping of these interactions, various experimental and computational techniques have been developed. In the experimental genre, in vivo UV-crosslinking immunoprecipitation assays such as CLIP-HITS \cite{RN1102} and in vitro selection assays such as HT-SELEX \cite{RN1103} and RNAcompete \cite{RN1104} are among the most successful technologies. In general, specificity patterns obtained from these methods can be expressed as logo diagram for each RBP or as analytical scores for individual RNA sequences. Through structure elucidation techniques, binding mechanisms for many of these characterized RBPs, e.g., hnRNP, Nova and PAZ, have also been clarified \cite{RN1105,RN1106,RN1107}. However, despite such remarkable achievements, experimental assays are constrained by reactivity, detection and scalability limits. For instance, UV-crosslinking assays prefer uridine-rich sequences, because pyrimidines are more photoactivatable than purines \cite{RN1108}. Although arguably the chemical origin of these assayed specificities can be validated by ribonucleoprotein co-crystals, single or a few such co-crystals could hardly explain the genuinely ambiguous patterns on logo diagrams (e.g., specific to both U and A on the same position).

To this end, computational approaches can enhance experimental results. In this genre, the body of sampled experimental knowledge, assays and structures, can be refined to uncover previously mis-/un-acknowledged specificity patterns. Exemplary assay-based computational approaches, e.g., DeepBind and variants \cite{RN11}, can integrate and learn over assay data collected for an RBP to infer specificity pattern that is consistent with large-scale assays. There are also less explored structure- and sequence-based computational approaches \cite{RN1110,RN1111}. Typically, in these latter approaches, given a three-dimensional protein structure or its amino acid sequence, local protein sequence context among other structural information (e.g., solvent accessibility, secondary structure, hydrophobicity and electrostatic patches) can be extracted in units of residues and used to train models in reference to RNA-RBP structures in the Protein Data Bank (PDB). As such, the demand for experimental data to start with is relaxed from assay-based methods. However, due to the highly limited amount of available features, their predictive power is restricted to distinction of RNA-binding sites from non-sites, i.e., binary predictions made over locations or indices of protein residues without suggesting the preferred base/sequence nor any informative interaction modes (e.g., via backbone or base). Nevertheless, computational approaches are scalable and cost-efficient, thus are important complements to experimental techniques.

In this chapter, we introduce NucleicNet, a structure-based computational framework, which addresses topical challenges presented above: (i) we developed ways to learn efficiently from the PDB such that we can predict interaction modes for different RNA constituents -- Phosphate (P), Ribose (R), Adenine (A), Guanine (G), Cytosine (C), Uracil (U), and non-site -- and visualize them on any protein surface; (ii) NucleicNet requires no external assay input to derive logo diagrams consistent with assay data, including RNAcompete, Immunoprecipitation Assay, and siRNA Knockdown Benchmark; (iii) the logo diagrams or position weight matrices (PWMs) obtained from NucleicNet can be used to score the binding potential of individual RNA sequences; (iv) NucleicNet can generalize across different families of RBPs and be potentially used to identify new RBPs and their binding pockets/preferences. Our pipeline is founded upon the FEATURE vector framework \cite{RN1112}, which encodes physicochemical properties on protein surfaces as high-dimensional feature vectors. This rich vector space not only has covered most features developed in other programs, but can also account for subtle differences in local topologies via its discrete radial distribution setup. Importantly, learning from these high dimensional feature space is nontrivial, therefore a deep residual network is proposed and trained for this purpose. 

On the other hand, from the structured prediction aspect, the interaction between RBP and RNA is challenging in the following ways. Firstly, people have not tried to study the interaction details between them from the learning angle. As a result, we need to define this problem from scratch. Secondly, investigating two 3D structures at the same time can be very difficult. Because of the multiple binding pockets and the rotation and orientation issue, the searching space for finding out the optimal conformation with the lowest binding energy from the physicochemical point of view is almost infinite.
Furthermore, brute-force solutions for this 3D problem can lead to extremely high-dimension outputs, which means we would encounter the curse of dimension. Thirdly, the structural data are very limited, especially for the interaction between two molecules. As we will discuss below, after filtering the redundancy, we only have 483 RBP-RNA complex structures. Finally, how to incorporate the structural information and the local physiochemical information into the deep learning model remains to be an open research direction. Essentially, in our framework, we decompose this hard structured prediction problem into a large number of subproblems. After solving those subproblems with a deep learning model, based on those sub-solutions, we use a hidden Markov model (HMM) to reconstruct the final solution for the original problem. Below, we discuss the method and results in detail to illustrate the above ideas.


\begin{figure}[!hpbt]
\centering
\includegraphics[width=1.0\textwidth]{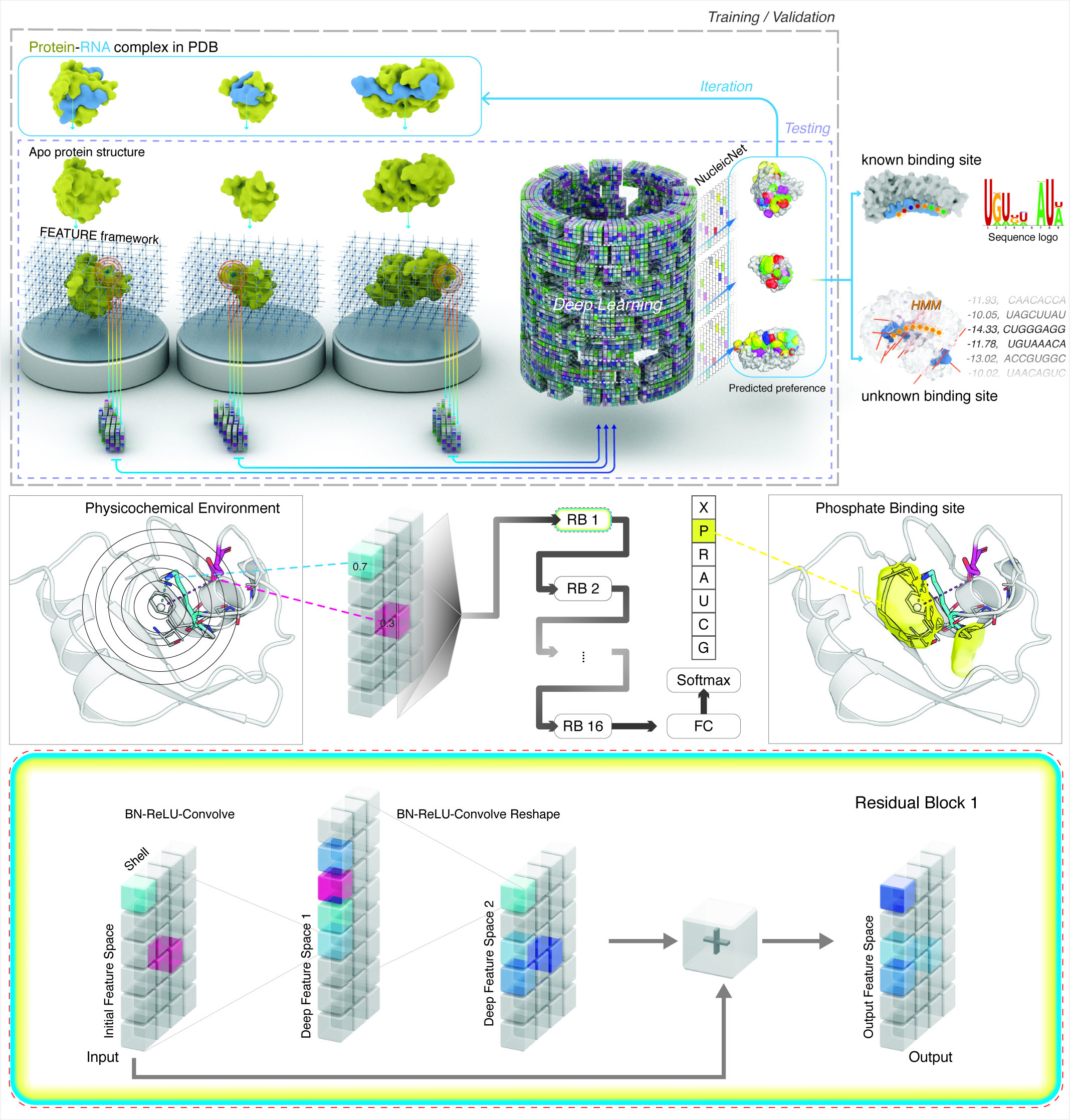}
\caption{Overview of NucleicNet. Top panel: training strategy and utilities of NucleicNet. Middle panel: physicochemical environment accession and introduction on the residual network. Bottom panel: the principle operations in a residual network.
}
\label{fig:overview}
\end{figure}

\section{Methods}

\subsection{Overview of NucleicNet} 
\begin{figure}[!hpbt]
\centering
\includegraphics[width=1.0\textwidth]{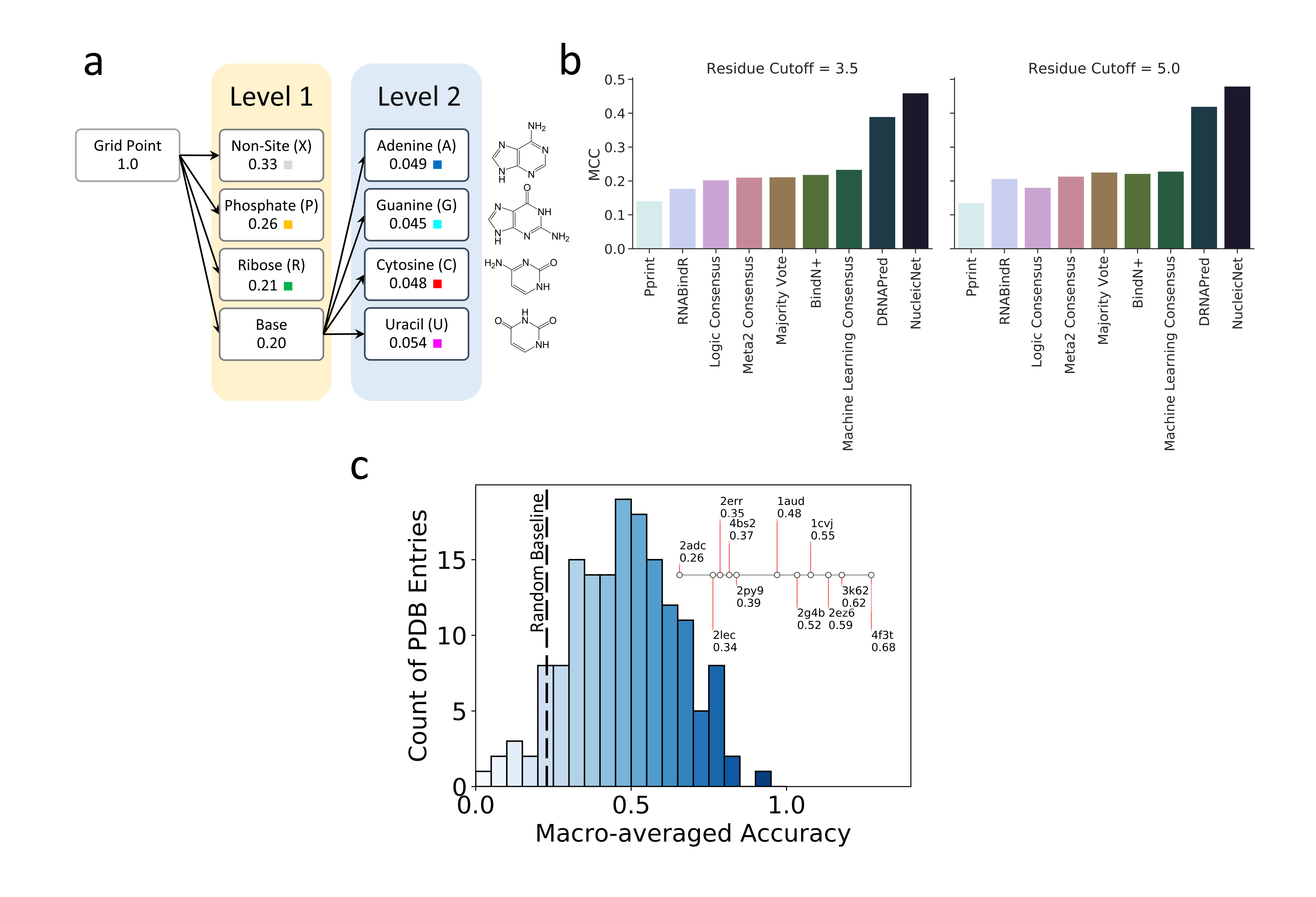}
\caption{Data statistics and performance of NucleicNet. 
}
\label{fig:performance}
\end{figure}

In NucleicNet, our goal is to predict on each location (grid point) of a protein's surface, whether the physicochemical environment presented on-site is fit to bind with an RNA and, if affirmative, the binding preference to each type of RNA constituent -- Phosphate (P), Ribose (R), Adenine (A), Guanine (G), Cytosine (C) and Uracil (U) -- that binds to the location. Computationally, we cast the problem as a supervised seven-class classification problem. Accordingly, we formulate the end-to-end training of NucleicNet as follows (Figure~\ref{fig:overview} top panel). First, surface locations on ribonucleoprotein complexes are retrieved from the PDB and typified as 7 classes that correspond to the bound RNA constituents and non RNA-binding site (X). Corresponding physicochemical environment on each location is then characterized using the FEATURE program \cite{RN1112} (Figure~\ref{fig:overview} middle panel). Next, a deep residual network is trained to associate each physicochemical environment with one of the 7 classes in a hierarchical manner (Figure~\ref{fig:performance}a). Finally, parameters of the network are optimized through standard backpropagation of the categorical cross entropy loss. Note that training data are entirely derived from three-dimensional structures in the PDB, i.e., we used no training data from external assays. Once training is completed for NucleicNet, raw surface characteristics extracted with FEATURE on surface location of query protein can then be fed forward to infer binding preference for each class on a location-by-location basis.

One strength that distinguishes our approach from related work is that not only binding sites of all 6 classes of RNA constituents are predicted and visualized on the surface of protein, but also, at the same time, these detailed results can be assimilated into logo diagrams or scoring interface for RNA sequences. As such, outcomes from the feed forward module are packaged into three utility modules: a Visualization module that indicates top predicted RNA constituents as a surface plot (Figure~\ref{fig:structure}a-c), a Logo Diagram module that generates the logo diagram when the RNA binding pocket on the protein surface is known (Figure~\ref{fig:logo}a-h), and a Scoring Interface module to apprehend binding score for a query RNA sequence (Figure~\ref{fig:logo}a-h, \ref{fig:exp}a-b), which can predict the most likely RNA sequence and the corresponding binding pocket on any query protein (Figure~\ref{fig:structure}a-c). The latter two modules can be summarized as a hidden Markov model (HMM), which incorporates both the locations of the bases and the geometric constraints for feasible RNA sequences. The Visualization module is used to compare our predictions with structural biology experiments. The Logo Diagram and Scoring modules are used to compare our predictions with in vivo or in vitro assay data.


In the following subsections, we shall give a summary of our methods. First, we describe how we represent the protein surface with grid points and physicochemical properties. We define how relevant non-redundant locations, corresponding to positive and negative examples of RNA-binding sites, are labeled and drawn from the PDB. Then, we describe how physicochemical environments on those locations are perceived by the FEATURE program, which formulates inputs for our deep learning network. Next, we examine the learning strategy and model architecture of NucleicNet to predict the binding class from those physicochemical environments. Finally, we explain how to infer the letter RNA sequences from the NucleicNet predictions.

\subsection{Dataset Construction}

\noindent {\bf Relevant Non-redundant Locations on Protein-RNA Complexes}

The surface locations that are 2.5 to 5.0 \angstrom{} away from any protein residue are established by three-dimensional coordinates of grid points on a cubic lattice spaced at 1 \angstrom{}. Relevant locations are selected from the surface locations by considering the topology of the local protein surface and also their bound RNA constituent labels. Non-redundant locations are retrieved by removing grid points associated with homologous proteins from the determined relevant locations. This strict strategy in collecting a relevant non-redundant dataset assures that the training and testing datasets are disjoint under cross validation (Figure~\ref{fig:performance}) and that the dataset does not carry prior information with respect to proteins.

To determine relevant surface locations, all ribonucleoprotein structures are retrieved from NPIDB \cite{RN1137}, an up-to-date server hosting RCSB Protein Data Bank (PDB) structures classified by their bound nucleotides (e.g. RNA-, DNA- or D/RNA). To define surface locations (surface grid points) on these PDB structures, Fpocket \cite{RN1138}, an alpha-sphere based external program, is adopted to mark out grid points on both buried and solvent-exposed protein surfaces. To provide positive examples of RNA constituent binding sites, geometric centroids of heavy atoms from each constituent are labeled. Surface grid points within 3 \angstrom{} of these labeled centroids and at most 5 \angstrom{} away from protein are considered positive relevant locations; each positive relevant location is labeled by a bound RNA constituent. Next, we consider locations where RNA-binding is unlikely. These negative relevant locations are provided by surface grid points selected randomly from space excluded by volumes within 3 \angstrom{} of any RNA atoms as well as alpha spheres from Fpocket. Note that the number of positive and negative relevant locations are balanced at ratio $2:1$ after the removal of redundant locations.

To remove redundant locations, data collected from the PDB are saturated with redundancy. Multiple copies of the same RNA-binding protein chain can exist within the same PDB entry due to the formation of homo- or hetero-multimeric complexes. Homologous chains can also be shared among different PDB entries dedicated to different bound RNA sequences, quality of resolved proteins and mutants, etc. We define the former situation as internal redundancy and the latter as external redundancy. Often, these homologous chains can share, to large extent, common RNA-binding configurations and physicochemical environments. Using redundancy-inclusive data for training and testing could introduce large bias to the evaluation and overstate the generalizability power of a model. Therefore, redundancy must be removed from the data.

To remove external redundancy, PDB entries are clustered into groups where each entry is linked with another that shares at least one RNA-bound chain with ≥90\% BLASTClust sequence homology; for each cluster, the PDB entry with the best global resolution is selected. In this way, 483 valid PDB entries becomes 158 cluster and each cluster contributes only one entry to the dataset. In addition, if the selected entry contains multiple copies of the same protein/RNA chains (i.e., internal redundancy), only grid points adhering to the best locally resolved RNAs are retained; grid points adhering to homologous protein chains are also discarded. Local resolution is defined by the average of B-factors on atoms of RNAs; grid points are assigned to adhere the closest RNA/protein residue. Note that the remaining non-redundant grid points are characterised in presence of the internal-redundant protein chains to preserve the intact physicochemical environment. In total, around 280k data points are compiled from the valid PDB entries; two-thirds of which are positive examples. The data points are randomly split into three disjoint folds that disallow both external and internal redundancy, even though members of the same BLAST group can exist within in the same training fold to maximize availability of training data. Note that testing is performed on data points contributed only by the representative member of each BLAST group, where in all three folds, there are 80k such data points.

\vspace{\baselineskip}
\noindent {\bf Capturing Physicochemical Environments with FEATURE}

RNA-protein interactions are maintained by physical forces and properties (e.g., electrostatics, hydrophobicity, solvent accessibility, etc.), but the origin and strengths of these interactions are determined by a varied spatial arrangement of chemical components and atoms on the protein surface (e.g., charged residues, hydrogen bond donors/acceptors, etc.). These complicated topological features, which we summarized as physicochemical environments, can be maneuvered into a feature vector, and by leveraging the power of deep learning, to predict RNA binding partners on protein surface locations -- this is the foundation that underlies our NucleicNet method.

In this work, the FEATURE vector framework \cite{RN1112} is adopted to perceive physicochemical environments on three-dimensional protein surfaces. Previously, this framework have been applied to predict cation \cite{RN1139,RN1140,RN1141,RN1142} and ligand/fragment binding sites \cite{RN1143,RN1144,RN1145}. In those studies, it has been shown as an effective implementation to describe similar binding sites shared by proteins with little structural or sequence resemblance. In contrast to other vector frameworks used by preceding structure-/sequence-based studies \cite{RN1146,RN1111}, where physical/structural features (at max 60 in total) are accounted in units of residue regardless of their spatial distribution, our physicochemical features are accounted in units of atoms and their discrete radial distribution over a location \cite{RN1112} (Figure~\ref{fig:overview} middle panel). As such, these features, 480 in total, preserve a much wider range of details (including atom types, elements, residues, functional groups, secondary structures, charges, hydrophobicity, solvent accessibility, etc. and, their radial distributions) than any other vector framework. This all-rounded information about physicochemical environments is indispensable for resolving subtle differences among RNA base- and backbone-binding sites (Figure~\ref{fig:performance}a, \ref{fig:structure}a-c). It has allowed us not only to tell the spatial region of RNA-binding as in other previous studies, but also to classify these binding sites into 6 different RNA constituents and deduce specificity towards the RNA bases.

To summarize, after obtaining a set of labeled relevant non-redundant locations, their protein-related physicochemical environment is then characterized under the FEATURE framework in absence of nucleic acid, solvent, substrate and ions. Hence, each of these locations is annotated by a FEATURE vector and a label that indicates the binding class, and our NucleicNet is trained to predict the label from the FEATURE vector.

\subsection{Hierarchical Classification of Physicochemical Environments}

In NucleicNet, our goal is to predict on each location of a protein surface, whether the physicochemical environment presented on-site is fit to bind with an RNA and, if affirmative, the most likely type of RNA constituent that binds to the location. This is a multi-class classification problem for which end-to-end training is possible, where the 7 attainable classes are Phosphate (P), Ribose (R), Adenine (A), Guanine (G), Cytosine (C), Uracil (U) and Non-binding site (X). However, as positive examples of backbone constituents (P and R) are 4 to 5 times more abundant than that of the nucleobases (A, U, C, G), straightforward deep learning model training suffers from the serious class imbalance problem \cite{RN1147} (Figure~\ref{fig:performance}a). To alleviate the situation, we therefore adopt a hierarchical classification scheme (Figure~\ref{fig:performance}a) that balances the data. In the first level, the grid point is classified by a 4-class coarse model, where attainable classes are Base, Ribose, Phosphate and Non-site, producing a normalized multi-label 4-class score vector. The training of this model requires merging data-points annotated with A/U/C/G to Base. This alleviates the class imbalance problem. To distinguish the four bases A/U/C/G, a second level classifier is compiled, which does not suffer from the class imbalance problem. A final normalized multi-label 7-class score vector is produced by multiplying the second level outcome (also normalized) with the Base prior from the first level. Consequently, based on such hierarchy, two models are built for the entire problem: one for predicting 4 coarse classes and the other for distinguishing the four bases. The model architecture common to learners at both levels will be introduced in the next subsection.

\subsection{Model Architectures}

Architectures of neural networks have been evolving along the development of the deep learning field. From the legendary AlexNet \cite{RN69} to cutting-edge architectures, such as residual networks (ResNet) \cite{RN1149} and generative adversarial nets (GAN) \cite{RN1150}, each of these architectures was designed to push forward the limit of prediction accuracy and resolve specific problems encountered in training on specific categories of data. In this work, considering the complexity of the problem and the convergence rate of the model, ResNet is chosen as our basic unit architecture due to its ability in handling the gradient vanishing problem, which obstructs extensive training of baseline multi-layer convolutional neural network models when deep networks are compiled. Our model is comprised of 16 residual blocks, a fully connected (FC) layer and a final Softmax layer to make a 4-class probability prediction. The residual blocks are considered as the feature extractor and the FC-Softmax is the classifier. In total, 32 convolutional layers are compiled, where each residual block contains 2 convolutional layers. The input tensor from the FEATURE program is of the shape $1*6$ shells $*80$ physicochemical properties. In the convolutional operation, a shared filter of size $1*2*80$ slides across the input, generating an inner product at each position as an intermediate output, which then goes through batch normalization (BN) and element-wise non-linear activation, in our case, rectified linear units (ReLUs) \cite{RN1129}, to produce the intermediate output. The use of the BN layer mitigates the internal covariate shift problem \cite{RN1151}. In total, 80 filters are used. Note that to enable a consistent size in the output tensor ($1*6*80$), the input tensor is zero-padded. In each residual block, an identical shortcut is added to allow learning of the residual between the input and the second intermediate output. The output from the final residual block is later flattened and fed to a fully connected layer to make 4-class probability prediction in the final Softmax layer. All parameters in the network are optimized, with weight decay, under Adam using categorical cross-entropy as the loss function. Training is implemented with TensorFlow. In general, it takes 4 days to train the model at all levels on a Titan X GPU. We also compared alternatives to ResNet in the NucleicNet predictor e.g. shallow machine learning methods and neural networks that do not consider spatial information. We find that on the grid level prediction, the proposed model, NucleicNet, outperform all the shallow methods as well as other deep learning architectures under the same experimental setting. Alternative machine learning strategies were also considered, e.g., MAX-AUC \cite{RN1152} and ensemble learning with data sampling \cite{RN1153,RN1154}, though issues in run time and overfitting were experienced.

\subsection{Scoring from Deep Learning Outputs}

\noindent {\bf Obtaining Sequence Logo with Predetermined Base Locations}

The feed-forward module of NucleicNet annotates each grid point with a normalized score vector that indicates predicted binding probability with respect to the 7 attainable classes on that location. For RBPs with predetermined ribonucleoprotein structures (e.g., those compared with the RNAcompete assay in Figure~\ref{fig:logo}), sequence logo diagrams can be easily generated by considering location $i$ of centroid for the corresponding nucleobase. As such, the NucleicNet score vectors predicted on grid points within 3 \angstrom{} of each base centroid are averaged to produce an averaged binding probability $p_i$. Information content $E_i$  on each base position $i$ is then accounted in terms of the following equation, where $p$ is the averaged binding probability on base position $i$ for class $c$:

\begin{gather}
E_i = \log_27+\sum_{c=\{AUCGPRX\}}p_i(c)\log p_i(c).
\end{gather}

A sequence logo diagram can then be generated by proportioning the information content $E_i$  according to $p_i(c)$. Class P, R and X, corresponding to Phosphate, Ribose and Non-RNA Binding Sites, are omitted from the logo diagram. Note that a gap is automatically assigned when location $i$ is $\geq$5\angstrom{} away from the protein.

\vspace{\baselineskip}
\noindent {\bf Scoring RNA Letter Sequence for Ago2}
\label{chapter6_sub:hmm}

Similar to the idea of applying position weight matrix scores (PWM scores) to study DNA sequence-specific binding of transcription-factors \cite{RN1155}, the results of NucleicNet for individual protein surfaces can be summarized as an equation $Q$ to score an arbitrary RNA letter sequence input: 

\begin{gather}
Q = \max \sum_{i}^{N} \log_2(p_i(b)T_{i,i+1}).
\end{gather}

This equation, which we refer to as a fixed hidden Markov model (HMM), is comprised of an emission probability $p_i$ and a transition probability $T_{i,i+1}$. Our goal is to assimilate NucleicNet outputs via $p_i$ and $T_{i,i+1}$ and to consider geometric constraints put forward by the covalent bond network and the torsional space of genuine RNA strands. The hidden states are locations indexed by $i$ of bases relevant to a continuous RNA strand bound to the RBP with the letter sequence of length $N$.The emission probability $p_i(b)$, referring to the binding probability of base b on the RNA sequence, is obtained by averaging the NucleicNet output within 3 \angstrom{} of the base location $i$. Note that $p_i(b)$  here is normalized among the bases. The transition probability $T_{i,i+1}$ refers to the transition probability between bases $i$ and $i+1$ on a continuous RNA strand from 5$^{\prime}$ to 3$^{\prime}$ end. In case, base locations are predetermined by ribonucleoprotein co-crystals, transitions between consecutive bases as well as their locations $i$ are certain, then $T_{i,i+1}=1$ and $p_i(b)$ can be deduced by averaging the NucleicNet output on locations just as we generate logo diagrams. The equation $Q$ is then reduced to an ordinary PWM scoring function; RNA string sequences of length $N$ are then evaluated by sliding across the co-crystal-native RNA strand locations to obtain a maximum in $Q$, which is implemented to calculate the NucleicNet score for comparison with the RNAC score in Figure~\ref{fig:logo}a-h.

We also investigate the situation where base locations referring to a continuous RNA strand are unknown but NucleicNet predicted RNA-binding sites are clearly directed by a phosphate-ribose backbone (e.g., in RNA-guided situations, for instance, Ago2 in Figure~\ref{fig:exp}). In this case, score $Q$ cannot be easily generated as in the case of co-crystals because those hidden locations and their transition probabilities $T_{i,i+1}$ are unknown, even though $p_i(b)$ can still be calculated from the NucleicNet outputs around location $i$ once locations are approximated. In the next paragraph, we outline how these unknowns can be efficiently estimated by aligning top predicted binding sites of RNA constituents with a conformational library of RNA trinucleotides. In this case, the score $Q$ can then be obtained by maximizing over all possible $i,i+1$ transition paths, when an RNA letter sequence of length N is enquired.

A continuous RNA strand may be considered as a transition graph between locations of consecutive bases, where base identities can be expressed by an emission probability $p_i(b)$  on each node indexed by a location i referring to the location of a base b on an RNA strand bound to an RBP. In case, where these locations are hidden, the transition probability $T_{i,i+1}$ is unknown. However, these transitions are certainly constrained, irrespective of the strand length, by the covalent bonds and the torsional space of the RNA \cite{RN1156,RN1157}. Therefore, they can be estimated by screening a database of RNA geometries that are tolerated by series of predicted RNA-binding sites on the RBP surface. In particular, for cases where predicted RNA-binding sites are clearly directed by a phosphate-ribose backbone (e.g., in Ago2 where the RBP is known to work in an RNA-guided manner), this trail of backbone binding sites and intermittent base binding sites are visually indicative for a continuous RNA strand. To efficiently screen out binding sites relevant to a continuous RNA strand in this case, top 10\% of binding sites reported by NucleicNet are aligned with a non-redundant library of trinucleotide conformations adopted from previous publication \cite{RN1156}. This library was compiled from ribonucleoprotein complexes in the PDB by binning over the pseudo-torsional space of RNA backbones \cite{RN1156}, from which, the $15^{\circ}$-bin library containing 296 conformers is chosen for our purpose. To compile a comprehensive trinucleotide conformer library, the $15^{\circ}$-bin library is permuted to cover all the $4^3$ possible trinucleotide sequences in atomic details for each conformer; the resultant 18944 trinucleotide conformers are optimized briefly under a AMBER99SB-ILDN force field \cite{RN1158} to assure proper geometry. Finally, these trinucleotides are reduced to centroids of their RNA constituents (nodes) resulting in some 9-nodes coarse-grained models ready to be aligned with the top binding sites. The clique-alignment process is done with a Bron-Kerbosch algorithm \cite{RN1159}, where only $\geq$7-cliques that show no atomic clash with the protein are retained. The 7-clique is chosen such that transition between consecutive $i,i+1$ bases (i.e., 2 Base nodes on the 9-node model) must be guided by at least 5 backbone constituents. These criteria assure that the proposed binding sites are geometrically feasible. To systematically assess how these aligned 3-mers can contribute to a continuous strand, we formulate the problem as a fixed HMM. Hypothetical base locations are the hidden states. To propose these locations, the Euclidean space covered by the aligned Base nodes is partitioned into multiple Voronoi cells seeded by k-means centers. To express the identity of the base, these Voronoi cells, each representing a hypothetical base location, are characterized by emission probabilities $p_i$  averaged from grid points within 3 \angstrom{} of a k-means center. Then, transitions, regarding consecutive bases within the same aligned clique, between different Voronoi cells are counted and symmetrized as an estimate of transition probability $T_{i,i+1}$. In case of Ago2, since it is ascertained that the 5$^{\prime}$ location is situated in the Mid domain \cite{RN1120}, a certain starting probability of one is assigned to a cell located in the Mid domain that is furthest away from any other cells. The 5$^{\prime}$ to 3$^{\prime}$ direction of transition is then ascertained by the ranking distance to this starting Voronoi cell; direction of the edge on the transition graph allows only transition from a high rank to a low one. With $p_i$ and $T_{i,i+1}$ affixed, score $Q$ can then be calculated using the equation presented above.

\section{Results}
\subsection{Evaluation from Structural Perspectives} 
Various reliable ground truths can be extracted from known structure information of ribonucleoproteins structures deposited in the PDB. First, we start with distinguishing RNA-binding residues from non-RNA-binding ones, i.e., a binary classification. This is the a classical problem tackled by most computational predictors on protein-RNA interaction \cite{RN1111}. Although NucleicNet was never trained on such a binary classification task, we converted our 7 class prediction into the binary one to compare with the state-of-the-art methods because there is no existing work on predicting detailed binding preference for RNA constituents as NucleicNet does. In general, a protein residue is considered RNA-binding in a co-crystal if at least one of its atoms is within a certain distance from atoms of the RNAs. In a recent review \cite{RN1111}, both 3.5 \angstrom{} and 5.0  \angstrom{} cutoffs were considered. The benchmark RNA\_T dataset \cite{RN1111} proposed therein, which consists of 175 RNA-binding protein chains, was generated by clustering protein chains with respect to their sequence and structural similarities, where annotations of RNA-binding residues were transferred among similar chains to alleviate effects of strand truncations \cite{RN1111}. Based on this ground truth, we benchmarked NucleicNet with a broad range of state-of-the-art predictors based on sequence information (Figure~\ref{fig:performance}b). To assign a binary label (site or non-site) on each protein residue using our NucleicNet predictor that works on grid points over the protein surface, score vectors on 30 grid points closest to a protein residue were taken to vote for 2 coarse classes, namely `RNA-binding site' and `non-site'; the 6 finer classes (that correspond to individual RNA constituents) are considered `RNA-binding site'. Testing benchmark proteins \cite{RN1111,RN1113} are omitted from training. At both aforementioned ranges of distance cutoffs, NucleicNet outperforms all available methods \cite{RN1113,RN1114,RN1115,RN1116} (Figure~\ref{fig:performance}b). This therefore demonstrates basic utility of NucleicNet as a tool to predict general RNA-binding sites.

\begin{figure}[!hpbt]
\centering
\includegraphics[width=1.0\textwidth]{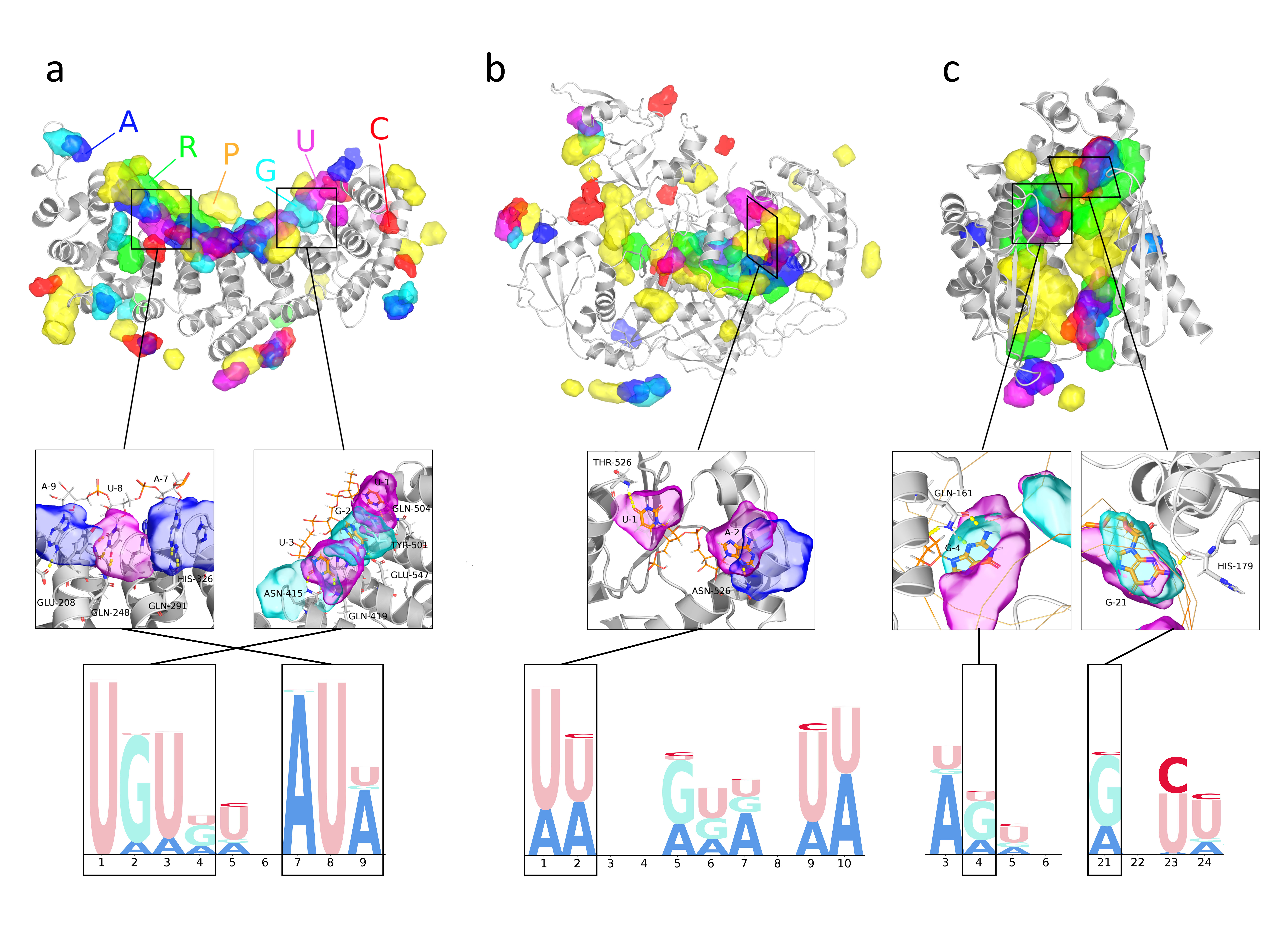}
\caption{NucleicNet prediction captures detailed binding motion determined by structural biology experiments. 
}
\label{fig:structure}
\end{figure}

\begin{figure}[!hpbt]
\centering
\includegraphics[width=1.0\textwidth]{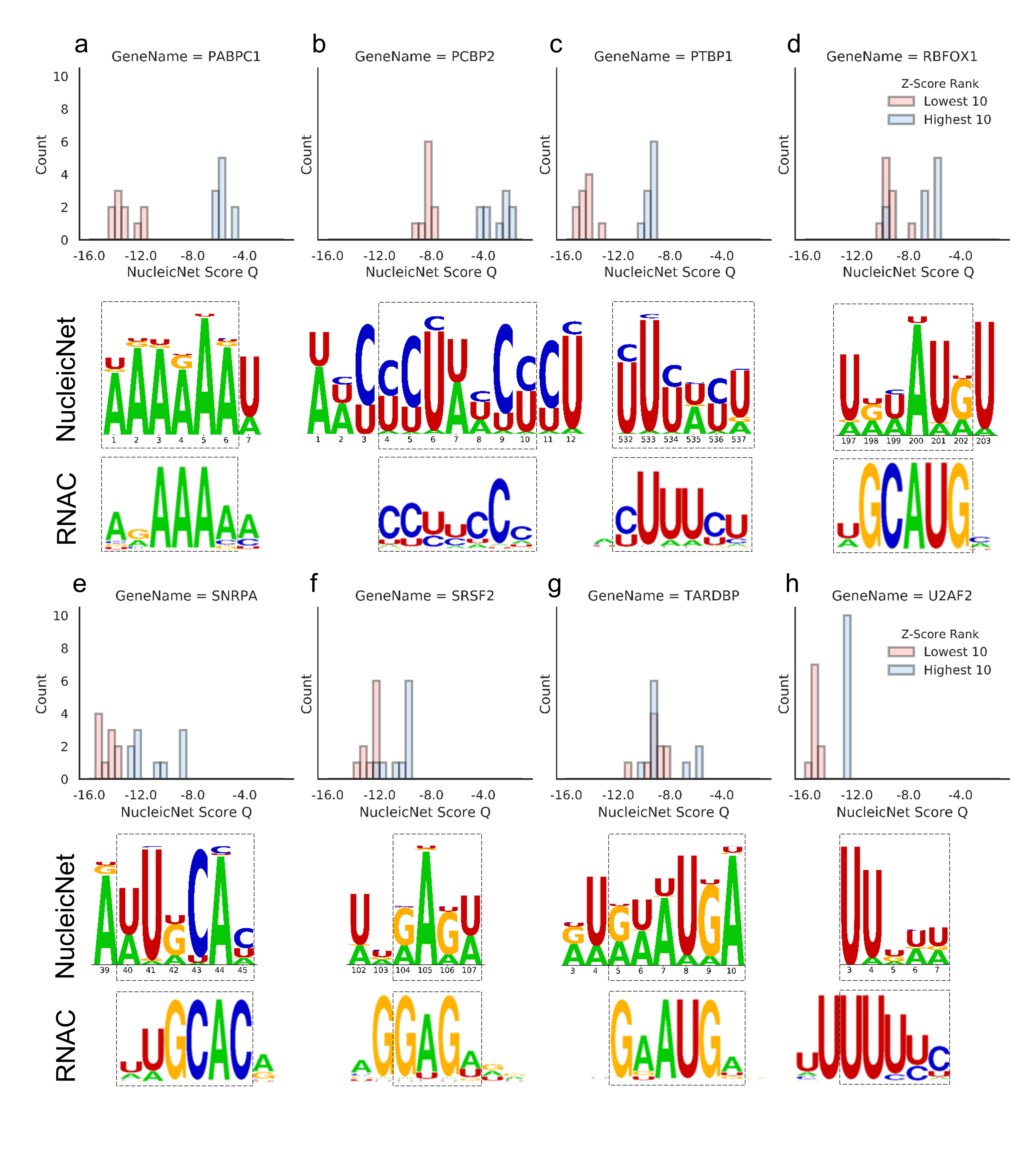}
\caption{Comparing scores and logo diagrams of NucleicNet with those obtained from RNAcompete (RNAC) assay.}
\label{fig:logo}
\end{figure}

Next, we evaluate on NucleicNet's ability to retrieve binding sites for the 6 detailed RNA constituents proposed; this includes `Phosphate' (P), `Ribose' (R), `Adenine' (A), `Guanine' (G), `Cytosine' (C), and `Uracil' (U). A three-fold cross-validation was performed over a carefully selected and curated non-redundant dataset from all protein-RNA complex structures from PDB, which consists of 158 complex structures, resulting in about 280,000 grid points in the dataset. We divided the 158 proteins into three folds. Each time, two folds of them were used for training and one fold for testing. Between folds, BLASTClust sequence homology of ≥90\% was disallowed. Notice that the granularity of this cross-validation is individual proteins, instead of grid points, which eliminates bias in the size of proteins. Table~\ref{tab:stat_cv} reports the performance in terms of AUROC, F1-score, Precision and Recall for each class. For the bases (A/U/C/G), an AUROC of 0.66 can be achieved in average. Remarkably, the power to differentiate sites and non-sites is recapitulated in an AUROC of 0.97. Categorical accuracy with respect to each protein is also calculated. A distribution of the accuracy score is shown in Figure~\ref{fig:performance}c; proteins covered in case studies (Figure~\ref{fig:structure}a-c, \ref{fig:logo}a-h) are marked out with their PDBID on the inset line diagram to indicate their performance, which shows that the accuracy of the case studies spreads over a wide range. In general, a median accuracy of 49\% is achieved in the non-redundant 3-fold cross validation (c.f. random baseline 23\%). This proof-of-principle analysis therefore demonstrates that NucleicNet can learn from a diverse structural database of physicochemical environment and generalize to unseen RBPs to recall potential binding RNA constituents, provided that structure of the elucidated protein is largely intact and contains relevant RNA-binding domains.  


\begin{table}[!hpbt]
\caption{Statistics of performance in cross validation of the non-redundant dataset from PDB.}
\label{tab:stat_cv}
\centering
\resizebox{\textwidth}{!}
{
	\begin{tabular}{cccccccc} 
	\hline
	Metrics & NonSite & Phosphate & Ribose & Adenine & Guanine & Uracil & Cytosine\\
	\hline\hline
	AUROC & 0.97 & 0.93 & 0.84 & 0.67 & 0.67 & 0.65 & 0.66\\ 
	\hline
	F1-score(macro) & 0.90 & 0.70 & 0.63 & 0.47 & 0.38 & 0.48 & 0.32 \\
	\hline
	Recall(macro) & 0.88 & 0.82 & 0.63 & 0.46 & 0.38 & 0.45 & 0.37 \\
	\hline
	Precision(macro) & 0.92 & 0.61 & 0.63 & 0.48 & 0.38 & 0.51 & 0.29 \\
	\hline
	F1-score(micro) & 0.90 & 0.70 & 0.64 & 0.47 & 0.41 & 0.48 & 0.32\\
	\hline
	Recall(micro) & 0.88 & 0.81 & 0.64 & 0.47 & 0.40 & 0.46 & 0.36\\
	\hline
	Precision(micro) & 0.92 & 0.61 & 0.64 & 0.48 & 0.41 & 0.51 & 0.29\\
	\hline
	\end{tabular}

}

\end{table}

\subsection{Reproducing Spatial Patterns of RNA-binding Sites}
One strength that structure-based methods offer is their potential to reveal and visualize binding sites on protein surfaces. While previous structure-based methods concern only binary classifications (sites and non-sites), our method can illustrate further on all six common RNA constituents -- `Phosphate' (P), `Ribose' (R), `Adenine' (A), `Guanine' (G), `Cytosine' (C) and `Uracil' (U). We demonstrate this unique power of our method via three exemplary RBPs: Fem-3-binding-factor 2 (FBF2, PDB Entry 3k62, Figure~\ref{fig:structure}a), Human Argonaute 2 (hAgo2, PDB Entry 4f3t, Figure~\ref{fig:structure}b) and Aquifex aeolicus Ribonuclease III (Aa-RNase III, PDB Entry 2ez6, Figure~\ref{fig:structure}c). FBF2 is an example from RBPs that interact directly with single-stranded RNA (ssRNA) motifs through base contacts, while the hAgo2 is an example from RBPs that functions in a RNA-guided manner through backbone or non-WC edge contacts. The third example, Aa-RNase III, involves double-stranded RNA binding domain (dsRBD). In Figure~\ref{fig:structure}, we indicate top predicted binding sites on these proteins for each binding class using our visualization module. In all cases, predictions were made on the protein structure after removing RNAs from the ribonucleoprotein complex. These proteins and their homologues were all excluded from the training process. In Figure~\ref{fig:structure} middle panel, we show that strong preference for nucleobases are mostly found at places where nucleotides interact explicitly with protein residues when superposed on a ribonucleoprotein structure. In Figure~\ref{fig:structure} lower panel, sequence logo diagrams were generated by averaging the NucleicNet score at the nucleobase locations on the long native RNA strand. In all cases, we show that NucleicNet has reproduced the detailed binding specificity captured by structural biology experiments.

\paragraph{Fem-3-binding-factor 2 (FBF2).} The PUMILIO/Fem-3-binding-factor (PUF) family of RBPs are important post-transcriptional regulators. In a typical PUF-mRNA interaction, the PUM-HD domain, common among all PUFs, will bind to the 3$^{\prime}$ untranslated region of mRNA that contains a conserved UGUR sequence motif. The strong sequence specificity is mediated through direct interactions (aromatic stacking and hydrogen bonds) made between protein surface residues and RNA nucleobases. The Fem-3-binding-factor 2 (FBF2) is one of the best-characterized PUF family proteins. In Figure~\ref{fig:structure}a middle panel, we show that interacting surface indicated by NucleicNet at Q504/Q419/N415/E542/Y501 and Q248/Q291/E208/H326 of FBF2 largely involves hydrogen bond donors or acceptors on the PUM-HD repeats. The respective sequence logo diagram derived from these locations (Figure~\ref{fig:structure}a lower panel) indicates a strong sequence preference at base 1-4 and 7-8 that is consistent with the 5$^{\prime}$-UGUR and downstream A7-U8 pattern reported previously. In addition, NucleicNet also correctly captures the modest preference for A or U (A $>$ U $>$ G) at base 9 consistent with the consensus reported by yeast three-hybrid assays \cite{RN1117,RN1118}, even though the crystal-bound native base at that position is a C. This therefore suggests that NucleicNet is able to reveal underlying sequence specificity patterns unseen in crystal structures and in the absence of third-party assay data.

\paragraph{Human Argonaute 2 (hAgo2).} Human Argonaute 2 (hAgo2) is an exemplary RBP that operates in an RNA-guided manner, where the guiding RNA strand can be a small interfering RNA (siRNA) or a micro RNA (miRNA). In cells, both of these RNAs pre-exist as a duplex of complementary single-strands. However, during assembly of the RNA-induced silencing complex (RISC), often one of the strands is preferentially loaded into hAgo2 to guide cleavage of the target RNAs. This asymmetric behavior is heavily affected by small changes in RNA sequences of the precursor duplex \cite{RN1119}. Two attributing factors were identified: 1) weakening of the base pair at one of the 5$^{\prime}$ ends, this decides which strand will unwind at its 5$^{\prime}$ end and subsequently enter the RISC complex \cite{RN1119}; (2) guiding RNA-hAgo2 interactions at base 1 (Figure~\ref{fig:structure}b middle panel) and the non-Watson-Crick edges of base 2-8 (the seed region) (Figure~\ref{fig:structure}b), these interactions are hypothesized to lower the enthalpic cost of RISC assembly \cite{RN1120,RN1121,RN1122}. However, compared to intensive studies on target RNA recognition by the RISC complex, the second factor, concerning RISC assembly that correlates loading and knockdown efficiency with guiding RNA-protein interaction, is much less explored.

\begin{figure}[!hpbt]
\centering
\includegraphics[width=1.0\textwidth]{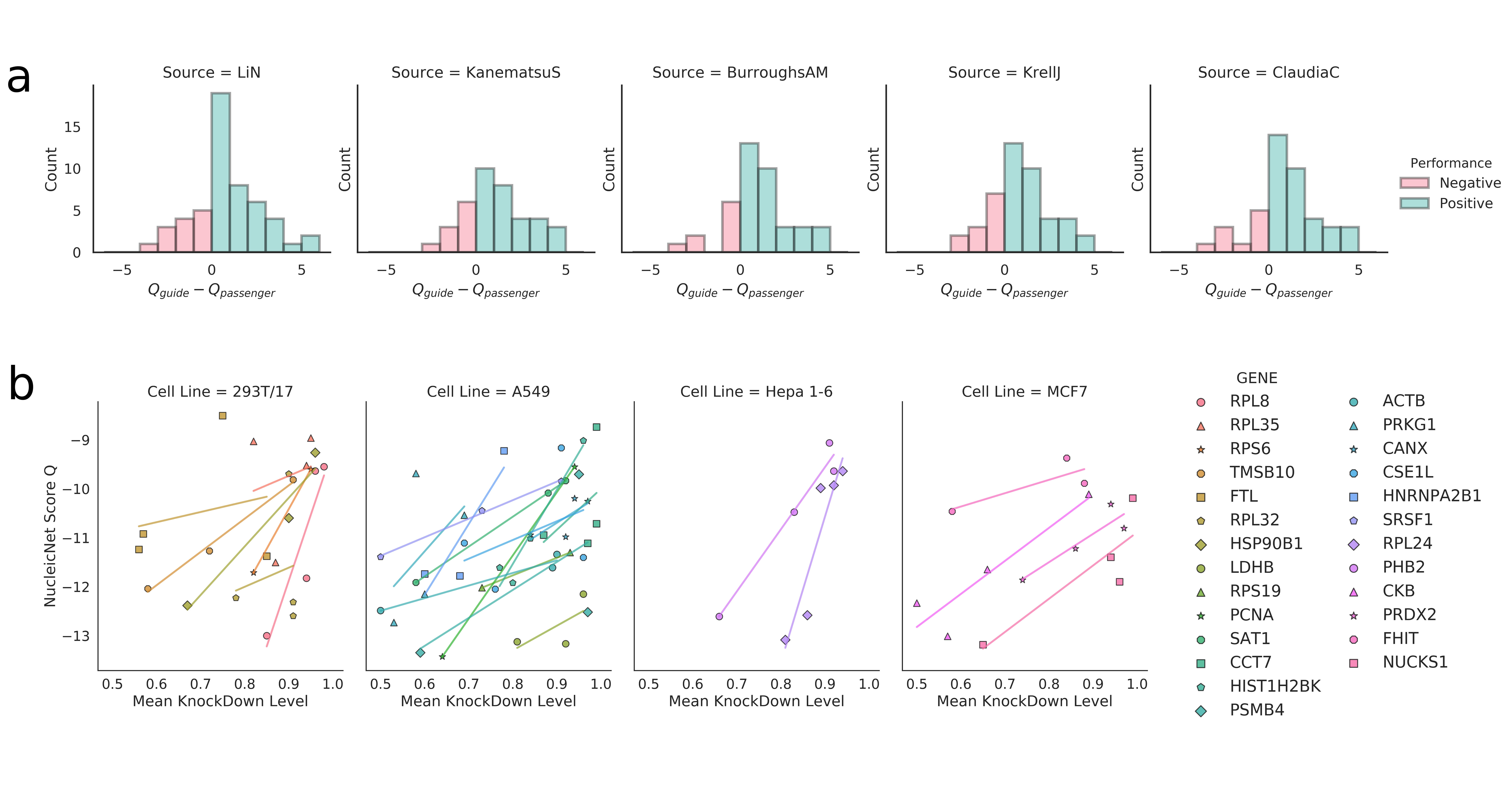}
\caption{NucleicNet predictions agree with in vivo experiments on Ago2.}
\label{fig:exp}
\end{figure}

In Figure~\ref{fig:structure}b upper panel, we show that binding sites of the guide strand, including the phosphate-ribose backbone around PAZ and N domains, are correctly captured by NucleicNet. Specifically, in Figure~\ref{fig:structure}b middle and lower panels, we focus on the 5$^{\prime}$-end binding pocket on the Mid domain and show that NucleicNet correctly predicts a strong U-binding pocket (U $>$ A $>>$ C/G) at base 1 and a U/A binding pocket (U $=$ A) at base 2. The first preference on base 1 and its order are well supported by structural evidence and NMR titration experiments performed using nucleoside monophosphates (mimics of the 5$^{\prime}$ end, UMP (0.12 mM) $>$ AMP (0.26 mM) $>>$ CMP (3.6 mM) / GMP (3.3 mM)) \cite{RN1120}. For other binding preferences in the seed region, only structural evidence is available and it is scattered among different PDB entries containing different seed sequences. For example, in the PDB entry 4f3t, A2 and G5 interact with N562 and Q757 respectively; in PDB entries 5js1/5t7b, U2 interacts with N562. These results are consistent with the logo diagram provided by NucleicNet (Figure~\ref{fig:structure}b lower panel). We show later in Figure~\ref{fig:exp}a-b that these NucleicNet predictions are supported by immunoprecipitation experiments and knockdown assays affirming that guide loading efficiency and sequence-protein interactions are correlated.

\paragraph{Aquifex aeolicus Ribonuclease III (Aa-RNase III).} Double-stranded RNA binding domain (dsRBD) is a domain that widely occurs among double-stranded-RNA-specific endoribonucleases, including the Aa-RNase III presented here. Originally, recognition of RNAs in dsRBDs were thought to be shape-dependent rather than sequence-specific. However, recent structural evidence confirms that this domain can recognize bases by interacting with the minor groove \cite{RN1123}. In Figure~\ref{fig:structure}c middle and lower panels, we show that NucleicNet has correctly predicted two strong G-binding sites concentrated around H179 and Q161, corresponding to the first $\alpha$ helix and the loop between $\beta$ strands 1 and 2 of the dsRBD, which agree well with the existing co-crystals. 

\subsection{Validation with In Vitro Assay Data}
To validate NucleicNet on RBPs that directly recognize RNA motifs on its surface, we compare the NucleicNet score with scores obtained from the RNAcompete assay (RNAC) \cite{RN1124,RN1125}. RNAC is a large-scale in vitro experiment that uses the epitope-tagged RBP to competitively select RNA sequences from a designed pool. For each RBP, 7-mer RNA-binding profiles obtained can be summarized as a Z-score for the individual RNA sequence or as a PWM by aligning the top 10 scoring sequences. Higher Z-score indicates better binding. We tested NucleicNet on all the RBPs for which both RNAC data and PDB structures are available (PABPC1, PCBP2, PTBP1, RBFOX1, SNRPA, SRSF2, TARDBP, and U2AF2). In all cases (Figure~\ref{fig:logo}a-h, Table~\ref{tab:stat_perf}), a Welch's t-test is performed and shows that NucleicNet is capable of differentiating between the top and bottom 10 sequences indicated by RNAC Z-scores with a positive test-statistics and p-value $<$ 0.005 except for TARDBP, where its RNAC binding profile is specific to a single sequence. In all cases, NucleicNet is capable of differentiating the sequences, although it was never trained on any assay data. This therefore suggests that the NucleicNet score is predictive and is suitable to complement selection assays.

Interestingly, NucleicNet is able to predict binding preference that is beyond structural biology information in PDB. For example, all the three PDB entries for protein PTBP1 (PDBID: 2adc, 2adc and 2ad9) are bound with the RNA sequence CUCUCU, which deviates from the RNAC suggested sequence YUUUYU (Table~\ref{tab:stat_perf}). This suggests that single or few PDB co-crystal structures may not inform about RNA binding preference comprehensively. However, by integrating with other PDB data through training, NucleicNet predicts a suggested sequence of UUUWYU in reasonable agreement with the RNAC sequence (Figure~\ref{fig:logo}c), which indicates its ability to make predictions that are not present in the training data. Accordingly in these cases, NucleicNet can have low accuracy scores with respect to PDB structural data. (accuracy 0.26 as in Figure~\ref{fig:performance}c inset 2adc) Another example is protein RBFOX1, for which there are only two deposited PDB entries 2err (with RNA sequence UGCAUGU) and 2n82 (with RNA sequence GGCAUGA). Even so, NucleicNet can correctly predict U/A at the first position with a dominant U, which is in agreement with the RNAC suggested sequence (Figure~\ref{fig:logo}d).


\begin{table}[!hpbt]
\caption{Statistics of performance and suggested sequences from NucleicNet and RNAcompete (RNAC). Best matching suggested sequences between RNAC and NucleicNet are underlined. R: A/G, M: A/C, Y: C/T, H: A/C/T, W: A/T, D: A/G/T, and N: A/C/G/U.}
\label{tab:stat_perf}
\centering
\resizebox{1.0\textwidth}{!}
{
	\begin{tabular}{|c|c|c|c|c|c|c|c|c|} 
	\hline
	For Figure~\ref{fig:logo} & a & b & c & d & e & f & g & h\\
	\hline\hline
	Gene name & PABPC1 & PCBP2 & PTBP1 & RBFOX1 & SNRPA & SRSF2 & TARDBP & U2AF2\\ 
	PDBID & 1cvj & 2py9 & 2adc & 2err & 1aud & 2lec & 4bs2 & 2g4b \\
	RNAC ID & 155 & 44 & 269 & 168 & 71 & 72 & 76 & 79 \\
	RNAC sequence & \underline{ARAAAA}M & \underline{CCYYCCH} & H\underline{YUUUYU} & \underline{WGCAUG}M & \underline{WUGCAC}R & G\underline{GAGW}D & \underline{GAAUGD} & UU\underline{UUUYC} \\
	Predicted sequence & \underline{AAAAAA}W & WHC\underline{YCUWHCY}CU & \underline{UUUWYU} & \underline{URHAUG}U & A\underline{WUGCAH} & WN\underline{GAGW} & RU\underline{RWAUGA} & \underline{UUDWW} \\
	PDB sequence & \underline{AAAAAA}A & AAC\underline{CCUAACC}CU & \underline{CUCUCU} & \underline{UGCAUG}U & A\underline{UUGCAC} & UG\underline{GAGU} & GU\underline{GAAUGA} & \underline{UUUUU} \\
	PCC  & 0.81 & 0.70 & 0.73 & 0.27 & 0.74 & 0.32 & 0.77 & 0.72 \\
	t-test statistics & 20.7 & 16 & 25.3 & 5.2 & 6.2 & 7 & 1.7 & 20.2 \\
	t-test P-value & $6.10e-13$ & $1.90e-9$ & $6.70e-13$ & $2.30e-4$ & $4.90e-5$ & $3.90e-6$ & $1.10e-1$ & $8.30e-9$\\
	\hline
	\end{tabular}

}

\end{table}

\subsection{Validation with In Vivo Assay Data}
As aforementioned, small changes in the sequence at the 5$^{\prime}$ end of guiding RNA (base 1-8) can lead to variable consequences in RISC assembly and thereafter affect the siRNA knockdown efficiency. Therefore, knowing how guide-hAgo2 interaction and loading efficiency are correlated is crucial towards the development of efficient RNA-induced silencing tools. To assess NucleicNet's ability in predicting asymmetry in gRNA loading, we compared the NucleicNet score $Q$ with quantitative results from two types of in vivo experiments -- immunoprecipitation assay and siRNA knockdown; $Q$ is derived from analysis of a hAgo2 structure (PDBID: 4f3t) and alignment with a trinucleotide conformation library.

\paragraph{Evaluation on the Ago2-RIP-Seq experiment.} We show that $Q$ can differentiate between guide and passenger sequences from the same precursor miRNA duplexes determined by Ago2 IP followed by small RNA sequencing from different cell lines, namely four from human (acute monocytic leukemia THP-1 \cite{RN1126}, colon cancer DLD \cite{RN1127}, colon cancer HCT116 \cite{RN1128} and T cell leukemia \cite{RN1129}) and one from mouse (neuroblastoma N2a \cite{RN1130}). In each dataset, a strand is considered the `guide' in the duplex when its reads per million (RPM) supersedes its complement by at least 2 orders of magnitude in an Ago2-RIP-Seq experiment (Ago2-RNA Immunoprecipitation and Sequencing) \cite{RN1131}. Duplexes with the guide strand having less than 25 RPM are also discarded resulting in a total of 222 duplexes under evaluation. For each dataset, a histogram of the NucleicNet score difference $Q_{guide}-Q_{passenger}$ between the guide and the passenger strands of each duplex is produced (Figure~\ref{fig:exp}a). A positive difference means that the guide is predicted more favorably than the passenger in binding according to NucleicNet analysis, which is the desired result. In summary, 76\% of the tested duplexes show positive differences. To quantify statistical significance of these differences, a paired T-test and a Wilcoxon signed rank test were conducted. Both tests survived p-value $<$ 0.005 criteria in all datasets confirming NucleicNet's ability in predicting small RNA asymmetry defined from an in vivo setup.

\paragraph{Evaluation on the siRNA knockdown experiment.} In siRNA knockdown experiments, different guide sequences with different loading efficiency can affect RISC assembly, therefore their silencing efficiency could be different \cite{RN1119,RN1132}. Here we evaluate how well the guide-hAgo2 interactions predicted by NucleicNet can explain these differences. In this regard, we collected knockdown benchmarks for shRNA registered on the Broad Institute RNAi Consortium from the website of a distributor\footnote{http://www.sigmaaldrich.com/life-science/functional-genomics-and-rnai.html} and tested for their correlations with the NucleicNet score. To accommodate for heterogeneity in cell lines and target genes, regression analyses were done separately on each entity and were restricted to entities that contain more than one data-points (i.e., different shRNA sequences at base 1-8) (Figure~\ref{fig:exp}b). Entities with the range of knockdown level narrower than 0.1 were excluded as trends could not be seen. In summary, 127 data points were used for evaluation, covering 37 genes in total; 90 data points (26 genes) show positive correlations with the NucleicNet score (Figure~\ref{fig:exp}b), whereas 37 data-points (11 genes) show negative correlation. Although many factors can affect knockdown efficiencies, our results suggest that sequence preferences in guide strand loading is one of them and therefore should be considered in future siRNA designs.

\section{Discussion}
Experimental assays and assay-based computational approaches are quintessential starting points to understand RNA-binding properties of proteins. However, apart from identifying RNA sequence motifs, little can be inferred about the chemistry of base-protein interactions, i.e., the origin of specificity, because atomic and topological details of the RBPs are excluded from analysis. Arguably, this gap of understanding can be filled by elucidating more ribonucleoprotein co-crystals. Nevertheless, even as structural elucidation techniques become more standardized and collections of co-crystals accumulate, efficient ways to exploit this vast abstract structural knowledge have yet to be realized. In this work, by perceiving local physicochemical environment through a deep residual network, we show that meaningful predictions about RNA-binding sites and interaction modes of RNA constituents can be deduced in a pure structure-based computational framework. More importantly, our results show that these learnings on structures can be applied to compare with state-of-the-art in vitro and in vivo experimental assay data, suggesting an ability to capture genuine RNA-binding interactions with verifiable biological implications.

Meanwhile, with this framework, NucleicNet, we show an example of using deep learning to handle the structured prediction in 3D space, whose output is the protein-RNA binding preference landscape along the protein surface. As we discussed above, when solving this hard problem, we might encounter the following challenges. Firstly, since they are high-dimension objects, we may encounter the curse of dimension and the tremendous search space. Secondly, the data are very limited. Furthermore, we need to incorporate the property of this structured prediction problem into the model. As we have discussed above and in Section \ref{chapter2_sec:ideas}, we used the following ideas to resolve the task. Firstly, instead of solving this problem straightforwardly and finding out the optimal conformation of the interaction complex, we wanted to predict the RNA chemical group binding preference landscape along the protein surface. However, this problem is still complicated. So we decomposed this problem into numerous subproblems, obtaining the binding preference landscape by predicting the binding preference distributions of all the grid points on the protein surface. Then, we used an HMM to aggregate the predictions into the solution for the original structured prediction problem. With such a decomposition, the data size can be boosted significantly, which enabled the deep learning method for this problem. Regarding the deep learning model, taking the local physicochemical information as input, the model convolved across different shells in the FEATURE framework. By doing so, we forced the model to incorporate the local structural information and the problem-specific property implicitly. 

On the other hand, we have to admit that our method is pioneering research in this field. There are a few limitations within our framework. Firstly, we omitted some useful structural information, such as RNA-RNA interactions,  base-stacking, base-pairing, and protein dynamics, due to the lack of data. Secondly, the downstream HMM for aggregating the predictions of the grid points from the deep learning model is not perfect, as we discussed in Section \ref{chapter6_sub:hmm}. We need to refine this part in the future further.

In the next chapter, we will present a project of predicting detailed enzyme functions, which is both a hierarchical classification problem and a multi-label classification problem, with deep learning methods.

\chapter{DEEPre: Sequence-based Enzyme EC Number Prediction by Deep Learning}
\label{chapter_deepre}

\section{Chapter Introduction}

Enzymes, an essential kind of proteins in the human body, catalyzing reactions \textit{in vivo}, play a vital role in regulating biological processes. The dysfunction of certain enzymes can cause serious metabolic diseases. For example, the deficiency of alpha-galactosidase, which hydrolyses the terminal alpha-galactosyl moieties from glycolipids and glycoproteins, would cause the Fabry disease, resulting in full body pain,  kidney insufficiency, and cardiac complications \cite{RN30}. The deficiency of DNA repair enzymes, which recognize and correct the physical damage in DNA, can cause the accumulation of mutations, which may further lead to various cancers \cite{RN31}. To investigate the causation of such diseases, an indispensable step of finding a way to cure them, it is crucial to understand the function of the related enzymes first. The most straightforward and accurate way of doing such investigation is conducting biological experiments. However, experimentally determining the enzyme function is both time-consuming and labor-intensive. Even worse, for a new query enzyme without any background information, biologists have little clue on how to set up the experiments. In this context, computational methods emerged to assist biologists in determining enzyme function and guiding the direction of setting up the validating experiments.

\begin{figure}[!tpb]
\centerline{\includegraphics[width=0.55\textwidth]{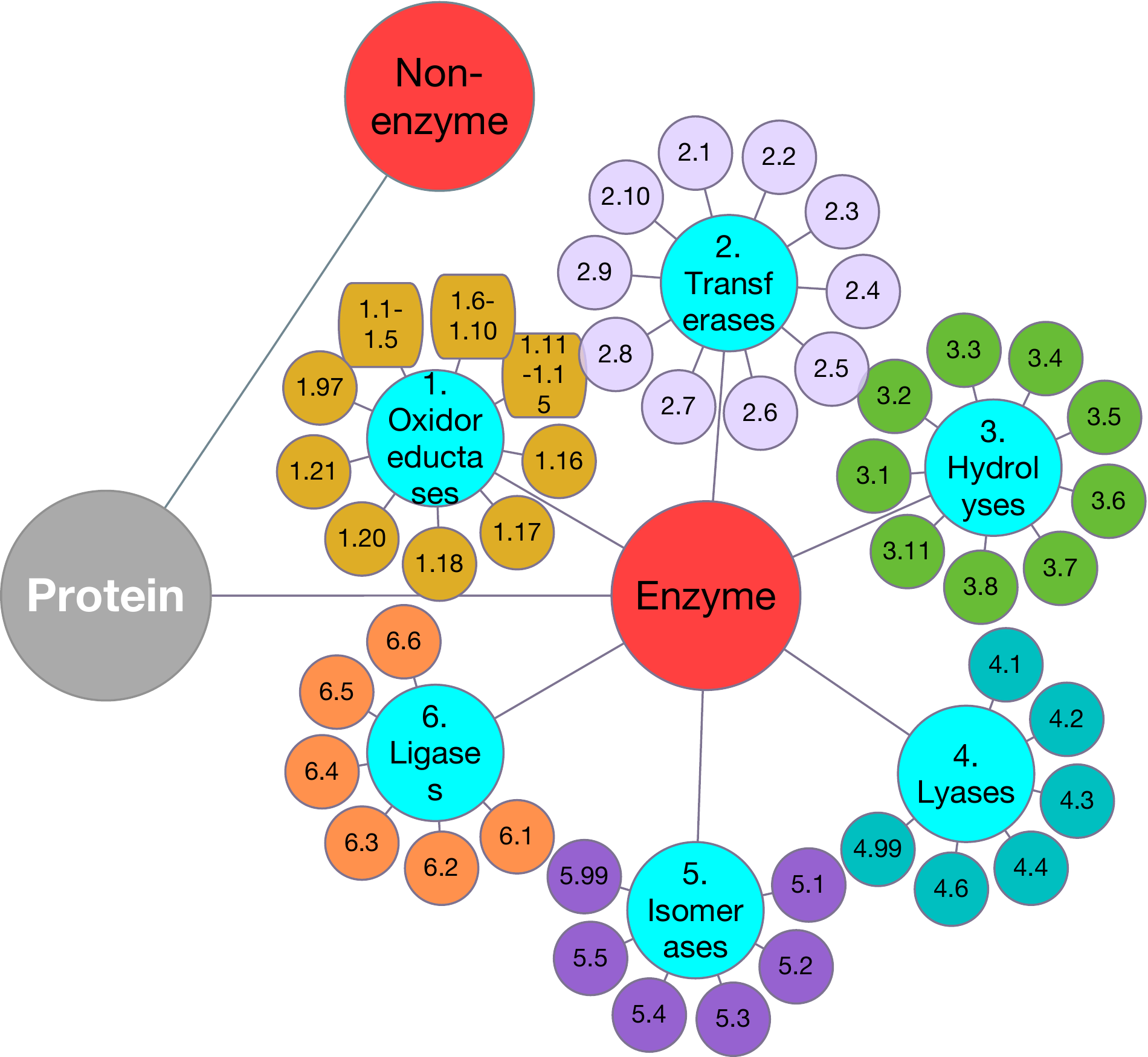}}
\caption{Illustration of the first three levels of the EC system.} \label{fig:01}
\end{figure}

According to SWISS-PROT \cite{RN2} (released on September 7, 2016), among the 539,566 manually annotated proteins, 258,733 proteins are enzymes. Such a large number of enzymes are usually classified using the Enzyme Commission (EC) system, the most well-known numerical enzyme classification scheme, which specifies the function of an enzyme by four digits. This classification system has a tree structure. As shown in Figure~\ref{fig:01}, after the root of the tree, there are two main nodes, standing for enzyme and non-enzyme proteins, respectively. The enzyme main node extends out six successor nodes, corresponding to the six main enzyme classes: (1) oxidoreductases, (2) transferases, (3) hydrolases, (4) lyases, (5) isomerases, and (6) ligases, represented by the first digit. Each main class node further extends out several subclass nodes, specifying the enzyme's subclasses, represented by the second digit. With the same logic, the third digit indicates the enzyme's sub-subclasses and the fourth digit denotes the sub-sub-subclasses. Take type II restriction enzyme, which is annotated as EC 3.1.21.4, as an example, the ``3'' denotes that it is an hydrolase; the ``1'' indicates that it acts on ester bonds; the ``21'' shows that it is an endodeoxyribonuclease producing 5-phosphomonoesters; the ``4'' suggests that it is a Type II site-specific deoxyribonuclease. Figure~\ref{fig:01} illustrates the first three levels of the EC system. By predicting the EC numbers precisely, computational methods can annotate the function of enzymes.

A number of computational methods have already been proposed to determine the enzyme function by predicting enzyme EC numbers. There have been three main research directions of this problem since \cite{RN103}, when machine learning methodologies and sequence information were used to investigate the problem for the first time. Firstly, because it is commonly believed that structures determine function, some researches, such as \cite{RN67, RN34, RN77, RN59, RN66}, focused on predicting the enzyme function by predicting the structure of the enzyme first. After obtaining the structure, they scan the database or the library, whose entries' EC numbers have already been determined and validated by experiments, and assign the EC number of the template with the most similar structure to the query. However, structure prediction is still relatively immature and time-consuming. Besides, since both the structure prediction step and the EC number prediction step can cause errors, the accumulated error would have a negative effect on the final prediction result. Secondly, the common assumption that enzymes with high sequence similarity tend to have similar functionality leads to a number of studies utilizing sequence similarity. For example, \cite{RN25, RN26} introduced programs  that are capable of performing genome-level enzyme function prediction. \cite{RN99, RN100, RN101} described a server using combined approaches: in addition to sequence similarity, it also incorporates PROSITE and PFAM database information. Although this category of methods is widely used in practice, they are unable to make a prediction when encountering a sequence without significant homologies in the current database. Thirdly, extracting features from the sequence and classifying the enzyme using machine learning algorithms is the most extensively studied direction. SVM-Prot \cite{RN85, RN46, RN94} uses SVM and composition, translation, and distribution (CTD) features to predict enzyme function. Other methods using SVM include \cite{RN65, RN49, RN90, RN80, RN48, RN47, RN64}. After long time investigation \cite{RN53,RN50,RN84,RN61,RN63}, \cite{RN54} proposed EzyPred, using pseudo position-specific scoring matrix (Pse-PSSM) and functional domain (FunD) as features and optimized evidence-theoretic $k$-nearest neighbor (OET-KNN) as classifiers, to predict the enzyme main classes and subclasses. \cite{RN60,RN62,RN45} also use KNN and its variants to perform prediction. \cite{RN56,RN78} utilize KNN to do multifunctional enzyme main class prediction. \cite{RN58} explored the performance of using random forest and a bunch of sequence-derived features. \cite{RN51} is based on neural network. Although this direction has already been studied for over 15 years with a number of softwares and servers available, few of them combines the procedure of feature extraction and classification optimization together. Instead, previous studies rely heavily on manually crafted features, and consider feature extraction and classification as two separate problems. In spite of the success of such methods, with the rapid expansion of the known enzyme sequences, the manually designed features are very likely to be a suboptimal feature representation which may be unsustainable in the omic era.


In addition to that, we also need to handle multi-functional enzymes, which can have multiple sets of EC numbers. Such a task is not appropriately solved by the previous methods. On the other hand, from the structured prediction aspect, we can encounter the following challenges when solving this enzyme function prediction problem. The biggest challenge, as we have discussed extensively in this thesis, is the data problem. For the mono-functional enzymes, we only have around 20K data samples, but we have around 200 classes if we consider the most refined classification.
Furthermore, the data are highly imbalanced. For example, in class 1.1.*.*, we can have 651 sequences, while in class 1.20.*.*, we only have 10 sequences. Secondly, we need to incorporate the structure of this problem into the algorithm design. There are two pieces of information that we should consider, both of which are from the labeling space. The first one is the hierarchical structure of the EC labeling space, which is explicit. The second one is the multi-labeling information for the multi-functional enzymes, which is implicit. 


In conquering the aforementioned limitations and challenges, 
here we propose a novel level-by-level prediction approach based on deep learning, only utilizing the sequence information. The enzyme sequences are represented by two kinds of raw encoding, sequence-length-dependent encoding, such as raw sequence one-hot encoding and PSSM, and sequence-length-independent encoding, such as functional domain encoding. Those two kinds of representations are combined into a deep learning model with a novel architecture to perform dimensionality uniformization, feature selection and classification model training simultaneously. Instead of training just one such deep learning model, we built multiple deep learning models following the hierarchical structure of the EC system, with one model for each internal node of the labeling tree. We also utilized hierarchical transfer learning to facilitate the information flow among those deep learning models, which can help them understand the labeling structure of the problem. Such a level-by-level classification strategy and transfer learning idea can also alleviate the data issue. Furthermore, when handling the multi-functional enzyme function prediction \cite{zou2019mldeepre}, we incorporated the multi-labeling information into the loss function, which can help the model to learn the problem structure implicitly. Below, we will discuss the method and results in detail.

\section{Methods}
%

\subsection{Sequence Representation}
The deep learning framework explained in Section \ref{model} eliminates the necessity of performing manual dimensionality uniformization and building complex, manually-designed features, which are unlikely to sustain the increasing amount and complexity of data, by conducting feature reconstruction and classifier training simultaneously. Therefore, we use the following raw features, constructed from the input sequence directly, to represent the sequences. Based on their dimensionality, they can be classified into two categories, sequence-length-dependent features and sequence-length-independent features. The first four features described below belong to the former while the last one belongs to the latter.

\paragraph{Sequence one-hot encoding.} To preserve the original sequence information, we use one-hot encoding as the first raw representation of the input sequence. This encoding uses one 1 and nineteen 0s to represent each amino acid. For example, A is encoded as $\irow{1&0_{1}&...&0_{19}}$, while C is encoded as $\irow{0_{1}&1&0_{2}&...&0_{19}}$. For each input protein sequence, the one-hot encoding produces an $L$ by 20 matrix, where $L$ represents the sequence length, with each row representing a specific spot and each column representing the appearance of a certain amino acid. For those sequences with undetermined amino acid at a particular spot, a vector with 20 0s is used to represent that special position.

\paragraph{Position Specific Scoring Matrix.} To provide the evolutional information to the training model, we deploy PSSM as the second sequence representation, which could be obtained in the following way.
\begin{enumerate}
	\item For an input query sequence, we use BLAST search against a database to generate a multiple sequence alignment (MSA) formed by the sequences with high similarity to the query sequence.
	\item From the MSA, we can produce the profile, listing the frequencies of each amino acid at each sequence spot in the alignment. Using the profile and the prior knowledge of amino acid substitutability, we can obtain the PSSM of the first iteration.
	\item The PSSM would be used as the substitution matrix used in BLAST to perform another iteration of BLAST. The newly detected sequences obtained in this iteration are added into the MSA to update the PSSM.
	\item The previous two steps are repeated until there is no new sequence or the procedure reaches the predefined iteration times.
\end{enumerate}

The PSSM from the last step is the evolutional sequence representation. In practice, we used PSI-BLAST \cite{RN95} from BLAST+ \cite{RN19} with three iterations, E-value being 0.002, against SWISS-PROT (released on May 11, 2016).

\paragraph{Solvent accessibility.} Solvent accessibility describes the openness of a local region. Because such information is unavailable directly from the database, we use DeepCNF \cite{RN1} to predict it. Taking the protein sequence as the input, DeepCNF outputs the possibilities of each amino acid of the sequence being in the state of buried, medium or exposed, respectively. The three states are defined by two solvent accessibility thresholds. Buried is defined as less than 10\%; exposed is defined as more than 40\%; and medium is defined within the range of 10\% and 40\%. This encoding produces an $L$ by 3 matrix. More details could be referred to \cite{RN1}.

\paragraph{Secondary structure one-hot encoding.} An amino acid could be in one of the three main secondary structure states, alpha-helix, beta-sheet, and random coil, which indicate the protein's local folding information. Similar to solvent accessibility, we take advantage of DeepCNF \cite{RN1} to predict the secondary structure of a given sequence, whose result is an $L$ by 3 matrix, each row of which shows the possibility of the amino acid folding into alpha-helix, beta-sheet or random coil, respectively. The details could be referred to \cite{RN1}.

\label{funcd}
\paragraph{Functional domain.} Usually, a protein sequence contains one or several functional domains, which provide distinct functional and evolutional information. Pfam \cite{RN20} is a collection of such functional domains, each represented by an HMM. Searching against the database and encoding in the following way generates the functional domain encoding used in our model.

\begin{enumerate}
\item Pfam has its default searching engine as HMMER \cite{RN97}. For each protein sequence, we use HMMER, with the inclusion E-value threshold as 0.01, to search against Pfam (Pfam 30.0 Released on July,1 2016), which contains 16,306 entries.
\item We employ a 16,306 dimensional vector to encode the searching result. If the $i$-th entry in the database is reported as hit, 1 appears on the corresponding position of the vector, otherwise it is 0.
\end{enumerate}

As a result, the functional domain encoding of a protein sequence would be:
\begin{equation}
\textbf{F}_{FuncD} =
		\begin{bmatrix}
         I_{1} & I_{2} & ...  & I_{i} & ... &  I_{16,306},
        \end{bmatrix}
        \label{eq:01}\vspace*{-10pt}
\end{equation}
where
\begin{equation}
\textit{I}_{i} = \begin{cases}1, \text{the $i$-th entry in Pfam reported as hit,}\\0, \text{otherwise.}\end{cases}
\end{equation}

\subsection{Classification Model} \label{model}
The enzyme function prediction problem has a tree-structured label space, which makes it a typical hierarchical classification problem. To solve this kind of problems, we propose a level-by-level prediction framework, building a model for each internal label node. The model contains two main components, namely, the problem-specific feature extractor, which is able to perform dimensionality uniformity and feature extraction, and the classifier. Such a novel, end-to-end model can perform feature selection and classifier training simultaneously in a virtuous circle, making it more likely to achieve high performance.

\vspace{\baselineskip}
\noindent {\bf Level-by-level Strategy}

\begin{figure}[!htpb]
\centerline{\includegraphics[width=0.95\textwidth]{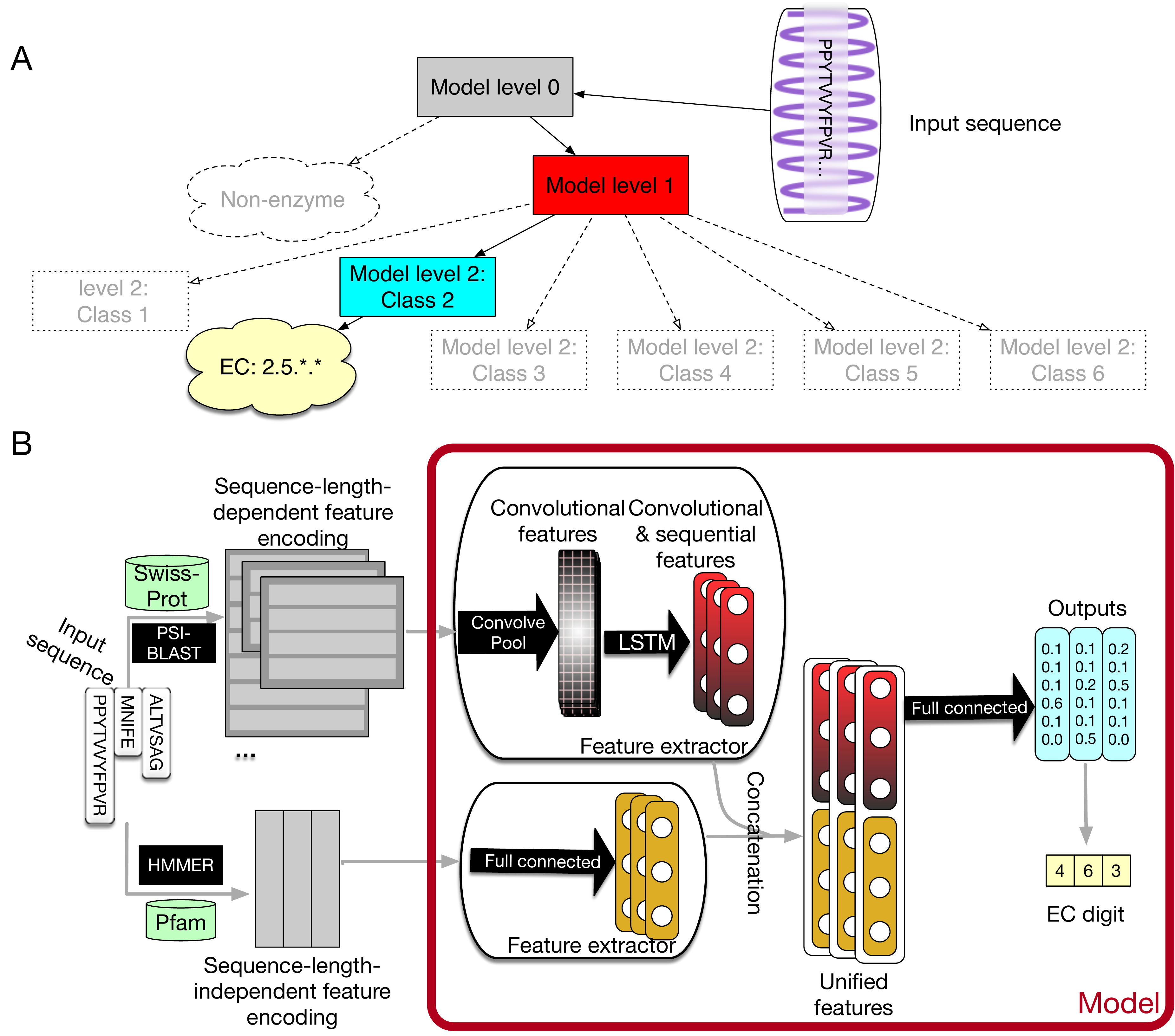}}
\caption{The prediction strategy and the deep learning model used for enzyme function prediction.
} \label{fig:02}
\end{figure}

Because of the relative small size (22,168 data points are assigned to 58 classes until the second digit) and, even worse, the extreme imbalance property (for example, the $NEW$ dataset contains 22,168 sequences belonging to non-enzyme while only 10 sequences belonging to subclass 1.20) of the data, this problem can be very difficult if we do not have additional information. Fortunately, we can take advantage of the labeling space structure when designing the method.
Particularly, a level-by-level prediction strategy is used. That is, given a sequence, the trained model would firstly predict whether it is an enzyme or not. If it is an enzyme, the model will further predict the first digit, which indicates its main class. Knowing the main class, our algorithm will choose the trained model for that specific main class and further predict the second digit, that is, the subclass. Corresponding to the label hierarchy, we build one model for determining whether the input is an enzyme, one model to determine the first digit, six models to determine the second digit. This prediction strategy could be referred to Figure~\ref{fig:02}(A).


\vspace{\baselineskip}
\noindent {\bf Deep Neural Network Model}

For each level of prediction, we build an end-to-end model based on several deep neural network components. In terms of the sequence-length-dependent features, such as PSSM, we build a feature extractor exploiting the convolutional neural network component to extract convolutional features from the input map and, after that, a recurrent neural network component, comprised of long short-term memory (LSTM) cells, to extract sequential features from the output of the previous component. As for the sequence-length-independent feature, i.e., the functional domain encoding, which is a vector, we use a fully-connected component to perform dimensionality reduction and feature extraction. We further employ a fully-connected component to combine those different pieces of information together, followed by a softmax layer for classification. The structure of the model could be referred to Figure~\ref{fig:02}(B). It should be noted that our default model uses only three input features, sequence one-hot encoding, PSSM and functional domain encoding. We encode local features, i.e., secondary structure and solvent accessibility, later to evaluate the importance of local information. During training, the training error is back-propagated to each component. The error would guide the convolutional neural network component and recurrent neural network component to perform an end-to-end feature selection, weighing more on the feature which would improve the final performance while weighing less on unimportant features automatically. At the same time, the weight of other components would be adjusted simultaneously to adopt the change. Such coupling effect of feature extraction and classifier training optimizes the performance dramatically.

When training the second digit prediction models, we adopted the pre-train and fine-tune techniques. Since the limited number of data is further divided into six parts corresponding to the six main classes, the amount of data belonging to each main class is unable to produce a model with the ability to extract features and being generalized well. To solve this issue, we pre-train the convolutional neural network component and recurrent neural network component by using all the training data. Then, for training each second digit prediction model, we fix the parameters of those components and only fine-tune the fully connected components using the specific subset of the training data. Notice that, this transfer learning idea, together with the level-by-level prediction strategy, can help the models understand the labeling structure of this problem, facilitating the information flow within the labeling space. In other words, we incorporate the problem structure into the methodology design, which is beneficial for solving this structured prediction problem. 

In practice, we used TensorFlow \cite{RN98} as the framework to construct the deep neural network. To alleviate the overfitting issue, we utilized weight decay, dropout \cite{RN21}, and batch normalization \cite{RN72}. We chose adaptive moment estimation (Adam) as the optimizer \cite{RN102}. With two Pascal Titan X cards, it took around 4 hours to obtain a well-trained model.


\begin{figure}[h!]
\begin{center}
\includegraphics[width=0.9\textwidth]{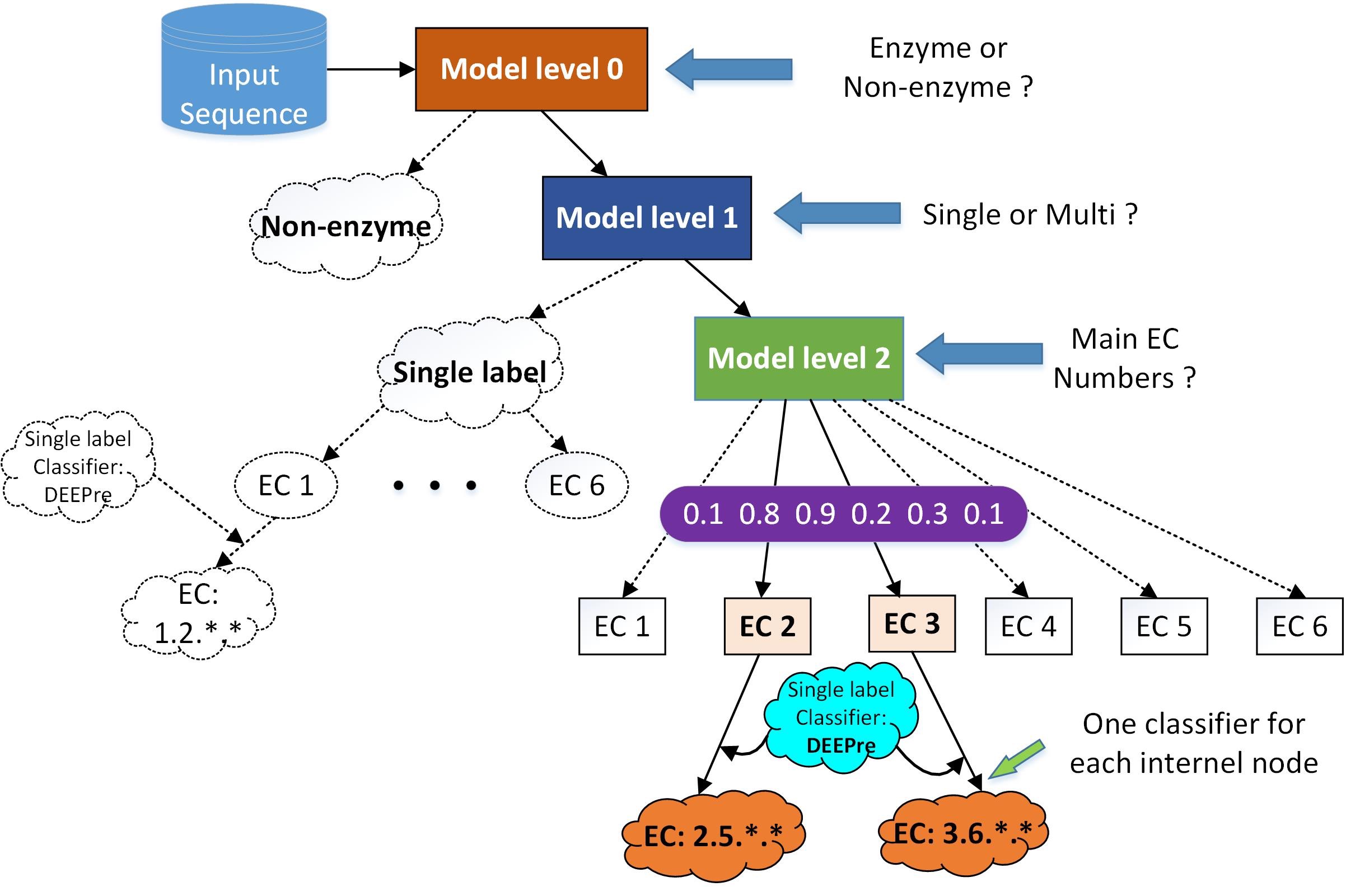}
\end{center}
\caption{The hierarchical classification strategy combining DEEPre and mlDEEPre.}
\label{fig:hierarchical}
\end{figure}

\subsection{mlDEEPre}
We further extend DEEPre from mono-functional enzyme function prediction to multi-functional enzyme function prediction, resulting in mlDEEPre \cite{zou2019mldeepre}. As shown in Figure~\ref{fig:hierarchical}, mlDEEPre has two levels. Given an enzyme sequence, the first level predicts whether the enzyme is a mono-functional enzyme or a multi-functional one. If the sequence is a multi-functional enzyme, the second level of mlDEEPre will predict the main classes of the enzyme's multi-functions. The first level of mlDEEPre essentially handles a binary classification problem while its second level deals with a multi-label classification problem. 

\paragraph{Multi-label: loss function.} Deep learning methods are often suitable for multi-label classification. As shown in Figure~\ref{fig:02}, the model's last layer has multiple nodes, whose outputs correspond to the predicted probability of each label. If we use the model to perform single label classification, we will find the label with the highest probability score and assign the query with that label. When we use the model to perform multi-label prediction, we can still use the predicted probabilities. However, we need to change the way of assigning labels. Instead of assigning the label with the highest probability, we want to assign the labels whose probability score is higher than a certain threshold so that multiple labels can be predicted. On the other hand, when we train the model, we also need to consider the multi-label information in the training data. One of the most straightforward way of incorporating such information is to modify the loss function accordingly and make the model know that we are performing multi-label prediction using a multi-label dataset. In terms of such a threshold and the loss function, we adopt the idea from \cite{RN467}, \textit{i.e.}, BP-MLL. We introduce the loss function in this section in detail and discuss the threshold in the next session. 

Formally, denote the $i^{th}$ enzyme instance as $x_i$, and its corresponding label vector as $D_i$. Each element of $D_i$ is a binary value, which indicates whether that enzyme instance belongs to a certain class. We use $d_i^j$ to denote that element, where $j\in[\,1,6]\,$ for our problem. 
If $d_i^j$ is 1, the enzyme $x_i$ belongs to the class $j$, 0 otherwise. As for a classification problem, the most intuitive way to define the global error of the network is to measure the distance between the predicted labels and the real labels of the training set:
\begin{equation} \label{eq:1}
E=\sum_{i=1}^{m}E_i,
\end{equation}
where $E_i$ represents the network error on the instance $x_i$ and $m$ is the size of the training data. For a multi-label classification problem, we can define $E_i$ as below:
\begin{equation} \label{eq:2}
E_i=\sum_{j=1}^{Q}{(\,l_i^j-d_i^j)\,}^2,
\end{equation}
where $l_i^{j}$ and $d_i^{j}$ are the output from the network and the true label of $x_i$ on the class $j$, respectively; Q is the total number of classes, which is $6$ in our problem. Using Eq. \ref{eq:2}, we can incorporate the multi-label information into the model to a certain degree since all the label information is considered in that loss function. However, the loss function in Eq.\ref{eq:2} assumes that class labels are independent, which ignores any relationship between different class labels. In reality, one of the most straightforward relationships between labels is that labels in $L_i^{true}$ should have higher ranks than those not in $L_i^{true}$, where $L_i^{true}$ is the set of labels that instance $x_i$ has. Accordingly, we can use the following function as the loss which considers the rank relationship between labels: 
\begin{equation} \label{eq:3}
E=\sum_{i=1}^{m}E_i=\sum_{i=1}^{m}\frac{1}{|L_i^{true}||\overline{ L_i^{true}}|}\sum_{(\,k,q)\,\in L_i^{true}\times\overline{ L_i^{true}}}e^{(\,-(\,l_i^k-l_i^q)\,)\,},
\end{equation}
where $\overline{ L_i^{true}}$ is the complementary set of $L_i^{true}$, that is, the label set which the instance $x_i$ does not have, and $|\bullet|$ is the cardinality of a set. From the equation, we can find that $(\,l_i^k-l_i^q)\,$ measures the difference between the outputs of the network on the labels belonging to the training instance and the ones not belonging to it, which is further fed to the exponential function. When $l_i^q$ happens to be much larger than $l_i^k$, which causes large discrepancy, the exponential function can penalize the error severely. By minimizing Eq.\ref{eq:3}, we can make the model output much higher values for the true labels while very small values for the labels that the training data do not have. Thus, labels in $L_i^{true}$ have higher ranks than those not in $L_i^{true}$, which is consistent with our goal.

\paragraph{Multi-label: threshold.} When we use the model, to determine and assign the labels, there should be a threshold $t(x)$, which is applied to the output of the deep learning model, so that we predict the labels as $L_i^{pred}=\{j|l_i^{j} >t(x),j\in[\,1,6]\}$. A straightforward and natural solution of the threshold function is to set $t(x)$ as a constant. However, that constant threshold does not consider the difference between different data points. To solve the problem, \cite{NIPS2001_1964} proposed an excellent idea to incorporate the information of each single data point into the threshold, which replaces the constant with a linear function $t(x_i)=\mathbf{w}^\intercal \cdot l(x_i)+b$, where $l(x_i)$ is the output of the network on the instance $x_i$. In this way, each data point can have its own threshold, which is more flexible than a constant. To obtain the threshold function, we need to solve the following problem:
\begin{equation} \label{eq:4}
t({x}_i)=argmin_t(\,|\{k|k\in L_i^{true},l_i^k \leqslant t\}|+|\{q|q\in \overline{L_i^{true}},l_i^q \geqslant t\}|)\, .
\end{equation}
If the solution of Eq. \ref{eq:4} is not unique and the solution composes a segment, the middle value of the value range is chosen as the threshold. For example, assume the real label and predicted label set of $x_i$ are \{1, 1, 0, 0, 0, 0\} and \{0.9, 0.8, 0.3, 0.1, 0.1, 0.1\}, when $0.3<t<0.8$, $|\{k|k\in L_i^{true},l_i^k \leqslant t\}|+|\{q|q\in \overline{L_i^{true}},l_i^q \geqslant t\}|$ always takes the minimum value as 0. Consequently, we choose the middle value of $(0.3, 0.8)$, which is 0.55, as the threshold. In BP-MLL, the solution of the threshold equation can be obtained through the linear least square method. 

To sum up, after we have a well-trained model and the threshold function parameters, and when we need to use the model to perform prediction, firstly, we feed the test instance to the trained network and get the outputs $l(x_i)$. Secondly, we calculate the threshold using $t(x_i)=\mathbf{w}^\intercal \cdot l(x_i)+b$ and apply the threshold to the output of the model, obtaining the predicted labels for the enzyme instance $x_i$.

\paragraph{DEEPre and mlDEEPre.} Although DEEPre is designed for mono-functional enzyme function prediction, it is very flexible, being able to predict the detailed function of an enzyme from the first level or the second level. For example, if we have already known that an enzyme has the follow incomplete EC number: 1.-.-.-, we can run DEEPre from the second level to fulfill the missing digits. Taking into consideration the enzyme's feature representation and the fact that the query sequence is an Oxidoreductase, we run the model trained specifically for the enzymes with the first EC digit as 1. With such flexibility, we can combine mlDEEPre and DEEPre to predict the detailed functionality of multi-functional enzymes easily. Using mlDEEPre, we can predict the main classes of those multi-functional enzymes, such as 2.-.-.- and 3.-.-.-. Feeding the sequence and the main classes annotation to DEEPre, we can fill in the missing digits for each incomplete annotation of a multi-functional enzyme. The idea of combining DEEPre and mlDEEPre is illustrated in Figure~\ref{fig:hierarchical}. Starting from a protein sequence, we first use level 0 of DEEPre to predict whether the protein is an enzyme or not. If yes, we use mlDEEPre's first level to predict whether the enzyme is a mono-functional enzyme or a multi-functional enzyme. If that is a mono-functional enzyme, we will further run DEEPre to get full annotation of that enzyme. If not, we will run the second level of mlDEEPre to predict the main classes of the enzyme. For each function, we run DEEPre to obtain the full annotation. Considering that most multi-functional enzymes have multiple EC number annotations for its different functions diverging in the first digit, our method is efficient and reliable under most circumstances.

\section{Results}

\subsection{Experimental Setting}

\noindent {\bf Datasets}

We adopted four datasets in the experiments. The first dataset is a widely used one from \cite{RN54}, constructed from the ENZYME database (released on 01-May-2007), with 40\% sequence similarity cutoff. More details of that dataset can be referred to \cite{RN54}. This dataset is denoted as the $KNN$ dataset in the rest of the chapter.

Following the same rule of constructing the $KNN$ dataset, we constructed a larger dataset using up-to-date databases. 
This larger dataset would be referred to as the $NEW$ dataset in the rest of this chapter.

Other than $KNN$ and $NEW$, which were used as the benchmark for cross-fold validation, we also need another dataset to further test the generalization power of the proposed method. This can be done by training the model on one dataset, and testing it on an independent and non-overlapping dataset, to avoid being overfitted on a particular dataset. Thus, the third dataset, the benchmark from \cite{RN67}, was used for cross-dataset validation. This non-homologous dataset was collected from PDB, satisfying two requirements: (1) the pair-wise sequence similarity is below 30\%, and (2) there is no self-BLAST hit within the database. All enzymes in this dataset have experimentally determined 3D structures. To avoid overlaps between the training and testing datasets, sequences contained in both dataset were removed from the latter one, which reduced the size of the dataset from 318 to 284. This benchmark is referred to as the $COFACTOR$ dataset in the following. Table~\ref{Tab:01} summarizes the three datasets.

\begin{table}[!hbt]
  \centering
  \caption{The $KNN$ and $NEW$ datasets summary. 
  }\label{Tab:01}
{\begin{tabular}{@{}lccc@{}}
\toprule
Dataset & $KNN$ Dataset & $NEW$ Dataset & $COFACTOR$ Dataset\\
\midrule
Source & \cite{RN54} & Self-constructed & \cite{RN67}\\
Enzymes & 9,832 & 22,168 & 284\\
Non-enzymes & 9,850 & 22,168 & -\\
\bottomrule
\end{tabular}}
\end{table}

Furthermore, we also used another dataset for the multi-functional enzyme function prediction. This dataset is from \cite{che2016identification}, which provides us with 4,076 multi-functional enzymes. More statistics of it are in Tables \ref{tab:multi_data}, \ref{tab:low_similarity_multi_data}. 

\begin{table}
\caption{Dataset II: 4,076 multi-labeled enzymes. This table shows the number of multi-functional enzymes in the dataset with different EC main class combinations. }
\label{tab:multi_data}
\centering
\resizebox{1.0\textwidth}{!}
{
\begin{tabular}{@{} |c|ccccccccccccc|ccc|c| @{}}
\hline
Number of classes  & \multicolumn{13}{c|}{2} & \multicolumn{3}{c|}{3} & 4  \\
\hline
\hline
\multirow{4}{*}{EC numbers} & \multirow{2}{*}{1} & \multirow{2}{*}{1} & \multirow{2}{*}{1} & \multirow{2}{*}{1} & \multirow{2}{*}{2} & \multirow{2}{*}{2} &\multirow{2}{*}{2} &\multirow{2}{*}{2} &\multirow{2}{*}{3} &\multirow{2}{*}{3} &\multirow{2}{*}{3} &\multirow{2}{*}{4} &\multirow{2}{*}{4} & 1 & 1 & 1 & 1 \\
 & & & & & & & & & & & & & & 2 & 3 & 4 & 2 \\
 & \multirow{2}{*}{2} & \multirow{2}{*}{3} & \multirow{2}{*}{4} & \multirow{2}{*}{5} & \multirow{2}{*}{3} & \multirow{2}{*}{4} & \multirow{2}{*}{5} & \multirow{2}{*}{6} & \multirow{2}{*}{4} & \multirow{2}{*}{5} & \multirow{2}{*}{6} & \multirow{2}{*}{5} & \multirow{2}{*}{6} & \multirow{2}{*}{4} & \multirow{2}{*}{6} & \multirow{2}{*}{5} & 3 \\
  & & & & & & & & & & & & & &  &  &  & 4 \\
\hline
Number & 147 & 841 & 63 & 37 & 1148 & 235 & 38 & 131 & 622 & 22 & 4 & 308 & 34 & 215 & 10 & 211 & 10 \\ 
\hline
\end{tabular}
}
\end{table}

\begin{table}
\caption{Dataset 
\uppercase\expandafter{\romannumeral2}: 1,085 multi-labeled enzymes with 65\% sequence similarity cut-off.}
\label{tab:low_similarity_multi_data}
\centering
\resizebox{1.0\textwidth}{!}
{
\begin{tabular}{ cccccccc }
\toprule
Multifunctional enzymes & EC 1 & EC 2 & EC 3 & EC 4 & EC 5 & EC 6 & Total \\
\midrule
Name & Oxidoreductase & Transferase & Hydrolase & Lyase & Isomerase & Ligase & \\
\midrule
Before redundancy & 1534 & 1924 & 2657 & 1698 & 616 & 179 & 4076 \\
\midrule
After CD-HIT & 386 & 503 & 689 & 473 & 137 & 52 & 1085 \\
\bottomrule
\end{tabular}
}
\end{table}

\vspace{\baselineskip}
\noindent {\bf Compared Methods}

For the cross-fold validation, in which training and testing are based on different parts within the same dataset, we compared our method with five other methods, including two state-of-the-art methods, EzyPred \cite{RN54} and SVMProt \cite{RN94}, and three baseline methods. One of the baseline methods uses SVM with the raw features used in our model; another baseline method uses SVM with Pse-PSSM; and the last baseline method uses the traditional neural network with our raw features. Due to the unchangeable database of EFICAz \cite{RN101} and COFACTOR \cite{RN34}, we did not include them in the cross-fold validation comparison. However, we performed cross-dataset validation, where the training and testing are on different datasets, to compare our method with EzyPred, SVM-Prot, COFACTOR, and EFICAz. Regarding mlDEEPre, for mono-function prediction, we compared our method with Pse-ACC \cite{chou2006}, ACC \cite{che2016identification}, EnzML \cite{RN56} and SVM. For multi-function prediction, we compared our method with ML-KNN \cite{zhang2007ml}, BR-KNN \cite{spyromitros2008empirical}, IBLR-ML \cite{cheng2009combining}, GM \cite{RN78}, and SVM-NN \cite{amidi2017automatic}.

\begin{figure}[!htpb]
\centerline{\includegraphics[width = 0.9\textwidth]{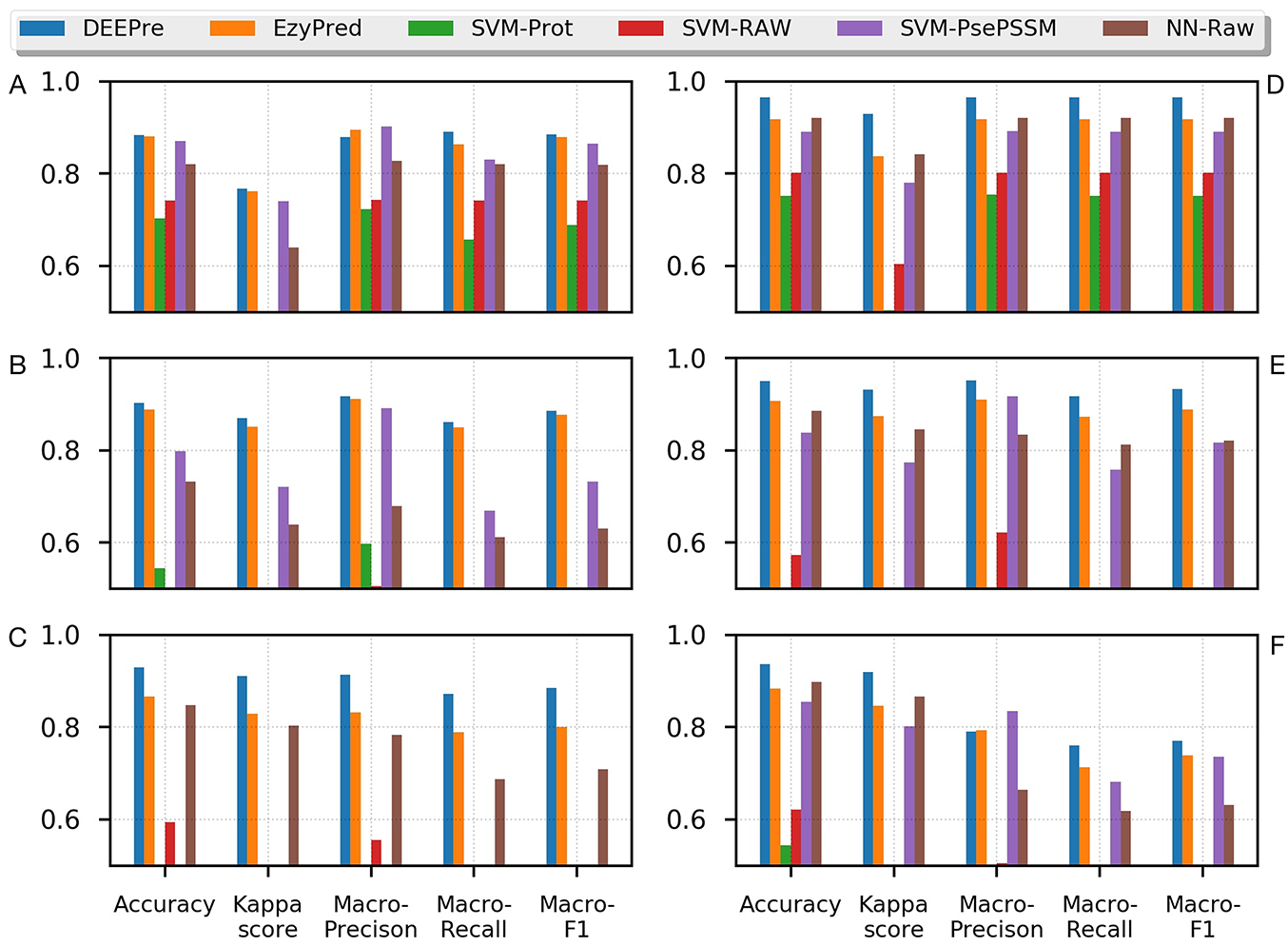}}
\caption{Cross-fold validation performance comparison. 
} \label{fig:03}
\end{figure}

\vspace{\baselineskip}
\noindent {\bf Evaluation Criteria}

For the enzyme or non-enzyme prediction, since it is a binary classification problem, we used accuracy, Cohne's Kappa Score \cite{RN24}, precision, recall and $F_1$ score to evaluate the classifiers' performance. For other predictions, since they are multi-class classification problems, we will report accuracy, Cohen's Kappa Score, Macro-precision, Macro-recall and Macro-$F_1$ score.
%

\subsection{Cross-fold Validation}

\noindent {\bf Overall performance}

Figure~\ref{fig:03} shows the 5-fold cross validation results. Our method almost always outperforms the other methods in both the $KNN$ dataset and the $NEW$ dataset across the five criteria and across the three hierarchical levels of prediction. As for the $NEW$ dataset, DEEPre outperforms the other five methods consistently in level 0 and level 1 prediction across the five criteria. As for the level 2 prediction, the only criterion that DEEPre does not improve over the existing methods is the Macro-Precision, which is an unweighted average of precision of each label. The appearance of very small classes (for example, subclass 1.20 only has 10 enzymes) in the second level prediction might be the reason for this result. In terms of the $KNN$ dataset,  although the smaller dataset makes the improvement of DEEPre over the other methods in level 0 prediction less significant, it still outperforms the other methods in level 1 and level 2 classification greatly.

\begin{figure}[!h]
\centerline{\includegraphics[width=0.9\textwidth]{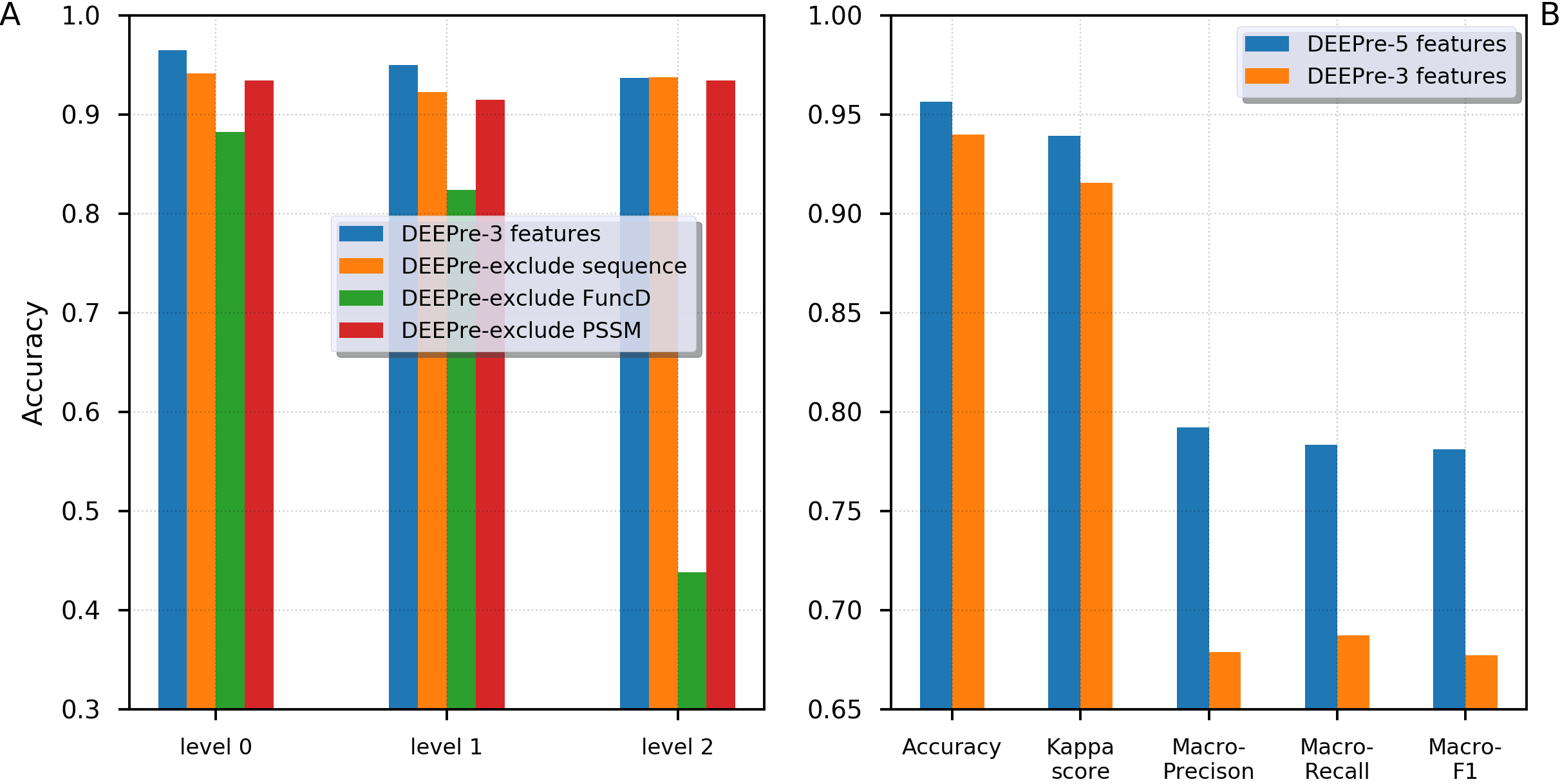}}
\caption{Feature importance analysis for the DEEPre framework.
} \label{fig:04}
\end{figure}

\vspace{\baselineskip}
\noindent {\bf Local Features Improve Performance}

It is believed that both global features and local features determine the function of a protein. For detailed function, local information can weigh even more in determining it. The features extracted by the convolutional component and the recurrent component from PSSM and sequence raw encoding could be considered as global features while the functional domain encoding can be considered as a local feature. We removed the three input raw encoding one by one when performing experiments and show the comparison of their performance on the $NEW$ dataset in Figure~\ref{fig:04}(A). Clearly, as the level goes deeper, the importance of functional domain is evidently increasing, which demonstrates the well-recognized hypothesis. To further prove it, we designed another experiment, in which we inputted more local feature encoding, including secondary structure and solvent accessibility, into our model. Figure~\ref{fig:04}(B) shows the performance comparison of this model and the previous model in level 2 prediction. The additional local features further improve the performance of our model, with accuracy improved by 1.8\% while Macro-precision, Macro-recall and Macro-F1 score improved by at least 11\%.

\vspace{\baselineskip}
\noindent {\bf Third Digit and Fourth Digit Prediction}

Using the same framework described above, we can also predict the enzyme's third digit, which represents its sub-subclass, on the $NEW$ dataset. The accuracy across all the sub-subclasses is 0.9415; the Kappa score is 0.8918; the macro-precision is 0.8942; the macro-recall is 0.8578; and the macro-F1 score is 0.8665. Regarding the fourth digit prediction, more data are needed to perform normal machine learning training-and-testing procedure. For example, within the sub-subclass 1.1.1 in the $NEW$ dataset, there are 188 classes. Each of those classes has less than 40 enzyme sequences, with 175 classes having less than 10 enzyme sequences. Using the current dataset with such distribution can lead to unreliable results.

\subsection{Cross-dataset Validation}
\begin{figure}[!h]
\centerline{\includegraphics[width=0.75\textwidth]{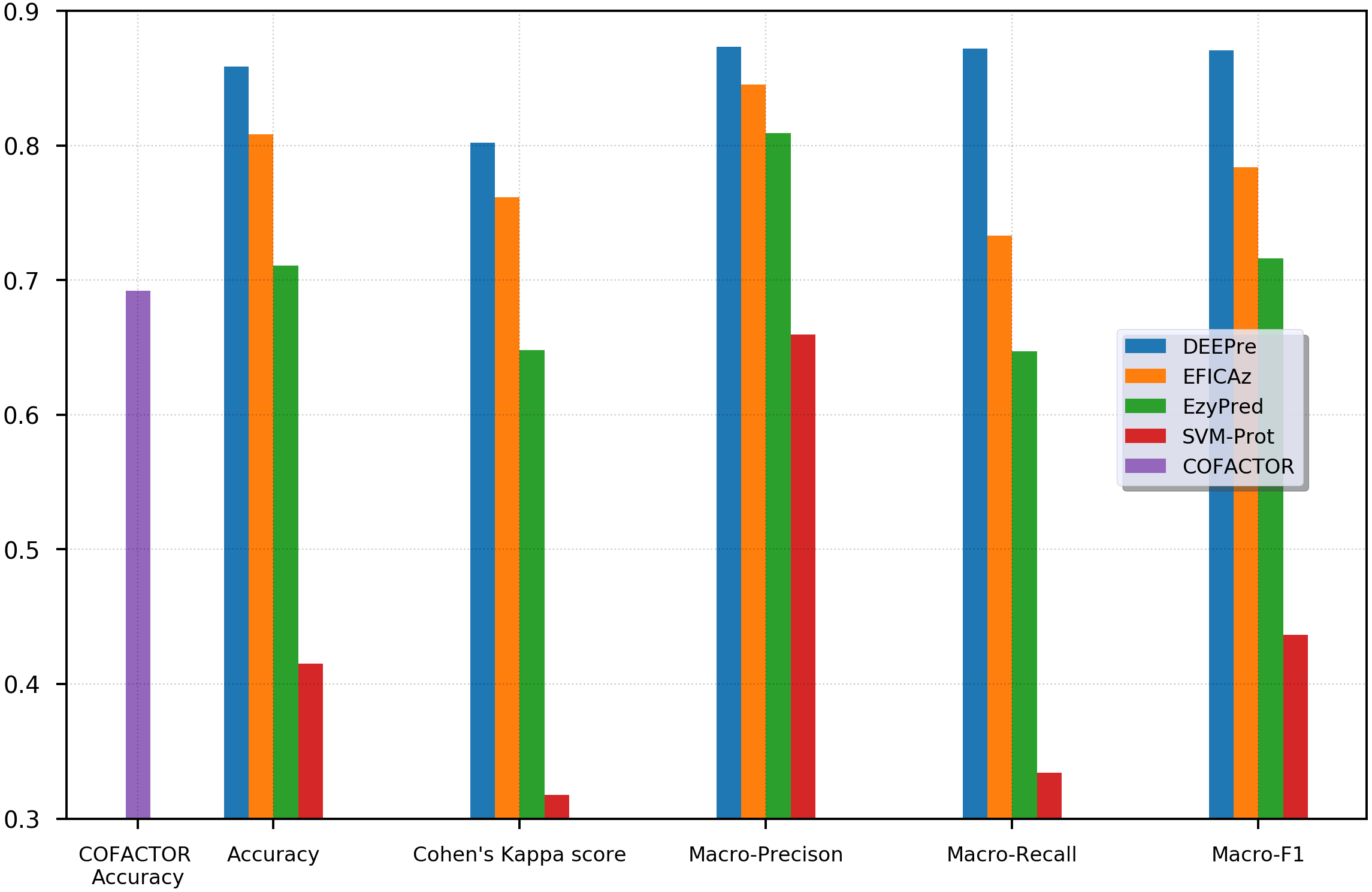}}
\caption{The performance comparison of different servers on predicting the main class of the $COFACTOR$ dataset. 
} \label{fig:05}
\end{figure}

In this experiment, we directly compared the performance of different servers in predicting the main class, that is, the first digit, of an enzyme. We used the $COFACTOR$ benchmark dataset, which is proved to be a difficult dataset in the enzyme function prediction field \cite{RN67}, as the test dataset. Firstly, we eliminated the sequences in the COFACTOR benchmark data which overlap with the DEEPre's training database ($NEW$) by 40\% sequence similarity filtering, reducing the data size from 318 to 284, to ensure that there is no bias in the DEEPre's results. Then we inputted the remaining sequences to each server manually and collected the prediction results. For COFACTOR, since it is quite time-consuming to run the server, about 4 hours to obtain the result for one query, we report the results from the original paper. As shown in Figure~\ref{fig:05}, DEEPre outperforms the other servers consistently across the five criteria, improving the accuracy by at least 6\% over the other servers, including COFACTOR. This is significant because COFACTOR requires 3D structures of enzymes whereas DEEPre only requires the sequence information. On the other hand, we should admit that we have changed the original $COFACTOR$ dataset to some extent to reduce the overlap between the training and testing sets, which might explain some of the performance difference between COFACTOR and DEEPre. We should also notice that all of those five servers have different training datasets, although those training datasets highly overlap with each other and each method was optimized on its corresponding dataset.

\begin{figure}[!t]
\begin{center}
\includegraphics[width = 0.7\textwidth]{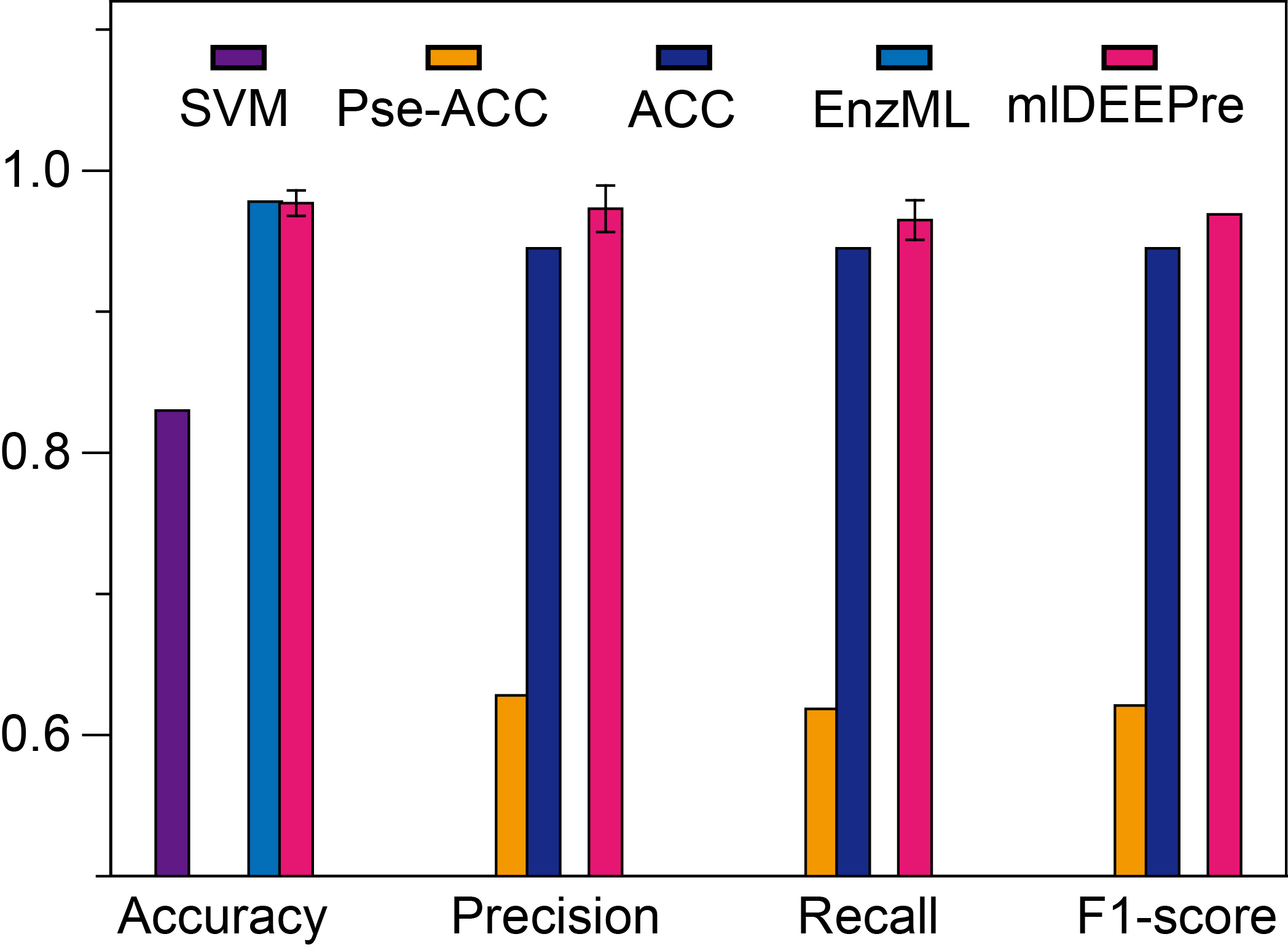}
\end{center}
\caption{The mono-functional enzyme VS multi-functional enzyme classification testing performance of different models. Performance lower than 0.6 are not shown in the figure.}
\label{fig:mono-results}
\end{figure}

\begin{figure}[!t]
\begin{center}
\includegraphics[width = 0.7\textwidth]{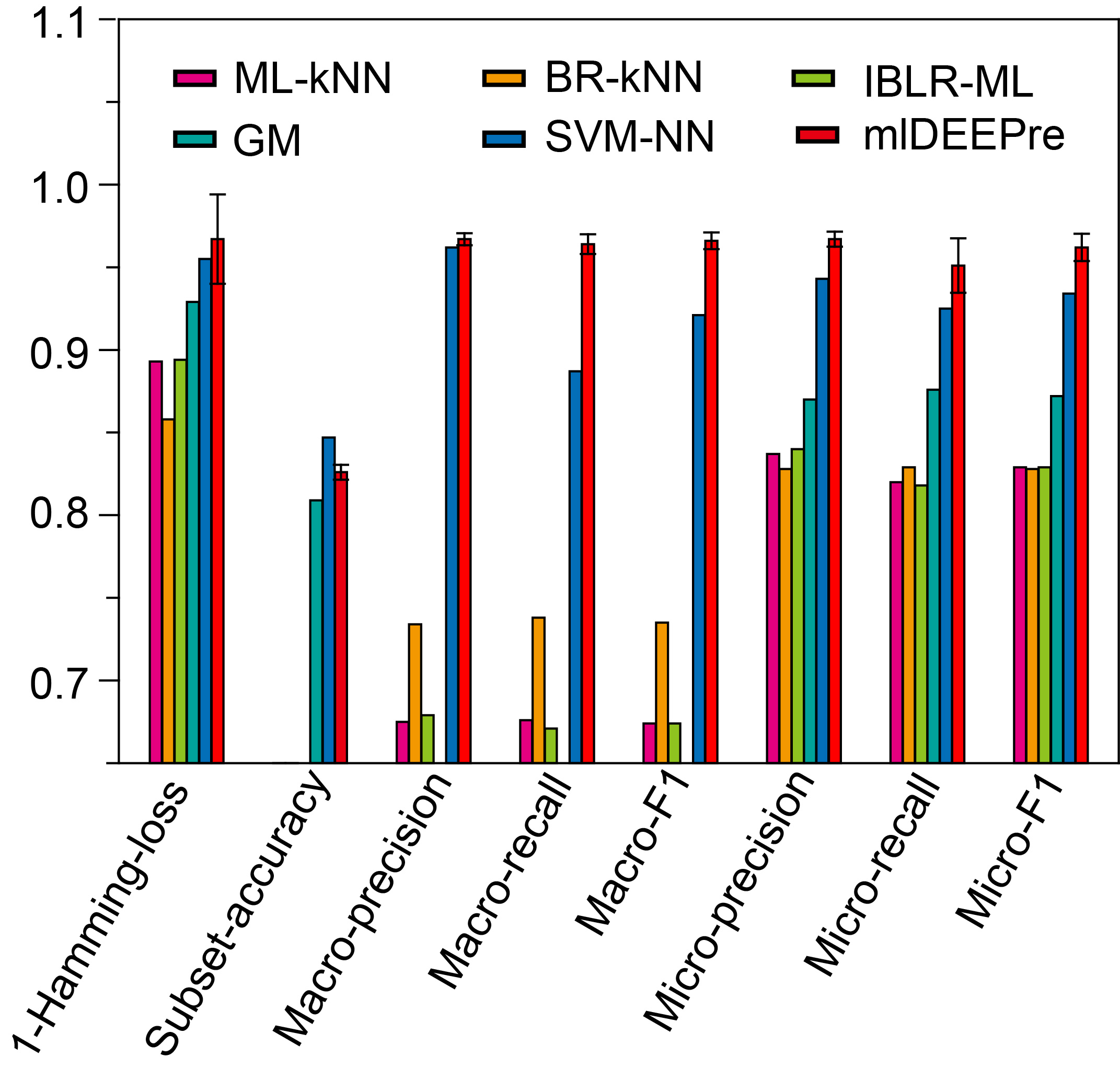}
\end{center}
\caption{The multi-functional classification testing performance of different models. Performance lower than 0.65 are not shown in the figure.}
\label{fig:multi-results}
\end{figure}

\begin{table}[!t]
\caption{The multi-functional classification performance of mlDEEPre on dataset II.}
\label{tab:multi-results}
\centering
\setlength{\tabcolsep}{10pt}
\begin{tabular}{@{} cccc @{}}
\toprule
Hamming-loss & Subset accuracy & Macro-precision & Macro-recall \\
\midrule
3.3 +/- 0.4\%  & 82.6 +/- 2.7\% & 96.7 +/- 0.3\% & 96.4 +/- 0.6\% \\
\midrule
\smallskip
Macro-F1 & Micro-precision  & Micro-recall & Micro-F1 \\
\midrule
 96.5 +/- 0.5\% & 96.7 +/- 0.4\% & 95.1 +/- 1.6\%  & 96.2 +/- 0.8\% \\
\hline
\end{tabular}
\end{table}

\subsection{mlDEEPre Performance}
In this paragraph, we describe the performance of the proposed method in predicting whether an enzyme is a mono-functional or multi-function enzyme. The training and testing datasets used in this work are shown in Table \ref{Tab:01} and Table \ref{tab:low_similarity_multi_data}. It is worthy pointing out that the data are imbalanced, with 22168 mono-functional enzymes and 1085 multi-functional enzymes. In this work, we employed penalized models to overcome the imbalance, forcing the model to pay more attention to the multi-functional class. We ran the model 30 times, each time with $70\%$ of all the data as training data and $30\%$ as testing data. We show the comparison results, both the average and the standard deviation, in Figure~\ref{fig:mono-results}. As suggested by Figure~\ref{fig:mono-results},our method can outperform all the other methods consistently across different criteria. Besides, our method is very stable, with the standard deviation of accuracy being as low as $0.09$.

Similar to the above experiments, we also ran the model 30 times, each time with $70\%$ of all the data as training data and $30\%$ as testing data, and evaluated mlDEEPre on the main classes prediction, obtaining the performance results shown in Table \ref{tab:multi-results} and Figure~\ref{fig:multi-results}.
According to Hamming-loss, the proposed multi-label model, mlDEEPre, predicts  97.6\% of all the actual main classes in the test dataset correctly, with the corresponding standard deviation being $2.7\%$, which outperforms all the other methods. Furthermore, we also compared our method with the other methods using other criteria. Although, because SVM-NN is good at predicting those rare class labels caused by imbalanced training samples (only 52 sequences belonging to class 6), the performance of SVM-NN (84.7\%) is slightly better than that of mlDEEPre (82.6\%) and GA (80.8 \%) in subset accuracy, mlDEEPre performs better than all the other methods in term of all the other criteria, including Macro-precision, Macro-recall, Macro-F1, Micro-precision, Micro-recall, and Micro-F1.

\subsection{Case Study}
Glutaminase is a phosphate-activated enzyme, which catalyzes the first step of glutaminolysis, hydrolysing glutamine into glutamate \cite{RN36}. The alternative splicing of its messager RNA results in its three isoforms, with isoform 1 and isoform 3 being capable of catalyzing while isoform 2 lacking the catalytic activity \cite{RN35}. To validate our model's ability to distinguish the different functionality of different isoforms, we obtained the sequences of the three Glutaminase isoforms from the UniProt and inputted them into our model. Our model predicted that isoform 1 and isoform 3 of Glutaminase are hydrolases acting on carbon-nitrogen bonds, being consistent with the experimental results. Our model also recognized isoform 2 as non-enzyme, which is consistent with the experimental result as well.

Aurora kinases B is a key enzyme regulating chromosomal segregation during mitosis, ensuring correct chromosome alignment and segregation as well as chromatin-induced microtubule stabilization and spindle assembly \cite{RN37}. Its over-expression may cause unequal distribution of genetic information, resulting in aneuploid cells, which can become cancerous \cite{RN38}. Aurora kinases B has five isoforms resulted from alternative splicing. Four of them have roughly equal length with high similarity, while isoform 3, having high expression in the metastatic liver with no expression in the normal liver, is only half of the length of the ``canonical'' isoform (142 amino acids v.s. 344 amino acids). Despite its much shorter length, the isoform does not lose its functionality. To further validate our model's ability of handling isoforms' functionality prediction, we collected the sequence of the five isoforms from the database and inputted them into our model. Our model's result is consistent with the experimental results. Particularly, our model predicted the functionality of the isoform 3 successfully, despite its sequence's large difference from the ``canonical'' sequence.

We also checked the performance of other four different servers on these two case studies. Among the five compared methods, only our method and EzyPred produced correct predictions for both cases.

\section{Discussion}

In this chapter, we discussed how to use deep learning to resolve the enzyme detailed function prediction problem. Although the outputs of this task are relatively simple compared to those of the other tasks in this thesis, this is a relatively standard structured prediction problem. There are two structures we need to consider when designing the method. The first one is the hierarchical structure of the EC labeling space. And the second one is the multi-labeling information for the multi-functional enzymes, which is implicit. To handle the first one, we used a novel level-by-level classification strategy that follows the hierarchical structure of the EC system.
Furthermore, we used transfer learning to facilitate the information flow within the labeling space, forcing the deep learning models to learn the structure of the labeling space implicitly. Regarding the second one, we used a novel multi-labeling loss function when training the deep learning models. Such a loss function can consider the dependency between different classes. Although this structured information is not explicitly presented, the loss function can incorporate it into the model implicitly. 

In the next chapter, we will go from a single molecule to system biology and demonstrate an example of predicting the disease genes, whose target is a graph. 


\chapter{PGCN: Disease Gene Prioritization by Disease and Gene Embedding through Graph Convolutional Neural Networks}
\label{chapter_pgcn}

\section{Chapter Introduction}
\label{sec:intro}
The last decade has seen a rapid increase in the adoption of whole-exome sequencing in the clinical diagnosis of genetic diseases \cite{Feero2014}.
However, the success rate of such genome-based diagnostics still remains far from perfect, with reported yields for a range of Mendelian diseases ranging from $\sim$20 to $\sim$50\% \cite{Taylor2015,Retterer2016}.
This relatively low yield is largely attributed to a considerable difficulty in differentiating disease-causing variants from a large pool of rare genetic variants that are not pathogenic and do not play roles in the expression of the disease phenotype \cite{MacArthur2014,RN867}.
To efficiently detect pathogenic variants and to improve the diagnostic rate of the genome-based approach, it is essential to have disease gene prioritization that substantially reduces the number of candidate causal variants and ranks them for further interrogations based on the association of the corresponding genes with the disease phenotype.

A number of computational methods have been developed to tackle the disease gene prioritization problem \cite{RN838,RN848}, and have been shown to be useful. For example, Endeavour \cite{RN814,tranchevent2008ndeavour,RN867} was able to associate \textit{GATA4} with congenital diaphragmatic hernia \cite{yu2013variants}; GeneDistiller \cite{RN820} discovered the role of \textit{MED17} mutations in infantile cerebral and cerebellar atrophy \cite{RN831}.
Based on the underlying computational techniques, existing disease gene prioritization methods can be categorized into five types. 
The first type is filter methods \cite{franke2004team,bush2009biofilter,RN835,RN853}, which sift the candidate list of genes into a smaller one according to the properties that associated genes should have. 
The second type of methods is based on text mining \cite{RN812,RN821,RN868,onto2vec,opa2vec}. 
Such methods score the candidate genes using the co-occurrence evidence with a certain disease from the literature. Thus, these methods can only detect associations that are already known. 
The third type is similarity profiling and data fusion methods \cite{RN814,RN816,RN823,RN829,RN830,RN845,RN860,RN859,RN858,RN867,RN877}. 
This is the dominant type in the disease gene prioritization community and includes the famous Endeavour method \cite{RN814}. These methods are based on the idea that similar genes should be associated with similar sets of diseases and vise versa. The similarity measurement can be defined using different data sources, such as Gene Ontology (GO) or the BLAST score. After obtaining the similarity scores from each data source, such methods apply data fusion to aggregate those scores into a global ranking. The fourth type is network-based methods \cite{RN838,RN836,RN844,RN847,RN846,RN851,RN837,singh2013prediction,rao2018phenotype}.
Such methods represent diseases and genes as nodes in a heterogeneous network, in which the edge weight represents their similarities. 
The last type is based on matrix completion techniques in recommender systems \cite{natarajan2014inductive,RN881}. These methods represent the disease gene association as an incomplete matrix and solve the disease gene prioritization problem by filling the missing values of the matrix. This category of methods has been shown to be the state-of-the-art \cite{RN881}. 

Despite the advances of the existing efforts, they have the following bottlenecks. Firstly, the similarity-based methods, which are rooted in the ``guilt-by-association'' principle, often fail to handle new diseases whose associated genes are completely unknown \cite{RN881}. Secondly, although the performance of the network-based methods is reasonable, they are biased by the network topology and cannot easily integrate multiple sources of information about genes and diseases \cite{RN848}. Thirdly, matrix completion methods assume and look for a weighted linear relationship between genes and diseases, which, in reality, is most likely to be highly nonlinear \cite{RN827}. In addition, most of the existing methods rely heavily on manually-crafted features or pre-defined rules of data fusion. Therefore, the disease gene prioritization problem remains elusive. 
On the other hand, the recent success of graphical models and deep learning in bioinformatics \cite{RN878,RN140, RN141,kim2018riddle,xia2018deerect} suggests the possibility to systematically incorporate multiple sources of information in the heterogeneous network and learn the highly nonlinear relationship between diseases and genes. 

In this chapter, we propose a novel disease gene prioritization method, PGCN, based on graph convolutional neural networks (GCNs) \cite{RN864,RN863,RN873,RN878}. Starting from a heterogeneous network which is composed of a genetic interaction network, a human disease similarity network, and a known disease-gene association network, with the additional information about genes and diseases from multiple sources, our method first learns embeddings for genes and diseases through graph convolutional neural networks, by considering both the network topology and the additional information of diseases and genes. Such embeddings are fed into an edge decoding (edge prediction) model to predict disease gene associations. Although we describe our method in two steps, our model can be trained in an end-to-end manner so that the model can learn the embedding and the decoding jointly. Usually, this disease gene prioritization problem is not solved as a structured prediction problem, but here, we solve it from the structured prediction aspect. We predict a relatively complete network by considering both the raw incomplete network and the side information of the nodes. The confidence scores of the predicted edges can be used for prioritization. Next, we will present how to formulate the original task as a link prediction problem and how to use GCNs to incorporate the graph structure information and the side information to resolve the task eventually in detail.

\section{Methods}
\label{sec:methods}
\begin{figure}[!t]
  \centerline{\includegraphics[width=100mm]{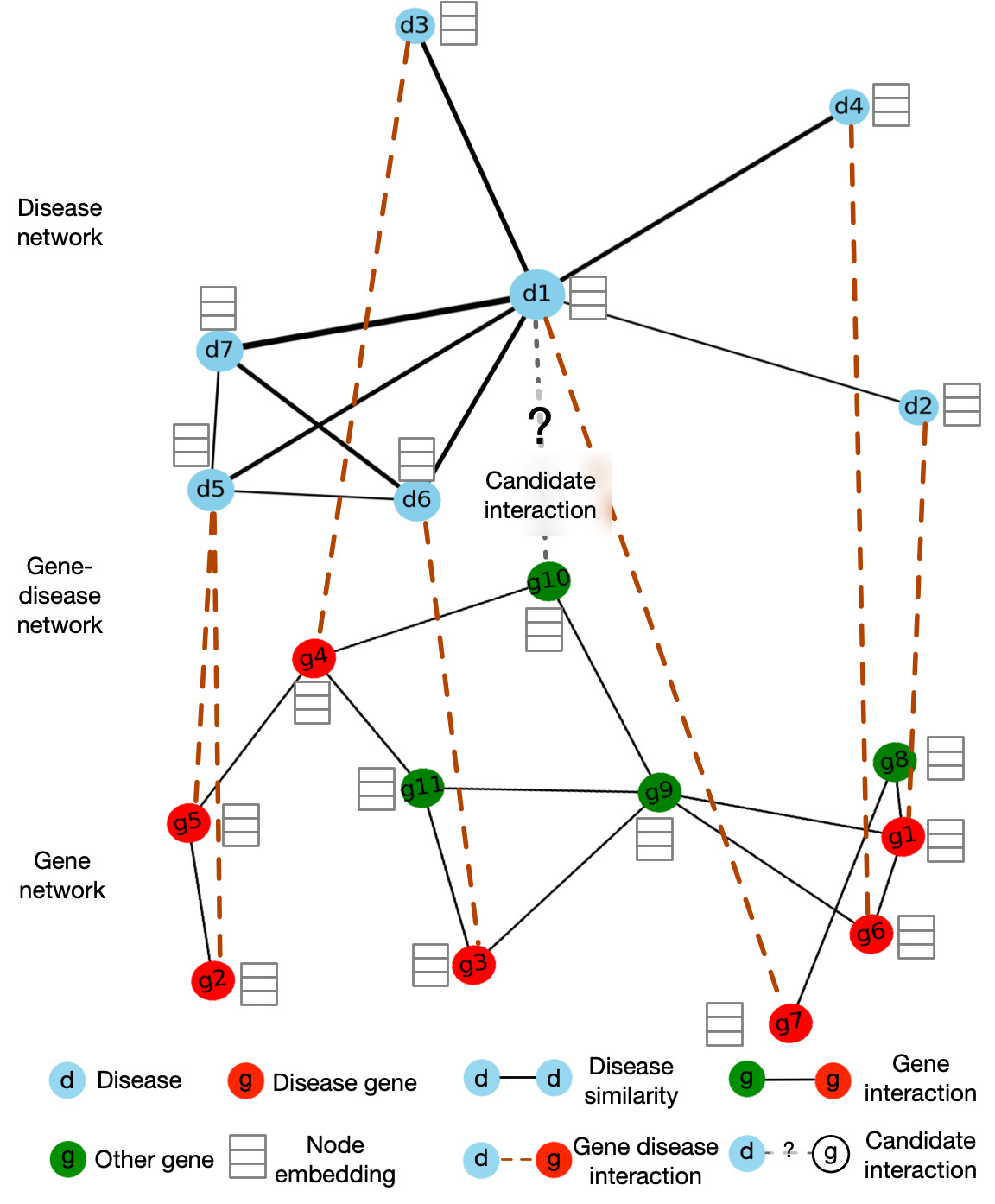}}
  \caption{Disease gene prioritization as a link prediction problem. Our goal is to predict the missing links given the heterogeneous network and additional raw representations of the nodes (diseases and genes).}
  \label{fig:link_prediction}
\end{figure}

\begin{figure}[!t]
  \centerline{\includegraphics[width=160mm]{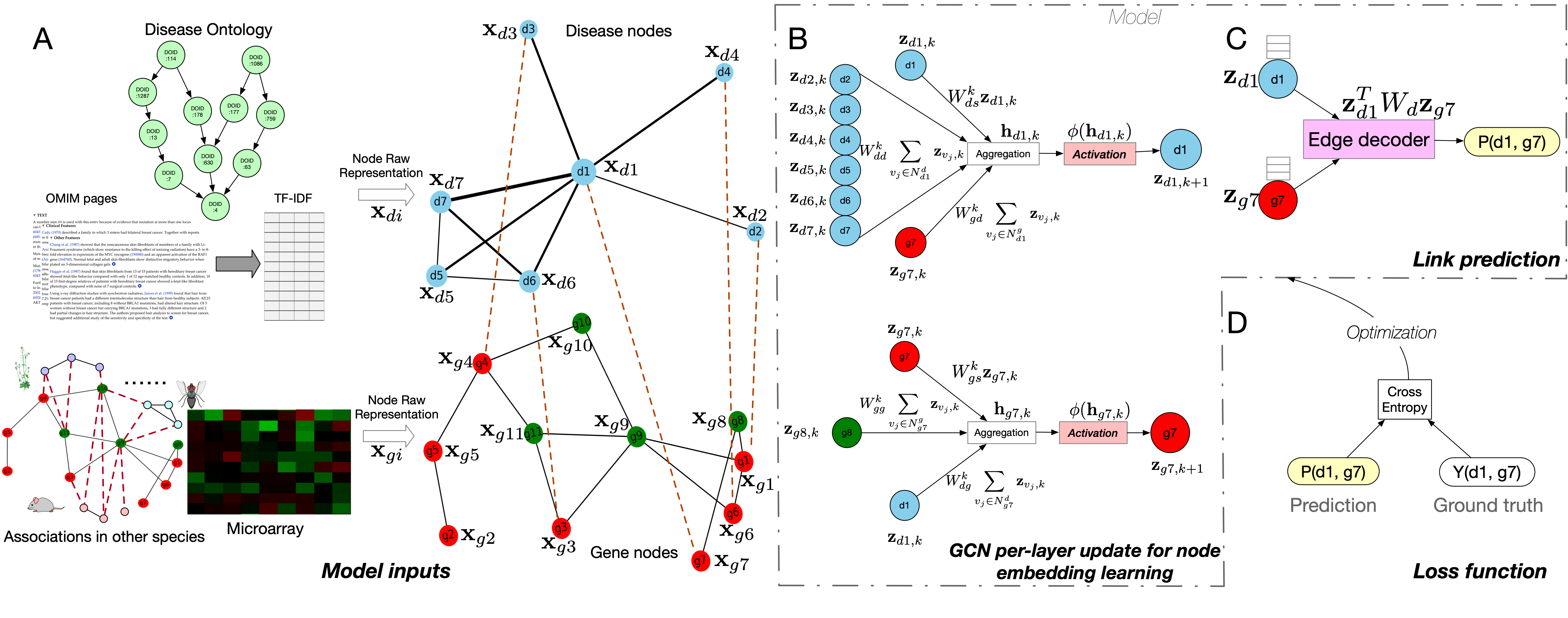}}
  \caption{Overview of the proposed method. (A) The input of our model contains two components, the heterogeneous network and the additional information for the nodes. (B) Examples of one layer of the graph convolutional neural network update for learning node embeddings. (C) The link prediction model. (D) The cross-entropy loss as the loss function for the end-to-end training.}
  \label{fig:gcn}
\end{figure}

In our work, we cast the disease gene prioritization problem as a link prediction problem. Unlike the previous studies \cite{natarajan2014inductive,RN881} which solve the problem with matrix factorization, 
we propose a novel method based on graph convolutional neural networks. We compile the disease similarities, genetic interactions, and disease-gene associations into a multi-nodal
heterogeneous network, as shown in Figure~\ref{fig:link_prediction}. In this network, the potential disease-gene associations can be considered as missing links and our goal is to predict those links \cite{link_prediction_2005,RN879}. 
The overview of our method is shown in Figure~\ref{fig:gcn}. The core idea of our method is to learn the nodes' latent representations (embeddings) from their initial raw representations (information encoded from different sources), considering the graph's topological structure and the nodes' neighborhood, after which we make predictions using the learned embeddings with the edge decoding model. Both the embedding model and the decoding model are trained in an end-to-end manner so that each model is optimized while being regularized by the other one.
In the following sections, we introduce each component of the proposed method in detail.

\subsection{Disease Gene Prioritization as a Link Prediction Problem}
\label{sub:link_prediction}
Recent studies \cite{natarajan2014inductive,RN881} have formulated the disease gene prioritization problem as a matrix completion problem and applied the recently developed methods in recommender systems, resulting in better performance than the previous state-of-the-arts. Although we also consider the problem as a recommender system problem, we treat the entire data structure as a heterogeneous network (Figure~\ref{fig:link_prediction} and Section \ref{sub:network}). Each node represents a disease or a gene, and each edge represents one specific kind of interaction. In addition, each disease or gene is supplemented with additional information from different data sources (Section \ref{sub:data_sources}). Our goal is to predict the potential links between disease nodes and gene nodes, whose link strength can be used for prioritization. Compared with the matrix factorization methods, our formulation can capture the nonlinear relationship between diseases and genes. Compared to the traditional network-based methods, our method is able to integrate the information from different sources in a systematic and natural way.

The core component of our method is the graph convolutional encoder (Section \ref{sub:node_embedding}), which can learn the embeddings from the nodes' neighborhood, node-specific information, and the topology of the heterogeneous network. The central problem for learning embeddings from graph data is to propagate and transform information. As shown in Figure~\ref{fig:gcn} (A), the entire graph starts from a heterogeneous network, with each node containing information from different sources. In the graph convolution model, each node's neighboring nodes define the computational graph of its local neural network, i.e., its own neural network architecture. Although the local computational graphs can be different for different nodes, the same operations share the same parameters and activation functions, which specify how the information is shared and propagated across the computational graph. Since we parameterize the graph convolution operation with a fully-connected neural network (Figure~\ref{fig:gcn} (B)), the model can seamlessly integrate information from different sources. The embeddings are fed into the link decoding model (Section \ref{sub:edge_prediction}). Thus, the proposed method can achieve problem-specific data integration systematically, whose parameters are learned from the data in an end-to-end manner.

\subsection{Network Compiling and Node Information}

\noindent {\bf Network Compiling}
\label{sub:network}

The network in our model (Figure~\ref{fig:link_prediction}) is a heterogeneous network containing three components: the gene network, the disease similarity network, and the disease-gene network. The disease-gene network is built from the OMIM database (November 26, 2017), with the associations being the links. After preprocessing, this network contains 12331 genes, 3215 diseases, and 3988 disease-gene associations. 

As for the gene network, we used HumanNet from \cite{RN836}. This large-scale functional gene network was constructed by considering multiple sources of information, including human mRNA co-expression, protein-protein interactions, protein complex, and comparative genomics information. In total, it incorporates 21 genomics and proteomics datasets from four species. Compared to the network built from single dataset, such as protein-protein interaction networks, it has higher accuracy and genome coverage \cite{RN836}. The usefulness of HumanNet in disease gene prioritization has been proved by previous studies \cite{singh2013prediction,natarajan2014inductive}. In summary, our gene network is composed of 12331 genes and 733836 edges with positive weights. More details about the network can be found in \cite{RN836}.

We used the MimMiner from \cite{van2006text} as the disease similarity network. This network was built from the text mining analysis on the OMIM database. For each disease, the anatomy and disease sections of the medical subject headings were used to extract terms from OMIM, whose frequencies were used as the feature vectors of the disease. After further refinement, the feature vectors were used to compute the pairwise similarities between the disease, which resulted in the MimMiner network. Although in the construction process, it did not involve gene information, the similarities were shown to be positively correlated with a number of measures of gene function. This network has also been used as a feature input in the previous disease gene prioritization methods \cite{singh2013prediction,natarajan2014inductive}. When setting the similarity threshold as 0.2, we obtain a disease similarity network with 3215 diseases and 645945 edges. 

\vspace{\baselineskip}
\noindent {\bf Data Sources for Node Raw Representation}
\label{sub:data_sources}

In contrast to the other network-based methods, our model can naturally incorporate additional information of the nodes from different sources. In our implementation, we incorporated the following data sources, although our method is generic and can take any source of information for diseases and genes.

As shown in Figure~\ref{fig:gcn} (A), we incorporated two kinds of additional information for the disease nodes. The first data source is Disease Ontology (DO) similarity. After collecting the ontology for the disease nodes, we calculated a similarity matrix for those diseases using the Resnik pairwise similarity \cite{Resnik_similarity} with the best-match average (BMA) strategy \cite{wang2007new}. For each disease, we took the corresponding row of this matrix as an additional feature vector for this node. 
The second data source is the clinical text from OMIM webpages. We collected the Clinical Feature and Clinical Management sections from the OMIM webpages for each disease, and we removed the most frequent and most rare words. Then, we counted the frequency of each unique word in the corpus related to each disease. To remove the bias of the relatively frequent words, we applied the TF-IDF scheme to the term frequency matrix and obtained the corresponding row as the feature vector for a disease. Finally, the two vectors were concatenated as the additional information for the disease. 

We also used two kinds of features as the additional information for the gene nodes. Following the strategy from \cite{natarajan2014inductive}, we collected the microarray measurement of gene expression level in different tissue samples from BioGPS and Connectivity Map. Since some genes are missing in the probes, we obtained 4536 features for 8755 genes. It is well-known that samples from the same cell type of different individuals tend to have a similar expression pattern, which results in redundant information in the obtained feature matrix. To eliminate the redundancy and reduce the dimensionality, we applied principle component analysis (PCA) on the features and used the first 100 eigenvectors as the feature representations from the gene expression microarray. The second type of additional information for genes is derived from gene-phenotype associations of other species. Following the previous studies \cite{singh2013prediction,natarajan2014inductive}, we used the phenotypes from eight species. As a result, we obtained eight matrices, whose rows represent different genes and columns represent the phenotypes of different species. We concatenated those gene-phenotype matrices together with the microarray matrix along the gene dimension, resulting in the additional information of the genes.

\subsection{Node Embedding with Graph Convolution}
\label{sub:node_embedding}
In this section, we introduce how we obtain the embeddings using graph convolutional neural networks, taking into consideration both the network topology, nodes' neighborhood, and the additional information of the nodes. Formally, given a graph $\mathcal{G} = (\mathcal{V}, \mathcal{E})$, where $\mathcal{V}$ represents the set of nodes and $\mathcal{E}$ represents the set of edges, with the adjacent matrix as $\bf{A}$, we denote $\textbf{x}_{i} \in \mathbb{R}^{m_{i}}$ as the additional information of the node $i \in \mathcal{V}$. Note that in our method, the value of $m_{i}$, which represents the dimension of the additional feature vectors, can be different for different kinds of nodes, i.e., gene nodes and disease nodes. The goal of embedding is to map each node to a vector $\textbf{z}_{i} \in \mathbb{R}^{c}$, where $c << m_{i}$, considering the information contained in $\bf{A}$ and $\{\textbf{x}_{i}\}_{i=1}^{|\mathcal{V}|}$.

The central problem of learning embedding with graph convolutional neural networks is to learn how to transform and propagate information (the additional information and intermediate embeddings of each node) across the entire network. In our method, the GCN module defines the information propagation architecture (the local computational graph) for each node using the node's neighborhood in the graph $\mathcal{G}$. In terms of the parameterization of the local computational graph, which defines how the information is propagated and shared, the parameters and weights are shared across all the local computational graphs built from $\mathcal{G}$, with the assumption that within the same graph $\mathcal{G}$, the way of sharing and propagating information should be the same. As a result, for a given node, each layer of graph convolutional neural networks aggregates and transforms the information (feature representations) from its neighbors and applies the same transformation to all part of the network, which is illustrated in Figure~\ref{fig:gcn} (B). If there is only one layer of graph convolution, the embedding will only aggregate information from its first-order neighbors. Thus, stacking $N$ layers of graph convolutional layers can make the embedding effectively convolve information from its $N$-order neighbors explicitly. Besides, when we stack more than one graph convolutional layers, the information of each single node can start broadcasting to the entire network implicitly, whose affect depends on the network topological structure (size, connectivity \textit{etc.}). By using multiple convolutional layers, we are able to learn the embedding of nodes, considering the network topology, local neighborhoods, and additional information of the nodes. 

Formally, in each layer, for each node, the information aggregation and transformation model takes the following form:
\begin{equation}
\begin{gathered}
\textbf{h}_{i, k} =  \sum_{l} \sum_{j \in \mathcal{N}_{i}^{l}} c_{i,j} W_{l}^{k} \textbf{z}_{j, k} + W^{k}_{t_{i}, s} \textbf{z}_{i, k},
\label{eq:aggre}
\end{gathered}
\end{equation}

\begin{equation}
\begin{gathered}
\textbf{z}_{i, k+1} = \phi(\textbf{h}_{i, k}),
\label{eq:act}
\end{gathered}
\end{equation}

\noindent where $\textbf{z}_{i, k} \in \mathbb{R}^{c_k}$ is the hidden representation of node $i$ in the $k$-th graph convolutional layer and $c_k$ is the dimensionality of that hidden representation;  $\textbf{h}_{i, k}$ represents the feature vector which has aggregated the information from the $k$-th layer hidden representations of the node's neighbors; $l$ represents the link type, \textit{i.e.}, genetic interaction, disease-disease similarity, or disease-gene association; $\mathcal{N}_{i}^{l}$ are the neighbors of $i$, which are linked by the link type $l$; $W_l^k$ is the weight parameter related to the link type $l$, such as $W_{dg}^k$, $W_{gd}^k$, $W_{dd}^k$ and $W_{gg}^k$ in Figure~\ref{fig:gcn} (B); $c_{i,j}$ is the normalization constant, inspired by \cite{RN878}, which is defined as $c_{i,j} = 1/\sqrt{|\mathcal{N}_{i}||\mathcal{N}_{j}|}$; $W^{k}_{t_{i}, s}$ is the weight parameter preserving the information from the node itself, where $t_{i}$ indicates the type of the node; $\phi$ is the non-linear activation function, which is usually chosen as rectified linear unit (ReLU). Note that the above aggregation and transformation formulas are related to the neighbors of a certain node, which means that the computational graph architecture can be different for nodes with different local neighborhood structures. We show examples of two very different computational graphs for nodes ${d1}$ and ${d7}$ in Figure~\ref{fig:gcn} (B). On the other hand, although the computational graphs can be different, the parameters are only related to the link type, not related to the node neighborhoods, which means that the parameterization is shared across the entire graph.

In our method, we use summation as the information aggregation method in the GCN model. With different information aggregation methods, it can result in different GCN variants. However, no matter which method we choose, the aggregation and transformation layer converts the hidden representation of node $i$ in layer $k$, $\textbf{z}_{i, k}$, into the hidden representation in the next layer as $\textbf{z}_{i, k+1}$. We use the output of the last graph convolutional layer, $\textbf{z}_{i, N}$, as the final embedding for that node, $\textbf{z}_{i}$. Naturally, the input of the first convolutional layer is the original feature vector of each node (Section \ref{sub:data_sources}). Formally, $\textbf{z}_{i, 0} = \textbf{x}_{i}$.

\subsection{Edge Prediction from Embeddings}
\label{sub:edge_prediction}
In this section, we introduce how to reconstruct edges in the network with the embeddings learned from GCN. We use the bilinear decoder with the following form as the the edge decoder:
\begin{equation}
\begin{gathered}
P({di}, {gj}) = \sigma(\textbf{z}^T_{di} \textbf{W}_d \textbf{z}_{gj}),
\label{eq:edge_encoder}
\end{gathered}
\end{equation}
where $\textbf{z}^T_{di} \in \mathbb{R}^c$ is the learned embedding of a disease node ${di}$; $\textbf{z}_{gj} \in \mathbb{R}^c$ is the learned embedding of a gene node ${gj}$; $ \textbf{W}_d \in \mathbb{R}^{c*c}$ is the trainable parameter matrix, which models the interaction between each two dimensions of $\textbf{z}^T_{di}$ and $\textbf{z}_{gj}$; $\sigma$ is the sigmoid function which converts the output value of the edge decoder to the range of $(0, 1)$, as a probability value. This edge decoder is illustrated in Figure~\ref{fig:gcn} (C). Note that, similar to the graph convolutional neural network model, the parameters of the bilinear decoder model are also shared across different gene-disease pairs, which can effectively reduce the risk of overfitting. 

Taking the GCN model and the edge decoder model together, we have the following trainable parameters: (1). The link-type-specific and layer-specific convolutional weight parameters $W_l^k$, which suggest how to aggregate and transform information from the node's neighbors. (2). The node-type-specific and layer-specific weight parameters $W_{t,s}^k$, which indicate how to preserve and transform nodes' self-information from one layer to the next. (3). The weight parameters of the bilinear edge decoder model, $\textbf{W}_d$, which model the interaction between two dimensions of the input embeddings of two nodes. As shown in Figure~\ref{fig:gcn} (B) and (C), the GCN model and the edge decoder model can be combined together to form an integrated model, which takes the raw representation of two nodes and output interaction probability. Consequently, the entire model and all the parameters can be trained in an end-to-end manner.

\subsection{Model Hyper-parameters}
\label{sub:model_para}
In this section, we introduce the hyper-parameters that we chose when building and training the model.

First, we used the cross-entropy loss as the loss function to train the entire model, which has the following form:
\begin{equation}
\begin{gathered}
L({di}, {gj}) = -\log P({di}, {gj}) - \mathbb{E}_{{gn} \sim \mathcal{P}({gj})} \log (1 - P({di}, {gn})),
\label{eq:loss_pgcn}
\end{gathered}
\end{equation}
where ${di}$ and  ${gj}$ form an edge in the training data. That is, the ground truth value $Y({di}, {gj}) = 1$ in Figure~\ref{fig:gcn} (D). By using the cross-entropy loss, we want the model to assign the probabilities for the observed training edges as high as possible while assigning low probabilities for the random edges. Following the previous studies \cite{RN866,RN878}, we used negative sampling to achieve this, which is illustrated by the last term in Eq. (\ref{eq:loss_pgcn}). For each existing edge $({di}, {gj})$, which is a positive sample, we sampled a random edge $({di}, {gn})$ by randomly choosing the second node ${gn}$, which follows the sampling distribution $\mathcal{P}$. Considering all the edges, we have the final cross-entropy loss of the model as:

\begin{equation}
\begin{gathered}
L = \sum_{({di}, {gj}) \in \mathcal{E}_{dg}} L({di}, {gj}),
\label{eq:loss_final}
\end{gathered}
\end{equation}

\noindent where $\mathcal{E}_{dg}$ represents all the edges connecting diseases and genes. As we discussed in the previous sections, we trained the model in an end-to-end manner, where the loss function gradient is back-propagated to the parameters in both the GCN model and the edge decoding model. This end-to-end training strategy is more likely to find problem-specific, effective models and embeddings, which has been proved by previous studies \cite{RN140,RN141,umarov2019promoter,zou2019mldeepre}. 

In terms of implementation, we set the number of layers as 2, with the dimension of the hidden representation as 64 and the final embedding dimension as 32. We trained the model using Adam optimizer, with the learning rate as 0.001. To reduce overfitting, we used the combination of dropout on the hidden layer unites with the dropout rate as 0.1, and the legendary weight decay method. We initialized the model's parameters using Xavier initializer.
During training, we fed mini-batch of edges to the model, with the batch size as 512. This can reduce the memory requirement and serve as an additional regularizer that further alleviates overfitting. In total, we trained the model for 300 epochs. With the help of a Titan Xp card, we finished the training of a model in 10 hours.

\section{Results}
\label{sec:results}

In this section, we show the performance of the proposed method and the five state-of-the-art methods. We first present the experimental settings in detail, introducing the five competing methods. After that, we show the performance of all the methods on recovering missing associations, and on discovering associations for novel genes and/or diseases that are not seen in the training. Finally, we demonstrate the effectiveness of the proposed method by investigating the predicted associations on breast cancer and the biological meaning of the learned embeddings.

\subsection{Experimental Setting}
\label{sub:exp_setting}

\noindent {\bf Evaluation Dataset and Experiments}

We built the dataset from the OMIM database (November 26, 2017). After preprocessing, we constructed a dataset with 12331 genes, 3215 diseases, and 3988 associations. Comprehensive experiments were designed to evaluate the performance of the proposed method. Firstly, we assessed the overall ability to recover the known disease-gene associations using the standard cross-fold validation strategy. During the experiments, we randomly hid 10\% associations as the testing set and used the remaining 90\% as the training set. This experiment mimics the situation in which partial knowledge about a disease is known (i.e., some associated genes are known) and we want to complete the knowledge by finding out other associated genes. The results are shown in Section \ref{sub:cross_fold}. However, this task is neither the most practically important nor the most challenging one for disease gene prioritization. In reality, researchers are more interested in predicting associations for diseases and/or genes that are not known before. To mimic such situations, we further designed three experiments. The first one is to predict associations for singleton genes \cite{singh2013prediction}, which means that the gene has only one associated disease and is not included in the training set (Section \ref{sub:new_gene}). The second one is to predict associations for new diseases. We excluded all the associations for certain diseases from the training set and challenged different methods to recover these associations (Section \ref{sub:new_disease}). In the third experiment, we tested the performance of different methods on recovering novel associations, which are defined as the ones that both the disease and the gene are absent in the training set (Section \ref{sub:new_associations}). Finally, we showed a case study of our predictions for breast cancer in Section \ref{sub:case_study}. 

\vspace{\baselineskip}
\noindent {\bf Compared Methods}
\label{sub:comp_methods}

Five state-of-the-art methods for disease gene prioritization are included in the comparison. 
The first one is Katz \cite{singh2013prediction}, which is a typical network-based method. It computes the node similarity based on the network topology. The similarity matrix is then used to make predictions for disease gene associations.
The second one is Catapult \cite{singh2013prediction}, another network-based method. It combines supervised learning with social network analysis, and has been shown to be the state-of-the-art network-based method \cite{singh2013prediction,natarajan2014inductive}.
This method deploys a biased support vector machine (SVM) as the classifier while the features are derived from random walks in the heterogeneous gene-trait network. It outperformed the previous network-based methods, such as PRINCE and RWRH, significantly.
The third one is a very recent network-based method, the Graph Convolution-based Association Scoring (GCAS) method \cite{rao2018phenotype}. This method used GCN as a pure network analysis tool which can perform information propagation on the similarity and association networks. Our method differs from GCAS in that we use GCN to integrate information from different sources and learn embeddings specifically for this problem, which are particularly suitable for the downstream edge prediction task.
The fourth one is the Inductive Matrix Completion (IMC) method \cite{natarajan2014inductive}, which introduced the matrix completion method into the disease gene prioritization field for the first time. It constructs features from genes and diseases from multiple sources, ranging from gene expression array to disease similarity networks. It then learns low-rank latent vectors for diseases and genes which can explain the observed disease-gene associations, taking into consideration features using a linear model. The learned latent vectors are then used for making further predictions.  
The last one is the very recently developed GeneHound method \cite{RN881}. It also utilizes the matrix completion method but combines Bayesian approach with matrix completion, which takes the disease-specific and gene-specific information as the prior knowledge. This method has been shown to outperform the legendary Endeavour method significantly \cite{RN881}.

\vspace{\baselineskip}
\noindent {\bf Evaluation Criteria}
\label{sub:evaluation_criteria}

We used the following criteria to evaluate our method and the competing methods: Area Under the Receiver Operating Characteristic curve (AUROC), Area Under the Precision-Recall Curve (AUPRC), Boltzmann-Enhanced Discrimination of ROC (BEDROC), Average Precision at $K$ (AP@$K$), and Recall at $K$ (R@$K$) score. AUROC is a commonly used criterion in machine learning, which computes the area under the ROC curve. In the disease gene prioritization problem, it can be interpreted as the probability of a true disease-associated gene is ranked higher than a false one selected randomly in a uniform distribution.
Similar to AUROC, AUPRC computes the area under the precision-recall curve. 
BEDROC, proposed to solve the ``early recognition'' problem, can be interpreted as the probability of a disease-associated gene being ranked higher than a gene selected randomly following a distribution in which top-ranked genes have a higher probability to be chosen. The formal definition of BEDROC can be referred to \cite{truchon2007evaluating}. P@$K$ computes the precision of the prediction if we consider the top $K$ predicted associations. Recall at $K$ considers the recall score within the top $K$ predictions. Those five criteria can provide a comprehensive evaluation of the proposed method.

\begin{table}[t]
\caption{The overall performance of the six compared methods. Under each criterion, the method with the best performance is in bold and the second best is underlined.}
\label{tab:overall_result}
\begin{center}
 \begin{tabular}{||c c c c c||} 
 \hline
 Method & AUROC & AUPRC & AP@200 & BEDROC \\ [0.5ex] 
 \hline\hline
 PGCN & \textbf{0.877} & \textbf{0.896} & \textbf{0.976} & \textbf{0.987} \\ 
 GeneHound & \underline{0.805} & 0.793 & 0.831 & 0.908 \\
 IMC & 0.780 & \underline{0.809} & \underline{0.928} & \underline{0.965} \\
 GCAS & 0.614 & 0.623 & 0.753 & 0.813 \\
 Catapult & 0.597 & 0.657 & 0.783 & 0.884\\
 Katz & 0.557 & 0.596 & 0.595 & 0.790\\ [1ex] 
 \hline
\end{tabular}
\end{center}
\end{table}

\begin{figure}[!t]
  \centerline{\includegraphics[width=160mm]{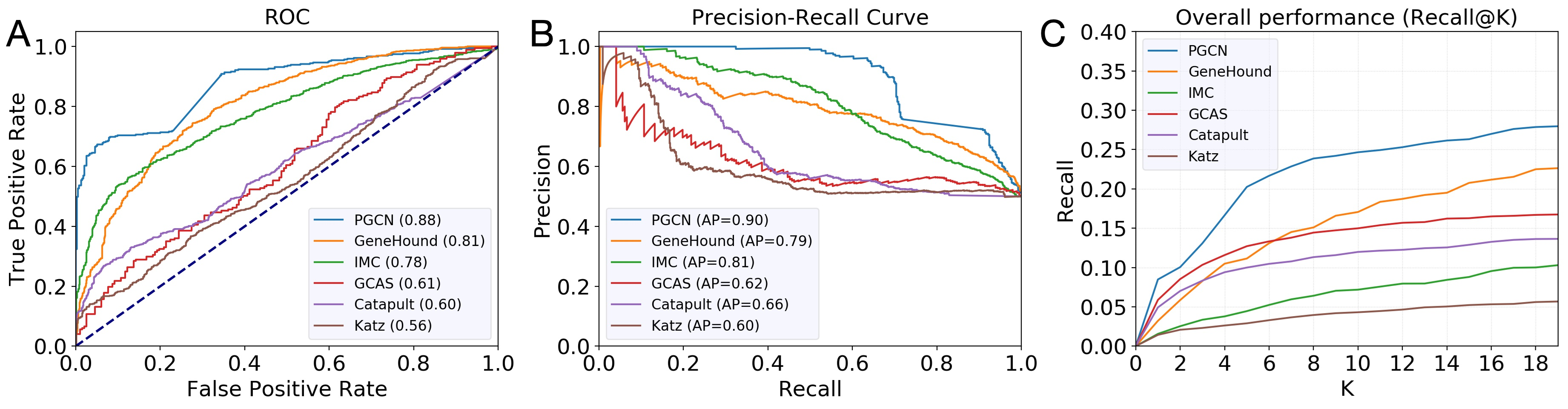}}
  \caption{Performance comparison of different methods. (A) ROC curves. (B) PRC curves. (C) Recall at $K$.}
  \label{fig:overall}
\end{figure}

\begin{figure}[!t]
  \centerline{\includegraphics[width=160mm]{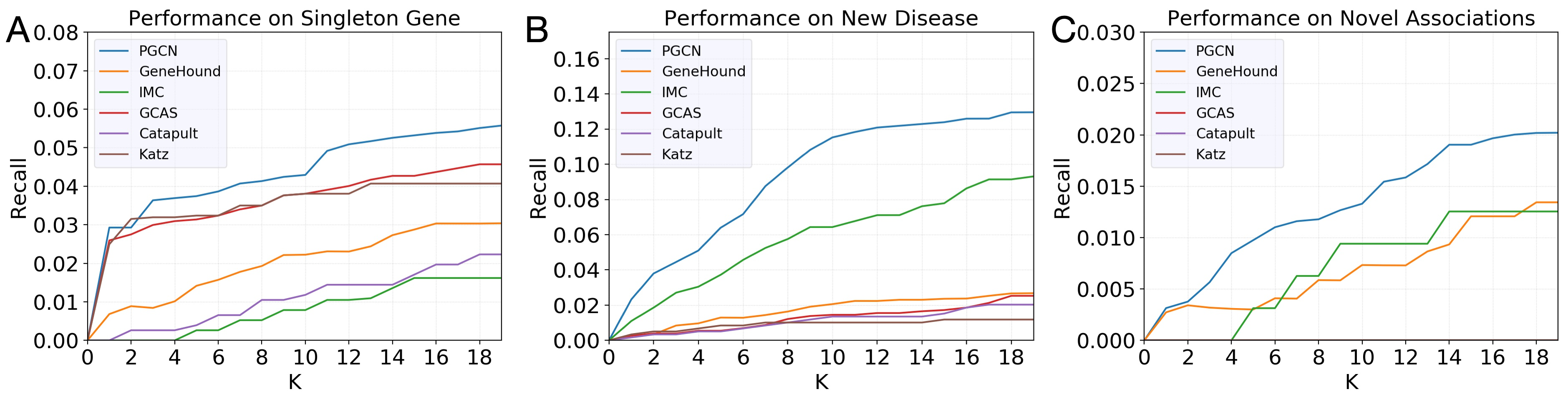}}
  \caption{Performance in terms of recall at $K$ of different methods on recovering associations for new genes and/or diseases. (A) Singleton genes association prediction. (B) New disease association prediction. (C) Novel association prediction.}
  \label{fig:ark_new}
\end{figure}

\subsection{Overall Performance}
\label{sub:cross_fold}
We randomly hid 10\% associations as the testing set and used the remaining 90\% edges as the training data to evaluate the overall performance of different methods on recovering the hidden associations. The performance of different methods is summarized in Table. \ref{tab:overall_result}. As shown in the table, the two matrix completion methods, GeneHound and IMC, can outperform the other three network-based methods, GCAS, Catapult and Katz, significantly across different criteria. The main reason is that they can take full advantage of the gene- and disease-specific information while the network-based methods are biased towards the network topology. On the other hand, the proposed method, PGCN, which can utilize both the network topology information and the additional information of the nodes in a systematic and natural way, can outperform all the state-of-the-art methods significantly and consistently across different criteria with a large margin. In terms of AUPRC, PGCN can outperform the second best method by around 10\%. We further show the ROC curves and the PRC curves in Figure~\ref{fig:overall} (A,B). It is clear that PGCN significantly outperforms all the state-of-the-art methods under all the false positive rates and all the recall values, which suggests that PGCN is an overall much better method. In disease gene prioritization, Recall at $K$ is also an important indicator because the top-ranked genes are candidates for further investigation. Figure~\ref{fig:overall} (C) shows the recall of different methods when different numbers of top predictions are considered. Interestingly, GCAS can perform quite well when $K$ is very small, compared to GeneHound, IMC, Catapult and Katz. Yet PGCN is clearly more sensitive than all the competing methods regardless of the number of top predictions to be considered. All the above results demonstrate that the proposed method can outperform the other methods in recovering the hidden associations between diseases and genes.

\subsection{Performance on Singleton Genes}
\label{sub:new_gene}
Following the idea of \cite{singh2013prediction}, we checked the performance of different methods on predicting the associations of singleton genes, which are defined as those genes with only one link in the database. In our experiment, the only links for the singleton genes were removed from training, which means that the methods needed to predict the associations ``from scratch". We used recall at $K$ to evaluate different methods, which is a difficult measurement because each test gene has one and only one true association. As shown in Figure~\ref{fig:ark_new} (A), PGCN consistently recovers the missing associations for singleton genes better than other methods. We also noticed that the network information is very important when $K$ is small (between 1 and 10) because the improvement of PGCN over the network-based method (e.g., Katz) is not large, which is consistent with the previous findings \cite{natarajan2014inductive}. However, as the number of top predictions being considered increases, the disease- and gene-specific information plays an increasingly important role, which leads to significantly better recall when $K$ is large.

\subsection{Performance on New Diseases}
\label{sub:new_disease}
Next, we evaluated the ability of different methods on predicting associations for novel diseases for which no associated genes are known. For a novel disease, all of its associations with genes were removed during training and different methods were challenged to recover those missing associations. This task is considerably less difficult in terms of recall than recovering the associations for singleton genes because a disease can be associated with more than one genes. At the same time, this task is practically important because it is directly related to the molecular diagnosis for human diseases. As shown in Figure~\ref{fig:ark_new} (B), IMC can outperform all the other previous methods with a large margin. The reason is that IMC is based on matrix completion techniques, which can effectively incorporate the disease-specific information \cite{natarajan2014inductive}. Our method, however, can not only incorporate disease- and gene-specific information, but also the known disease-gene associations in a unified framework. Furthermore, our method trains the disease and gene embeddings and link prediction in an end-to-end manner, and thus further significantly improves the performance over IMC. 

To further understand how our method works, we investigated a disease, atrioventricular septal defect-4 (AVSD4), for which we removed its only associated gene, \textit{GATA4}, during training and PGCN successfully recovered it with the highest score. The link between AVSD4 and \textit{GATA4} is built through another disease, ventricular septal defect-1 (VSD1), which is known to be associated with \textit{GATA4}. Our method detected the similarity between the two diseases, AVSD4 and VSD1, according to their embeddings learned by our method, which is illustrated in Figure~\ref{fig:embedding} (B). However, this similarity is very difficult to be detected because in the disease similarity network, the two diseases have a wrong similarity score of 0, which suggests that they are two completely irrelevant diseases. Therefore, all the network-based methods failed to predict the association between AVSD4 and \textit{GATA4}. Our method, on the contrary, systematically incorporates not only the network topology, but also the disease-specific information. In this particular case, the disease-specific information plays an important role in the disease embedding and thus PGCN was able to detect the similarity between the two diseases in the embedding space, which led to the correct prediction on the association between AVSD4 and \textit{GATA4}.

\subsection{Performance on Novel Associations}
\label{sub:new_associations}
We then evaluated the prediction performance of different methods for novel associations, which are defined to be the association between a disease and a gene, both of which have no association in the training set. This is the most stringent and challenging requirement. In order for a method to recover such associations, neither the disease end nor the gene end of the association can be directly used. The method must be powerful enough to effectively use the disease- and gene-specific information, and propagate the information through other diseases, genes, and their associations in the heterogeneous network. The results are shown in Figure~\ref{fig:ark_new} (C). As expected, the recall values of all the methods have a clear drop comparing to the two previous tasks. We found that the three network-based methods did not perform well in this task as they were unable to recall any true associations. We suspect that the main reason for this is that the definition of novel associations makes network propagation alone extremely difficult. To support this view, the two matrix completion methods, which can take advantage of the specific information of genes and diseases, performed much better than the network-based methods. Our method consistently outperforms all the competing methods, and the improvement increases with a larger $K$. 

\subsection{Ablation Study and Case Study}

\noindent {\bf Ablation Study}
\label{sub:data_input}

\begin{figure}
  \centerline{\includegraphics[width=90mm]{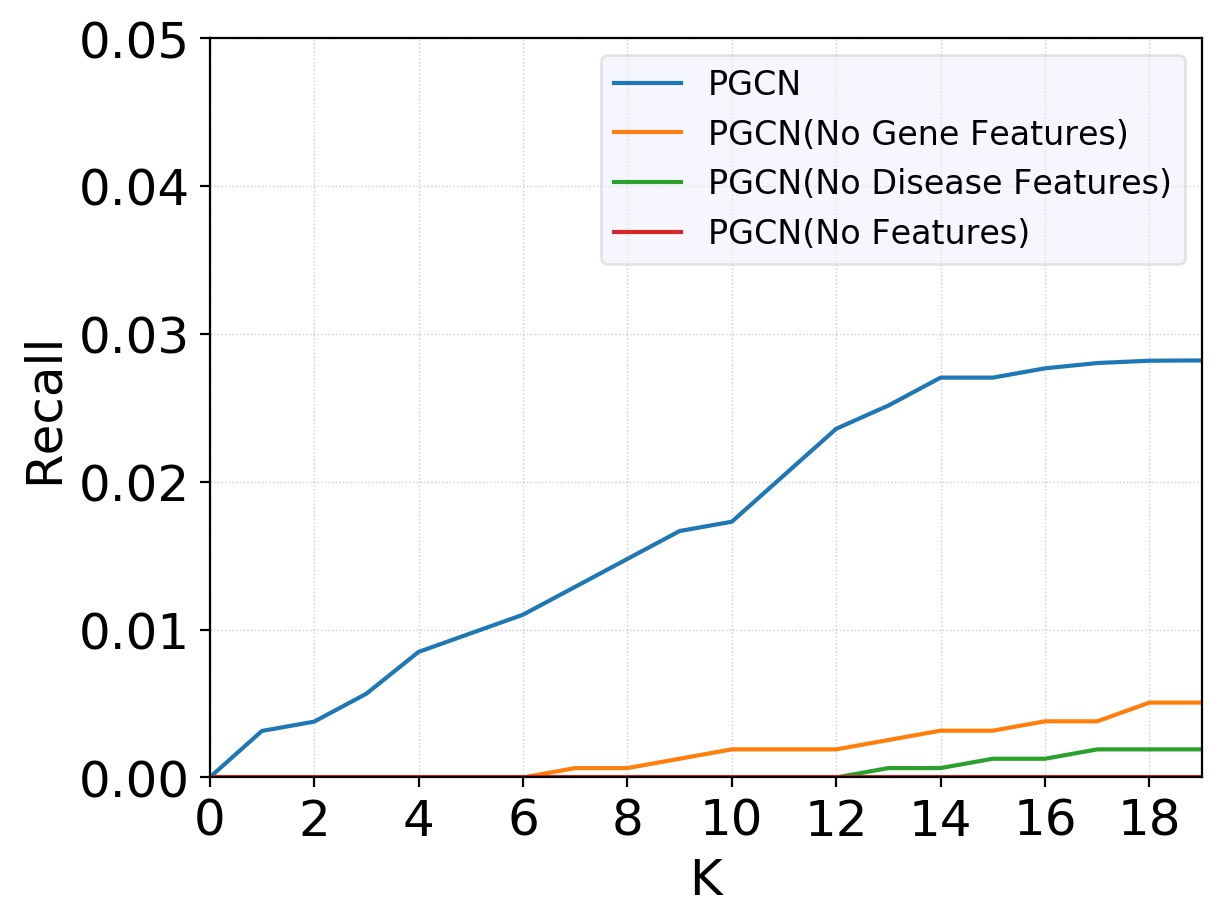}}
  \caption{The importance of the disease- and gene-specific information.}
  \label{fig:ark_feature}
\end{figure}

To expand the analysis for the importance of the disease- and gene-specific information, we further investigated its contributions to the prediction performance of our method.  Focusing on the novel association prediction task, we excluded the disease features, the gene features, and both of them, respectively, and evaluated the performance of the corresponding models. As shown in Figure~\ref{fig:ark_feature}, both the disease features and the gene features are very important for the proposed method. If we exclude either one of them, the performance will degrade significantly. If we exclude both of them, the model cannot recall anything when $K$ is in the range of $(1, 19)$. On the other hand, disease features are more important than the gene features as the model with the disease features begins to recall some true associations when $K=7$ while the model with the gene features begins to recall some true associations when $K=13$. The reason may be that the gene network we used is HumanNet, which is a very informative database that was built from multiple data sources. 

\vspace{\baselineskip}
\noindent {\bf Case Study}
\label{sub:case_study}

As a case study, we investigated the top 10 associations for breast cancer. Among these 10 genes, other than the four ground-truth breast cancer-related genes reported in the OMIM dataset, our model also predicted three interesting genes: \textit{Axin2}, \textit{TLR4}, and \textit{PTPRJ}, which were reported to be related to breast cancer. For example, \textit{Axin2} was found to be included in the \textit{Wnt/$\beta$-catenin/Axin2} pathway, which can regulate the breast cancer invasion and metastasis \cite{li2016c}; \textit{TLR4} was found to be overexpressed in the majority of the breast cancer samples and also related to the metastasis of breast cancer \cite{volk2014paclitaxel}; and \textit{PTPRJ} forms \textit{DEP-1/PTPRJ/CD148}, which is receptor-like protein tyrosine phosphatases (PTP), that was found to be mutated or deleted in human breast cancer \cite{spring2015protein}. These results suggest the potential application of our method on discovering new genes related to complex human diseases.

\begin{figure}
  \centerline{\includegraphics[width=80mm]{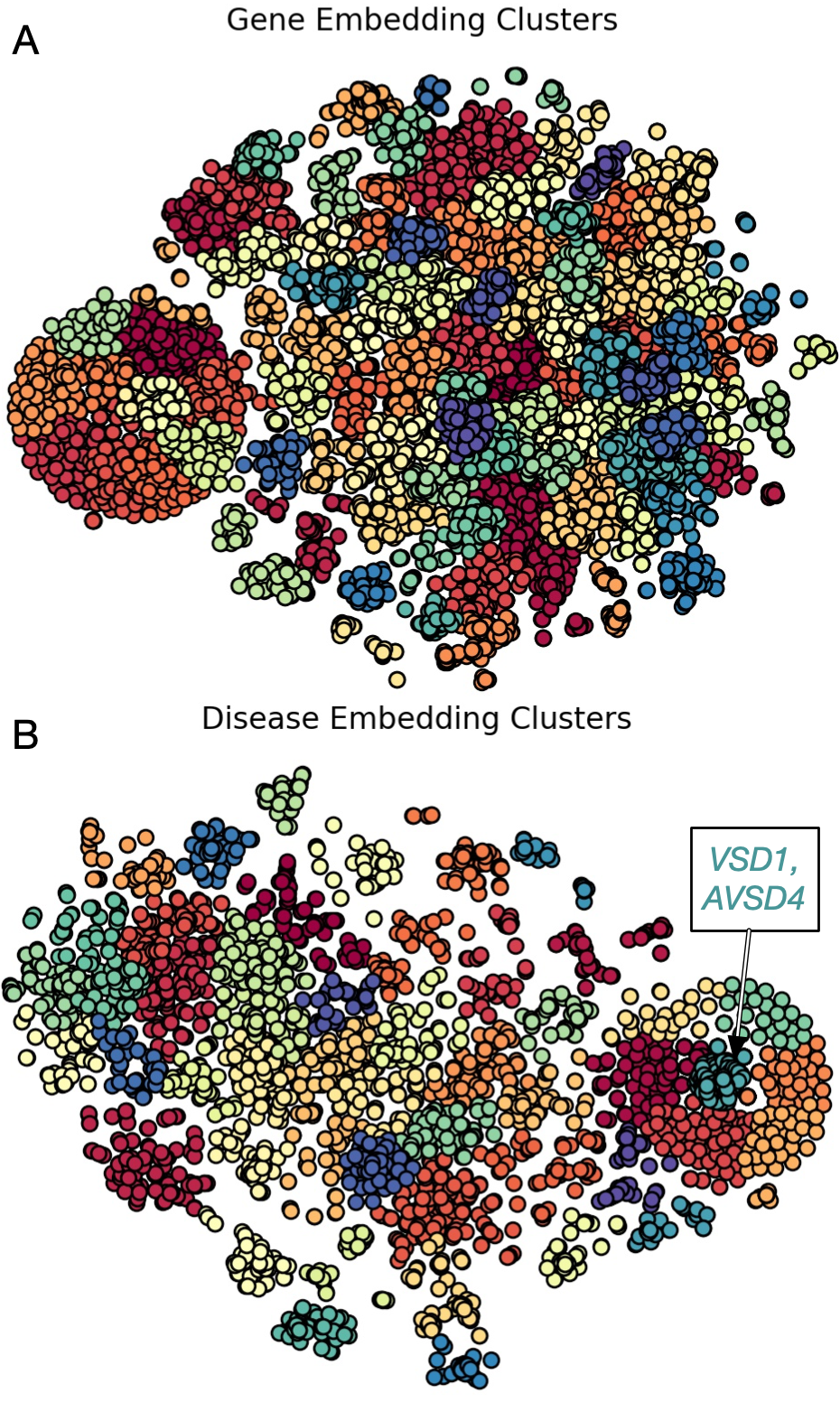}}
  \caption{Visualization of the clustering of embeddings in 2D space using t-SNE.}
  \label{fig:embedding}
\end{figure}

\subsection{Biological Meaning of Learned Embeddings}
\label{sub:embedding_visualization}

To gain insights into how the final embeddings represent the gene and the disease features, we mapped the 32-dimensional vector of each node into a 2-dimensional space using t-SNE \cite{maaten2008visualizing} for visualization (Figure~\ref{fig:embedding}).
From these 2-dimensional data, we observed that many points are located closely to some other points and they form clusters of a wide range of sizes in both the gene feature space and the disease feature space.
Since closely located data points suggest that the corresponding features are biologically similar in our embedding, we analyzed the extent to which closely located data points in the low dimensional space represent the biological association of the corresponding features.

To this end, we first clustered the data points into 100 groups for the gene node embedding and 50 groups for the disease node embedding using hierarchical clustering.
To analyze the functional association for gene features, we mapped genes in each cluster to biological pathways that they are associated with using the KEGG pathway data \cite{Kanehisa2000} and evaluated their statistical significance.
We found that all of the 100 clusters have statistically significant levels of association with biological function ($p < 0.05$; hypergeometric test).
Notably, the cluster which includes RPL3L over-represents the genes involved in the formation of the ribosome ($p < 10^{-82}$), while the one including H2AFX has a disproportionately large number of genes involved in the DNA repair response of systemic lupus erythematosus ($p < 10^{-39}$).  

For the analysis of the disease node embedding, we used the Human Phenotype Ontology (HPO) dataset \cite{Kohler2019} to associate each disease with corresponding HPO phenotypic abnormality terms.  
Similar to the gene-function association analysis, all of the 50 clusters are found to have statistically significant number of diseases with association to some sort of phenotypic abnormalities ($p < 0.01$).
In particular, we found that the cluster including Parkinson Disease, Late-onset (OMIM:168600) is enriched in diseases that are associated with slow movements ($p < 10^{-22}$), while the cluster with Neuropathy, Hereditary Sensory And Autonomic, Type II (OMIM:201300) over-represents genes associated with muscular hypotonia ($p < 10^{-10}$).

These results indicate the ability of our method to generate embeddings that preserve the gene and the disease associations that are critical to the disease gene prioritization task.
They also highlight the possibility to interpret the gene node and the disease node embeddings in a biologically meaningful way, which is essential to gain biomedical insights into novel disease-gene associations.

\section{Discussion}
Disease gene prioritization is usually not solved as a structured prediction problem in bioinformatics. Here, we creatively resolved it as a link prediction problem. Essentially, we wanted to predict a more complete graph based on a relatively incomplete graph and the side information of the nodes. To utilize both the graph structure information and the side information, we used a graph convolutional neural network to learn meaningful embeddings for all the nodes based on those two pieces of information. We could then use the embeddings to perform the disease gene prediction. By training the entire model in an end-to-end fashion, our method took full advantage of the training data. As we have discussed extensively in this thesis and further demonstrated by this project, incorporating the problem structure (the topological information of the graph in this task) and the prior knowledge (side information of the nodes) into the methodology design can help us resolve the problem much more efficiently and accurately.

Until now, we have illustrated our ideas of using deep learning to tackle the structured prediction problems in bioinformatics with six detailed examples. In the next chapter, we conclude this thesis by summarizing the main points of the thesis and pointing out the potential future research directions.


\chapter{Concluding Remarks}
\label{chapter_conclusion}

\section{Summary}
Because the computational tasks in bioinformatics are usually related to the involved real-life biological problems, structured prediction problems, whose targets are complex objects, are prevalent in this field. Due to the unique properties of the structured prediction problems in bioinformatics, we may encounter the following challenges when solving those tasks: 1) the lack of data; 2) tremendous search space; 3) incorporating problem structures into the methods; 4) interpretability. The previous methods for solving those problems in bioinformatics have various limitations as they did not consider and respond to those challenges sufficiently. We argue the following two ideas can be beneficial for handling those problems. Firstly, we can combine deep learning with other classic algorithms, such as PGMs, which model the problem structure explicitly. Secondly, we can design and train the problem-specific deep learning architectures or methods by considering the structured labeling space and problem structures, either explicitly or implicitly. 

In this thesis, we demonstrated and showcased our ideas with six concrete examples from four subfields of bioinformatics, including sequence analysis, structure prediction, function annotation, and system biology. Starting from the most fundamental scale, the sequence scale, in Chapter \ref{chapter_ds}, we discussed a deep learning-based simulator for modeling the entire pipeline of Nanopore sequencing \cite{li2018deepsimulator}. We used a novel deep learning model, which integrated the canonical time warping algorithm into the Bi-LSTM model, to simulate the 1D electrical signals and DNA reads in the sequencing technology. Then, we went from sequence to high-level structures. In Chapter \ref{chapter_dlbi}, we discussed a deep learning guided Bayesian inference framework, DLBI \cite{li2018dlbi}, for the super-resolved structure reconstruction of super-resolution fluorescence microscopy. This framework combined the strengths of stochastic simulation, deep learning, and Bayesian inference to solve the structured prediction problem efficiently. 
In Chapter \ref{chapter_e2efold}, we discussed a novel deep learning architecture, E2Efold \cite{chen2020rna}, for predicting RNA secondary structure, which is one of the oldest problem in bioinformatics. In E2Efold, we embedded the unrolled constrained optimization algorithm into the deep learning architecture, which helped us incorporate the problem structure efficiently and resolve the task accurately. After that, we moved one level forward further, going from structure to function. In Chapter \ref{chapter_nucleicnet}, we discussed a framework, NucleicNet \cite{lam2019deep}, for investigating the protein-RNA interaction. This 3D structured prediction problem is very challenging. We decomposed this hard problem into a large number of sub-problems, which could be solved efficiently with a deep learning model. Then we used an HMM to reconstruct the solution for the original problem from the deep learning outputs. 
In Chapter \ref{chapter_deepre}, we discussed a tool, DEEPre \cite{RN140,zou2019mldeepre}, for annotating the detailed enzyme function. We used a level-by-level prediction strategy, hierarchical transfer learning, and a novel multi-label loss function to incorporate the hierarchical labeling structure and the multi-class information into our method effectively. Finally, we went from a single molecule to system biology. In Chapter \ref{chapter_pgcn}, we discussed how to use deep learning to predict and prioritize disease genes \cite{li2019pgcn}. We creatively formulated it as a link prediction problem. And then, we used a graph convolutional neural network, which could utilize both the graph structure information and the side information of nodes, to resolve this structured prediction problem efficiently.

\section{Future Research Work}
Considering the success of using deep learning for handling the structured prediction problems in bioinformatics, as discussed above, we believe our ideas can be further extended to solve other difficult but influential problems in bioinformatics. Here, in this last section, we want to highlight several tasks from the last scale in Figure \ref{fig:thesis_overview}, that is, health-care. 

The health-care problems can be divided into three categories: monitoring, diagnosing, and curing. For monitoring, it is fascinating and useful to predict the future health condition of patients based on their current health condition, genetic information, and other auxiliary information \cite{jia2019estimating}. With such a prediction, we can make some preparations in advance in case of an emergency. This task is difficult, not only because we want to predict something in the future. The desired predicted targets can be very complicated, instead of just a predicted probability of the patients' condition becoming worse. For example, we may want to make multiple predictions for different time points during a period. The results of different time points should have some dependency and correlation. Furthermore, we need to resolve some ``what if'' questions---if the patient did not take medicine in the morning, what would happen during noon? The problem structure in such daily life scenarios can be much more complicated than that at the molecular level as human beings are much more flexible and unpredictable than the micro-environment. 

Regarding diagnosing, only after we have diagnosed the specific diseases that a patient has, can the doctor gives detailed treatment protocol. However, nowadays, the diagnosis of rare diseases can still be very time-consuming. For example, in Europe, on average, it will take us around nine years to make the correct diagnosis for rare diseases \cite{wright2018paediatric}. This task is challenging because, for rare diseases, the available data are internally highly limited. Assisted by transfer learning and few-shot learning, the system, which takes advantage of deep learning, multi-modal data, and doctor's prior knowledge, can be potentially useful for accelerating such a diagnosing process. The correlation between different diseases (the structure within the labeling space) can also be very informative, which we should consider when designing the methods. 

In terms of curing, there is an exciting research direction, that is, to design new drugs. This research direction is related to our previous research on investigating the interaction between two molecules \cite{lam2019deep}. Drugs can also interact with abnormal molecules. By binding against the molecules and preventing them from functioning, the drug can help us cure the diseases. The most challenging part of drug design is to perform decoding, that is, to recover the 2D/3D topological structure from the dense vector representation of the drug. Although the opposite encoding problem, that is, to learn the embedding of the drug which can preserve their physiochemical property and structural information, has been well studied, this decoding problem has been less explored. The integration between the deep generative models and combinatorial optimization algorithms should be helpful for this fascinating topic.

Although those problems are very challenging, based on the ideas discussed within this thesis and other further extensions, we believe we can make our contributions in solving the aforementioned influential tasks in the future, improving people's health and wellness.






\begin{onehalfspacing}
\renewcommand*\bibname{\centerline{REFERENCES}} 
\addcontentsline{toc}{chapter}{References}
\newcommand{\BIBdecl}{\setlength{\itemsep}{0pt}}
\bibliographystyle{IEEEtran}
\bibliography{References}
\end{onehalfspacing}

\appendix

\newpage

\begingroup
\let\clearpage\relax
\begin{center}
\vspace*{2\baselineskip}
{ \textbf{{\large APPENDICES}}} 
\addcontentsline{toc}{chapter}{Appendices} 
\end{center}


\refstepcounter{chapter}%
\chapter*{\thechapter \quad Summary of Publications}
\label{appendix_publications}

\textbf{Journal:}{($^{*}$Equal contribution, $^\dagger$Co-corresponding)}
\begin{benumerate}{30}
	\item \textbf{Y Li}$^{*}$, S Wang$^{*}$, C Bi, Z Qiu, M Li, X Gao, ``DeepSimulator1.5: a more powerful, quicker and lighter simulator for Nanopore sequencing''. \textit{Bioinformatics}. (2020). \href{https://academic.oup.com/bioinformatics/article/36/8/2578/5698265}{10.1093/bioinformatics/btz963} \cite{li2020deepsimulator1}
	\item T Zhang, \textbf{Y Li}, Y Li, S Sun, and X Gao. ``A Self-adaptive Deep Learning Algorithm for Accelerating Multi-component Flash Calculation.'' \textit{Computer Methods in Applied Mechanics and Engineering}. (2020). \href{https://www.sciencedirect.com/science/article/pii/S0045782520303923}{10.1016/j.cma.2020.113207}. \cite{ZHANG2020113207}
	\item H Li$^{*}$, S Tian$^{*}$, \textbf{Y Li}$^{*}$, R Tan, Y Pan, C Huang, Y Xu, and X Gao. ``Modern Deep Learning in Bioinformatics.'' \textit{Journal of Molecular Cell Biology}. (2020). \href{https://academic.oup.com/jmcb/article/doi/10.1093/jmcb/mjaa030/5861537}{10.1093/jmcb/mjaa030}. \cite{Li2020JMCB}
	\item Z Li, Y Li, B Zhang, \textbf{Y Li}, Y Long, X Zou, M Zhang, Y Hu, W Chen$^\dagger$, X Gao$^\dagger$. ``DeeReCT-APA: prediction of alternative polyadenylation site usage through deep learning''. \textit{Genomics, Proteomics \& Bioinformatics (GPB), 2020, accepted.} \cite{li2020deerect}
	\item J Lam$^{*}$, \textbf{Y Li}$^{*}$, L Zhu$^\dagger$, R Umarov, H Jiang, A Heliou, F Sheong, T Liu, Y Long, Y Li, L Fang, R Altman, W Chen$^\dagger$, X Huang$^\dagger$, X Gao$^\dagger$. ``A deep learning framework to predict binding preference of RNA constituents on protein surface''. \textit{Nature Communications}. (2019). \href{https://www.nature.com/articles/s41467-019-12920-0}{s41467-019-12920-0}. \cite{lam2019deep}
	\item G Jia, \textbf{Y Li}, H Zhang, I Chattopadhyay, A Jensen, D Blair, L Davis, P Robinson, T Dahlén, S Brunak, M Benson, G Edgren, N Cox, X Gao, A Rzhetsky. ``Estimating heritability and genetic correlations from large health datasets in the absence of genetic data''. \textit{Nature Communications}. (2019). \href{https://www.nature.com/articles/s41467-019-13455-0}{s41467-019-13455-0}. \cite{jia2019estimating}
	\item J Lei, G Sheng, P Cheung, S Wang, \textbf{Y Li}, X Gao, Y Zhang, Y Wang, X Huang. ``Two symmetric Arginine residues play distinct roles in Thermus thermophilus Argonaute DNA guide strand-mediated DNA target cleavage''. \textit{Proceedings of the National Academy of Sciences of the United States of America (PNAS)}. (2019). \href{https://www.pnas.org/content/early/2018/12/26/1817041116}{10.1073/pnas.1817041116}. \cite{lei2019two}
	\item \textbf{Y Li}, T Zhang, S Sun$^\dagger$, X Gao$^\dagger$. ``\href{https://arxiv.org/abs/1809.07311}{Accelerating Flash Calculation through Deep Learning Methods}''. \textit{Journal of Computational Physics (JCP)}. (2019). \href{https://www.sciencedirect.com/science/article/pii/S0021999119303596}{10.1016/j.jcp.2019.05.028}. \cite{li2019accelerating}
	\item \textbf{Y Li}, C Huang, L Ding, Z Li, Y Pan, X Gao. ``Deep learning in bioinformatics: introduction, application, and perspective in big data era''. \textit{Methods}. (2019). \href{https://github.com/lykaust15/Deep_learning_examples}{10.1016/j.ymeth.2019.04.008}. \textbf{Cover article of the Methods issue: Deep Learning in Bioinformatics}. \cite{li2019deep}
	\item Z Zou, S Tian, X Gao, \textbf{Y Li}. ``mlDEEPre: Multi-functional enzyme function prediction with hierarchical multi-label deep learning''. \textit{Frontiers in Genetics}. (2019) \href{https://www.frontiersin.org/articles/10.3389/fgene.2018.00714/full}{10.3389/fgene.2018.00714}. \cite{zou2019mldeepre}
	\item R Umarov, H Kuwahara, \textbf{Y Li}, X Gao$^\dagger$, V Solovyev$^\dagger$. ``Promoter analysis and prediction in the human genome using sequence-based deep learning models''. \textit{Bioinformatics}. (2019). \href{https://academic.oup.com/bioinformatics/advance-article/doi/10.1093/bioinformatics/bty1068/5270663?rss=1}{10.1093/bioinformatics/bty1068}. \cite{umarov2019promoter}
	\item U Hameed, C Liao, A Radhakrishnan, F Huser, S Aljedani, X Zhao, A Momin, F Melo, X Guo, C Brooks, \textbf{Y Li}, X Cui, X Gao, J Ladury, L Jaremko, M Jaremko, J Li, S, Arold. ``H-NS uses an autoinhibitory conformational switch to achieve environment-controlled gene silencing''. \textit{Nucleic Acids Research (NAR)}. (2018). \href{https://academic.oup.com/nar/advance-article/doi/10.1093/nar/gky1299/5266712}{10.1093/nar/gky1299}. \cite{shahul2018h}
	\item Zhihao Xia, \textbf{Y Li}, B Zhang, Z Li, Y Hu, W Chen$^\dagger$, X Gao$^\dagger$. ``DeeReCT-PolyA: a robust and generic deep learning method for PAS identification''. \textit{Bioinformatics}. (2018). \href{https://github.com/likesum/DeeReCT-PolyA}{10.1093/bioinformatics/bty991}. \cite{xia2018deerect}
	\item \textbf{Y Li}, R Han, C Bi, M Li, S Wang$^\dagger$, X Gao$^\dagger$. ``DeepSimulator: a deep simulator for nanopore sequencing''. \textit{Bioinformatics}. (2018). \href{https://github.com/lykaust15/DeepSimulator}{10.1093/bioinformatics/bty223}. \cite{li2018deepsimulator}
	\item V Kordopati, A Salhi, R Razali, A Radovanovic, F Tifratene, M Uludag, \textbf{Y Li}, A Bokhari, A AlSaieedi, A Raies, C Neste, M Essack, V Bajic. ``DES-Mutation: System for Exploring Links of Mutations and Diseases''. \textit{Scientific Reports}. (2018). \href{http://www.cbrc.kaust.edu.sa/des-mutation/home/index.php}{10.1038/s41598-018-31439-w}. \cite{kordopati2018mutation}
	\item \textbf{Y Li}$^{*}$, F Xu$^{*}$, F Zhang, P Xu, M Fan, L Li, X Gao$^\dagger$, R Han$^\dagger$. ``DLBI: Deep learning guided Bayesian inference for structure reconstruction of super-resolution fluorescence microscopy''. \textit{Bioinformatics}. (2018). \href{https://github.com/lykaust15/DLBI}{10.1093/bioinformatics/bty241}. \cite{li2018dlbi}
	\item S Wang, S Fei, Z Wang, \textbf{Y Li}, J Xu, F Zhao, X Gao. ``PredMP:a web server for de novo prediction of membrane protein''. \textit{Bioinformatics}. (2018). \href{http://www.predmp.com:3001/#}{10.1093/bioinformatics/bty684}. \cite{wang2019predmp}
	\item R Han, X Wan, L Li, A Lawrence, P Yang, \textbf{Y Li}, S Wang, F Sun, Z Liu, X Gao, F Zhang. ``AuTom-dualx: a toolkit for fully automatic alignment of dual-axis tilt series with simultaneous reconstruction''. \textit{Bioinformatics}. (2018). \href{http://ear.ict.ac.cn}{10.1093/bioinformatics/bty620}. \cite{han2019autom}
	\item R Han, \textbf{Y Li}, X Gao, S Wang. ``An accurate and rapid continuous wavelet dynamic time warping algorithm for end-to-end mapping in ultra-long nanopore sequencing''. \textit{Bioinformatics}. (2018). \href{https://github.com/realbigws/cwDTW}{10.1093/bioinformatics/bty555}. \cite{han2018accurate}
	\item \textbf{Y Li}, S Wang, R Umarov, B Xie, M Fan, L Li, X Gao. ``DEEPre: sequence-based enzyme EC number prediction by deep learning''. \textit{Bioinformatics}. (2017). \href{https://www.ncbi.nlm.nih.gov/pubmed/29069344}{10.1093/bioinformatics/btx680}. \cite{RN140}
	\item H Dai, R Umarov, H Kuwahara, \textbf{Y Li}, L Song$^\dagger$, X Gao$^\dagger$. ``Sequence2Vec: a novel embedding approach for modeling transcription factor binding affinity landscape''. \textit{Bioinformatics}. (2017). \href{https://www.ncbi.nlm.nih.gov/pubmed/28961686}{10.1093/bioinformatics/btx480}. \cite{RN141}
	\item S Wu, D Wang, J Liu, Y Feng, J Weng, \textbf{Y Li}, X Gao, J Liu, W Wang. ``The dynamic multisite interactions between two intrinsically disordered proteins''. \textit{Angewandte Chemie}. (2017). \href{https://www.ncbi.nlm.nih.gov/pubmed/28493424}{10.1002/anie.201701883}. \cite{wu2017dynamic}
	\item X Li, Q Tao, Y Fang, C Cheng, Y Hao, J Qi, \textbf{Y Li}, W Zhang, Y Wang, X Zhang. ``Reward sensitivity predicts ice cream-related attentional bias assessed by inattentional blindness''. \textit{Appetite}. (2015). \href{https://www.ncbi.nlm.nih.gov/pubmed/25681293}{10.1016/j.appet.2015.02.010}. \cite{li2015reward}
\end{benumerate}
\textbf{Conference:}{($^{*}$Equal contribution, $^\dagger$Co-corresponding)}
\begin{benumerate}{7}
	\item X Chen$^{*}$, \textbf{Y Li}$^{*}$, R Umarov, X Gao, L Song, RNA Secondary Structure Prediction By Learning Unrolled Algorithms, \textit{Eighth International Conference on Learning Representations (ICLR-20), \textbf{Oral}}. \cite{chen2020rna}
	\item X Chen, H Dai, \textbf{Y Li}, X Gao, and L Song. Learning to Stop While Learning to Predict. \textit{Thirty-seventh International Conference on Machine Learning (ICML-20)}. \cite{chen2020learning}
	\item L Ding, M Yu, L Liu, F Zhu, Y Liu, \textbf{Y Li}, L Shao. Two Generator Game: Learning to Sample via Linear Goodness-of-Fit Test. \textit{Thirty-third Conference on Neural Information Processing Systems (NeurIPS-19)}. \cite{ding2019two}
	\item L Ding, Z Liu, \textbf{Y Li}, S Liao, Y Liu, P Yang, G Yu, L Shao, X Gao. ``\href{https://www.aaai.org/Papers/AAAI/2019/AAAI-DingLizhong2.4563.pdf}{Linear Kernel Tests via Empirical Likelihood for High Dimensional Data}''. \textit{The Thirty-Third AAAI Conference on Artificial Intelligence (AAAI-19)}. \cite{ding2019linear}
	\item L Ding, S Liao, Y Liu, \textbf{Y Li}, P Yang, Y Pan, C Huang, L Shao, X Gao. ``\href{https://www.aaai.org/Papers/AAAI/2019/AAAI-DingLizhong1.4483.pdf}{Approximate Kernel Selection with Strong Approximate Consistency}''. \textit{The Thirty-Third AAAI Conference on Artificial Intelligence (AAAI-19)}. \cite{ding2019approximate}
	\item \textbf{Y Li}$^{*}$, F Xu$^{*}$, F Zhang, P Xu, M Fan, L Li, X Gao$^\dagger$, R Han$^\dagger$. ``\href{https://github.com/lykaust15/DLBI}{DLBI: Deep learning guided Bayesian inference for structure reconstruction of super-resolution fluorescence microscopy}''. \textit{The Twenty-Sixth Conference on Intelligent Systems for Molecular Biology (ISMB-18)}. \cite{li2018dlbi}
	\item R Han, \textbf{Y Li}, X Gao, S Wang. ``\href{https://github.com/realbigws/cwDTW}{An accurate and rapid continuous wavelet dynamic time warping algorithm for end-to-end mapping in ultra-long nanopore sequencing}''. \textit{The Seventeenth European Conference on Computational Biology (ECCB-18)}. \cite{han2018accurate}
\end{benumerate}

\textbf{Under Review \& Preprint}{($^{*}$Equal contribution, $^\dagger$Co-corresponding)}
\begin{benumerate}{6}
	\item G Jia$^{*}$, \textbf{Y Li}$^{*}$, X Zhong, K Wang, M Pividori, R Alomairy, A Esposito, H Ltaief, C Terao, M Akiyama , K Matsuda, D Keyes, H Im, T Gojobori, Y Kamatani, M Kubo, N Cox, X Gao, A Rzhetsky. ``Genome-wide Association Mapping of Geometric Space of Human Disease''. \textit{Nature Genetics, under review}.
	\item \textbf{Y Li}, Z Xu, W Han, H Cao, R Umarov, A Yan, M Fan, H Chen, L Li, P Ho, X Gao. ``HMD-ARG: hierarchical multi-task deep learning for annotating antibiotic resistance genes.'' \textit{Microbiome, under review}.
	\item C Bi, L Wang, B Yuan, X Zhou, Y Li, S Wang, Y Pang, X Gao, Y Huang, M Li. Long-read Individual-molecule Sequencing Reveals CRISPR-induced Genetic Heterogeneity in Human ESCs. \textit{bioRxiv}, 2020. \cite{bi2020long}
	\item \textbf{Y Li}, Z Li, C Huang, L Ding, Y Pan, X Gao. ``\href{https://github.com/lykaust15/SupportNet}{SupportNet: solving catastrophic forgetting in class incremental learning with support data}''. \textit{arxiv.org/abs/1806.02942}. \cite{li2018supportnet}
	\item \textbf{Y Li}, L Ding, X Gao. ``\href{https://github.com/lykaust15/NN_decision_boundary}{On the decision boundary of deep neural network}''. \textit{arxiv.org/abs/1808.05385}. \cite{li2018decision}
	\item \textbf{Y Li}, H Kuwahara, P Yang, L Song$^\dagger$, X Gao$^\dagger$. ``\href{https://www.biorxiv.org/content/10.1101/532226v1}{PGCN: Disease gene prioritization by disease and gene embedding through graph convolutional neural networks}''. \textit{bioRxiv 532226}, 2019. \cite{li2019pgcn}
\end{benumerate}


\endgroup

\end{document}